\pdfoutput=1
\documentclass[fleqn,usenatbib]{mnras}    

\usepackage[T1]{fontenc}
\usepackage{ae,aecompl}

\DeclareRobustCommand{\VAN}[3]{#2}
\let\VANthebibliography\thebibliography
\def\thebibliography{\DeclareRobustCommand{\VAN}[3]{##3}\VANthebibliography}

\usepackage{graphicx}	
\usepackage{amsmath}	
\usepackage{amssymb}	
\usepackage{pdflscape}	

\let\oldAA\AA
\renewcommand{\AA}{\text{\normalfont\oldAA}}

\usepackage{newtxtext,newtxmath}  

\usepackage{enumitem}

\usepackage{placeins}  
\usepackage{caption}    

\usepackage{xcolor}

\hypersetup{pdfauthor={G. Bihain},
               pdftitle={Search of nearby resolved neutron stars among optical sources},
               pdfkeywords={stars: neutron, pulsars: general, white dwarfs, solar neighbourhood},
               bookmarksnumbered=true}

\newcommand\rnaas{Res. Notes Am. Astron. Soc.}             
\newcommand\rpph{Rep. Prog. Phys.}                
\newcommand\phlb{Phys. Lett. B}                          
\newcommand\cse{Comput. Sci. Eng.}                    

\title[Search of nearby resolved neutron stars]{Search of nearby resolved neutron stars among optical sources}

\author[G. Bihain]{Gabriel Bihain$^{1,2}
$\thanks{E-mail:gabriel.bihain@aei.mpg.de}
\\
$^{1}$Max Planck Institute for Gravitational Physics (Albert Einstein Institute), Callinstrasse 38, D-30167 Hannover, Germany\\
$^{2}$Leibniz Universit{\"a}t Hannover, D-30167 Hannover, Germany\\
}

\date{Accepted 2023 April 24. Received 2023 March 27; in original form 2022 August 17}

\pubyear{2023}

\begin{document}
\label{firstpage}
\pagerange{\pageref{firstpage}--\pageref{lastpage}}
\maketitle

\begin{abstract}
Neutron stars are identified as pulsars, X-ray binary components, central objects of supernovae remnants, or isolated thermally emitting sources, and at distances beyond 120 pc. A population extrapolation suggests 10$^3$ objects within that boundary. Potentially, neutron stars could continuously emit gravitational waves at sensitivity reach of present instrumentation. As part of our Search for the Nearest Neutron Stars ``Five Seasons'' project, we search for nearby resolved neutron stars. Based on expected fluxes and magnitudes of thermally cooling neutron stars and pulsars, we selected sources in \textit{Gaia} DR3. The sources have $G$-band absolute magnitudes $M_G>16$~mag, parallax signal-to-noise ratios greater than two, and colours $G_{BP}-G<0.78$ and $G-G_{RP}<0.91$~mag for power-law emitters of flux $F_{\nu} \propto \nu^{-\alpha_{\nu}}$ with spectral indices $\alpha_{\nu}<3$. The photometric region overlaps with that of white dwarfs, in confluence with most known pulsars in binaries having white dwarf companions. We looked for counterparts in gamma-ray, X-ray, ultraviolet, radio, optical, and infrared catalogues. We find about two X-ray-, 15 ultraviolet-, one radio probable counterparts, and at least four sources with power-law profiles at the ultraviolet--optical(--infrared). Because the sources have $G\gtrapprox20$~mag, we rely on \textit{Gaia} DR3 single-source parameters. We identify possible binaries based on photoastrometric parameters, visual companions, and flux excesses. Some emission components suggest small thermal radii. Source types, neutron star content, and properties require further inquiry. 
\end{abstract}




\section{Introduction}


Neutron stars are extremely compact objects resulting from the collapse of (i) OB transition type stars of roughly 8--20 solar masses or (ii) accreting white dwarfs. If rotating and axially asymmetric, these compact objects are expected to emit continuously gravitational waves of very small amplitudes. The wave frequency relates to the rotation frequency, and because ground-based gravitational wave detectors are most sensitive to the 10$^{0}$--10$^4$~Hz frequency range \citep{2005MNRAS.359.1150P}, targets would have to be rapid rotators. This is fulfilled for example by 0--1~Gyr young pulsars that still have high spin energy or by millisecond pulsars, spun up by accretion. When magnetic fields or magnetic spin-down torques are weak and insignificant, high rotation frequencies are expected to be sustained over long times. Continuous gravitational wave searches are progressively increasing in sensitivity \citep[recent results can be found e.g. in][]{2021ApJ...921...80A, 2021ApJ...922...71A, 2022PhRvD.105h2005A, 2022PhRvD.106f2002A, 2022PhRvD.106j2008A, 2022ApJ...932..133A, 2022ApJ...935....1A, 2021ApJ...923...85A, 2021ApJ...909...79S, 2023arXiv230304109S, 2021ApJ...906L..14Z, 2022ApJ...929L..19C, 2023PhRvX..13b1020D, 2022ApJ...925....8M}.

Young and millisecond pulsars represent a low percentage of the estimated $10^8$--$10^9$ neutron stars in the Galaxy, most of which being expected to be very dim electromagnetically. For a mean inter neutron-star distance of 10~pc \citep{2010A&A...510A..23S, 2023arXiv230304714P}, there would be $\sim$10, 1000, and 8000 neutron stars within 20, 100, and 200~pc, respectively. RX~J1856.5$-$3754 at $d = 123^{+11}_{-15}$ pc \citep{1996Natur.379..233W,2010ApJ...724..669W} is the nearest known exemplar, and it is isolated, X-ray dim, and thermally emitting. As other known thermally emitting neutron stars \citep{2020MNRAS.496.5052P}, it is expected to cool down quickly over time. Because neutron stars would be numerous within 100--200~pc, some could be still detectable electromagnetically from thermal or non-thermal emission \citep[see e.g.][]{2022MNRAS.510..611T}. These would also be promising targets for continuous gravitational wave searches. Knowing the position of the nearest one would allow for a targeted gravitational-wave search that is more sensitive than broad all-sky searches.

The combination of very high effective temperatures and small radii ($\sim$10~km) implies that the youngest neutron stars have very high X-ray-to-optical flux ratios, with practically no visible optical counterparts. This led to extensive searches using X-ray sources as reference, in particular with the \textit{ROSAT} All-Sky Survey \citep[RASS;][]{1999A&A...349..389V,2016A&A...588A.103B}, to look for optical counterparts, despite the lower X-ray spatial resolution, and identify the sources. Some of these were indeed found to be thermally emitting neutron stars \citep[see review by][]{2000PASP..112..297T}. Over the last decades, wide-area optical surveys have become deeper and spatially sharper, as for example the Global Astrometric Interferometer for Astrophysics (\textit{Gaia}) mission survey \citep{2016A&A...595A...1G}, providing in addition trigonometric parallaxes and proper motions, and the Dark Energy Spectroscopic Instrument (DESI) Legacy Imaging Surveys \citep{2019AJ....157..168D}. Moreover, several pulsars have been detected in the optical with a significant component of non-thermal origin \citep[see e.g.][]{2021MNRAS.502.2005Z}. This motivates us, as part of our Search for the Nearest Neutron Stars Five Seasons project, to revisit the search, by using instead faint optical sources as reference and identifying neutron star candidates.


\section{Methods}

The search methods are as follows:
\begin{enumerate}[label=(\roman*),align=left,itemindent=9pt]
\item First, we consider the evolution in luminosity and effective temperature of an isolated thermally emitting neutron star, anchored to an accurately observed object (RX~J1856.5$-$3754), and we compute its emission profile at different ages. We consider also measurements at the ultraviolet--optical--(infrared) of several known thermally emitting neutron stars and pulsars. For all of these sources, we compute synthetic fluxes and magnitudes at different survey bandpasses.
\item Next, we select \textit{Gaia} sources with parallaxes and photometry closest to the expected photometry.
\item Then we prospect the multiband detectability of neutron stars among these sources.
\item Finally, we crossmatch the sources with catalogue sources at other wavelengths and verify them in images. We discard unambiguously identified sources and keep potential candidates for searches of continuous gravitational waves.
\end{enumerate}



\subsection{Expected photometry of resolved neutron stars}\label{sect:expphot}

After formation, neutron stars are expected to cool mainly through neutrino emission from the core and then through thermal radiation from the surface. The transition time-scale from neutrino- to photon cooling depends on the interior composition and is theoretically estimated to be in the range 0.01--0.1 Myr \citep[see e.g.][]{1999PhyU...42..737Y,2004AdSpR..33..523Y,2006NuPhA.777..497P}. Observations of young pulsars indicate a decrease and then an increase in the X-ray and optical to spin-down luminosity ratios, that is of the capacity in converting rotational energy to X-ray and optical radiation, for increasing spin-down age. The change occurs at $\tau\approx0.01$~Myr and could be associated with the transition to photon cooling \citep{2006AdSpR..37.1979Z,2013MNRAS.435.2227Z}. For what follows, we assumed that the cooling is mainly thermal at $\ga$0.1~Myr. During thermal cooling, the surface temperature and luminosity can be approximated by $T\propto t^{-1/2}$ and $L\propto t^{-2}$ \citep{1979PhR....56..237T, 2017A&A...608A.147P}, as it roughly depicts, at ages of 0.1--10~Myr, the distribution of thermally emitting neutron stars in a log--log plane, such as presented in \citet{2020MNRAS.496.5052P}. This allows us to estimate the main observables as a function of time:
\begin{itemize}
\item[-] the surface temperature
\begin{equation}\label{eqn:tevol}
T= T_0 (t/t_0)^{-1/2}
\end{equation}
where $T_0$ and $t_0$ are the present temperature and age,
\item[-] the thermal luminosity
\begin{equation}\label{eqn:BBlum}
L=4 \pi R^2 \sigma T_\mathrm{eff}^4 = L_0 (t/t_0 )^{-2}
\end{equation}
where $R$ is the emitting radius, $\sigma$ the Stefan--Boltzmann constant, $T_\mathrm{eff}$ the effective temperature, and $L_0$ the present luminosity,
\item[-] the bolometric magnitude
\begin{equation}\label{eqn:Mbol-d}
m_\mathrm{bol} = M + 5~\log(d [\mathrm{pc}]/10)
\end{equation}
where the absolute magnitude
\begin{equation}\label{eqn:Mbol-lum}
M = - 2.5~\log(L/\mathrm{L_\odot}) + \mathrm{M_\odot},
\end{equation}
with L$_\odot = 3.826\times10^{33}$~erg~s$^{-1}$ and M$_\odot = 4.755$~mag for the Sun, and $d$ is the distance, and
\item[-] the flux
\begin{equation}\label{eqn:fluxlum}
F=L/(4 \pi d^2 ).
\end{equation}
\end{itemize}

We assumed an emission profile resembling that of RX~J1856.5$-$3754, as a simplified version of emission profiles of known thermally emitting isolated neutron stars, and because this object is the nearest known neutron star, has an accurate distance measured from trigonometric parallax, and was frequently observed for observational calibration and object comparison at X rays. The template profile is composed of (i) a very soft X-ray blackbody, (ii) ultraviolet and optical fluxes greater than the Rayleigh--Jeans tail of the X-ray blackbody, and (iii) optionally, X-ray emission from accretion of interstellar matter. 

We note that RX~J1856.5$-$3754 and the other known X-ray dim thermally emitting isolated neutron stars would have very strong magnetic fields, similar to those of strong magnetic-field pulsars but weaker than magnetars \citep{2019RPPh...82j6901E}. We assumed that this affects only weakly the thermal luminosity \citep{2020MNRAS.496.5052P} and their potentialities for gravitational wave detection. We note also that the most luminous and youngest thermally emitting neutron stars are associated with supernova remnants at $\gtrapprox1$~kpc, and have X-ray luminosities as high as 10$^{34.5}$~erg~s$^{-1}$ at $\sim$1 kyr \citep{2020MNRAS.496.5052P}, thousand times brighter than RX~J1856.5$-$3754. Finally, we note that millisecond pulsars emit typically less thermally than RX~J1856.5$-$3754, where the low thermal emission is maintained through re-heating (from accretion or internal processes) or from hot polar caps \citep{2020MNRAS.496.5052P,2021PhR...919....1B}. However, compared to less young thermally emitting neutron stars, some isolated millisecond pulsars could turn out to be brighter (see below).

\begin{figure}
	\includegraphics[width=\columnwidth]{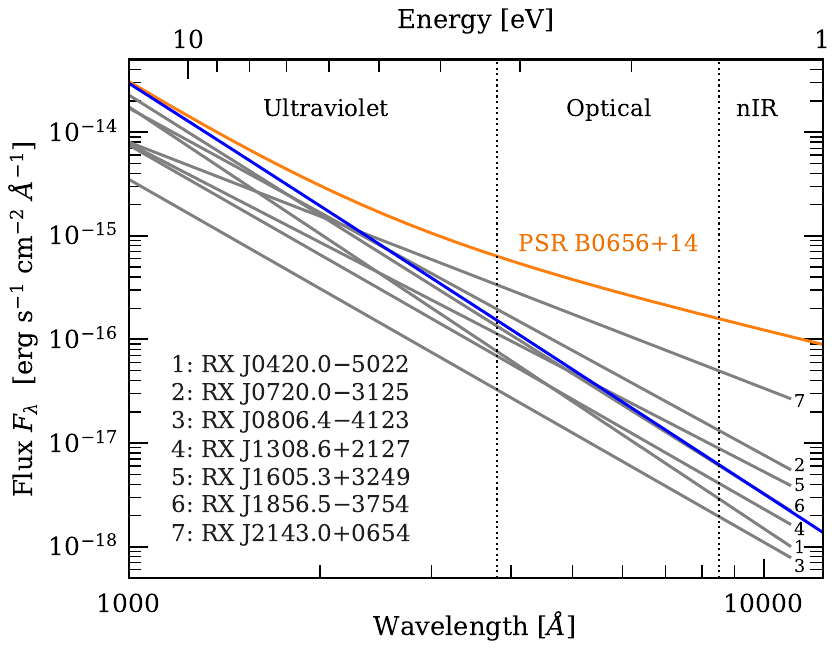}
    \caption{10~pc unabsorbed spectra at 1000--11000~$\AA$. The ultraviolet--optical fits from \citet{2011ApJ...736..117K} for the seven known X-ray dim isolated neutron stars are represented by the grey lines. The tail of the X-ray blackbody fit from \citet{2003A&A...399.1109B} for RX~J1856.5$-$3754, multiplied by eight, is shown by the blue line, which is very close to the grey line for this object. The fit from \citet{2021MNRAS.502.2005Z} for PSR B0656+14 is represented by the orange line.}
    \label{sfd_m7}
\end{figure}

\begin{figure*}
    \includegraphics[width=\textwidth]{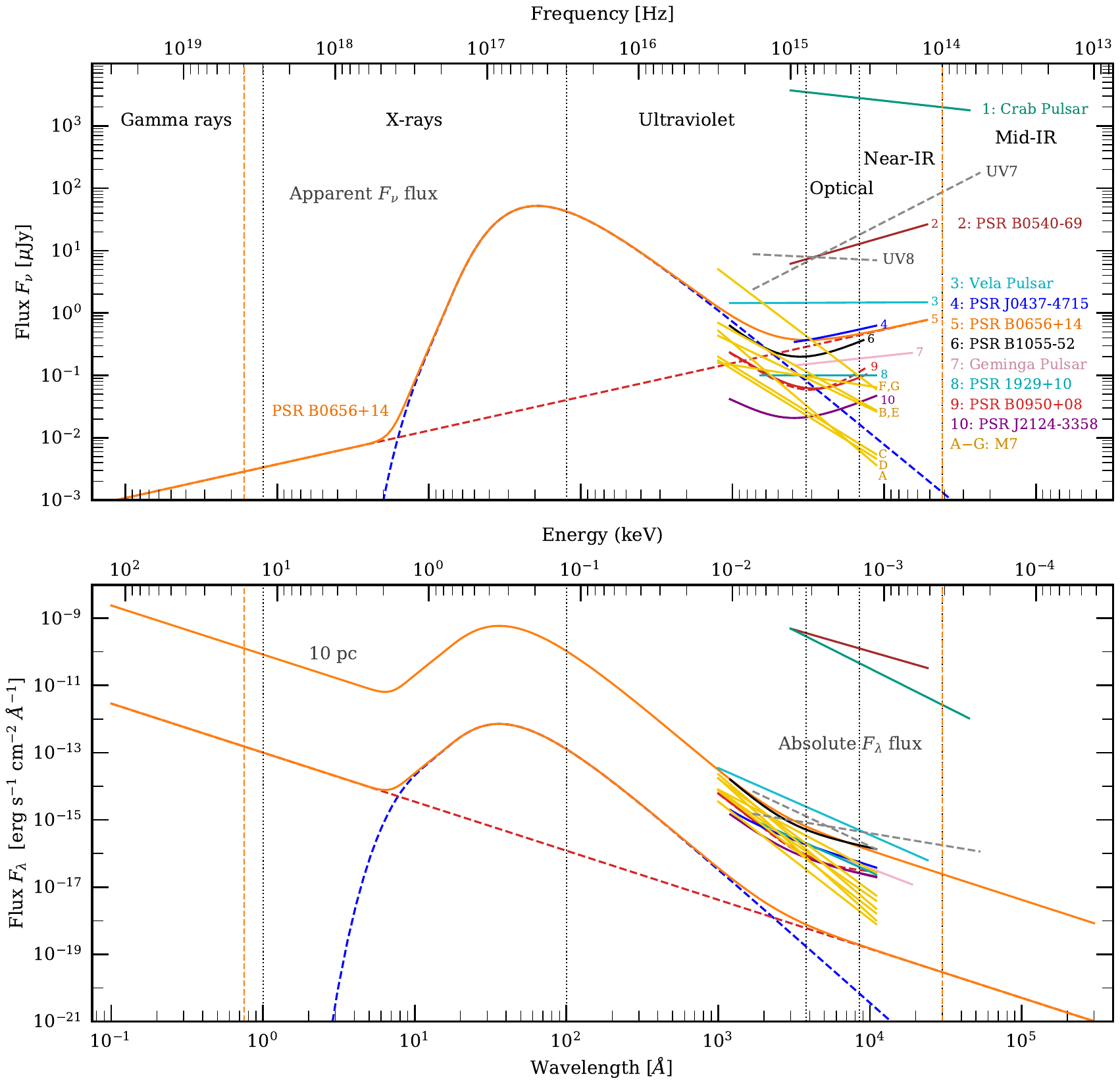}
    \caption{Unabsorbed spectra in frequency ({\sl top panel}) and wavelength ({\sl bottom panel}) of the model fit adopted for PSR~B0656+14, based on observed fluxes in the range delimited by the vertical orange dashed lines \citep[see Section~\ref{sect:expphot} and ][]{2021MNRAS.502.2005Z}. The blackbody component, power-law component, and sum of both are represented by the blue dashed, red dashed, and orange solid lines, respectively. The other solid lines represent fits to unabsorbed fluxes of other neutron stars, as observed ({\sl top panel}, annotated with numbers and names), and scaled to 10~pc based on their known distances ({\sl bottom panel}). These neutron stars are the Crab~Pulsar (green), PSR~B0540$-$69 (brown), the Vela~Pulsar (cyan), PSR~J0437$-$4715 (blue), PSR~B1055$-$52 (black), the Geminga~Pulsar (pink), PSR~1929+10 (turquoise), PSR~B0950+08 (red; the dashed line is for the maximal spectral index), PSR~J2124$-$3358 (purple), and the known X-ray dim thermally emitting isolated neutron stars (M7, yellow) RX~J0420.0$-$5022 (A), RX~J0720.0$-$3125 (B), RX~J0806.4$-$4123 (C), RX~J1308.6+2127 (D), RX~J1605.3+3249 (E), RX~J1856.5$-$3754 (F), and RX~J2143.0+0654 (G). Example of power-law components of the UV7 and UV8 candidates are represented by the grey dashed lines.
    }
    \label{sfd}
\end{figure*}

We adopted for RX~J1856.5$-$3754 an unabsorbed X-ray blackbody of temperature $kT_{X}^{\infty} = 63.5\pm0.2$~eV ($736\,882\pm2321$~K), corresponding to a luminosity $L^\infty=10^{31.63}$~erg~s$^{-1}$ \citep{2003A&A...399.1109B} and an emitting radius $R^{\infty} = 4.51$~km for the distance $d=123$~pc. We assumed an unabsorbed ultraviolet--optical flux excess of factor eight relative to the blackbody Rayleigh--Jeans tail ($F_{\lambda}\propto\lambda^{-\alpha_{\lambda}}$ with a spectral index $\alpha_{\lambda}=4$, or $F_{\nu}\propto\nu^{-\alpha_{\nu}}$ with $\alpha_{\nu}=-2$), so that it matches the unabsorbed fit from \citet{2001A&A...378..986V}. It agrees also with the unabsorbed fit from \citet{2011ApJ...736..117K}. The excess corresponds to a $-$2.26~mag brighter magnitude. From \citet{2011ApJ...736..117K}, the seven known thermally emitting isolated neutron stars have flux excesses of factor 5--12 at 4700~$\AA$ and $\alpha_{\lambda}=3.2-4.2$ at the ultraviolet--optical, except RX~J2143.0+0654, which has a flux excess of factor 50 and $\alpha_{\lambda}=2.5$. As shown in fig.~5 of that study, the hotter these objects are at X rays (e.g. compared to RX~J1856.5$-$3754), roughly the smaller their ultraviolet--optical $\alpha_{\lambda}$ spectral indices and the shallower the slopes, implying less blue colours. This diversity of optical power-law slopes motivated us to adopt for RX~J1856.5$-$3754 a blackbody at X rays and a power law in the optical rather than two blackbodies adjusting the X-ray and optical fluxes. Figure~\ref{sfd_m7} illustrates the spectral flux density distributions (hereafter spectra, or more generally speaking, spectral energy distributions -- SEDs) of the seven objects, scaled to 10~pc using their distances of $1.2$--$4.1\times10^2$~pc compiled in \citet{2020MNRAS.496.5052P} and assuming validity of extrapolation at 8500--11\,000~$\AA$. Their apparent and absolute spectra are also shown in Fig.~\ref{sfd}.

For RX~J1856.5$-$3754, the unabsorbed RASS 0.1--2.4 keV flux is of $1.5\times10^{-11}$~erg~s$^{-1}$~cm$^{-2}$ \citep{1996Natur.379..233W} and translates to $2.27\times10^{-9}$ erg~s$^{-1}$~cm$^{-2}$ at 10~pc; from the blackbody, we retrieve a flux of $2.67\times10^{-9}$~erg~s$^{-1}$~cm$^{-2}$, assuming a 100 per cent transmission, which is fairly in agreement. We adopted a present age of 2 Myr ($t_0$), in between the kinetic age of 0.42~Myr \citep{2013MNRAS.429.3517M} and the spin-down age of 3.8~Myr \citep{2008ApJ...673L.163V}. We assumed that the ultraviolet--optical flux excess is slowly damped over time, by a factor of $1/\log((2+t/t_{0})^2)$ and starting at $1.16 \times t_0$. At 400--1000~$\AA$, we assumed a linear transition in logarithmic scale from the X-ray blackbody to the flux excess. For simplicity, we assumed a constant emission radius and that the surface temperature relates with time as in Equation~(\ref{eqn:tevol}) even beyond 10~Myr.

Finally, we adopted an optional, 0.1--2.4 keV X-ray accretion luminosity of 10$^{28}$~erg~s$^{-1}$ of very low-density gas \citep[table 2 in][]{1993ApJ...403..690B}, as in the interstellar surroundings of the Sun, also known as the local cavity or Local Bubble, delimited by dense clouds at 40--180~pc and other neighbouring cavities \citep[see e.g.][]{2010A&A...518A..31V}. This accretion luminosity could be activated at any time during the thermal evolution of the neutron star. 

We then computed the emission profiles for ages of 0.01, 0.05, 0.1, 0.2, 1, 2, 5, 10, 20, 50, 120, 300, 600, 1000, and 5000 Myr. At ages $>$0.1~My, the traced-back luminosities are consistent with the observed luminosities of thermally emitting neutron stars. At younger ages of 0.01 and 0.05~Myr, the traced-back ones are up to $10^{36}$~erg~s$^{-1}$ and unrealistically higher than those observed ($10^{31}-10^{33}$~erg~s$^{-1}$, as in fig.~2 of \citealt{2020MNRAS.496.5052P}). 

Similarly as in \citet{2022MNRAS.510..611T}, we considered the theoretical surface thermal evolution of a pulsar with rotochemical heating and an initial spin-period of 1~ms (having the hottest temperature over time) presented in \citet{2019PhLB..795..484H}, which is the same with or without dark matter heating. In this case, we did not add any flux factor for any optical excess-mission, for simplicity. We assumed a neutron star radius of 11~km when computing the thermal luminosities.

We included also isolated pulsars that are firmly identified and measured photometrically in the ultraviolet--optical--(infrared). In general, isolated pulsars present (i) power-law emission associated to the magnetosphere, in the wavelength range between high-energy photons and radio waves, and sometimes also (ii) hot thermal emission of the neutron-star surface or caps. First we considered PSR B0656+14 (0.11~Myr, 288~pc), which is extensively characterized \citep{2021MNRAS.502.2005Z}. From the $N1o$ fit with the G2BB + PL spectral model, it has (i) a power law of photon index $\Gamma=1.54$ and normalization $PL_\mathrm{norm}=1.97\times10^{-5}$~photons~keV$^{-1}$~cm$^{-2}$~s$^{-1}$ at 1~keV, and (ii) a blackbody for the surface ($kT^\infty=68$~eV) and another for the hot polar caps ($kT^\infty=134$~eV). The bolometric thermal luminosity is of $\log(L^\infty)=32.5$~erg~s$^{-1}$. Figure~\ref{sfd} shows the spectra in frequency (top panel) and wavelength (bottom panel) of the unabsorbed model fit, where we omitted the Gaussian absorption line at 547~eV, which does not affect significantly the X-ray part. We verified that the relative difference of SED $(\lambda F_{\lambda} - \nu F_{\nu})/(\lambda F_{\lambda})$ is zero over the energy range (standard deviation $\sigma<10^{-15}$) and thus that the representations are equivalent. This figure shows that, at optical wavelengths, the power-law component has a flux that is about 10 times the flux extrapolation of the X-ray blackbody, and a slope that is less blue than those of the seven known thermally emitting isolated neutron stars (see also Fig.~\ref{sfd_m7}), implying photometric colours that are redder, albeit still close to neutral.

We considered the dereddened blackbody and power-law fit to ultraviolet--optical photometry (1200--9000~$\AA$) of the PSR~B1055$-$52 radio pulsar (0.535 Myr, $\sim$350~pc) obtained by \citet{2010ApJ...720.1635M}. The power-law component ($F_{\nu}\propto\nu^{-\alpha_{\nu}}$ with $\alpha_{\nu}=1.05\pm0.34$) dominates in the optical at $>$3000~$\AA$, and we assumed validity of extrapolation at 9000--11\,000~$\AA$. We considered also the dereddened blackbody and power-law fit to ultraviolet--optical photometry of the intrinsically fainter PSR B0950+08 radio pulsar (17.5 Myr, 262~pc) obtained by \citet{2017ApJ...850...79P}. Although this fit is less certain ($\alpha_{\nu}=1.21\pm0.42$) and we assumed validity of extrapolation at 9500--11\,000~$\AA$, it suggests an even redder colour. (Assuming a spectral index of $\alpha_{\nu}=1.63$, its synthetic $G-G_{RP}$ and $G_{BP}-G_{RP}$ colours would be redder by $<$0.1 and $<$0.2~mag.) 

Next, we considered the $\alpha_{\nu}=0$ dereddened fit to ultraviolet--near-infrared photometry ($\approx$1000--24\,000~$\AA$) of the Vela Pulsar (0.011~Myr) obtained by \citet{2013ApJ...775..101Z}, for its distance of about 285~pc \citep{2003ApJ...596.1137D}. Then, we considered the $\alpha_{\nu}=0.26$ dereddened power-law fit at 3290--19\,000~$\AA$ for the Geminga Pulsar (0.342~Myr) from \citet{2006A&A...448..313S}, for its distance of 0.25~kpc \citep{2012ApJ...755...39V}. We considered the $\alpha_{\nu}=0$ dereddened power-law fit at 1890--4750~$\AA$ for PSR~1929+10 (3.11~Myr) from \citet{2002ApJ...580L.147M}, for its distance of 0.31~kpc \citep{2012ApJ...755...39V}, and assuming validity of extrapolation up to 11\,000~$\AA$. Also, we considered the $\alpha_{\nu}=-0.27$ dereddened power-law fit at 3000--45\,000~$\AA$ for the Crab Pulsar (1054~kyr) from \citet{2009A&A...504..525S}, for its distance of 2~kpc \citep{1973PASP...85..579T}, and we considered the $\alpha_{\nu}=0.70$ dereddened power-law fit at 3000--24\,000~$\AA$ for PSR B0540$-$69 (1700~kyr; in the Large Magellanic Cloud at 48.97~kpc) from \citet{2012A&A...544A.100M,2019ApJ...871..246M}. The Crab Pulsar and PSR B0540$-$69 are so young and bright that we did not consider their absolute magnitudes ($M_G=3.5$ and 2.8~mag) for the search of nearby neutron stars.

Finally, we included the non-accreting, recycled millisecond pulsars PSR~J0437$-$4715 \citep{2004ApJ...602..327K, 2012ApJ...746....6D,2019MNRAS.490.5848G} and PSR~J2124$-$3358 \citep{2017ApJ...835..264R}. PSR~J0437$-$4715 (6.7 Gyr) is the nearest known millisecond pulsar, at 156.3$\pm1.3$~pc. Its gamma-ray to ultraviolet emission distinguishes clearly from the optical--infrared emission of its 4000~K white dwarf companion, which is bright in \textit{Gaia} with $M_G=14.4$~mag. We considered the CutPL+3BB fit obtained shortwards of 3300~\AA~\citep{2012ApJ...746....6D}, composed of a cutoff power law of photon index $\Gamma=1.562\pm0.013$ ($\alpha_{\nu}=\Gamma-1=0.562$, similar to that of PSR B0656+14), extending from gamma-rays to X-rays and possibly the ultraviolet, and three consecutive X-ray--ultraviolet blackbodies. Extrapolating the fit to the optical, the power-law contribution dominates the contribution of the Rayleigh--Jean tail of the coolest of the three blackbodies, implying higher fluxes and redder colours in the optical. In the case of PSR~J2124$-$3358 ($\sim$11~Gyr, $\sim$410~pc), it has two ultraviolet and one optical photometric measurements in the range 1200--7000~$\AA$ \citep{2017ApJ...835..264R}, whose values can be associated with thermal and magnetospheric emissions, respectively. We adopted the blackbody+power-law fit with $R_\mathrm{bb}=12$~km, $T_\mathrm{bb}=11.72\times10^4$~K, and $\alpha{_\nu}=1$, and we assumed validity of extrapolation up to 11\,000~$\AA$. In Fig.~\ref{sfd}, we show the corresponding apparent and absolute spectra for these neutron stars.

We then derived synthetic fluxes and magnitudes at different instrumental bandpasses for the above emission profiles and power-law profiles of different spectral indices, using the \textsc{pyphot}\footnote{\url{https://mfouesneau.github.io/pyphot/}} package from Morgan Fouesneau in the \textsc{python} programming language. We adopted the $G_{BP}$, $G$, and $G_{RP}$ bandpasses of the \textit{Gaia} Early Data Release 3 \citep[EDR3;][]{2021A&A...649A...1G}, which span the wavelength range 3300--10\,500~$\AA$. Given that we focus on small distances $<$50--120~pc, we assumed in a first approximation no significant interstellar extinction and reddening.

\begin{figure*}
	\includegraphics[width=\textwidth]{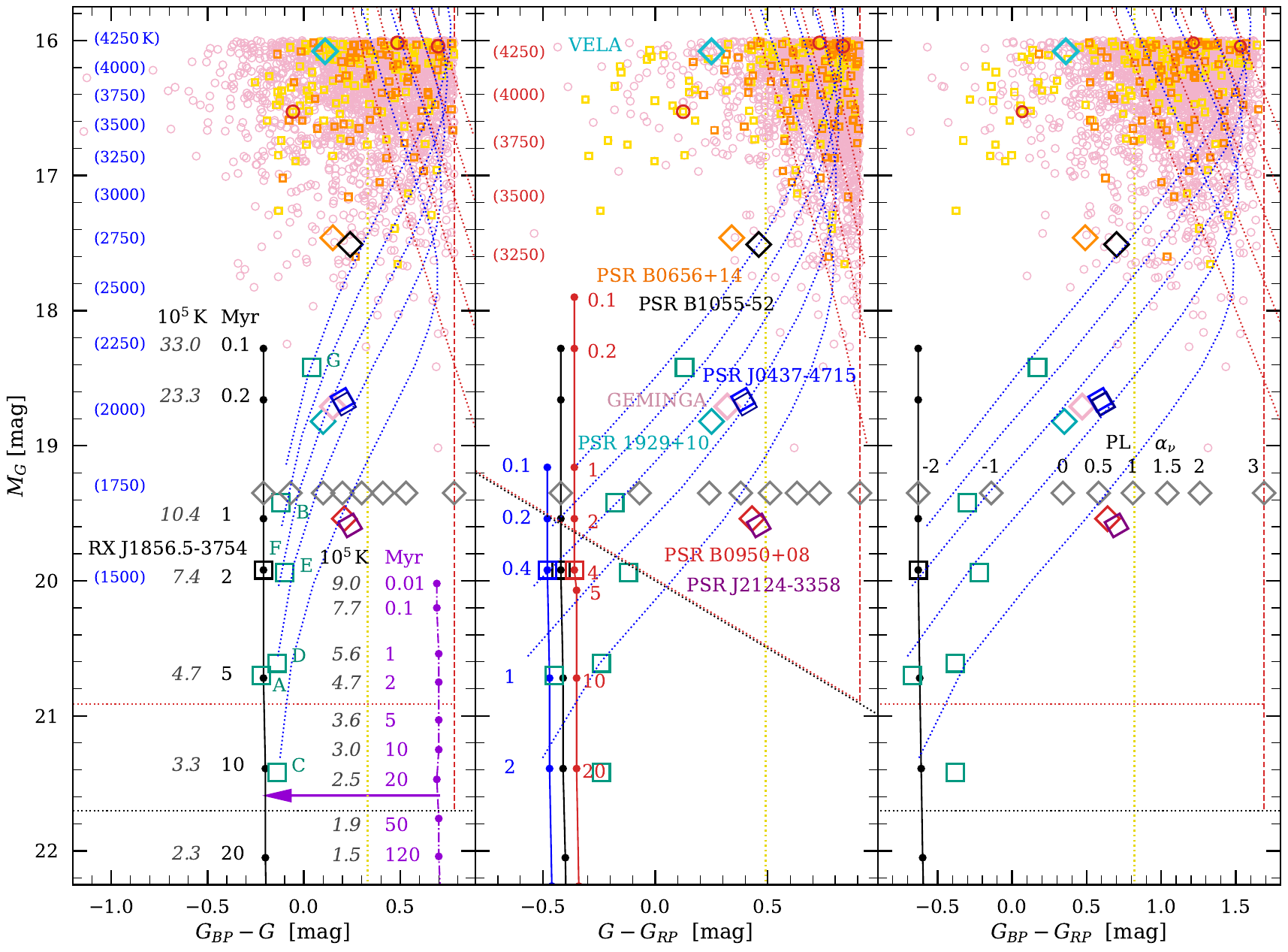}
    \caption{$M_G$ versus $G_{BP}-G$, $G-G_{RP}$, and $G_{BP}-G_{RP}$ colour--absolute magnitude diagrams. In the middle panel, the thermal cooling tracks with ultraviolet--optical excess, anchored to RX~J1856.5$-$3754 for its fiducial, kinetic, and spin-down ages of 2, 0.4, and 4 Myr, are represented by the black, blue, and red solid lines, respectively. The later two tracks are offset in colour by $-$0.06 and 0.06~mag for clarity. RX~J1856.5$-$3754 is indicated by the black open square and labelled F in the left-hand panel. Other X-ray dim, thermally emitting isolated neutron stars are represented by green squares and labelled in the left-hand panel: RX~J0420.0$-$5022 (A), RX~J0720.0$-$3125 (B), RX~J0806.4$-$4123 (C), RX~J1308.6+2127 (D), RX~J1605.3+3249 (E), and RX~J2143.0+0654 (G). The thermal model with rotochemical heating for a pulsar of initial period of 1~ms is shown by the violet dash--dotted line (offset in colour by +0.9~mag for clarity); effective temperatures in 10$^5$~K and ages in Myr are indicated left to the track. Only \textit{Gaia} sources with $M_G>16$~mag and less red than the colour selection boundaries (red vertical dashed lines) are shown, as pink circles. Those identified as probable white-dwarf photometric candidates in GCNS or else are indicated by yellow or orange small squares ($P_\mathrm{WD}>0.5$; see Appendix~\ref{app:method1} for details) and those part of resolved wide binaries are indicated by red circles (the other components have $M_G<16$~mag). The Vela, PSR B0656+14, PSR~B1055$-$52, Geminga, PSR~1929+10, and PSR B0950+08 pulsars are represented by the cyan, orange, black, pink, turquoise, and red diamonds, respectively. The PSR~J0437$-$4715 and PSR~J2124$-$3358 millisecond pulsars are represented by the blue and purple slightly tilted diamonds. A pulsar such as PSR~J0437$-$4715 but with an insignificant thermal emission of $T_\mathrm{eff}<10^5$~K (dark-blue diamond) has practically the same location as PSR~J0437$-$4715. The grey diamonds represent pure power-law sources of $\alpha_{\nu}=-2$, $-1$, 0, 0.5, 1, 1.5, 2, and 3, respectively; their $M_G$ magnitude is arbitrary. The expected colours of the Sun are represented by the yellow vertical dotted lines. The lower grey dotted lines correspond to the maximum $M_G$, $M_{BP}$, and $M_{RP}$ magnitudes of \textit{Gaia} DR3 sources with $snr_{\pi}>2$, most of which have red colours and are not shown in the plots. The red dotted lines indicate the combined limit with the colour selection criteria. Theoretical cooling sequences of white dwarfs with pure hydrogen and pure helium atmospheres (blue and red dotted curves) are shown for masses of 0.4, 0.6, 0.8, 1.0, and 1.2 M$_\odot$ from top to bottom. H and He effective temperatures for 0.8~M$_\odot$ are indicated in parenthesis in the left-hand and middle panels.}
    \label{cmd_gaia}
\end{figure*}

Possibly some isolated neutron stars emitting as pulsars might have even redder ultraviolet--optical colours than the known pulsars. We can adopt a colour upper bound from radio pulsars, for which many radio observations are available, although the spectral slopes at radio and optical might differ. At low radio frequencies (down to $\sim$100 MHz), spectral indices peak at about $\alpha_{\nu}=1.7$ and values are typically lesser than 3 \citep{2001A&A...368..230K, 2015MNRAS.453..828K}. We assumed thus in the optical an upper bound for the spectral slope of $\alpha_{\nu}=3$, which implies colours $G_{BP}-G=0.78$, $G-G_{RP}=0.91$, and $G_{BP}-G_{RP}=1.69$~mag.


\subsection{Selection of \textit{Gaia} sources with parallaxes}
\label{sec:gaia_plx} 

First, we considered the location of neutron stars in the $M_G$ versus $G_{BP}-G$, $G-G_{RP}$, and $G_{BP}-G_{RP}$ colour--absolute magnitude diagrams, which are shown in Fig.~\ref{cmd_gaia}. In the three panels, the black solid line represents the thermal cooling track with ultraviolet--optical excess, anchored to RX~J1856.5$-$3754 (black open square, labelled F in the left-hand panel) at the present age $t_0=2$~Myr. In the middle panel, the blue and red tracks are for present ages $t_0=0.4$ and 4 Myr, in which cases the $G$-band absolute magnitudes at $<$2 Myr offset by up to +1.5 and $-$0.4~mag. The other X-ray dim, thermally emitting neutron stars are represented by green squares: RX~J0420.0$-$5022 (A), RX~J0720.0$-$3125 (B), RX~J0806.4$-$4123 (C), RX~J1308.6+2127 (D), RX~J1605.3+3249 (E), and RX~J2143.0+0654 (G). The Vela (0.011~Myr), PSR B0656+14 (0.11~Myr), PSR~B1055$-$52 (0.535 Myr), Geminga (0.342~Myr), PSR~1929+10 (3.11~Myr), and PSR B0950+08 (17.5~Myr) pulsars are represented by the cyan, orange, black, pink, turquoise, and red diamonds, respectively. The PSR~J0437$-$4715 (6.7 Gyr) and PSR~J2124$-$3358 (11~Gyr) millisecond pulsars are represented by the blue and purple slightly tilted diamonds. As a curiosity, we note that their potential colours are very close to the expected colours of the Sun, $G_{BP}-G=0.33$, $G-G_{RP}=0.49$, and $G_{BP}-G_{RP}=0.82$~mag \citep{2018MNRAS.479L.102C} delineated by the yellow vertical dotted lines. RX~J2143.0+0654 is 1.6~mag brighter at $M_G$ than RX~J1856.5$-$3754. Given its spin-down age of 3.7~Myr \citep{2020MNRAS.496.5052P}, its trace-back to $\sim$0.1 Myr would lead to an even brighter $M_G$. Also, the absolute magnitudes of the cooling tracks anchored to RX~J1856.5$-$3754 are for an emitting radius at infinity of 4.51~km; for a neutron star emitting from its whole surface, that is with an emitting radius about thrice larger, the magnitudes would be about 2.5~mag brighter (cf. equations~(\ref{eqn:BBlum}) and (\ref{eqn:Mbol-lum})). Besides this, PSR B0656+14, PSR~B1055$-$52 and the Vela Pulsar are also brighter, the last one having $M_G=16.1$~mag. Finally, equal-brightness neutron star binaries would have absolute magnitudes brighter by 0.75~mag. For these reasons, for the search of neutron stars older than 0.01-0.1~Myr, it seemed appropriate to us to select \textit{Gaia} source with $M_G>16.0$~mag. 

We queried \textit{Gaia} sources with $M_G>16.0$~mag (\texttt{phot\_g\_mean\_mag} + 5 $\log_{10}$(\texttt{parallax}) $-$ $10 > 16$) and parallax signal-to-noise $snr_{\pi}>2$ ($\texttt{parallax\_over\_error}>2$). \textit{Gaia} DR3 has 90 and 50 per cent completenesses at about $G=20.5$ and 21.0~mag, based on a comparison \citep{2021A&A...649A...6G} with the Panoramic Survey Telescope and Rapid Response System (Pan-STARRS; PS1) survey \citep{2016arXiv161205560C}. The query yielded 75\,073 sources. Details on the flux uncertainties and sample limits in absolute- and apparent magnitudes are given in Appendix~\ref{app:method1}. The sources have parallaxes of $\pi>8.9$~mas and are in the $\pi>8$~mas \textit{Gaia} Catalogue of Nearby Stars\defcitealias{2021A&A...649A...6G}{GCNS}\citep[GCNS;][]{2021A&A...649A...6G}. Of these, 3312 and 71761 (4.4 and 95.6 per cent) are among the selected and rejected sources of the GCNS 100~pc (GCNS100pc) sample. The latter ones are either expected to be beyond 100~pc or to have a spurious astrometric solution. Because we are searching for sources that are outliers in terms of absolute magnitude, colour, and kinematics respect to known-population statistics and stellar mock catalogues used as priors for distance estimates, we relied first on the parallaxes and the absolute magnitudes derived directly from the parallaxes. We note that the absolute magnitude criterium is equivalent to the parallax criterium $\pi>10^{(26 -G)/5}$~mas. This can be converted to a distance criterium for sources with $snr_{\pi}$ well above 3, for which the distance inverse-proportionality to the parallax, $d=1/\pi$, tends to be valid \citep[see e.g.][]{2019AJ....158...20M}. In this case, $G=19$~mag implies $\pi>25$~mas and $d<40$~pc, and $G=21$~mag implies $\pi>10$~mas and $d<100$~pc. At these distances, neutron stars with transverse physical velocities of $v_\mathrm{t}\leq200$~kms/s have proper motions ($pm$) of $\mu\leq1055$ and 422 mas~yr$^{-1}$, since $\mu=(1000/4.74)\,v_\mathrm{t} / d$.

We then selected sources less red than the boundary colours defined by a power law of spectral index $\alpha_{\nu}=3$ (see above; vertical red dashed lines in Fig.~\ref{cmd_gaia}), with $G_{BP}-G<0.78$, $G-G_{RP}<0.91$~mag, and $snr_\mathrm{flux}\geq2$ at $G_{BP}$ and $G_{RP}$. It yielded 2464 sources (pink circles in Fig.~\ref{cmd_gaia}), which have $\pi=9.9-48.5$~mas and $pm=0.1-2354$~mas~yr$^{-1}$. In Appendix~\ref{app:method1}, we comment about two sources with very blue $G_{BP}-G$ colours of $-$2.7~mag, the corrected colour-excess factor (\texttt{phot\_bp\_rp\_excess\_factor\_corrected}), the \textit{Gaia} (E)DR3 bias at low fluxes, and the taking in of sources that are in fact redder than the colour values in the catalogue. The combination of the maximum $M_G$, $M_{BP}$, and $M_{RP}$ of the $M_G>16$~mag sample (grey dotted lines) and the colour criteria reduces the maximum $M_G$ to 20.91~mag at $G_{BP}-G$ and $G_{BP}-G_{RP}$ (red dotted lines). (We note that we initially started the study with \textit{Gaia} DR2, which provides much fainter absolute $G$-band magnitudes, closer to those of thermal cooling neutron stars; however, the fainter-$M_G$ \textit{Gaia} DR2 sources have spurious astrometric parameters, as described below). The maximum apparent magnitudes at $G_{BP}$, $G$, and $G_{RP}$ in the subsample are of about 21.7, 21.1, and 21.1~mag, as defined by different sources that are real, relatively isolated, and have magnitude errors smaller than about 0.15~mag. Finally, 4 and 29 per cent of the 2464 sources have $snr_{\pi}=2-3$ and $3-5$, and $M_G=16.0-16.5$ and $16.0-17.4$~mag, respectively.

Some nearby stars with measured parallax and proper motion are not in \textit{Gaia} DR3 or do not have parallax and proper motion in \textit{Gaia} DR3 \citep{2021A&A...649A...6G}. We thus checked the list from \citet{2021A&A...649A...6G} of 1258 sources missing in DR3. We anchored the power law with spectral index $\alpha_{\nu}=3$ to $M_G=16$~mag, and we saw that most sources have $M_H<14.8$~mag, brighter than this power law would permit at $H$ band, whereas the fainter ones are all classified as TY-type brown dwarfs. There are 124 sources without $JH$-band photometry that have $M_V$ from $-$1 to 14~mag and that are also too bright. The remaining sources are either bright multiple stars, low-mass stars, brown dwarfs, or Simbad entries of radial-velocity planets. Besides this, we queried \textit{Gaia} DR2 sources with the same parallax and colour selection criteria as above and, to limit verification, we considered only the 24 sources at $M_G=18.5-23.5$~mag that do not have parallax nor proper motion in DR3. From visual verification in optical archive images, we found that these sources are significantly blended within~arcseconds and faint, probably causing the spurious DR2 parameters, and therefore these sources can be discarded.

Fig.~\ref{cmd_gaia} shows that despite \textit{Gaia} DR3 allowing us in principle to probe absolute magnitudes as faint as $M_G\sim20.9$~mag in the colour-selected subsample, most of the colour-selected sources are very bright and red, and much brighter and redder than the known X-ray dim, thermally emitting isolated neutron stars. Redwards of the $G-G_{RP}$ selection boundary, the number of sources that are fainter increases progressively, up to $M_G\approx21.7$ at $G-G_{RP}\approx1.8$~mag (not shown in Fig.~\ref{cmd_gaia}). However in this first step, we consider only neutron stars that are resolved (as those described above) or that do not have excessively red companions. L and T-type brown dwarfs have $M_G\approx16-20$ and $>$20~mag, and $G-G_{RP}>1.55$ and $>$1.8~mag \citep{2018A&A...619L...8R,2019MNRAS.485.4423S}, and were discarded by the colour-selection criterium. Late-T or Y-type dwarfs, closer to the planetary-mass regime ($<$13~M$_\mathrm{Jup}$), have $M_G\gtrapprox21$~mag and even redder $G-G_{RP}$ colours. As unresolved companions, these could remain hidden in the photometry of the brightest colour-selected sources.

In Fig.~\ref{cmd_gaia}, probable white-dwarf photometric candidates from \citetalias{2021A&A...649A...6G} and other references are indicated by yellow and orange small squares, respectively (more information is given in Appendix~\ref{app:method1}). Also, the theoretical cooling sequences of white dwarfs with pure-hydrogen and pure-helium atmospheres\footnote{\url{https://www.astro.umontreal.ca/~bergeron/CoolingModels/}} \citep{2020ApJ...901...93B} are represented by blue and red dotted curves, for masses of 0.4, 0.6, 0.8, 1.0, and 1.2~M$_\odot$ from top to bottom. From these sequences, white dwarfs with $M_G>16$~mag would be older than 6.6, 8.1, 8.6, 7.6, and 5.0~Gyr (pure He) and 8.0, 10.3, 11.0, 9.8, and 6.4~Gyr (pure H), for masses of 0.4, 0.6, 0.8, 1.0, and 1.2~M$_\odot$. At the considered colours, those with pure-He atmospheres would always be younger than 10~Gyr, the approximate age of the Galaxy, whereas those with pure-H atmospheres would be older than 10~Gyr at $M_G>16.8$, 15.9, 15.6, 16.1, and 21.3~mag for masses of 0.4, 0.6, 0.8, 1.0, and 1.2~M$_\odot$, respectively.

The distribution of sources in the absolute-magnitude--colour diagrams of Fig.~\ref{cmd_gaia} can be explained in part as a faint and diffuse extension of the white dwarf sequence \citep{2022RNAAS...6...36S}, extension consisting of relatively old white dwarfs that are either cooler- \citep{2020MNRAS.499.1890M} or more massive \citep{2020ApJ...898...84K,2021MNRAS.503.5397K}. Because degenerate objects tend to be dominated by the gravitational force, that is by their own weight, those that are more massive, have smaller radii and thus, for surface--light-emitting objects, fainter absolute magnitudes. The distribution of the sources could also be explained by the notion that neutron stars are (a) rare, given that these are probably one order of magnitude less frequent than white dwarfs, mainly because their higher-mass progenitor stars are less frequent, and (b) intrinsically faint, in particular the thermally emitting ones, expected to cool down and dim to $M_G\ga21$~mag in less than a half-dozen Myrs after formation.

\subsection{Prospects of multiband detectability of neutron stars among the \textit{Gaia} sources}
\label{sec:gaia_nsdetec} 

\begin{figure}
    \includegraphics[width=\columnwidth]{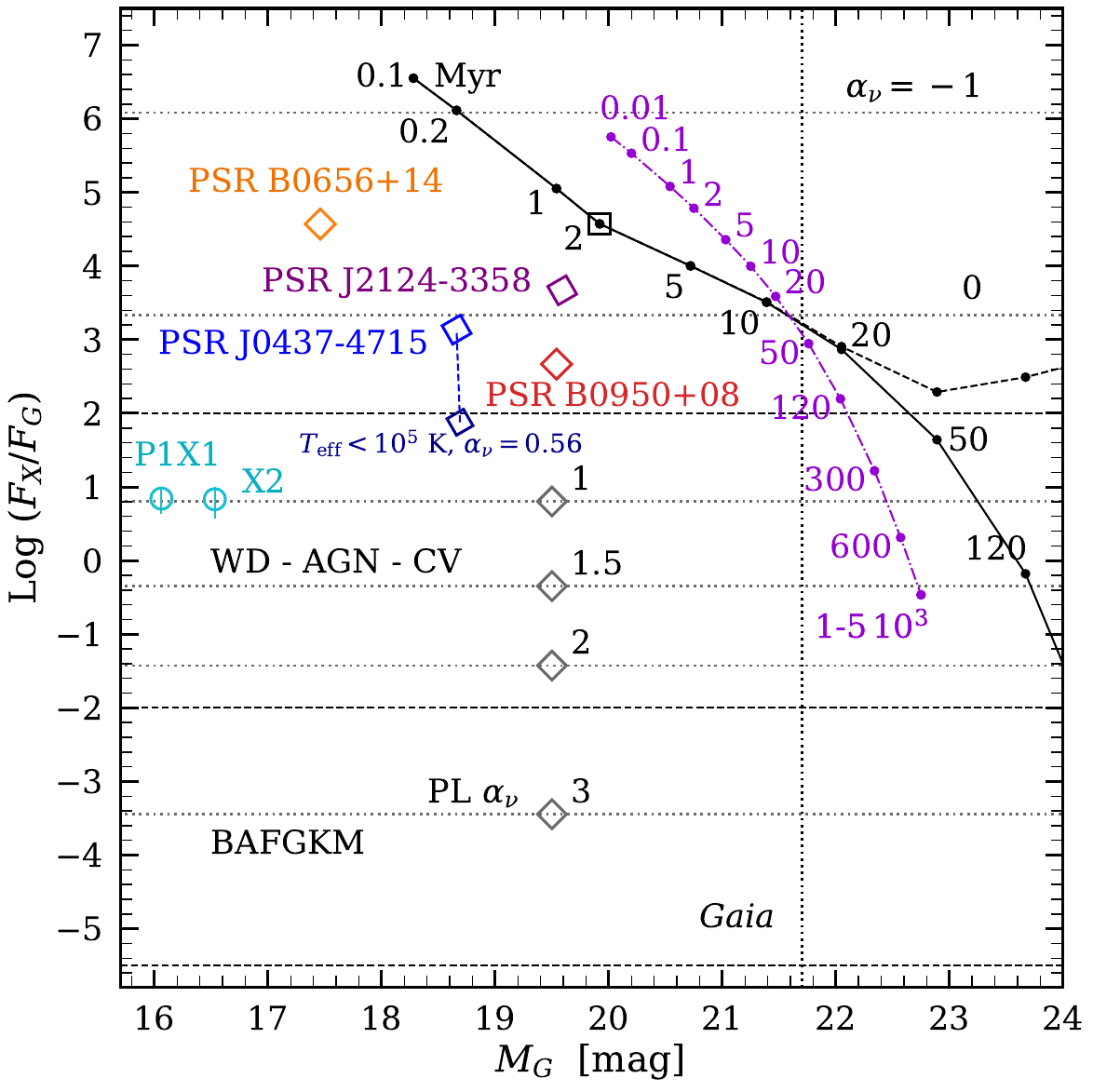}
    \caption{X-ray-to-optical flux ratio versus $G$-band absolute magnitude diagram. The thermal cooling tracks without and with interstellar accretion are represented by the black solid and dashed lines. Power-law sources are indicated by the grey dotted lines and the corresponding spectral indices. The P1X1 and X2 RASS crossmatch candidates are indicated by the cyan open circles. Same as in Fig.~\ref{cmd_gaia}.}
    \label{xg_vs_mg}
\end{figure}

\begin{figure*}
  \begin{minipage}[c]{0.695\textwidth}
    \includegraphics[width=\textwidth]{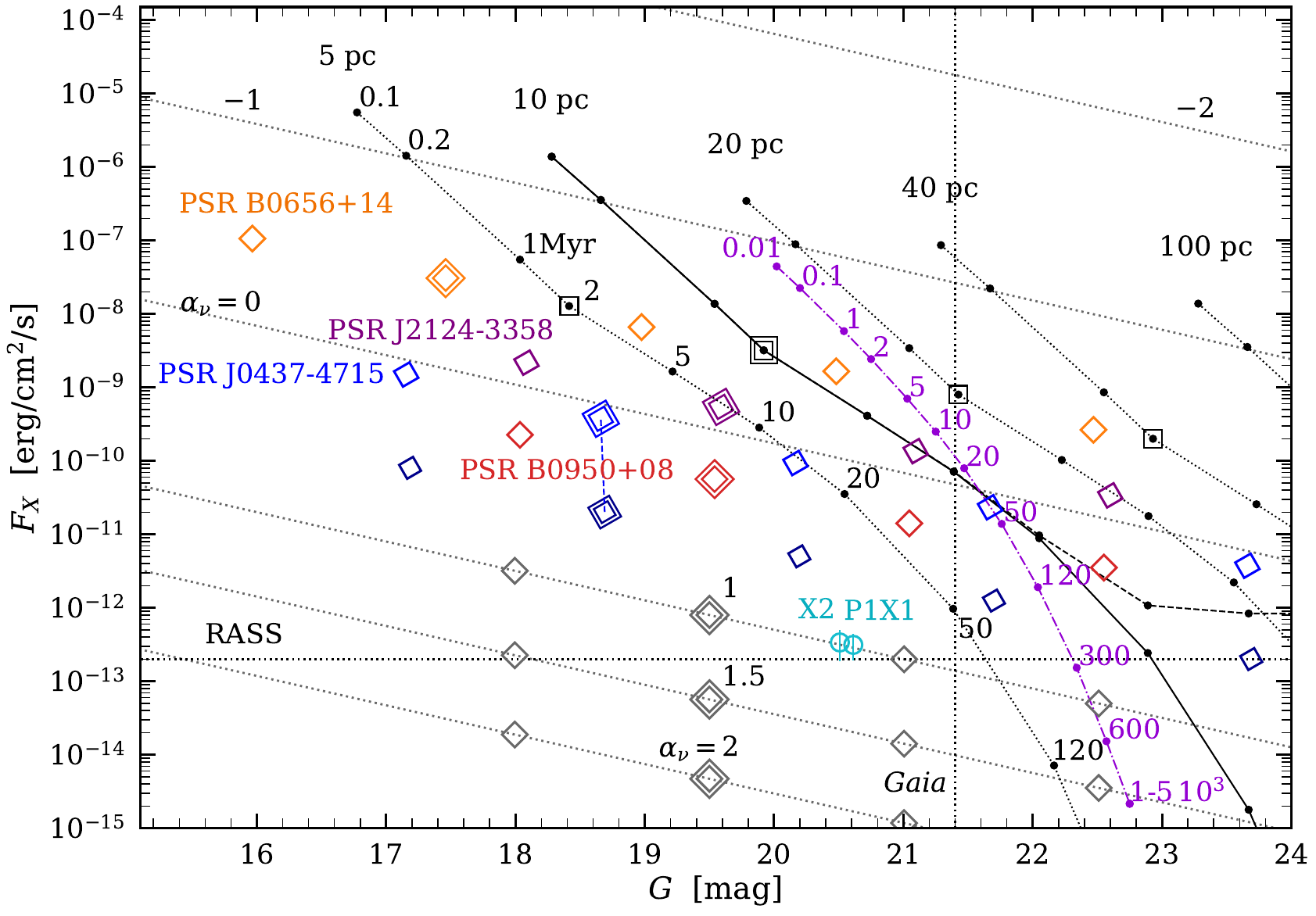}
  \end{minipage}\hfill
  \begin{minipage}[c]{0.265\textwidth}
    \caption{X-ray flux versus $G$-band magnitude diagram. Thermal cooling tracks at distances other than 10~pc are represented by black dotted lines. Double square or lozenge symbols are for 10~pc. The pulsar thermal track with rotochemical heating (violet dash--dotted line) is at 10~pc. The vertical dotted line at $G=21.4$~mag indicates the faintest apparent magnitude reached by \textit{Gaia} DR3 sources of $M_G>16$ and $snr_{\pi}>2$. As expected, the plot indicates that from 10 to 100~pc, the X-ray flux decreases by a factor 10$^{-2}$, and the $G$-band magnitude increases by 5~mag. Same as in Figs.~\ref{cmd_gaia} and \ref{xg_vs_mg}.
    } \label{x_vs_g}
  \end{minipage}
\end{figure*}

The SEDs of white dwarfs can be approximated by single blackbodies of $T_\mathrm{eff}\lesssim10^5$~K \citep[see e.g.][]{2019MNRAS.482.4570G}, whereas those of known neutron stars imply typically very hot blackbodies or power laws. Figure~\ref{xg_vs_mg} shows an X-ray-to-optical flux ratio versus $M_G$ diagram. The ratio is defined as $\log (f_{X}/f_{G}) =\log(f_X )+0.4G +5.10$, where the flux is in erg~s$^{-1}$~cm$^{-2}$ and the constant 5.10 stems from $-\log (f_G[\mathrm{Vega}])$, where $f_G[\mathrm{Vega}]$ is the flux of the Vega star without normalization by the filter transmission. The ratio is a variant of that defined by \citet{1988ApJ...326..680M}, and we assume schematically that white dwarfs, active galactic nuclei (AGN), and cataclysmic variables (CV) have log ratios of $-$2 to 2, and that BAFGKM-type stars have lesser ratios, below $-$2 \citep[for M dwarfs in \textit{eROSITA}, see][]{2022A&A...661A..29M}. Known neutron stars have the greatest X-ray-to-optical flux ratios, typically upped by thermal emission. For pulsars whose thermal emission is insignificant compared to the magnetospheric emission, the log ratio can be in the range [$-$2, 2], overlapping with that of white dwarfs, or even lower. In Fig.~\ref{xg_vs_mg}, this is illustrated with a pulsar such as PSR J0437$-$4715 but with $T_\mathrm{eff}<10^5$~K, represented by the bottom dark-blue lozenge at $\alpha_{\nu}=0.56$, linked with a blue dashed line to the blue lozenge of PSR J0437$-$4715. Considering PSR~B0656+14 instead, similarly, the ratio would decrease to that at $\alpha_{\nu}\approx0.55$. Lower ratios are illustrated with power-law sources of $M_G=19.5$~mag and $\alpha_{\nu}=1$, 1.5, 2, and 3. The same example sources are represented in Fig.~\ref{cmd_gaia}, where we note that removing the thermal component of PSR J0437$-$4715 does not change its optical photometry.

\begin{figure*}
   \includegraphics[width=\columnwidth]{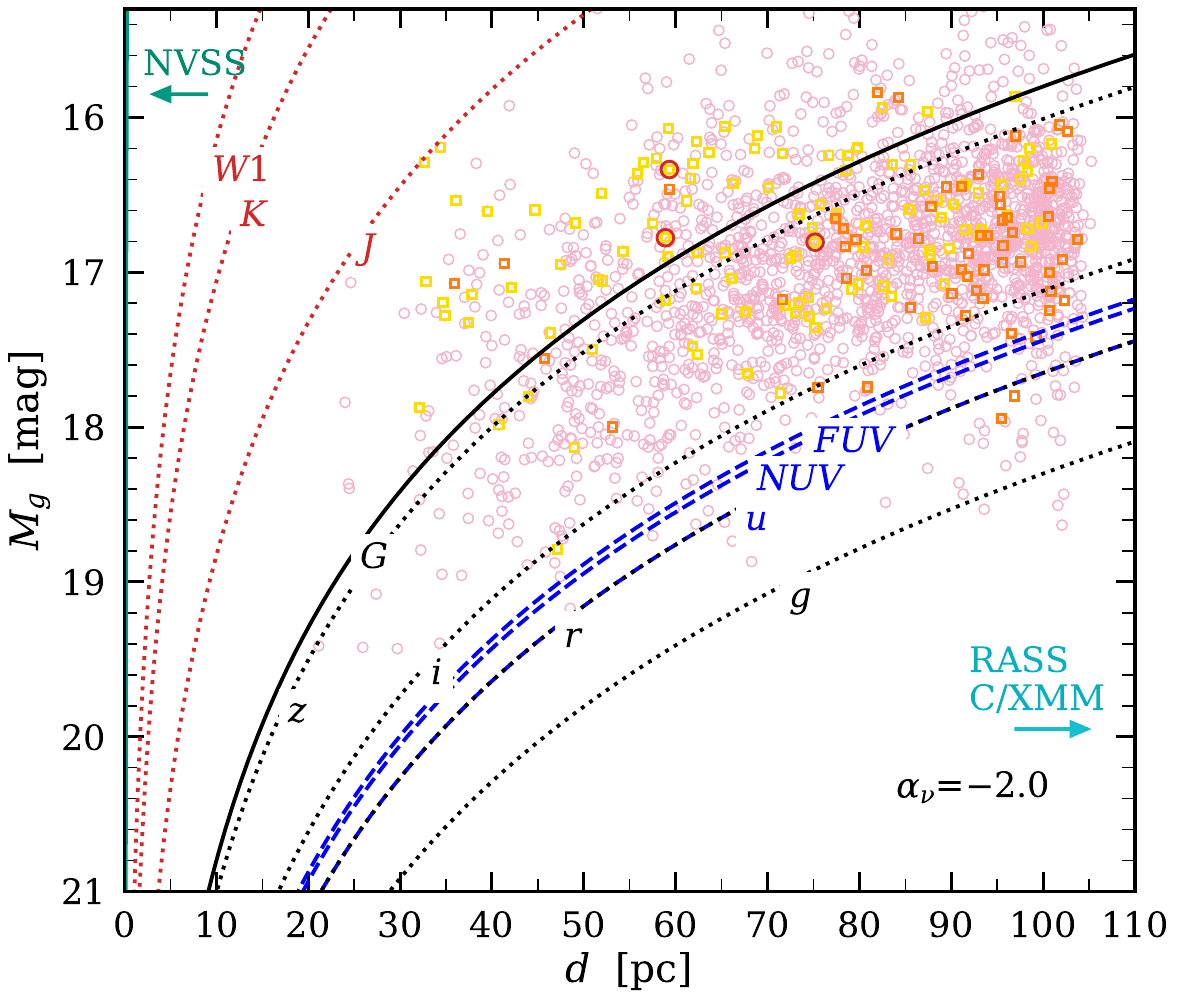}
   \includegraphics[width=\columnwidth]{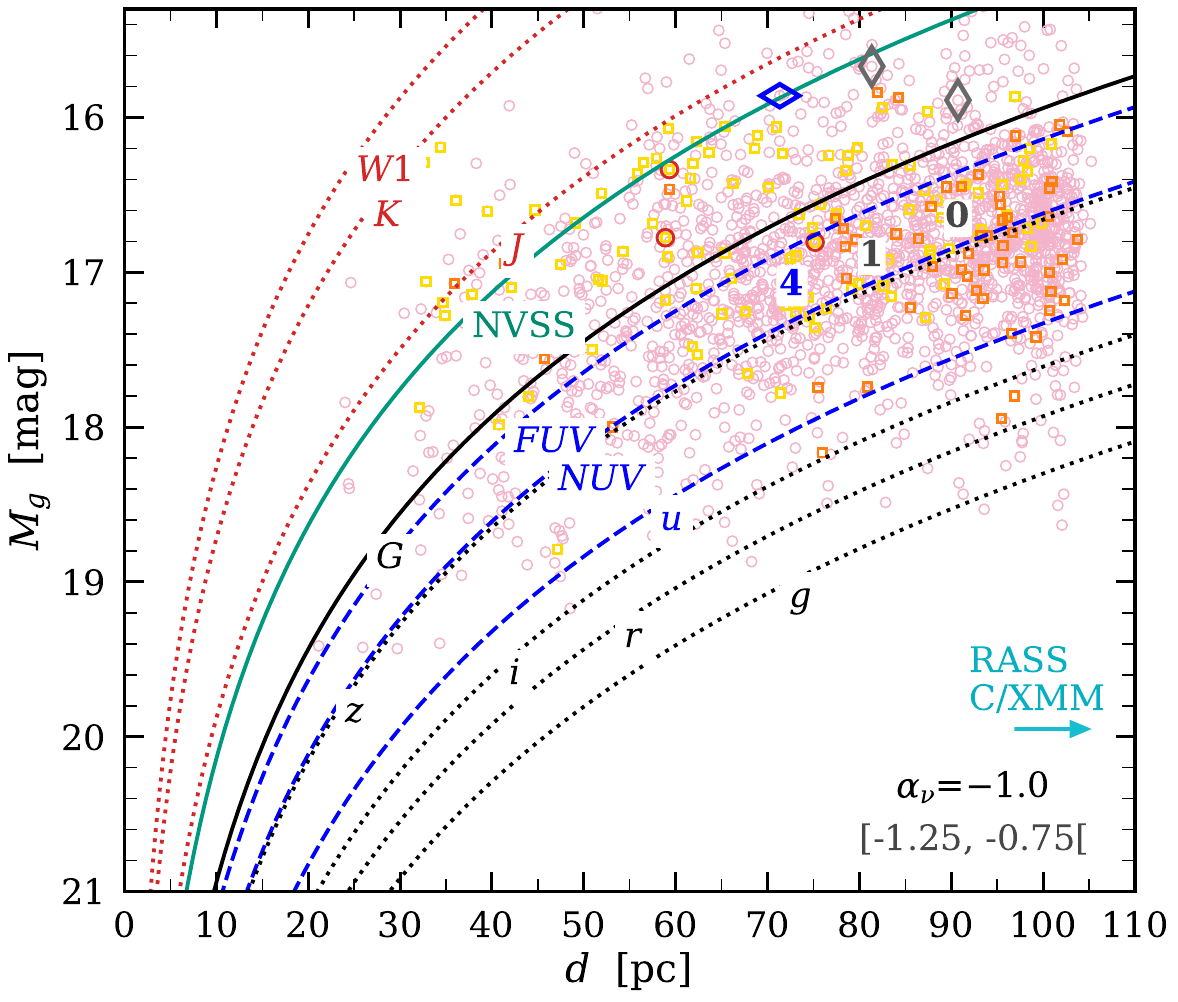}
   \includegraphics[width=\columnwidth]{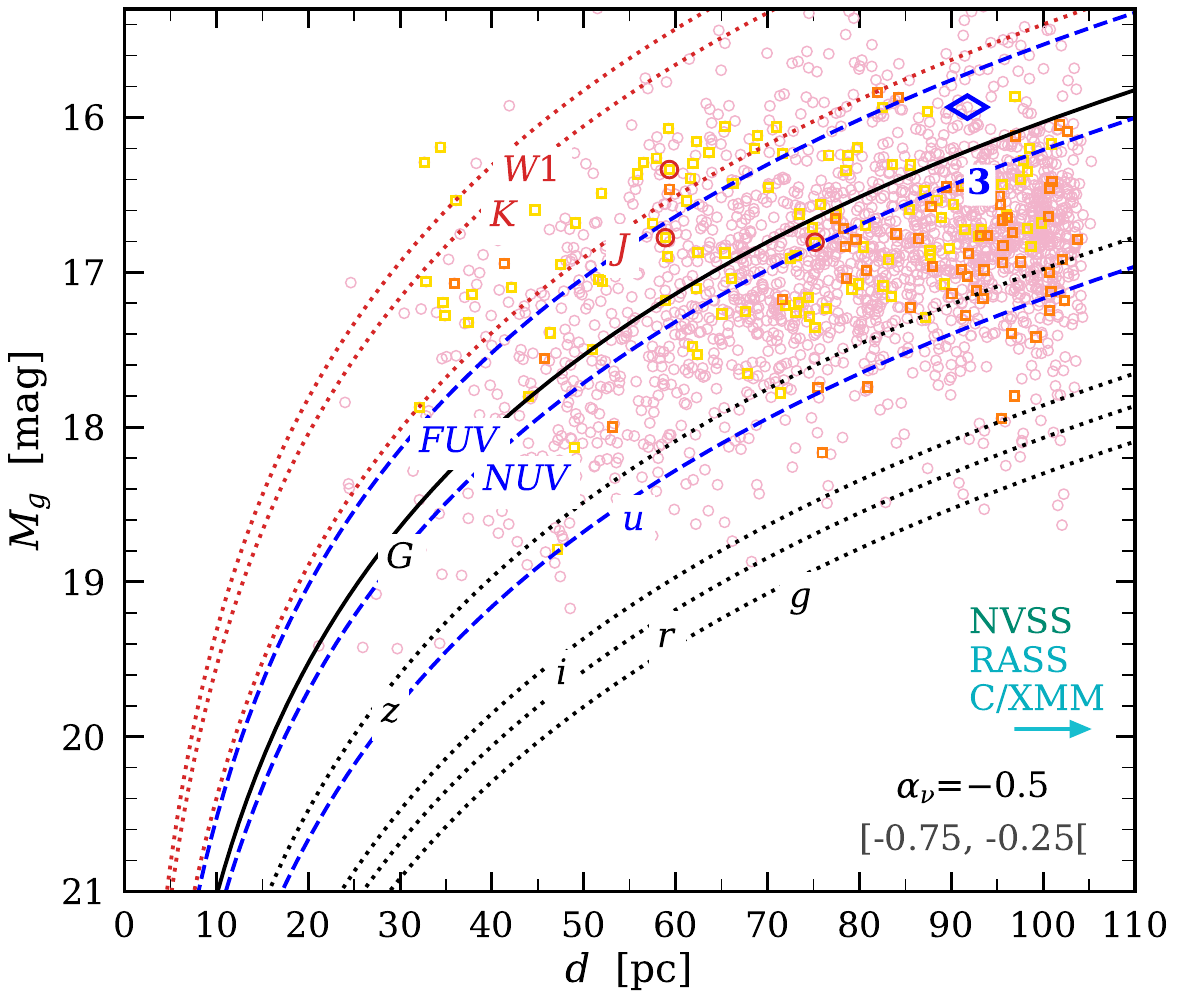}
   \includegraphics[width=\columnwidth]{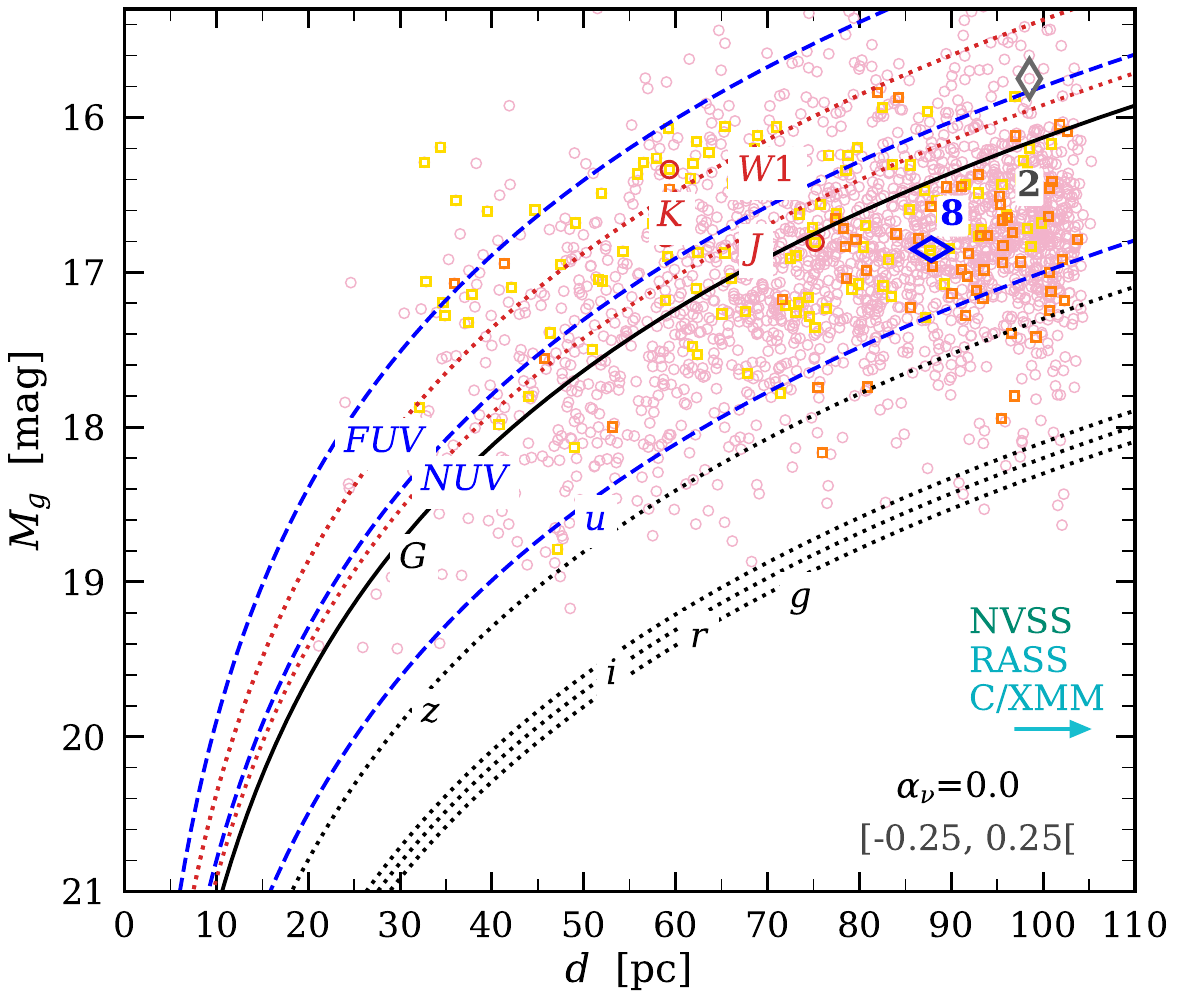}
    \caption{Detectability of nearby power-law emitters of spectral index $\alpha_{\nu}=-2$, $-$1, $-$0.5, and 0, as a function of their absolute $g$-band magnitude. Catalogue or survey limits are for (see text for details and references): \textit{X rays}: \textit{Chandra} 0.1--10.0~keV (or \textit{XMM}) and RASS 0.1--2.4~keV (cyan dashed and solid lines, respectively); \textit{Ultraviolet}: \textit{GALEX} $FUV$ and $NUV$ bands, and SDSS $u$ band (blue dashed lines); \textit{Wide optical}: \textit{Gaia} $G$ and PS1 $griz$ bands (black solid and dotted lines); \textit{Infrared}: WFCAM $JK$ and \textit{WISE} $W1$ bands (red dotted lines); \textit{Radio}: NVSS 1.4~GHz (green solid line). \textit{Gaia} sources are represented as in Fig.~\ref{cmd_gaia}, but using $d50$ distances and $g$-band magnitudes from PS1, else NSC, or else SDSS. Some ultraviolet and power-law sources of Tables~\ref{tab:uvcan}~and~\ref{tab:plcan} are represented by horizontal and vertical diamonds in the plots of nearest spectral index (spectral index ranges are indicated at the bottom), and in grey, blue, or red depending on whether their power-law component was determined mostly from the optical, the ultraviolet, or the infrared, respectively.} \label{mg_vs_dlim1}
\end{figure*}

The X-ray flux versus $G$-band apparent magnitude diagram of Fig.~\ref{x_vs_g} shows that \textit{Gaia} sources similar to known thermal-emitting neutron stars would be very bright at X-rays. We considered RX~J1856.5$-$3754, PSR~B0656+14, and PSR~J0437$-$4715, which are measured in RASS. We also considered PSR~1929+10, PSR~B0950+08, and PSR~J2124$-$3358, which are not detected in RASS, but when brought nearer to 100~pc, would be detected with fluxes of $10^{-13}-10^{-11}$~erg~s$^{-1}$~cm$^{-2}$. We adopted for PSR B0950+08 a 10~pc unabsorbed flux of $5.65\times10^{-11}$~erg~s$^{-1}$~cm$^{-2}$ in the RASS 0.1--2.4 keV energy band, from deep X-ray Multi-Mirror Mission \textit{(XMM)-Newton} observations \citep{2004ApJ...615..908B}, and for PSR J2124$-$3358 a 10~pc unabsorbed flux of $5.43\times10^{-10}$~erg~s$^{-1}$~cm$^{-2}$ (sum of the thermal and non-thermal contributions) in the X-ray soft 0.1$-$2 keV band (\citealt{2009ASSL..357.....B} and reference therein). Fig.~\ref{x_vs_g} indicates that neutron stars at $G<21.4$~mag, resembling (i) PSR~J0437$-$4715 or PSR~B0656+14 would be at less than about 40--50~pc, and those resembling (ii) RX~J1856.5$-$3754, PSR B0950+08, or PSR J2124$-$3358 would be at less than about 20~pc.

\begin{figure*}
   \includegraphics[width=\columnwidth]{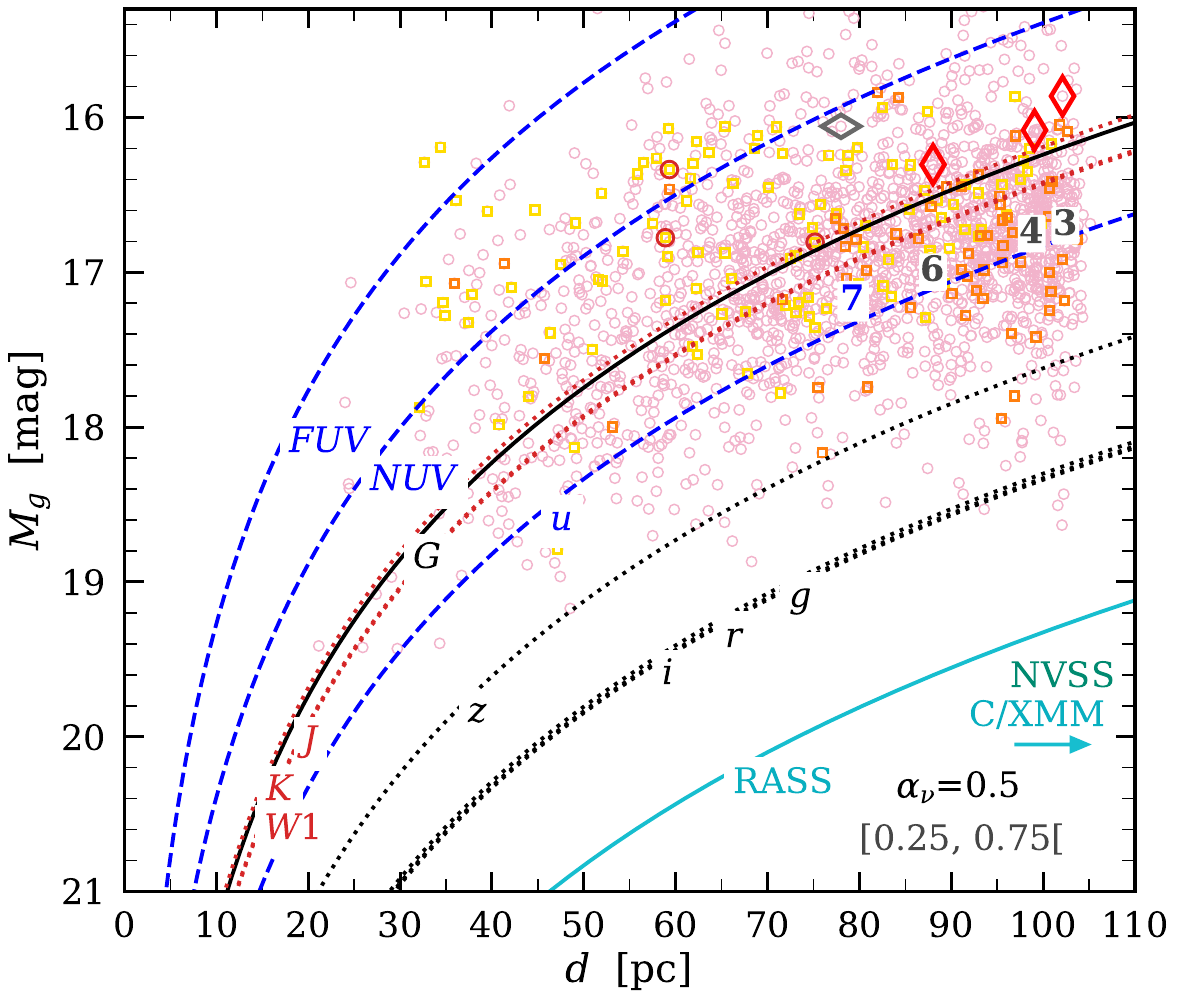}
   \includegraphics[width=\columnwidth]{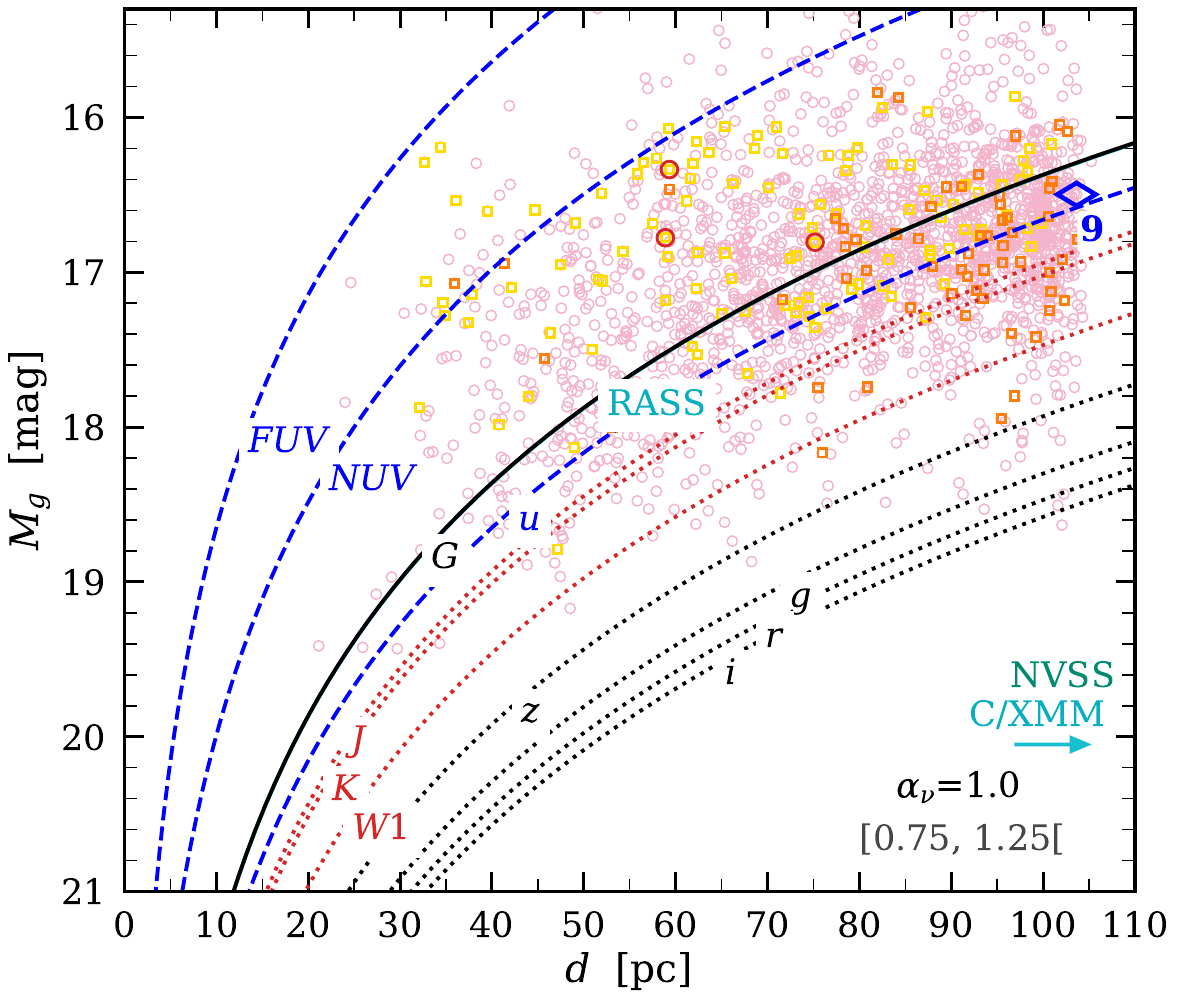}
   \includegraphics[width=\columnwidth]{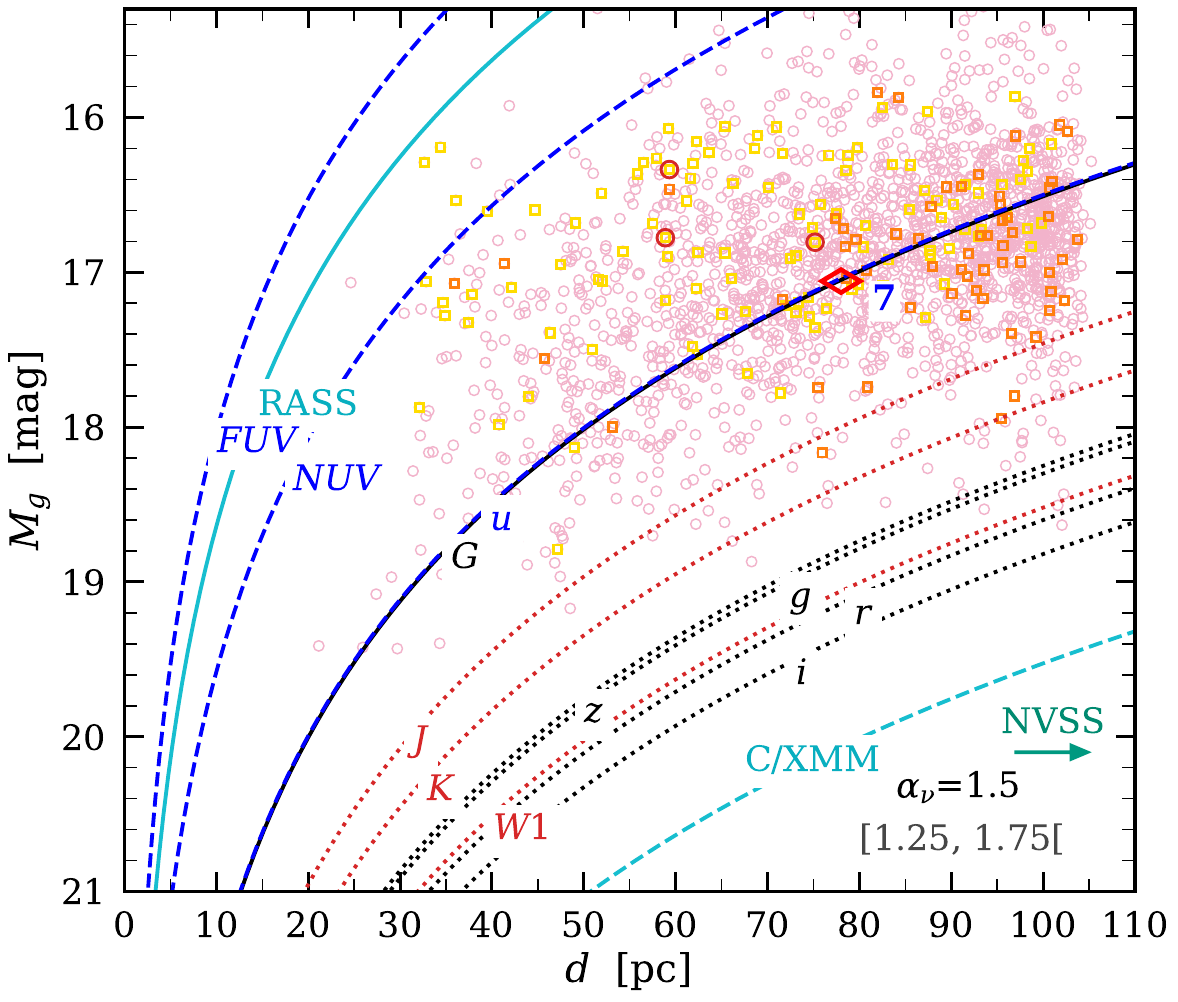}
   \includegraphics[width=\columnwidth]{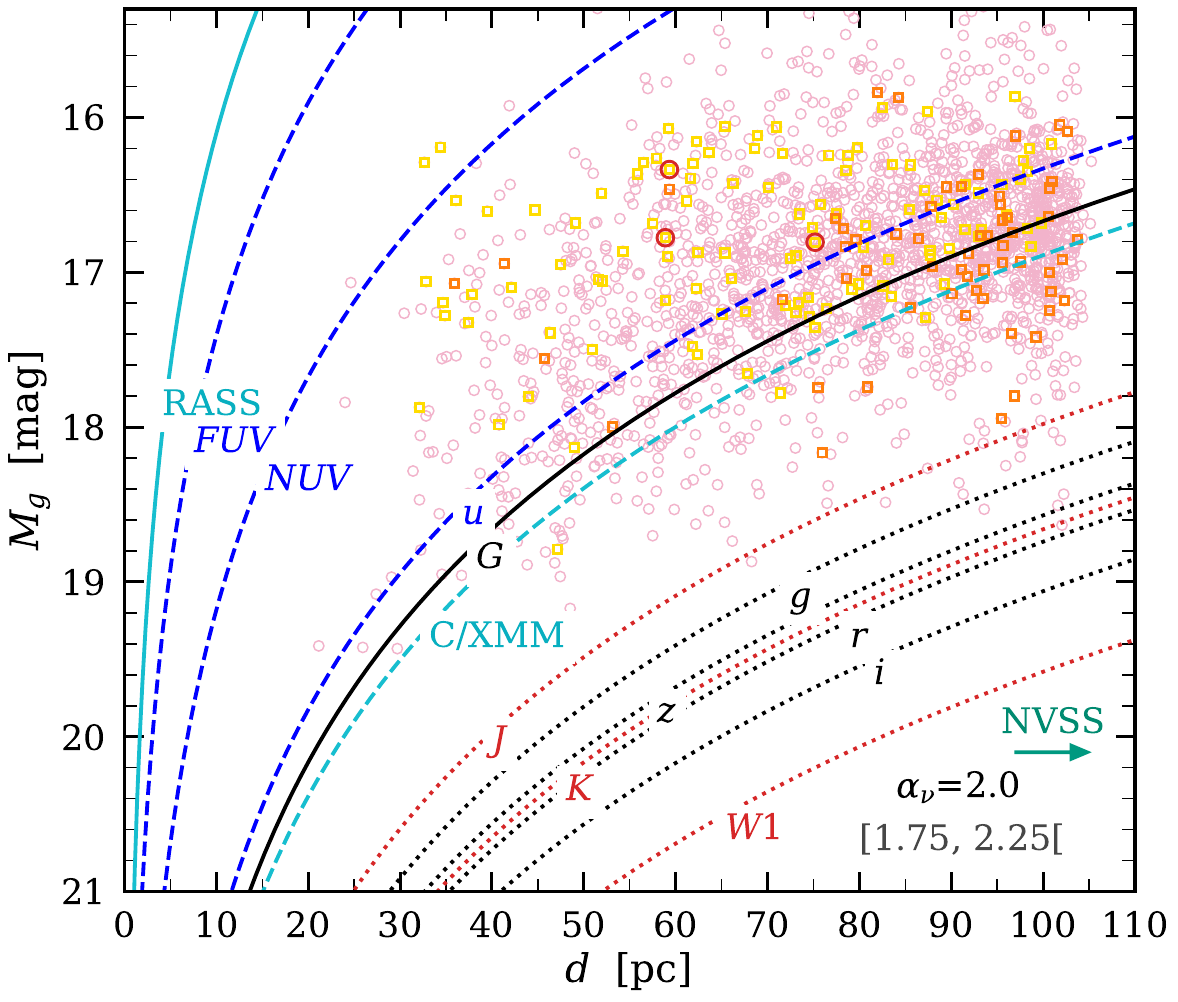}
   \includegraphics[width=\columnwidth]{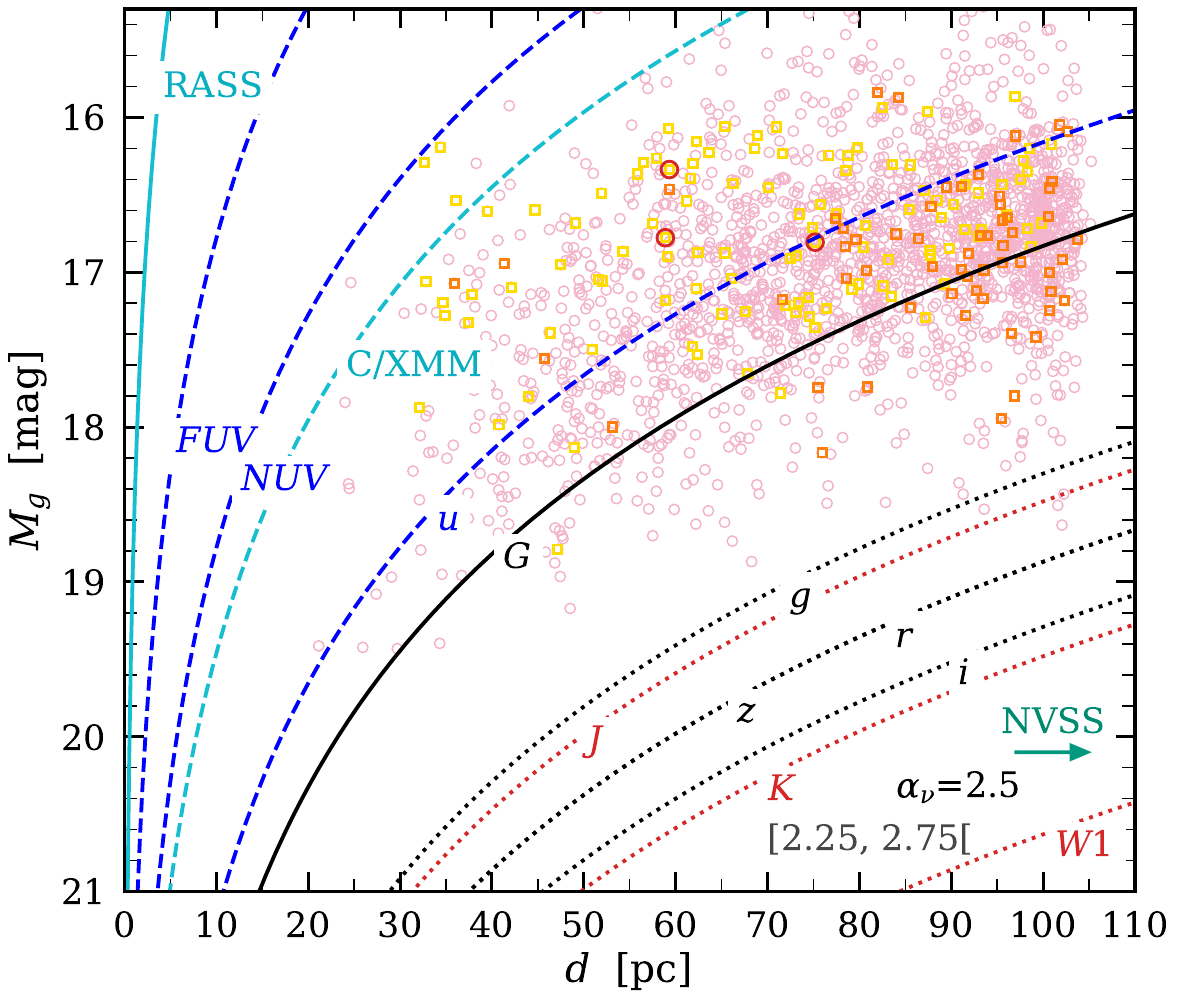}
   \includegraphics[width=\columnwidth]{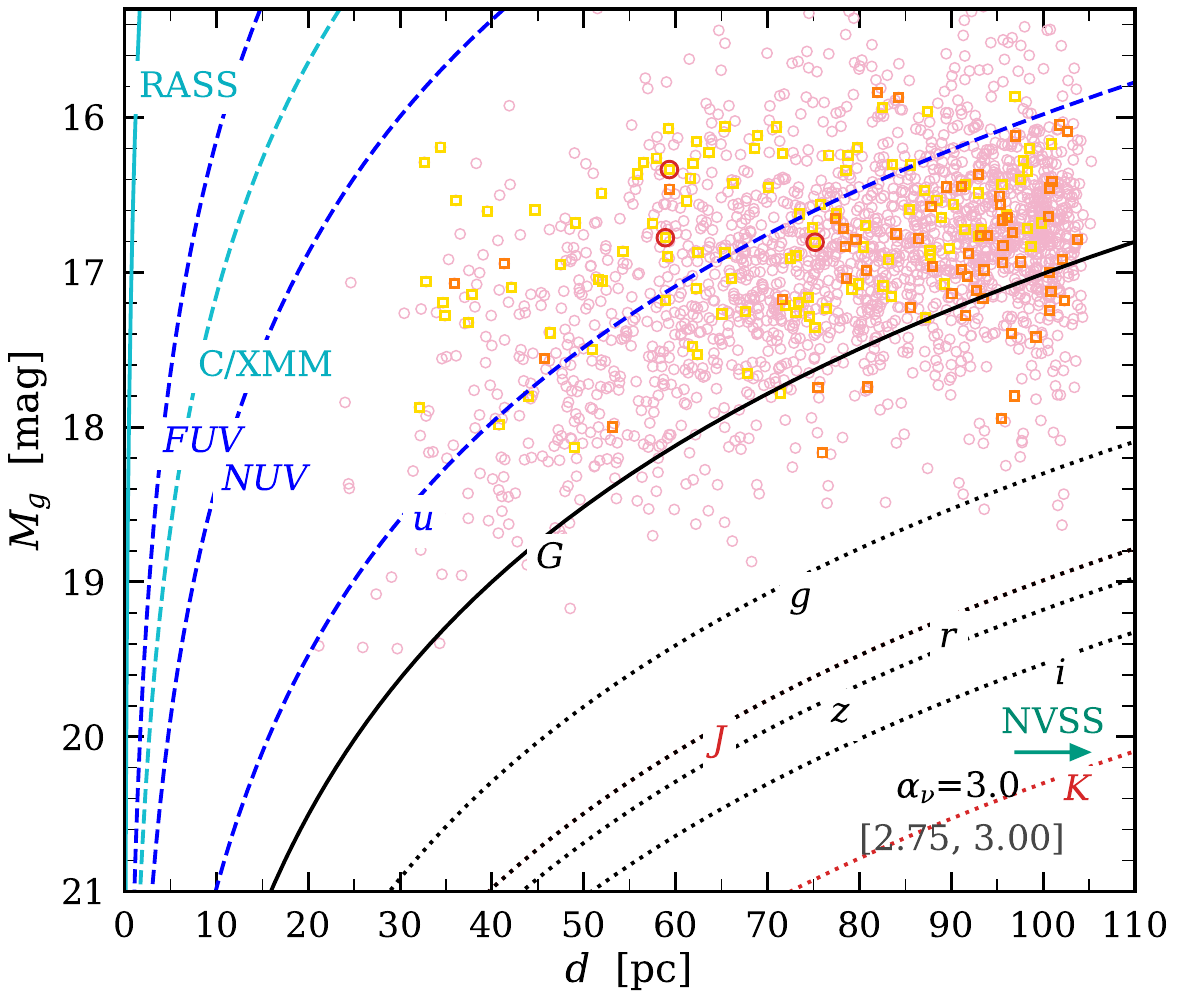}
    \caption{Same as Fig.~\ref{mg_vs_dlim1}, for power-law emitters of spectral index $\alpha_{\nu}=0.5$, 1, 1.5, 2, 2.5, and 3.} \label{mg_vs_dlim2}
\end{figure*}

On the other side, Fig.~\ref{x_vs_g} indicates that power-law-source neutron stars of $M_G=19.5$~mag and no significant thermal emission would elude RASS detection if these are at more than 20~pc for $\alpha_{\nu}=1$ and at 
more than 5~pc for $\alpha_{\nu}=1.5$ (assuming a single power law from X rays to the optical).

For a more precise estimation, we produced Figs.~\ref{mg_vs_dlim1} and \ref{mg_vs_dlim2}, which represent the maximum distances of detectability as function of PS1 $g$-band absolute magnitude, for different catalogues or surveys and power-law emitters of spectral index $\alpha_{\nu}=-2.0$, $-$1.0, $-$0.5, 0.0, 0.5, 1.0, 1.5, 2.0, 2.5, and 3.0. The PS1 $g$ band is much narrower than the \textit{Gaia} $G$ band and is appropriate to anchor power-law components identified in SEDs. From equation~(\ref{eqn:Mbol-d}), we have the dependence between the distance and magnitude limits:

\begin{equation}\label{eqn:dlim}
d_\mathrm{lim}= 10 \times 10^{0.2(m_\mathrm{f,lim}-M_\mathrm{f})},
\end{equation}

where we can express the absolute magnitude $M_\mathrm{f}$ at any filter as a function of the absolute magnitude $M_\mathrm{rf}$ at a reference filter (here PS1 $g$ band),

\begin{equation}\label{eqn:magf}
M_\mathrm{f}=M_\mathrm{rf}+(M_\mathrm{f,0}-M_\mathrm{rf,0}), 
\end{equation}

for an arbitrarily flux-scaled SED (subscript 0).

We adopted the following limits (usually point-source limits, see references for details): \textit{X rays}: catalogue deep depth of $10^{-15}$~erg~s$^{-1}$~cm$^{-2}$ for \textit{Chandra} at 0.1--10~keV, also applicable for \textit{XMM-Newton} \citep{2010ApJS..189...37E,2020A&A...641A.136W}, and survey limit of $2\times10^{-13}$~erg~s$^{-1}$~cm$^{-2}$ for RASS at 0.1--2.4~keV \citep{2016A&A...588A.103B}; \textit{Ultraviolet}: Galaxy Evolution Explorer (\textit{GALEX}) typical 5$\sigma$ detection depth of $FUV=19.9$ and $NUV=20.8$~mag \citep{2017ApJS..230...24B} and Sloan Digital Sky Survey (SDSS) 95 per cent completeness at $u=22.0$~mag \citep{2017ApJS..233...25A}; \textit{Wide optical}: \textit{Gaia} 50 per cent completeness at $G=21.0$~mag \citep{2021A&A...649A...6G}, PS1 stack 5$\sigma$ detection depth of $g=23.3$, $r=23.2$, $i=23.1$, and $z=22.3$~mag ($\sim$50 per cent completeness; \citealt{2016arXiv161205560C}); \textit{Infrared}: representative 5$\sigma$ detection depths of $J=20$ and $K=18.5$~mag (see references of near-infrared surveys in Section~\ref{sec:opticalir}), and Wide-field Infrared Survey Explorer (\textit{WISE}) 90 per cent completeness at $W1=17.7$~mag \citep{2021ApJS..253....8M}; \textit{Radio}: 1.4~GHz NRAO VLA Sky Survey (NVSS) 90 per cent completeness at 3.0~mJy \citep{1998AJ....115.1693C}. We note that the catalogue or survey limits are not necessarily homogenous over the covered sky regions.

We represented the \textit{Gaia} sources as in Fig.~\ref{cmd_gaia}, but using as distance the median of the posterior distance estimation from \citetalias{2021A&A...649A...6G} (\texttt{dist\_50}, heareafter $d50$), which is independent of colour (spectral type) or direction in the sky. The $d50$ of the 2464 sources is in the range 21--105~pc and of 80$\pm18$~pc in average. Also we used $g$-band magnitudes from PS1, else from the National Optical-Infrared Astronomy Research Laboratory (NOIRLab) Source Catalogue (NSC) \citep{2021AJ....161..192N}, or else from SDSS (see Section~\ref{sec:opticalir}). Some UV candidates with power-law components (Table~\ref{tab:uvcan}) and power-law profile sources (Table~\ref{tab:plcan}) are represented by horizontal and vertical diamonds, respectively, in the plots of nearest spectral index. In Section~\ref{sec:plbr}, we briefly discuss these sources in the diagrams.

The diagrams of Figs.~\ref{mg_vs_dlim1} and \ref{mg_vs_dlim2} adjust to modifications of source distances and search limits as follows. If a source is at a smaller or larger distance, it shifts to the left or to the right, respectively, as on detection-limit curves, implying that the detectability of the source is invariant to changes of the source distance (at fixed observed flux and negligible interstellar extinction). From equation~(\ref{eqn:dlim}), if another magnitude or flux limit (with a prime) is chosen for a given filter, the corresponding distance limit simply changes by a factor of

\begin{equation}\label{eqn:dlimfactor}
10^{0.2(m^{\prime}_\mathrm{f,lim}-m_\mathrm{f,lim})}=(F^{}_\mathrm{f,lim}/F^{\prime}_\mathrm{f,lim})^{0.5}, 
\end{equation}

implying that the curves visible in the plots can easily be used to compute other search limits.

Bright, power-law-source neutron-stars of $M_g=16.50$~mag and $\alpha_{\nu}=1$ (i.e. $M_G=16.13$~mag, from Table~\ref{tab:plcolours}) would be detected both in \textit{Gaia} and at X rays in RASS up to 94~pc (Fig.~\ref{mg_vs_dlim2}). However, those of $\alpha_{\nu}=1.5$, 2.0, 2.5, and 3.0 ($M_G=15.99$, 15.83, 15.67, and 15.49~mag) would be detected in \textit{Gaia} up to 100, 108, 116, and 128~pc, but not in RASS at $d>27$, 8, 3, and 1~pc, respectively.

At ultraviolet, because of the shallower depth of the \textit{GALEX} wide survey, the known pulsars that we considered can only be detected at less than 20--40~pc, as shown in the $NUV$- versus $G$-band magnitude--magnitude diagram of Fig.~\ref{uv_vs_g}. Power-law source neutron-stars of $M_G=19.5$~mag (with no significant thermal emission) would elude \textit{GALEX} detection if these are at more than about 10~pc for $\alpha_{\nu}\ge1$ (assuming a single power law from the ultraviolet to the optical). From Fig.~\ref{mg_vs_dlim2}, if these have $M_g=16.5$~mag, then for $\alpha_{\nu}=1$ these would be detected up to 50~pc, whereas for $\alpha_{\nu}=1.5$, 2.0, 2.5, and 3.0, these would elude $NUV$-band detection at $d>41$, 34, 29, and 24~pc, respectively.

To summarize, we see from Figs.~\ref{xg_vs_mg}, \ref{x_vs_g}, \ref{mg_vs_dlim1}, \ref{mg_vs_dlim2}, and \ref{uv_vs_g} that if young neutron stars or pulsars that are still very hot and thermally emitting are measured in \textit{Gaia}, then these would be detected at X rays in RASS, and to a lesser extent, at $NUV$ and $FUV$ bands in \textit{GALEX}. On the other hand, pulsar emitters with insignificant thermal emission and power-law slopes $\alpha_{\nu}\ge1.5$ would remain undetected in RASS and \textit{GALEX}, at strongly and weakly decreasing distances for increasing spectral indices, respectively. This quantitative reasoning has some overlap with the interpretation of \citet{1997MNRAS.287..293M} that some of the red (and faint) optical sources they identify, using optical $B$ and $R$-band photometry, as candidate counterparts to extreme-ultraviolet sources without RASS X-ray detection, could be nearby neutron stars.

Finally, we note that among the 3320 pulsars of the Australia Telescope National Facility (ATNF) Pulsar Catalogue version 1.67\footnote{\url{https://www.atnf.csiro.au/research/pulsar/psrcat/}} \citep{2005AJ....129.1993M}, 315 are in binaries with a known companion type: about 50 per cent have low-mass He-core white dwarf companions and 15 per cent have higher-mass CO- or ONeMg-core white dwarf companions, as shown in Fig.~\ref{porbsep}, for the pulsars with measured orbital periods and projected semimajor axes. There is a confluence of neutron stars with white dwarfs, though it has not been explored thoroughly at medium to large binary separations. Based on this confluence, pulsars are searched as companions to white dwarfs \citep[e.g. at radio wavelengths,][]{2021MNRAS.505.4981A}. If neutron stars are among our candidates and in binaries, then these will probably have white dwarf companions.

\subsection{Crossmatches}

Next, we crossmatch the coordinates of the 2464 \textit{Gaia} sources of the colour-selected subsample with the coordinates of sources of broad-optical, infrared, gamma-ray, X-ray, radio, ultraviolet, and literature catalogues. We verify the crossmatch candidates in archival images and catalogues using mainly the Aladin Interactive Sky Atlas tool. To better resolve the sources, especially for surveys of low spatial resolution, we query images in high-resolution mode in Aladin or in native pixel sizes with the embedded SkyView image selector.

\section{Results}

\subsection{Optical and infrared crossmatches}\label{sec:opticalir}

At the broad optical, we considered the PS1-, SDSS-, and NSC catalogues. Best-neighbour cross-identifications with PS1, SDSS, and the Two Micron All Sky Survey \citep[2MASS;][]{2006AJ....131.1163S} are provided in \citetalias{2021A&A...649A...6G} and are from the \textit{Gaia} (E)DR3 external crossmatch tables \citep[see description for \textit{Gaia} DR2 in][]{2019A&A...621A.144M}.

Cross-identifications with PS1 are available for 1419 of the 2464 \textit{Gaia} sources. Using Aladin, we noted that the sources with very large proper motions of $pm>200$~mas~yr$^{-1}$ typically have more than one entry in the PS1 DR1 VizieR catalogue (\citealt{2016arXiv161205560C}; CDS/II/349/ps1), corresponding to different trailing points defining the trajectory of the sources. Among the sources with $pm>100$~mas~yr$^{-1}$, for eight with and four without initial \textit{Gaia}--PS1 cross-identification, we could recover identifications with the widest ($grizy$) bandpass coverage and the best astrometric and photometric quality. Also, from a cone search within 1~arcsec of the \textit{Gaia} sources, we could find 21 additional cross-identifications, of which one has $pm=90$~mas~yr$^{-1}$ and the others have $pm=3$$-$20~mas~yr$^{-1}$, the latter ones being towards the Galactic plane and bulge and mostly in blends. In total, there are thus 1444 sources with PS1 photometry. Using the proper motions of the \textit{Gaia} sources, we propagated their coordinates at epoch 2016.0 to the epochs of the PS1 crossmatches, obtaining same-epoch angular separations of $0.10\pm0.11$~arcsec in average and smaller than 0.89~arcsec. Large separations appear typically for large proper motion sources or binary sources that are not resolved in PS1. We list the angular separations for sources of interest in Tables~\ref{tab:opt_ir} and \ref{tab:opt_ir2} (column `Sep'). We did not update to PS1 DR2, because DR1 photometry is included in DR2 and 84 DR1 sources, mostly with incomplete bandpass coverage, are missing in DR2.

Cross-identifications with SDSS DR13 \citep{2017ApJS..233...25A} are available for 150 of the \textit{Gaia} sources. The angular separations provided in the \textit{Gaia} DR3 -- SDSS external crossmatch table can easily be reproduced by propagating the coordinates to the epochs of the SDSS crossmatches. The angular separations are of $0.11\pm0.08$~arcsec in average and smaller than 0.61~arcsec. Using Aladin, we verified that all of the sources with $pm>100$~mas~yr$^{-1}$ and in the SDSS catalogue footprint are already cross-identified. We list the angular distances for sources of interest in Tables~\ref{tab:opt_sdss_nsc} and \ref{tab:opt_sdss_nsc2}. For sources having ultraviolet $u$-band fluxes, we also derived the angular separations at $u$-band using the coordinates $(\texttt{ra}+\texttt{offsetra\_u}/3600/\cos_\mathrm{deg}(\texttt{dec}), \texttt{dec}+\texttt{offsetdec\_u}/3600)$, where \texttt{offsetra\_u} and \texttt{offsetdec\_u} are the proper offsets in arcseconds provided in the SDSS catalogue.

For NSC DR2 \citep{2021AJ....161..192N}, we crossmatched using a search radius dependent of the proper motion: $(pm \times dyr)/1000 + \delta$~arcsec, where (i) $dyr=4$~yr is the maximum epoch difference, given that the NSC observation epochs are within 2012.0--2020.0~yr, and (ii) $\delta=1.5$~arcsec is the minimum search radius. For the resulting crossmatches, we then propagated the \textit{Gaia} coordinates to the individual NSC epochs and we recomputed the angular separations. As for all the other crossmatches in this study, we omit to formalize the negligible propagated uncertainties caused by the very small coordinate- (0.11--3.76~mas) and proper motion errors (0.23--4.38~mas~yr$^{-1}$) of the \textit{Gaia} sources; these are absorbed in the large search radii. Similarly as for PS1, 27 large proper motion sources have different trailing point measurements in NSC, and we selected the best (single or multiple) epochs of (complementing) widest bandpass coverage, whose \textit{Gaia}--NSC angular separations are not necessarily the smallest but are smaller than 0.17~arcsec. We obtained matches for 2102 \textit{Gaia} sources, with angular separations of $0.16\pm0.16$~arcsec in average and smaller than 1.42~arcsec. The largest separations correspond to multiple sources that are not resolved in the seeing-dependent data of the NSC object catalogue and may correlate with larger NSC values of semimajor axis radius, full width at half-maximum, and non-stellar classification. We list the angular distances for sources of interest in Tables~\ref{tab:opt_sdss_nsc} and \ref{tab:opt_sdss_nsc2}. At declinations $\delta>-29\degr$, the NSC DR2 photometry for the $grizY$ bands are calibrated on the PS1 photometry. At $\delta<-29\degr$, $griz$ band photometry are calibrated on ATLAS-Refcat2 \citep{2018ApJ...867..105T}. $u$-band photometry at $\delta<0\degr$ are calibrated on Skymapper DR1 \citep{2018PASA...35...10W}. Finally, $u$ and $Y$-band data are calibrated on model magnitudes at $\delta>0\degr$ and $<$$-29\degr$, respectively (see \citealt{2021AJ....161..192N} for details). For the SEDs, we assumed the same bandpass parameters as for SDSS $u$- and PS1 $grizy$ bands, for simplicity.

In the $JHK_\mathrm{s}$-band near-infrared catalogue of 2MASS, only one \textit{Gaia} source has a cross-identification, and it corresponds to a brighter and unrelated source at 1.75~arcsec. We considered thus deeper near-infrared catalogues. Similarly as for the crossmatch with NSC, we crossmatched using $\delta=1.5$~arcsec, with the Visible and Infrared Survey Telescope for Astronomy (VISTA) Hemisphere Survey (VHS) DR6 \citep[$dy=8$~yr;][]{2013Msngr.154...35M,2015A&A...575A..25S}, United Kingdom Infrared Telescope (UKIRT) Hemisphere Survey (UHS) DR1 \citep[$dy=3.621$~yr;][]{2018MNRAS.473.5113D}, UKIRT Infrared Deep Sky Survey (UKIDSS) DR11PLUS Large Area Survey (LAS), Galactic Cluster Survey (GCS), and Galactic Plane Survey \citep[GPS; $dy=10.671$~yr;][]{2007MNRAS.379.1599L}, VISTA Magellanic Survey (VMC) DR6 \citep[$dy=6.17$~yr;][]{2011A&A...527A.116C}, VISTA Variable in the Via Lactea Survey (VVV) DR5 \citep[$dy=6$~yr;][]{2010NewA...15..433M}, and VISTA Kilo-degree Infrared Galaxy (VIKING) Survey DR5 \citep[$dy=6.14$~yr;][]{2013Msngr.154...32E} catalogues. Among the UHS-, VVV, and in particular the VHS matches, there were multiple matches from different observing epochs and at close angular separations, also for very small proper-motion sources, and we applied a complex selection accounting for magnitude errors, error flags, multiband coverage, and angular separation. We obtain counterpart numbers of 1170 (VHS), 284 (UHS), 23 (LAS), 24 (GCS), 219 (GPS), 76 (VMC), 228 (VVV), and 15 (VIKING), implying $ZYJHK_\mathrm{s}K$-band photometry for 1957 \textit{Gaia} sources. From the VISTA and UKIRT catalogues, we used the $AperMag3$ magnitude, which is the default point source aperture corrected magnitude (2.0~arcsec diameter). The VISTA filters are similar to the UKIDSS Wide-Field Camera (WFCAM) filters, though the $K$ filter is shorter ($K_\mathrm{s}$); overall, magnitude differences are typically smaller than 0.10~mag \citep{2018MNRAS.474.5459G}. In Table~\ref{tab:opt_ir}, we list the $JHK_\mathrm{s}$-band magnitudes next to the PS1 $y$-band magnitudes, and we indicate $ZY$-band magnitudes in the footnotes. 

Also, similarly as for the crossmatch with NSC, we crossmatched using $\delta=1.75$~arcsec and $dy=6$~yr with the all-sky CatWISE2020 catalog, which is based on \textit{WISE} and \textit{NEOWISE} survey data at 3.4 and 4.6~$\micron$ ($W1$ and $W2$) obtained within 2010.0--2018.0~yr, and which has a 90 per cent completeness depth at $W1=17.7$ and $W2=17.5$~mag \citep{2021ApJS..253....8M}. There are 276 counterparts. We list the mid-infrared magnitudes for sources of interest in Tables~\ref{tab:opt_ir} and \ref{tab:opt_ir2}. Most of the \textit{Gaia} sources of the colour-selected subsample are not visible in unWISE images, because these are either too faint or unresolved in blends with brighter sources.

\subsection{Gamma ray crossmatches}\label{sec:gamma}

We crossmatched the J2016.0 coordinates of the \textit{Gaia} sources with the coordinates of the 6659 point and 78 extended sources of the {\sl Fermi} Large Area Telescope Fourth Source Catalogue \citep[4FGL-DR3;][]{2022ApJS..260...53A}, and within the 4FGL confidence and model positional error ellipses in the sky, respectively.

Within the 68 per cent confidence ellipses, there are five point-source matches, which we flag as $\gamma$\_a, $\gamma$\_b, $\gamma$\_c, $\gamma$\_d, and $\gamma$\_e. These have the following (\textit{Gaia} DR3,~4FGL) identifications, centroid separations and ellipse semimajor axes (in arcminutes), and total numbers of \textit{Gaia} sources in the ellipses:\\
- 4043369078205765888, J1758.5$-$3219; 7.48, 18.49; 203623,\\- 4040994030072768896, J1743.9$-$3539; 9.28, 21.11; 211462,\\- 5990998826943947520, J1606.2$-$4602; 18.49, 25.46; 78735,\\- 4144738037302227072, J1750.4$-$1721; 3.13, 3.85; 10530,\\- 2957941232274652032, J0524.4$-$2413; 17.07, 18.57; 6676.\\
The likelihood that these are true counterparts is of 4.91, 4.73, 12.7, 95.0, and 150 parts per million, respectively, that is extremely small. Nevertheless, we account for them exploratively and as reminders of the gamma-ray search perspective. Their \textit{Gaia} parameters are listed in Table~\ref{tab:all1}. At a later stage, we crossmatched within the 95 per cent confidence ellipses, because even the brightest 4FGL sources can have true counterparts that are beyond the 68 per cent confidence ellipses. There are 14 additional matches at 0.04--0.25~deg separation of point sources of 95 per cent confidence semimajor axes of 0.04--0.39~deg; to limit the scope of the paper, we did not study them further. Finally, there are 11 matches at 0.16--2.17~deg separation of gamma-ray extended sources of semimajor axes of 0.52--3.0~deg, which we did not consider further, because these are too extended or in crowded regions in the Magellanic Clouds. None of the above \textit{Gaia} sources has also an X-ray- (or radio) counterpart (considering the crossmatches we present in the next sections), which otherwise would have given it more weight.

\begin{figure}
   \includegraphics[width=\columnwidth]{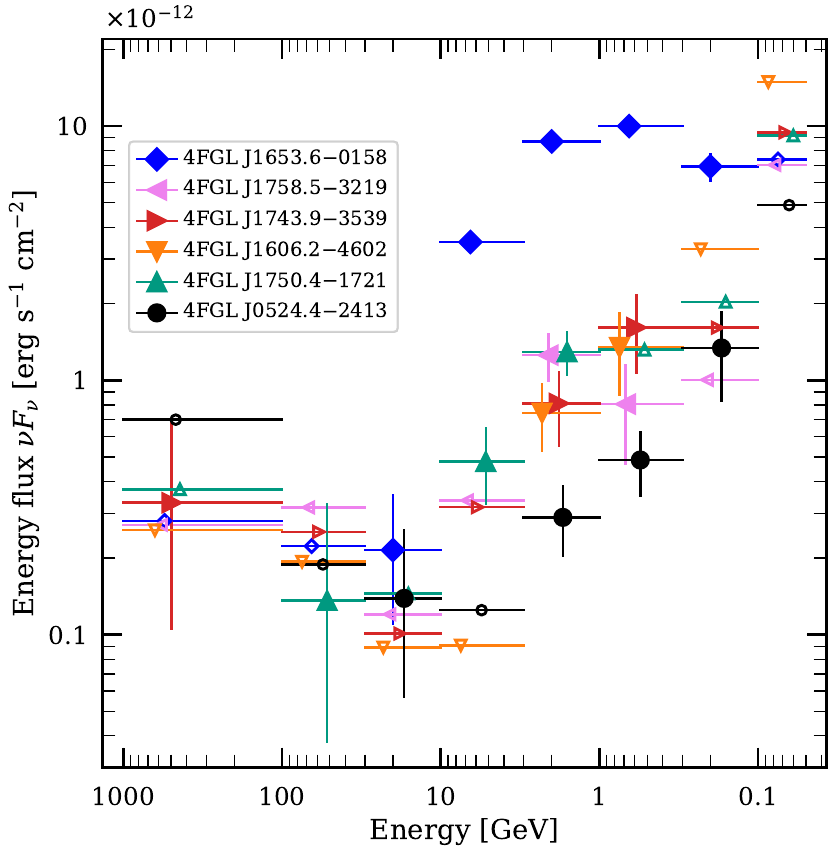}
    \caption{SEDs of the gamma-ray point-source crossmatches and the 4FGL J1653.6$-$0158 millisecond pulsar. Data points at each energy band are slightly offset horizontally to increase visibility. Smaller symbols without vertical error bars are upper limits.} \label{gamsed}
\end{figure}

\begin{table*}
	\centering
	\caption{X-ray crossmatches.}
	\label{tab:xcan}
	\footnotesize
	\begin{tabular}{l@{\hspace{1.2\tabcolsep}}r@{\hspace{1.2\tabcolsep}}r@{\hspace{1.2\tabcolsep}}r@{\hspace{1.2\tabcolsep}}r@{\hspace{1.2\tabcolsep}}r@{\hspace{1.2\tabcolsep}}r@{\hspace{1.2\tabcolsep}}r@{\hspace{1.2\tabcolsep}}r@{\hspace{1.2\tabcolsep}}r@{\hspace{1.2\tabcolsep}}r@{\hspace{1.2\tabcolsep}}r@{\hspace{1.2\tabcolsep}}r@{\hspace{1.2\tabcolsep}}r@{\hspace{1.2\tabcolsep}}r}
		\hline
Flag  &  $\mu_{\alpha^*}$ &  $\mu_{\delta}$ & $G$   &  $BP-G$ &$G-RP$  & $g-r$     & $r-i$       &  $i-z$   & P & Sep & L  & Rate  &  HR1  & HR2    \\
         &  \multicolumn{2}{c}{(mas~yr$^{-1}$)}            & (mag) &  (mag) &(mag)    & (mag)    & (mag)    & (mag)   &   & ($\prime\prime$) & (\%)  &      (cts s$^{-1}$)                &          &        \\
		\hline
P1X1   &     -0.2 &     -1.5 &  20.61 &    -0.33~$\pm$~0.08  &   0.64~$\pm$~0.17  &   -0.24 &   -0.21 &   -0.02 &   1 &  44.1 &   20.0 &   0.0290~$\pm$~0.0105 &    1.00~$\pm$~0.50 &    0.38~$\pm$~0.31 \\ 
X2     &     99.0 &   -164.9 &  20.51 &     0.19~$\pm$~0.08  &   0.56~$\pm$~0.09  &    0.63 &    0.13 &    0.13 &   2 &  47.7 &   16.7 &   0.0313~$\pm$~0.0135 &   -0.46~$\pm$~0.36 &    1.00~$\pm$~0.62 \\ 
X\_a   &      4.1 &     -1.7 &  20.81 &     0.25~$\pm$~0.13  &   0.88~$\pm$~0.10  &    0.68 &    0.41 &    0.10 &   2 &  35.0 &    1.6 &   0.3833~$\pm$~0.0504 &    0.95~$\pm$~0.04 &    0.21~$\pm$~0.09 \\ 
X\_b   &    -12.5 &     11.7 &  20.43 &     0.42~$\pm$~0.11  &   0.85~$\pm$~0.07  &    0.63 &    0.34 &    0.18 &   2 &  37.5 &    0.5 &   0.0484~$\pm$~0.0164 &    0.33~$\pm$~0.21 &   -0.04~$\pm$~0.28 \\ 
X\_c   &     -6.2 &      3.7 &  20.28 &     0.04~$\pm$~0.12  &   0.91~$\pm$~0.14  &    0.51 &    0.22 &    0.11 &   2 &  42.7 &    0.3 &   0.1066~$\pm$~0.0235 &    1.00~$\pm$~0.11 &    0.38~$\pm$~0.17 \\ 
X\_d   &     -4.1 &     -3.7 &  20.72 &     0.18~$\pm$~0.15  &   0.88~$\pm$~0.08  &    0.70 &    0.38 &    0.14 &   1 &  57.3 &    5.0 &   0.0182~$\pm$~0.0082 &    1.00~$\pm$~0.41 &   -0.89~$\pm$~0.26 \\ 
X\_e   &     -8.1 &     -5.8 &  20.67 &     0.15~$\pm$~0.17  &   0.78~$\pm$~0.13  &     --  &     --  &     --  &  -- &  54.3 &    1.0 &   0.0441~$\pm$~0.0189 &    1.00~$\pm$~2.44 &   -1.00~$\pm$~0.83 \\ 
              \hline
	\end{tabular}
\scriptsize
\vspace{-0.23cm}
\begin{flushleft}
$\mu_{\alpha^*}$ stands for $\mu_{\alpha} \cos \delta$. Sep is the separation (arcsec) at epoch 1991 between the \textit{Gaia} and 2RXS centroids. $L$ is a counterpart likelihood, as the inverse of the number of \textit{Gaia} sources at epoch 1991 within Sep+0.05~arcsec of the X-ray source. The $griz$-band photometric (P) colours are from PS1 (1) and NSC (2). Proper motion component errors are of 0.5--2.3 mas~yr$^{-1}$. From top to bottom, the names of the X-ray sources are: (1) 2RXS J003745.3+440747, (2) 2RXS J210253.8$-$583920, (3) 2RXS J171535.3$-$545013, (4) 2RXS J165504.0$-$304049, (5) 2RXS J165739.0$-$294945, (6) 2RXS J000819.5+423453, and (7) 2RXS J171902.3$-$603235. The 2RXS hardness ratios HR are defined as HR = (hard - soft)/(hard+ soft) with HR1~=~$(H - S)/(H+S)$ and HR2~=~$(V - M)/(V+M)$, where the $S$ (soft), $H$ (hard), $M$ (medium), and $V$ (very hard) bands correspond to 0.1--0.4, 0.5--2.0, 0.5--0.9, and 0.9--2.0~keV.
\end{flushleft}
\end{table*}

The five \textit{Gaia} sources of $\gamma$\_a--$\gamma$\_e are in crowded stellar fields; the first four are at $(|l|<11\degr,~|b|<5\degr)$, towards the Galactic Centre, and the fifth ($\gamma$\_e) is at $(l,b)=(227.1986,-29.3487)\degr$, at 1.6~arcmin of the centre of the M~79 globular cluster (NGC 1904), of 4.8~arcmin radius. The average detection significance over the 100~MeV--1~TeV energy range (\texttt{Signif\_Avg}) of $\gamma$\_a--$\gamma$\_e is of 3.9--7.2~$\sigma$ and their spectrum type is log-parabola, except for 4FGL~J0524.4$-$2413 that has a power-law spectrum type. In Fig.~\ref{gamsed}, we present their SEDs as a function of energy. Pulsar-type sources are identified in 4FGL as having curved- (log-parabola) or significantly curved (subexponentially cutoff power law) spectral shapes \citep{2022ApJS..260...53A}. In Fig.~\ref{gamsed}, we represent also the 4FGL J1653.6$-$0158 black-widow millisecond pulsar, which has the typical gamma-ray pulsar properties of a time-stable photon flux and a highly curved spectrum \citep{2020ApJ...901..156N,2020ApJ...902L..46N}. Two Fermi point sources have associations in the catalogue. 4FGL~J0524.4$-$2413 has a high probability association with the above mentioned M~79 globular cluster, because the cluster is in the 68 and 95 per cent confidence ellipses (the gamma-ray centroid and cluster centre are separated by 17.1~arcmin). 4FGL~J1606.2$-$4602 has a low confidence association with the ECC~G334.57+04.63 source of unknown type. The five \textit{Gaia} sources have red colours suggesting apparent temperatures of $\sim$3000--5500~K (see also the SEDs in Fig.~\ref{fsed_gamma_ir}, available as online supplementary material).

\subsection{X ray crossmatches}\label{sec:xrays}

At X-rays, we considered the second RASS source catalogue (2XRS; $2\times10^{-13}$~erg~s$^{-1}$~cm$^{-2}$; \citealt{2016A&A...588A.103B}), the \textit{XMM-Newton} catalogues of serendipitous sources from observations (4\textit{XMM} DR11; $1\times10^{-15}$~erg~s$^{-1}$~cm$^{-2}$; \citealt{2020A&A...641A.136W}) and stacked observations (4\textit{XMM} DR11s; \citealt{2020A&A...641A.137T}), the \textit{XMM-Newton} Slew survey catalogue (XMMSL2; $3\times10^{-12}$~erg~s$^{-1}$~cm$^{-2}$; \citealt{2008A&A...480..611S}), the \textit{Chandra} Source Catalogue \citep{2010ApJS..189...37E} Release 2.0 (CSC2.0; $1\times10^{-15}$~erg~s$^{-1}$~cm$^{-2}$), and the \textit{eROSITA} Final Equatorial Depth Survey (eFEDS) X-ray catalogue ($7\times10^{-15}$~erg~s$^{-1}$~cm$^{-2}$; \citealt{2022A&A...661A...1B,2022A&A...661A...3S}), where the sensitivities are in the 0.5--2.0 keV energy range and adopted from fig.~14 of \citet{2022A&A...661A...1B}.

For convenience, we crossmatched the \textit{Gaia} and X-ray sources in two steps. In the first step, we used a constant crossmatch radius, defined by the template expression $(pm_\mathrm{max}/1000)\,dyr + n\,\sigma_\mathrm{pos, rl} + \delta$~arcsec, where $pm_\mathrm{max}$ is the maximum proper motion in our \textit{Gaia} subsample, $dyr$ is the absolute value of the epoch difference between \textit{Gaia} DR3 and the farthest or representative epoch of the counterpart catalogue, $\sigma_\mathrm{pos, rl}$ is the maximum- or representatively large positional uncertainty in the counterpart catalogue, $n$ is an integer factor of $\sigma$, and $\delta$ is to allow for any unaccounted systematic- or source-centroiding uncertainty. In the second step, we required the crossmatches to have angular separations smaller than $(pm/1000)\,dyr+n\,\sigma_\mathrm{pos}+\delta$~arcsec, using the individual measurements.

For the \textit{Chandra} CSC2.0 catalogue, we used $dyr=16.25$~yr, $\sigma_\mathrm{pos, rl}$ as the maximum of the major radius of the 95 per cent confidence-level positional error ellipse, $n=2$, and $\delta=2$~arcsec. In the first step, 56 matches are found. In the second step, one remains. Gaia DR3 4108419965650416128 ($pm=4.8$~mas~yr$^{-1}$) is in a very dense field and at 64~arcsec of the faint ($SNR_\mathrm{X}=2.4$) 2CXO~J171032.0$-$281942 ($\sigma_\mathrm{pos}=43$~arcsec), whose centroid is at 4~arcsec of the optically bright CD$-$28~12887 star ($G=10.1$~mag). Also, within 65~arcsec of the X-ray source there are 916 \textit{Gaia} sources, implying a very small counterpart likelihood of $L=0.0011$.

For the \textit{XMM} catalogues, we used $dyr=|2000-2016|=16$~yr, $\sigma_\mathrm{pos, rl}$ as the maximum of the positional 1-sigma uncertainty in the respective \textit{XMM} catalogues, $n=3$, and $\delta=2$~arcsec. We find four matches, which are in dense stellar fields. We propagated the coordinates of the \textit{Gaia} sources using their proper motions ($pm<15$~mas~yr$^{-1}$) to the mean epochs of the respective XMM matches. Gaia DR3 4685846291048239744 is at $3.58 \times \sigma_\mathrm{pos}$ of 4XMM J004425.7$-$732635 ($\sigma_\mathrm{pos}=1.43$~arcsec) of low detection significance ($SC\_DET\_MIL=9.3$), which has seven \textit{Gaia} sources within $3.58 \times \sigma_\mathrm{pos}$. Gaia DR3 4124307015366667776 is at $3.03 \times \sigma_\mathrm{pos}$ of 4XMM J173204.2$-$170613 ($\sigma_\mathrm{pos}=2.83$~arcsec), which has 15 \textit{Gaia} sources within $3.03 \times \sigma_\mathrm{pos}$. Gaia DR3 6029397037233478400 is at $1.04 \times \sigma_\mathrm{pos}$ of 4XMM J170238.7$-$295445 ($\sigma_\mathrm{pos}=17.70$~arcsec), which has 79 \textit{Gaia} sources within $1.04 \times \sigma_\mathrm{pos}$ and has an extent of 37.20$\pm$4.71~arcsec. Finally, Gaia DR3 4124500048336584064 is at $1.41 \times \sigma_\mathrm{pos}$ of XMMSL2 J173139.7$-$165636 ($\sigma_\mathrm{pos}=10.69$~arcsec) of low detection significance ($DET\_ML\_B8=9.2$), which has 27 \textit{Gaia} sources within $1.41 \times \sigma_\mathrm{pos}$. The four matches have very small counterpart likelihoods of 0.143, 0.067, 0.013 and 0.037, respectively.

We proceeded similarly with the \textit{ROSAT} 2XRS catalogue, with $dyr=25.5$~yr, $n=3$, and $\delta=3$~arcsec. We adopt $\sigma_\mathrm{pos, rl}=\sigma_\mathrm{pos}$ as a representatively large value of 20~arcsec, based on the comparison of 2RXS- with SDSS positions \citep{2016A&A...588A.103B} and because there are no formal 2XRS positional errors available. In the first step, there are 26 matches. In the second step, seven remain, which are listed in Table~\ref{tab:xcan}. Also listed are (i) the angular separation (Sep) at epoch 1991 and (ii) the counterpart likelihood (L), as the inverse of the number of \textit{Gaia} sources at epoch 1991 within Sep+0.05~arcsec of the X-ray source.

\begin{table*}
	\centering
	\caption{Ultraviolet crossmatches.}
	\label{tab:uvcan}
       \footnotesize
	\begin{tabular}{l@{\hspace{1.2\tabcolsep}}r@{\hspace{1.2\tabcolsep}}r@{\hspace{1.2\tabcolsep}}r@{\hspace{1.2\tabcolsep}}r@{\hspace{1.2\tabcolsep}}r@{\hspace{1.2\tabcolsep}}r@{\hspace{1.2\tabcolsep}}r@{\hspace{1.2\tabcolsep}}r@{\hspace{1.2\tabcolsep}}r@{\hspace{1.2\tabcolsep}}r@{\hspace{1.2\tabcolsep}}l@{\hspace{1.2\tabcolsep}}r@{\hspace{1.2\tabcolsep}}r@{\hspace{1.2\tabcolsep}}r}
		\hline
Flag  &  $\mu_{\alpha^*}$ &  $\mu_{\delta}$ & $G$   &  $BP-G$ &$G-RP$  & $g-r$     & $r-i$       &  $i-z$   & P & Sep & L  & $FUV$$-$$NUV$      &  $NUV-G$    & $u-G$          \\
      &  \multicolumn{2}{c}{(mas~yr$^{-1}$)}            & (mag) &  (mag) &(mag)    & (mag)    & (mag)    & (mag)   & &  ($\prime\prime$)     &    & (mag)                         &  (mag)                  & (mag)          \\
		\hline
UV1    &    -15.8 &     -3.9 &  20.89 &   -0.35~$\pm$~0.13  &    0.35~$\pm$~0.15  &   -0.37 &   -0.25 &   -0.27 &   2 &   0.3 &  1  &    0.21~$\pm$~0.41  &   -0.31~$\pm$~0.23  &     --            \\ 
UV2    &     -1.4 &     -7.1 &  20.17 &   -0.25~$\pm$~0.08  &    0.71~$\pm$~0.10  &    0.26 &    0.11 &    0.02 &   1 &   1.6 &  1* &     --              &    1.98~$\pm$~0.34  &     --            \\ 
UV3    &     -1.5 &      2.6 &  20.89 &   -0.19~$\pm$~0.17  &    0.80~$\pm$~0.15  &   -0.19 &     --  &     --  &   2 &   2.2 &  1  &   -0.34~$\pm$~0.25  &   -0.66~$\pm$~0.15  &     --            \\ 
UV4    &      1.6 &     -9.0 &  20.23 &   -0.10~$\pm$~0.06  &    0.83~$\pm$~0.06  &   -0.13 &    0.34 &    0.22 &   1 &   0.7 &  1* &   -0.47~$\pm$~0.11  &   -0.78~$\pm$~0.06  &     --            \\ 
UV5    &     -2.6 &     -2.3 &  20.90 &   -0.05~$\pm$~0.16  &    0.47~$\pm$~0.26  &    0.30 &     --  &     --  &   2 &   3.0 &  0.5&     --              &    1.82~$\pm$~0.40  &     --            \\ 
UV6    &     -0.7 &    -13.4 &  20.12 &   -0.04~$\pm$~0.05  &    0.79~$\pm$~0.05  &    0.33 &    0.11 &    0.05 &   1 &   1.7 &  1  &     --              &    1.89~$\pm$~0.36  &     --            \\ 
UV7    &    -10.0 &      3.1 &  20.56 &   -0.03~$\pm$~0.10  &    0.81~$\pm$~0.10  &    0.08 &    0.22 &    0.00 &   2 &   1.9 &  1  &     --              &    1.34~$\pm$~0.35  &     --            \\ 
UV8    &     -5.9 &     14.7 &  20.86 &    0.15~$\pm$~0.26  &    0.88~$\pm$~0.15  &    0.47 &    0.88 &    0.52 &   1 &   1.9 &  1  &     --              &    0.93~$\pm$~0.35  &    0.11~$\pm$~0.13  \\ 
UV9    &      2.3 &      7.8 &  20.89 &    0.26~$\pm$~0.15  &    0.85~$\pm$~0.13  &    0.40 &    0.67 &    0.55 &   1 &   0.9 &  1  &     --              &    1.60~$\pm$~0.28  &    0.42~$\pm$~0.16  \\ 
UV10   &     -6.3 &     -9.5 &  20.89 &    0.26~$\pm$~0.17  &   -0.12~$\pm$~0.30  &   -0.08 &   -0.23 &    0.04 &   1 &   0.6 &  1  &   -0.41~$\pm$~0.37  &    1.07~$\pm$~0.26  &     --            \\ 
UV11   &    -11.7 &    -12.9 &  20.86 &    0.35~$\pm$~0.28  &    0.55~$\pm$~0.22  &    0.51 &    0.20 &    0.17 &   1 &   1.5 &  0.5&     --              &    1.10~$\pm$~0.32  &    1.96~$\pm$~0.55  \\ 
UV12   &     -4.2 &    -14.6 &  20.70 &    0.36~$\pm$~0.14  &    0.43~$\pm$~0.19  &    0.33 &    0.17 &    0.08 &   1 &   1.5 &  1* &     --              &    1.73~$\pm$~0.43  &    1.06~$\pm$~0.19  \\ 
UV13   &    210.3 &    149.6 &  20.19 &    0.48~$\pm$~0.09  &    0.53~$\pm$~0.07  &    0.99 &    0.24 &    0.14 &   1 &   0.1 &  1  &      --             &     --              &   -0.31~$\pm$~0.10  \\ 
UV14   &    195.0 &     40.2 &  20.51 &    0.49~$\pm$~0.15  &    0.81~$\pm$~0.08  &    0.89 &    0.39 &    0.13 &   1 &   0.0 &  1  &      --             &     --              &    1.17~$\pm$~0.30  \\ 
UV15   &    100.2 &   -158.2 &  20.75 &    0.58~$\pm$~0.16  &    0.69~$\pm$~0.11  &    0.73 &    0.09 &    0.24 &   1 &   0.6 &  1  &      --             &     --              &    0.40~$\pm$~0.19  \\ 
              \hline
	\end{tabular}
\scriptsize
\vspace{-0.23cm}
\begin{flushleft}
Proper motion component errors are of 0.5--2.7 mas~yr$^{-1}$. The $griz$-band photometric (P) colours are from PS1 (1) and NSC (2). Sep is the separation (arcsec) between the \textit{Gaia} and (i) \textit{GALEX} centroids at the \textit{GALEX} epochs for UV1--12, and (ii) SDSS DR13 $u$-band centroids at the SDSS epochs for UV13--15. $L$ is the counterpart likelihood, as the inverse of the number of unique \textit{Gaia}-, PS1-, and NSC sources within $3\times\sigma_{\mathrm{pos}}$ of the ultraviolet source at its epoch. * Likelihood shared with a visual companion at 0.6~arcsec (i) with separate Gaia parameters (UV2 in Table~\ref{tab:all2}) or (ii) only partly resolved in PS1 images (UV4, UV12), and whose relative separation is indicated in column `Vis.' of Table~\ref{tab:mul}.
\end{flushleft}
\end{table*}

Gaia DR3 387812285286341504 (P1X1; $d50=81$~pc) is in a sparse field and 44.1~arcsec south of the centroid of the faint 2RXS J003745.3+440747. Considering the faintness, uncertain centroid and spread ($\leq$30~arcsec), and pixel locations of the soft and hard counts of the RASS source, the \textit{Gaia} source could be a potential match. Also, the \textit{Gaia} source has low counts in the \textit{GALEX} GR6+7 $NUV$ and $FUV$-band images. Comparing these counts with those of GALEX J003744.6+440750 ($FUV=20.1\pm0.25$ and $NUV=19.61\pm0.13$~mag) 47.4~arcsec north, it would have $NUV\sim FUV\sim20.8\pm0.4$~mag, implying $FUV-NUV\sim0$ and $NUV-G\sim0.2$~mag. Among the seven 2RXS matches, P1X1 has the smallest proper motion ($pm\sim1.5\pm2.4$~mas~yr$^{-1}$) and the bluest $G_\mathrm{BP}-G$ and PS1 colours. Its blue $G_\mathrm{BP}-G$ and very red $G-G_\mathrm{RP}=0.64\pm0.17$~mag are reflected in its positive \texttt{phot\_bp\_rp\_excess\_factor\_corrected} of 0.58 (see comments on this parameter in Appendix~\ref{app:method1}). Its SED based on the PS1 photometry indicates an almost straight profile over the $griz$ bands, with some excess at $zy$ band (Fig.~\ref{fsed_pl_can1}; see Section~\ref{sec:pl} for details about SEDs and the fitting procedure). Among all the \textit{Gaia} sources of the colour-selected subsample that have PS1 $griz$-band photometry and a spectral-index fitting error smaller than 0.1, it is the only source with a spectral index lesser than zero, $\alpha_{\nu}= -0.86\pm0.07$, and has the highest value of the renormalised unit weight error (\texttt{ruwe}) for astrometry, of 7.6 (also the third greatest \texttt{ruwe} value in the colour-selected subsample). A value of \texttt{ruwe}~$\ga1.4$ is indicative of an ill-behaved astrometric solution and may indicate binarity or multiplicity \citep{2021A&A...649A...6G}. Within 44.15~arcsec of the 2RXS centroid and at epoch 1991, there are five \textit{Gaia} sources ($G=18.20-20.61$~mag), among which P1X1 is the only one with a negative $G_\mathrm{BP-G}$ colour (it has also the bluest $g-r$ colour among the PS1 sources in that area), is the faintest ($G=20.61$~mag), and its \texttt{ruwe} value is much greater than those of the four other sources (1.0). However GALEX J003744.6+440750 is also a potential match at 8.4~arcsec of the 2RXS centroid; it is relatively brighter in the optical (Gaia DR3 387812869401900672; $G=19.15$, $G_\mathrm{BP}-G=0.14$, $G-G_\mathrm{RP}=0.29$~mag) and mid infrared, and has a 100 per cent probability of being extragalactic as a quasi stellar object (QSO), in the Northern Extragalactic \textit{WISE} $\times$~Pan-STARRS (NEWS) catalogue \citep{2020A&A...644A..69K}.

Gaia DR3 6456175155513145344 (X2) is in a sparse field and 47.7~arcsec north of the uncertain centroid of the faint 2RXS~J210253.8$-$583920 source. The \textit{Gaia} source has $G=20.51$~mag, a large proper motion of $(99,-165)\pm0.5$~mas~yr$^{-1}$, and is at $d50=62$~pc. It is in GCNS100pc and is a probable white-dwarf photometric candidate  (\citetalias{2021A&A...649A...6G}; \citealt{2021MNRAS.508.3877G}). At less than 47.75~arcsec of the 2RXS centroid and at epoch 1991, there are six \textit{Gaia} sources ($G=12.04$ and 20.21--21.00~mag). Among these, X2 has the least red $G_\mathrm{BP}-G_\mathrm{RP}$ colour, of 0.75~mag. In the Dark Energy Survey (DES) DR1 catalogue \citep{2018ApJS..239...18A}, X2 corresponds to DES~J210255.98$-$583838.3, with colours of $g-r=0.79$, $r-i=0.23$, and $i-z=0.07$~mag suggesting a cool atmosphere. Within 5.5~arcsec, it has five DES neighbours and it is the second brightest. Its large proper motion can easily be seen in image tiles from 2013 to 2018, queried at the NOIRLab Astro Data Lab\footnote{\url{https://datalab.noirlab.edu}}. (We note that within 47.75~arcsec of the 2RXS centroid, there are 55 NSC sources, which would decrease the counterpart likelihood). The brightest \textit{Gaia} source is Gaia DR3 6456175048138766720 ($G=12.04$~mag, 25.8~mas~yr$^{-1}$, 350~pc; $NUV=17.84\pm0.04$~mag), at 29~arcsec of the 2RXS centroid. It could also be a potential match, because it is an EW--W~Ursae~Majoris-type eclipsing variable \citep{2012ApJS..203...32R}, a contact binary of orbital period of 0.3456~d \citep{2017MNRAS.469.3688D}.

For the possible P1X1 and X2 candidates, we converted the 2RXS count rates to fluxes, adopting a conversion factor of $1.08\times10^{-11}$ erg~s$^{-1}$~cm$^{-2}$ for an intrinsic hydrogen column density $N_H=0$ and a power-law emission of photon index $\Gamma=1$, and we represent them by cyan circles in Figs.~\ref{xg_vs_mg} and \ref{x_vs_g}.

The five remaining 2RXS matches are located in dense fields close to the plane or centre of the Galaxy, with very small counterpart likelihoods of 0.003--0.05. The relatively bright 2RXS J171535.3$-$545013 (X\_a) is associated, as an AGN candidate, with 2MASS 17153752$-$5450062 at 20~arcsec and the 2SXPS J171537.7$-$545007 \textit{Swift} X-ray point source \citep{2012ApJ...751...52E,2022A&A...657A.138T}. Similarly, the relatively bright 2RXS J165739.0$-$294945 (X\_c) can be associated with the 2SXPS J165739.6$-$294926 AGN candidate \citep{2022A&A...657A.138T}, at 20~arcsec and in the opposite direction to the X\_c \textit{Gaia} source.
 
For the eFEDS catalogue, we used $dyr=4$~yr, $\sigma_\mathrm{pos, rl}$ as the maximal positional uncertainty, $n=3$, and $\delta=3$~arcsec. We find one match. Gaia DR3 5763109404082525696 has a proper motion of ($-$114,~86)~mas~yr$^{-1}$, $G/G_\mathrm{BP}/G_\mathrm{RP}\approx20.4$~mag, $d50=62$~pc, and is a probable white-dwarf photometric candidate in GCNS100pc and \citet{2019MNRAS.482.4570G}. At the eFEDS epoch, it is at 28.0~arcsec of the centroid of the faint eFEDS J091001.1$-$022226 ($\sigma_\mathrm{pos}=16.3$~arcsec), of X-ray count rate of 0.044$\pm$0.020 cts/s. The crossmatch likelihood is of 0.5, because there is another \textit{Gaia} counterpart, closer at 11~arcsec and brighter in the optical ($G=17.6$~mag) and infrared (2MASS 09100156$-$0222173).

\subsection{Ultraviolet crossmatches}\label{sec:uv}

\begin{figure*}
  \begin{minipage}[c]{0.695\textwidth}
    \includegraphics[width=\textwidth]{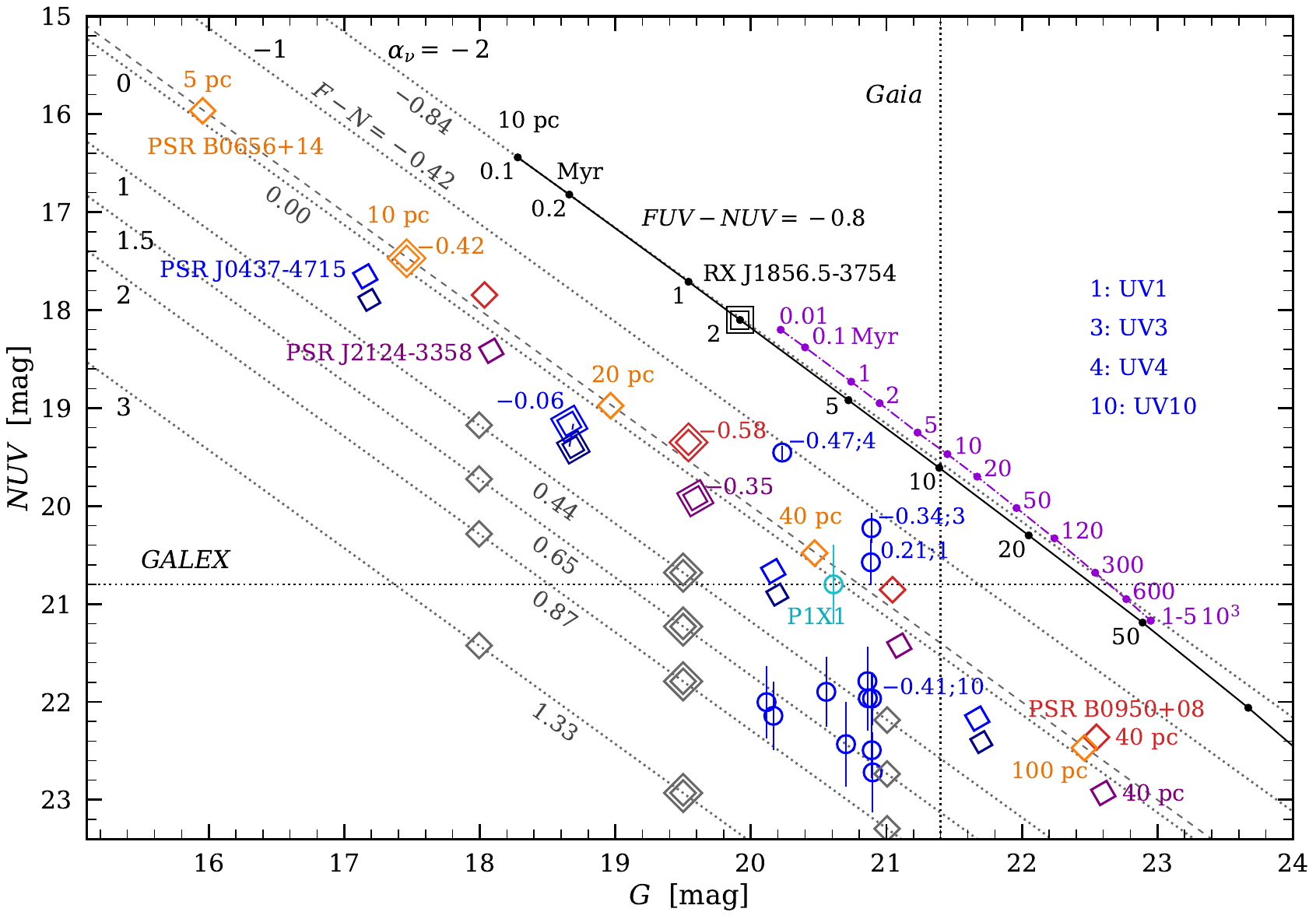}
  \end{minipage}\hfill
  \begin{minipage}[c]{0.265\textwidth}
    \caption{NUV- versus $G$-band magnitude diagram. The grey dashed line represents equal magnitudes. Synthetic $FUV-NUV$ colours of known neutron stars are indicated next to their 10~pc magnitudes. The neutron-star thermal cooling track is shown only at 10~pc, because the tracks at other distances are overlapping with it. The pulsar thermal track with rotochemical heating, also at 10~pc (violet dash--dotted line), is offset in $G$ band by +0.15~mag for clarity. The \textit{GALEX} crossmatches are indicated by blue open circles, annotated with measured $FUV-NUV$ colours. The oblique grey dotted lines are for power-law sources and are annotated with synthetic $FUV-NUV$ colours (more colour values are in Table~\ref{tab:plcolours}). Same as in Fig.~\ref{x_vs_g}.
    } \label{uv_vs_g}
  \end{minipage}
\end{figure*}

At the ultraviolet, we considered the \textit{GALEX} All-Sky Survey GUVcat\_AIS GR6+7 catalogue \citep{2017ApJS..230...24B}. We propagated the coordinates of the \textit{Gaia} sources to epoch 2008.0, midst of the \textit{GALEX} 2003--2013 epoch range. We then proceeded similarly as for the X-ray crossmatches, with $dyr=4.5$~yr, $\sigma_\mathrm{pos, rl}=3$~arcsec as the maximum positional uncertainty at $NUV$ and $FUV$ bands, $n=3$, and $\delta=3$~arcsec. The second step yields 38 matches. After visual inspection in archival images, we find 12 true or probable matches (Table~\ref{tab:uvcan}). These are all in sparse stellar fields, with typically no other \textit{Gaia} source within 10~arcsec of the \textit{GALEX} centroid (except for UV4 that has five other \textit{Gaia} sources). The 26 remaining matches were discarded, because, the ultraviolet sources (i) have brighter and closer optical sources that match their coordinates, or (ii) are too far off, given the relatively precise coordinates and proper motions.

The angular separations of the 12 matches at the respective \textit{GALEX} epochs are listed in Table~\ref{tab:uvcan}. Expressed in multiples of $\sigma_{\mathrm{pos, }NUV}$, these are of 0.4-1.0 (five sources), 1.0-1.4 (five sources), 2.4 (UV5), and 3.2 (UV3). Some ultraviolet positional scatter could be explained by the larger pixel size, detector field distortion, and centroiding at faint fluxes. The ultraviolet sources come from coadds (multiple visits), are classified as point source, and have artefact and source extraction flags of value 0 for the retrieved $FUV$ and $NUV$ measurements, except for three sources, which have non-zero artefact flags that are not among those causing real concern \citep{2017ApJS..230...24B}. Gaia DR3 6901051613248041216 (UV4) and Gaia DR3 4438623270475014656 (UV11) have both an $NUV$-band artefact flag \texttt{Nafl} of 17, that is of 1 and 16 indicating a detector bevel edge reflection and a bright star near the field edge, respectively. The first candidate is very bright at $FUV$- and $NUV$ bands and the latter candidate is only detected faint at $NUV$ band. Gaia DR3 837736394742229248 (UV9) has an $NUV$-band source-extraction flag \texttt{Nexf} $= 2$, though the source is well isolated.

In the $NUV$ versus $G$ magnitude--magnitude diagram of Fig.~\ref{uv_vs_g}, we show the 12 candidates, which have $G=20.1-20.9$~mag and are at $d50=60$$-$104~pc. Gaia DR3 5777941133043108992 (UV3) and UV4, are well above the equal-magnitude line (grey dashed line), and their location and blue $FUV-NUV$ colours are consistent with power law profiles of $\alpha_{\nu}\approx-1$. The nine sources well below the equal-magnitude line are located at  $NUV-G$ colours of power-law profiles of $\alpha_{\nu}=0.5$$-$2.0. Their profiles can be verified with fluxes at narrower bands than the $G$ band (and with the $FUV$ band flux in the case of UV10).

\begin{figure*}
    \includegraphics[width=\columnwidth]{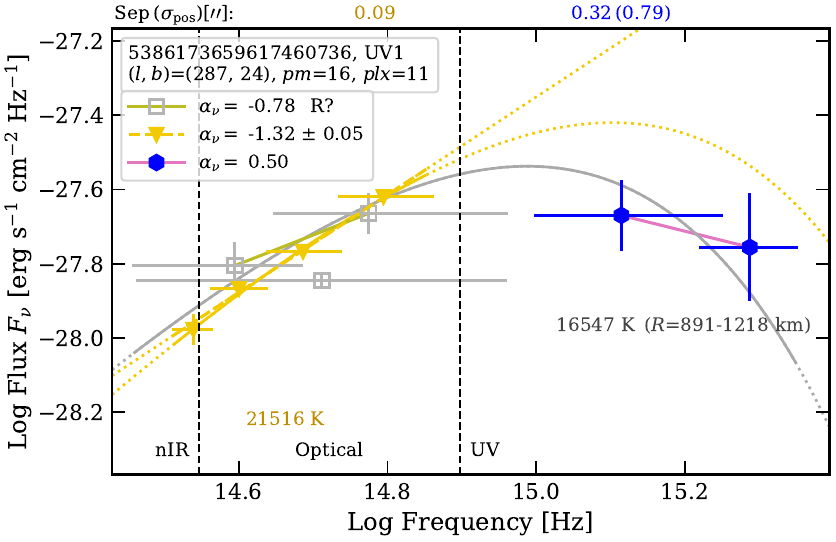}
    \includegraphics[width=\columnwidth]{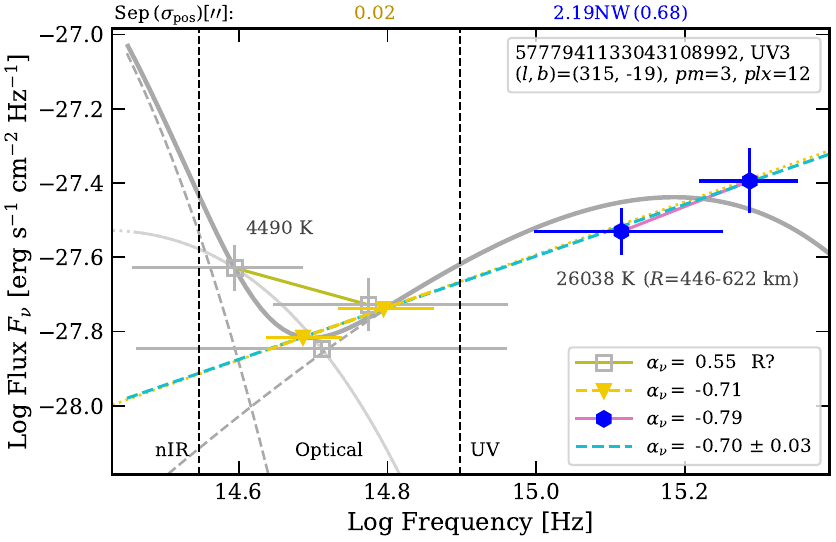}
    \includegraphics[width=\columnwidth]{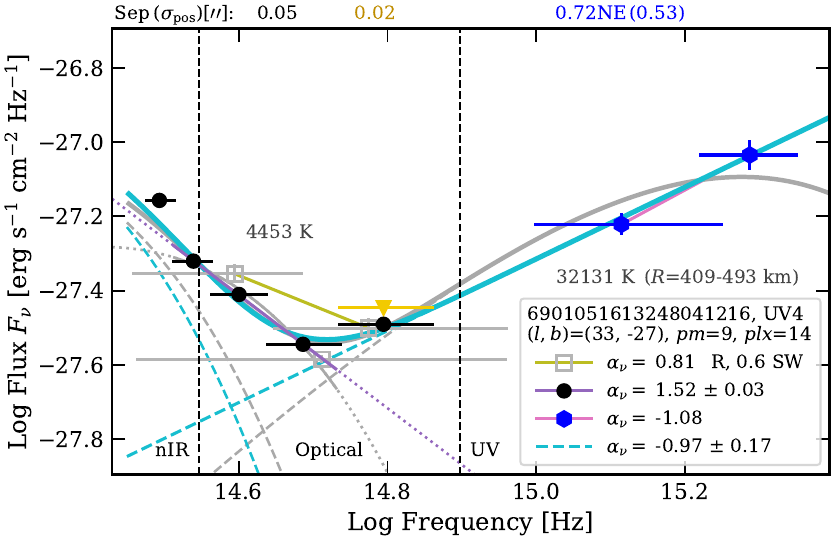}
    \includegraphics[width=\columnwidth]{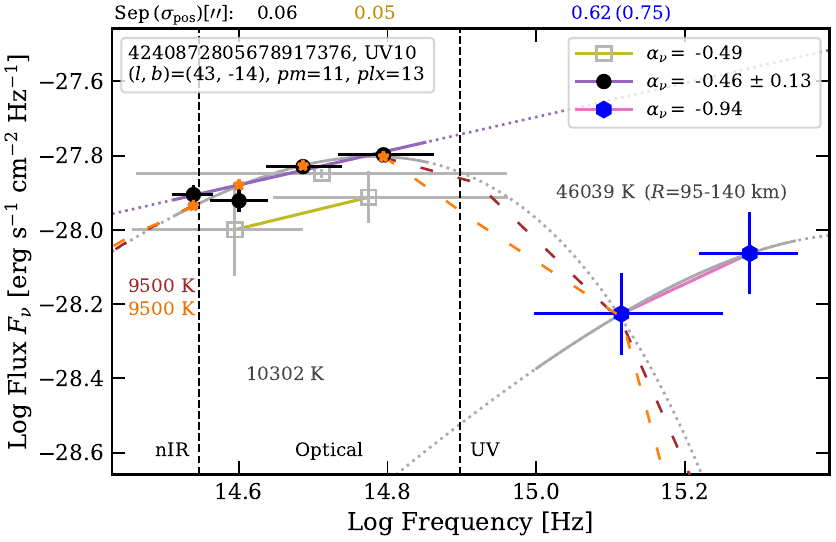}
    \caption{SEDs of the four candidates with {\it GALEX} $FUV$ and $NUV$ fluxes, UV1, UV3, UV4, and UV10. The \textit{Gaia} DR3 source ID, galactic coordinates in deg, proper motion in mas~yr$^{-1}$, parallax in mas, and photoastrometric- and visual-companion binary indicators (columns `Ast.' and `Vis.' in Table~\ref{tab:mul}) are indicated in the legend boxes. The $\Delta \log F_{\nu}$ ordinate span is of 1.2~dex, else if it is greater, it is explicited. In general, also for the next SED figures, \textit{Gaia}, PS1, SDSS, NSC, {\it GALEX}, and additional infrared fluxes are represented by grey squares, black dots, blue upward triangles, yellow downward triangles, blue hexagons, and red open circles, respectively. Above the top axis, we indicate the angular separations in arcsec between the counterpart and the \textit{Gaia} source at the mean observation epoch of the counterpart; at mid-infrared and ultraviolet, we also indicate the positional uncertainty ($\sigma_{\mathrm{pos}}$) in parenthesis, and the cardinal direction if the separation is larger than 3~$\sigma_{\mathrm{pos}}$ and 0.75~arcsec, at the mid-infrared, and larger than 1~$\sigma_{\mathrm{pos}}$ at the ultraviolet. Observed fluxes are fitted by power laws (coloured segments), blackbodies (curves), and theoretical He and H atmosphere white-dwarf grid points (brown and orange small stars and dashed lines). Effective temperatures indicated in the optical range are for the single blackbody fits (thin solid curves) `passing through' the optical fluxes. Example two-blackbody and blackbody+power-law fits are represented by grey and cyan thick lines, respectively (dashed lines for individual components and solid lines for their sums); in these cases, only the effective temperature of the hot blackbody component is indicated in the ultraviolet range. Thermal radii derived from the ultraviolet blackbody fits are indicated in kilometres and account for distance uncertainties (listed in Table~\ref{tab:mul}).}
    \label{fsed_uv1}
\end{figure*}

\begin{figure*}
    \includegraphics[width=\columnwidth]{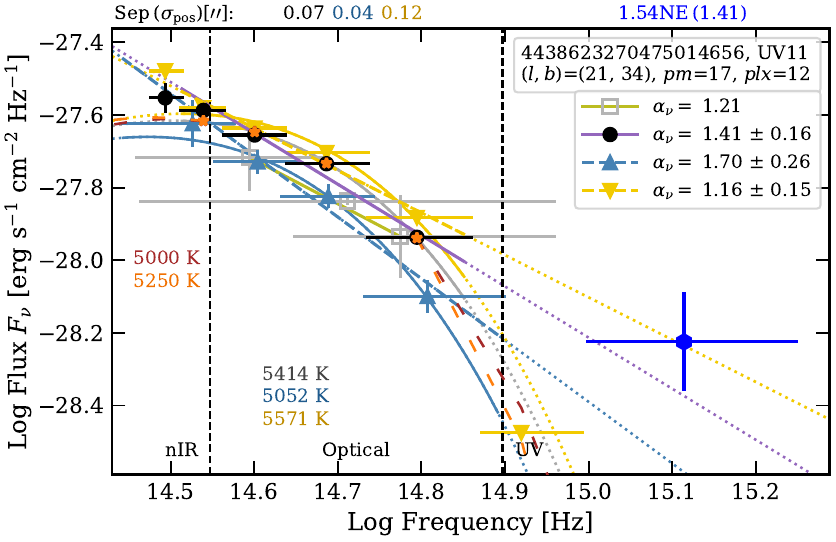}
    \includegraphics[width=\columnwidth]{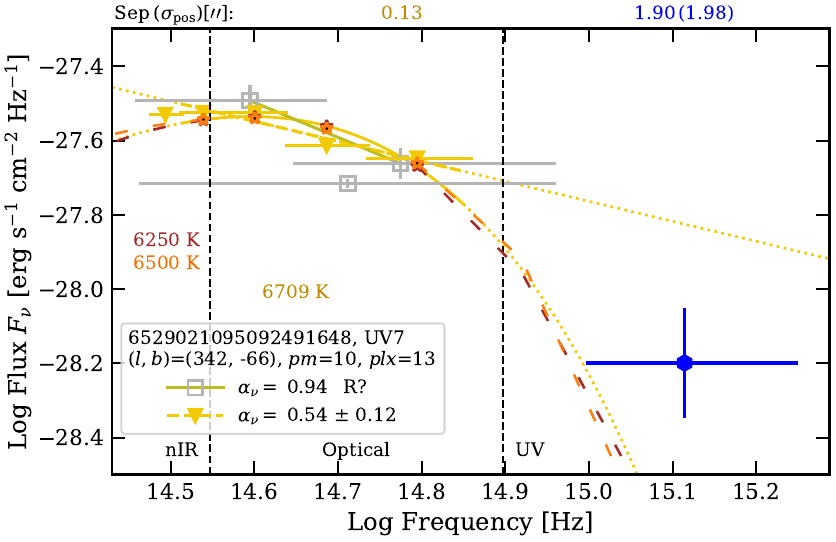}    
    \includegraphics[width=\columnwidth]{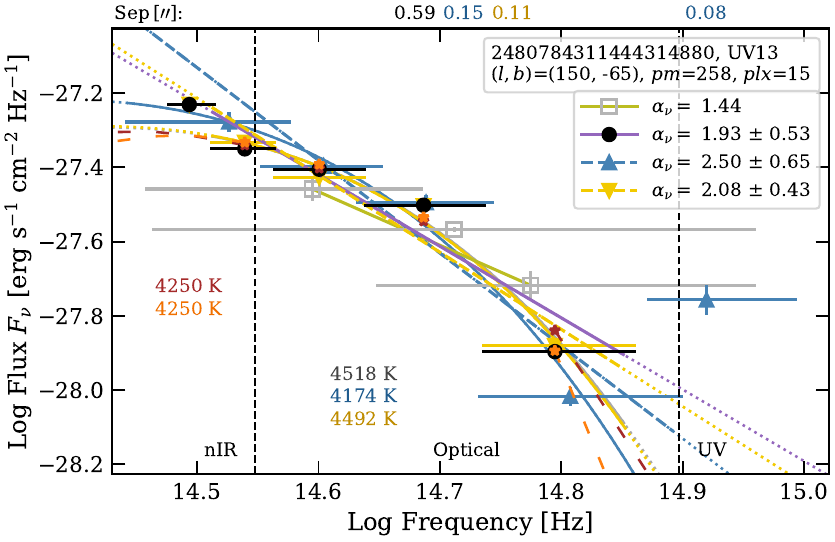}
    \includegraphics[width=\columnwidth]{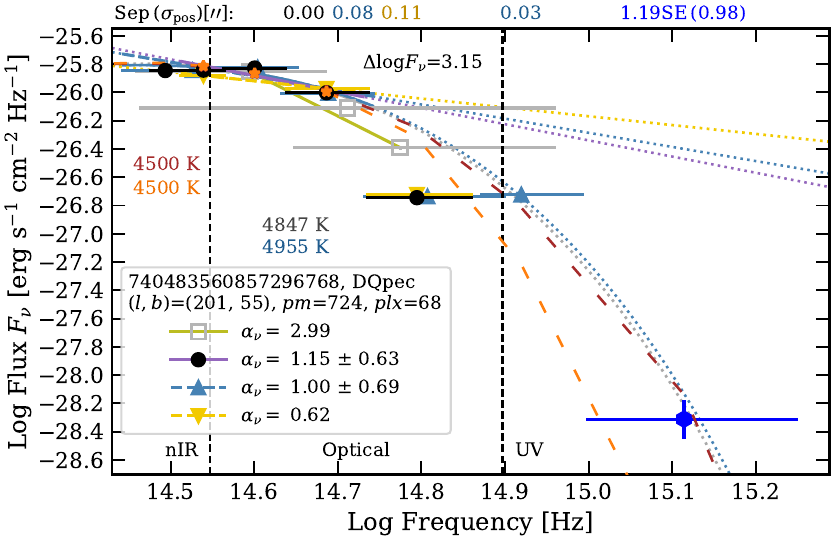}
    \caption{{\sl Upper left panel}: SED example of UV11 with significant $NUV$-band flux excess (UV8 and UV9 are shown in Fig.~\ref{fsed_uv8_uv9}). {\sl Upper right panel}: SED example of UV7 with low $NUV$-band flux excess (UV2, UV5, UV6, and UV12 are shown in Fig.~\ref{fsed_uv_ir} of the online supplementary material). {\sl Lower left panel}: SED example of UV13 with significant SDSS $u$-band flux excess (UV15 is shown in Fig.~\ref{fsed_uv_ir} of the online supplementary material). {\sl Lower right panel}: For comparison, SED of the LHS~2229 DQpec white dwarf (commented in Appendix~\ref{sec:serfind}), with a very low $NUV$-band flux. Same as in Fig.~\ref{fsed_uv1}.}
    \label{fsed_uv2}
\end{figure*}

In Fig.~\ref{fsed_uv1}, we show the SEDs of the four candidates having $FUV$ and $NUV$ fluxes, Gaia DR3 5386173659617460736 (UV1), UV3, UV4, and Gaia DR3 4240872805678917376 (UV10). These have distances $d50=101$, 92, 71, and 93~pc, respectively. None is a white dwarf photometric candidate in catalogues. The SED profile of UV1 (the source above and closest to the $NUV=G$ line in Fig.~\ref{uv_vs_g}) can be approximately fitted by a single hot blackbody, represented by the grey solid line. Alternatively, the profile could consist of (i) a component that is straight-up in the optical, peaking at $g$ band, and significantly suppressed at $u$ band, such as the profile of the Gaia~DR3~1528861748669458432 DC white dwarf (Fig.~\ref{wd_blue}), and (ii) another component  that is related to the ultraviolet emission. UV3 is well fitted at optical--ultraviolet by a power law of $\alpha_{\nu}=-0.7$; its bright infrared fluxes and red \textit{Gaia} colours indicate that it has a cool companion (the full SED is available in Fig.~\ref{fsed_uv_ir} of the online supplementary material). Similarly, UV4 is well fitted at optical--ultraviolet by a power law of $\alpha_{\nu}=-1.0\pm0.2$, summed with a cool blackbody (straight and leftmost cyan dashed lines, respectively, in Fig.~\ref{fsed_uv1}, where the cyan thick solid line is the sum). Assuming unbroken power laws, these are expected to be detected at X rays in 2RXS RASS, as indicated in Fig.~\ref{mg_vs_dlim1}, in the panels for $\alpha_{\nu}=-0.5$ and $-$1.0. However, these are not detected, possibly because the exposure times (as indicated by the 2RXS exposure map) are of only about 130 and 280~s, respectively, close to the minimum exposure time \citep{2016A&A...588A.103B}. Both sources could be binaries. UV3 has a very small proper motion, possibly related to it being binary, faint, or a QSO; it appears as a point-source in \textit{GALEX}. In the case of UV4, the cool component stems from a blended visual companion, 0.6~arcsec south-west of the \textit{Gaia} source. The ultraviolet emission is very bright, extends from the \textit{Gaia} source up to a few~arcsec in radius, and appears as a hot nebula or bubble. The blue-optical--ultraviolet fluxes of UV3 and UV4 are better fitted by power-laws than blackbodies, and the blackbody fits suggest hot temperatures near $\sim$30\,000~K. Finally, UV10 has fluxes indicating a hot optical source of $\sim$10\,000~K with a slightly lower ultraviolet emission. Assuming that the optical component, of blue slope, has a significantly suppressed $u$-band flux (as for Gaia~DR3~1528861748669458432 in Fig.~\ref{wd_blue}), and that the ultraviolet component stems entirely from another source that emits thermally, then the latter would have a temperature of $\sim$45\,000~K.

Gaia DR3 3150537594577224960 (UV8), UV9, and UV11 have significant $NUV$-band flux excesses. Their $d50$ distances are of 88, 104, and 96~pc and their proper motions of 16, 8, and 17 mas~yr$^{-1}$, respectively. None is a catalogue white dwarf photometric candidate. The SEDs of UV8 and UV9 (Fig.~\ref{fsed_uv8_uv9}) consist both of a (i) relatively bright, red, and ultra-cool component ($T_\mathrm{eff}<3250\pm230$~K) and (ii) a blue, power-law- or hot blackbody component; we discuss these in detail in Section~\ref{disc:pl}. UV11 is a cool source ($T_\mathrm{eff}\sim5250$~K) with a very low $u$-band flux compared to the $NUV$-band flux (Fig.~\ref{fsed_uv2}). In fact, the centroid of the \textit{GALEX} source ($\sigma_{\mathrm{pos}}=1.41$~arcsec; count spread of about 2~arcsec) is 1.5~arcsec north-east of the \textit{Gaia} source and 2.4~arcsec south-east of the centroid of an extended galaxy or nebula (SDSS J162911.33+055225.9), which are both clearly resolved and bright in a NOIRLab $u$-band stack image. The counterpart likelihood for UV11 is thus of 0.5.

The five remaining \textit{Gaia} sources with a \textit{GALEX} counterpart are Gaia DR3 6829498931265990656 (UV2), Gaia DR3 6585452983927662080 (UV5), Gaia DR3 6858721824326705536 (UV6), Gaia DR3 6529021095092491648 (UV7), and Gaia DR3 4465431356821050496 (UV12). These appear to have blackbody profiles peaking at the red-optical--near-infrared and a low $NUV$-band flux excess relative to the blackbody extrapolation (UV7 is shown as example in Fig.~\ref{fsed_uv2}; the others are shown in Fig.~\ref{fsed_uv_ir} of the online supplementary material). Their $d50$ distances are of 60--98~pc. UV2, UV6, and UV12 have high values of \textit{Gaia} photoastrometric parameters associated with binarity (see Table~\ref{tab:mul} and Section~\ref{disc:bin}). UV5 and UV7 have no and slight photoastrometric indication of binarity, respectively. The SED of UV7 has a clear power-law component when infrared fluxes are taken into account (Fig.~\ref{fsed_uv8_uv9}), which we discuss in detail in Section~\ref{disc:pl}. UV5 has a counterpart likelihood of 0.5, because the \textit{GALEX} centroid ($\sigma_{\mathrm{pos}}=1.25$~arcsec) is 3.04 east of it and 3.3~arcsec west of another optical source, NSC DR2 160974\_6762, which is 0.25~mag fainter at $g$-band.

Finally, we identified six SDSS sources with $u$-band flux excesses, using the selection criterium $u-g<g-r$, for $u$-band point-spread-function (PSF) magnitude errors smaller than 0.45~mag. UV8 is already commented above. Gaia DR3 2480784311444314880 (UV13), Gaia DR3 1900830823016998144 (UV14), and Gaia DR3 2807663865638075264 (UV15) have cool $griz$-band components of $\sim$4000--5000~K, large proper motions of about 260, 200,
and 190~mas~yr$^{-1}$, and $d50$ distances of 66, 82 and 90~pc, are well isolated in sparse fields, and are white dwarf photometric candidates in \citetalias{2021A&A...649A...6G} (UV13 and UV14 also in \citealt{2021MNRAS.508.3877G}). The SED of UV13 is shown in Fig.~\ref{fsed_uv2} and those of UV14 and UV15 are shown in Fig.~\ref{fsed_uv_ir} of the online supplementary material. UV15 appears to have (i) a slight flux dip at $i$ band compared to an overall curved $griz$-band profile, and (ii) a high mid-infrared, $W1$-band flux excess. UV13 and UV15 have significant $u$-band flux excesses, and UV14 has a low flux excess that agrees within 1--2~sigma with the blueward extrapolations of the PS1, SDSS, and NSC blackbody fits. The two remaining sources, Gaia DR3 430017946216189145 and Gaia DR3 4518304014850969088, have large $u$-band errors of 0.41--0.44~mag and their counts in the $u$-band image appear as noise compared to similar features around these sources.

\begin{table*}
\tiny
	\centering
	\caption{Radio crossmatches.}
	\label{tab:radiocan}
	\footnotesize
	\begin{tabular}{l@{\hspace{1.2\tabcolsep}}r@{\hspace{1.2\tabcolsep}}r@{\hspace{1.2\tabcolsep}}r@{\hspace{1.2\tabcolsep}}r@{\hspace{1.2\tabcolsep}}r@{\hspace{1.2\tabcolsep}}r@{\hspace{1.2\tabcolsep}}r@{\hspace{1.2\tabcolsep}}r@{\hspace{1.2\tabcolsep}}r@{\hspace{1.2\tabcolsep}}l@{\hspace{1.2\tabcolsep}}r@{\hspace{1.2\tabcolsep}}r@{\hspace{1.2\tabcolsep}}r@{\hspace{1.2\tabcolsep}}r@{\hspace{1.2\tabcolsep}}r}
		\hline
Flag    &  $\mu_{\alpha^*}$ &  $\mu_{\delta}$ & $G$   &  $BP-G$ &$G-RP$  & $g-r$     & $r-i$       &  $i-z$   & P & Sep & n & L   & Name   & $F_\mathrm{radio}$ & Band  \\
        &  \multicolumn{2}{c}{(mas~yr$^{-1}$)}            & (mag) &  (mag) &(mag)    & (mag)    & (mag)    & (mag)   &   &   ($\prime\prime$) &  ($\sigma_\mathrm{pos}$) &  (\%)   &    &  (mJy)      &   (MHz)   \\
		\hline
Rad1    &   -91.2  &    65.5 &  20.78 &   -0.07~$\pm$~0.12  &    0.52~$\pm$~0.22 &    0.37 &    0.47 &    0.05 &   1   &    5.6 &   1 &  100.0 & (1)  &        2.3~$\pm$~0.5 &   1400 \\ 
Rad\_a  &     38.8 &    18.1 &  20.87 &    0.13~$\pm$~0.29  &    0.41~$\pm$~0.22 &    0.48 &    0.28 &    0.05 &   1   &   13.2 &   3 &   25.0 & (2)  &        4.8~$\pm$~0.4 &   1400 \\ 
Rad\_b  &     -2.8 &   -15.7 &  20.56 &   -0.01~$\pm$~0.10  &    0.58~$\pm$~0.11 &   -0.01 &    0.10 &   -0.02 &   2   &    1.6 &   1 &    0.6 & (3)  &       36.0~$\pm$~7.2 &   4850 \\ 
Rad\_c  &    -14.1 &     7.2 &  20.10 &   -0.01~$\pm$~0.13  &    0.89~$\pm$~0.05 &    0.64 &    0.38 &   -0.02 &   1   &   11.6 &   2 &    5.0 & (4)  &        3.5~$\pm$~0.5 &   1400 \\ 
Rad\_d  &     -0.1 &     8.8 &  20.64 &   -0.10~$\pm$~0.34  &    0.77~$\pm$~0.12 &    1.15 &    0.65 &    0.40 &   1   &   11.8 &  20 &   16.7 & (5)  &       31.0~$\pm$~1.0 &   1400 \\ 
Rad\_e  &     -0.3 &    -3.9 &  20.93 &    0.60~$\pm$~0.20  &    0.88~$\pm$~0.12 &    1.11 &    0.38 &    0.55 &   1   &   10.3 &  12 &   20.0 & (6)  &       23.7~$\pm$~1.2 &   1400 \\ 
--      &   --     &         &  --    &    --               &   --               &   --    &   --    &    --   &  --   &   11.1 &   6 &   20.0 & (7)  &     173.1~$\pm$~19.2 &    150 \\ 
Rad\_f  &     -3.0 &    -1.7 &  20.87 &    0.38~$\pm$~0.08  &    0.54~$\pm$~0.08 &    0.88 &    0.52 &    0.24 &   1   &    7.6 &   1 &    6.7 & (8)  &           95~$\pm$~9 &   4850 \\ 
--      &   --     &         &  --    &    --               &   --               &   --    &   --    &    --   &  --   &   12.7 &  14 &    1.8 & (9)  &     227.9~$\pm$~6.71 &   1420 \\ 
--      &   --     &         &  --    &    --               &   --               &   --    &   --    &    --   &  --   &    6.3 &  11 &   14.3 & (10) &      279.1~$\pm$~9.1 &   1400 \\ 
--      &   --     &         &  --    &    --               &   --               &   --    &   --    &    --   &  --   &    4.2 &   1 &    2.6 & (11) &        560         &    408 \\ 
--      &   --     &         &  --    &    --               &   --               &   --    &   --    &    --   &  --   &    6.3 &  27 &   14.3 & (12) &         739~$\pm$~40 &    365 \\ 
--      &   --     &         &  --    &    --               &   --               &   --    &   --    &    --   &  --   &    6.0 &   3 &   14.3 & (13) &   1710.0~$\pm$~171.8 &    150 \\ 
--      &   --     &         &  --    &    --               &   --               &   --    &   --    &    --   &  --   &    9.1 &   3 &    6.7 & (14) &  2085.49~$\pm$~75.367&     74 \\ 
Rad\_g  &     -1.0 &    -2.8 &  20.88 &    0.18~$\pm$~0.21  &    0.28~$\pm$~0.26 &    0.88 &    0.33 &    0.16 &   1   &   12.1 &   4 &    3.0 & (15) &        8.1~$\pm$~1.2 &   1400 \\               \hline
	\end{tabular}
\scriptsize
\vspace{-0.23cm}
\begin{flushleft}
Proper motion component errors are of 1.1--2.8 mas~yr$^{-1}$. The $griz$-band photometric (P) colours are from PS1 (1) and NSC (2). Angular separation (Sep) is in arcseconds. Radio source name (catalogue, if distinct from the source name acronym):(1) NVSS J053905+820343, (2) NVSS J201013+540843, (3) PMN J0050$-$7235, (4) NVSS J165455$-$293302, 
(5) NVSS J174923$-$034832, (6) NVSS J075143$-$221953, (7) TGSSADR J075143.6 $-$221952 (GMRTAS150M), (8) GB6 J1927+2526, (9) CGPS J192720+252626 (CGPSNGPCAT), (10) NVSS J192721+252624, (11) B2.2 1925+25A (B2), (12) TXS 1925+253 (TEXAS), (13) TGSSADR J192721.0+252623 (GMRTAS150M), (14) VLSSr J192721.2+252627, and (15) NVSS J174425$-$144134. Catalogue references are given in the Master Radio Catalogue.
\end{flushleft}
\end{table*}

\subsection{Radio crossmatches}\label{sec:radio}

We crossmatched the list of \textit{Gaia} sources with the Master Radio Catalogue (as revisited on 2022 December 12), provided by the High Energy Astrophysics Science Archive Research Centre (HEASARC; \url{https://heasarc.gsfc.nasa.gov}) and that contains radio sources from many catalogues, using a crossmatch radius of $(pm/1000 \times 50) +15$~arcsec. There are 27 multiple radio matches for 12 \textit{Gaia} sources, all within 90~arcsec. Ten \textit{Gaia} sources have a match in images of the 1.4~GHz NRAO VLA Sky Survey (NVSS; epochs 1993--1997; \citealt{1998AJ....115.1693C}). We compiled the positional uncertainties and mean observation epochs of the radio sources in the different catalogues. We then propagated the \textit{Gaia} coordinates to these epochs and we recomputed the angular separations. Four \textit{Gaia} sources were discarded because their sky separations of 28--77~arcsec are too far off to the radio sources, of coordinate uncertainties of a few arcsec. In Table~\ref{tab:radiocan}, we list the matches of the remaining eight \textit{Gaia} sources, together with the separation at the radio epoch, the multiple $n$ of the radio positional-uncertainty ellipse ($\sigma_\mathrm{pos}$) required to include the match, and the counterpart likelihood in percent, as the inverse of the number of \textit{Gaia} sources located at the radio epoch in the enlarged ellipse.

\begin{figure*}
    \includegraphics[width=\columnwidth]{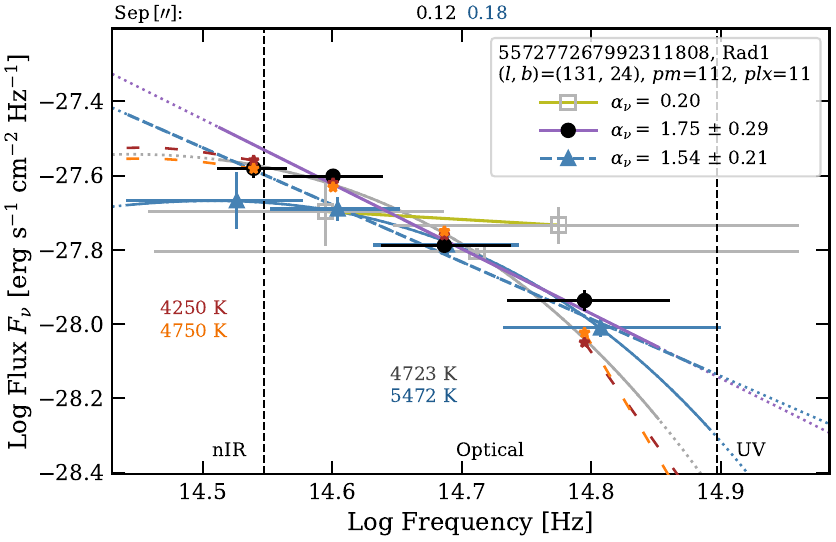}
    \includegraphics[width=\columnwidth]{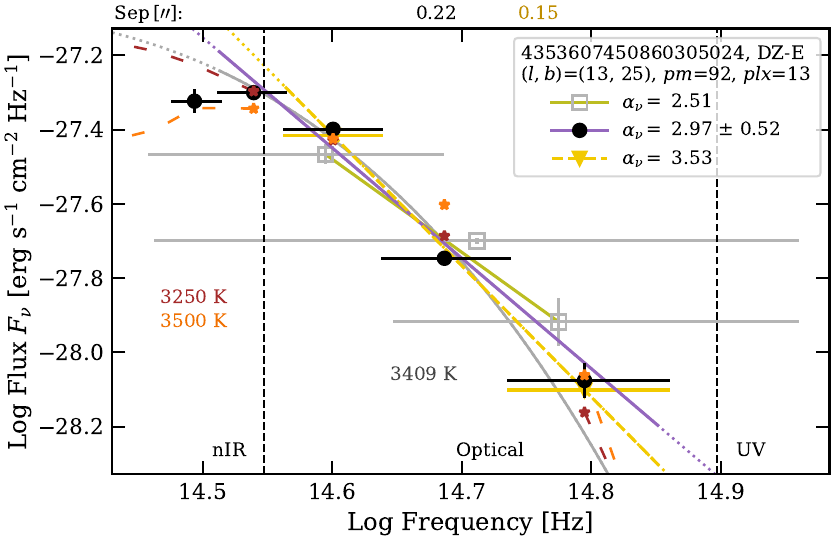}
   \caption{The PS1 SED of Rad1 ({\sl left-hand panel}) is similar to that of Gaia DR3 4353607450860305024 ({\sl right-hand panel}), a DZ white dwarf with an atmosphere enriched in lithium, sodium, potassium, and calcium ascribed to planetesimal accretion, and whose reduced flux at $r$ band is caused by intense sodium line absorption. Same as in Fig.~\ref{fsed_uv1}.}
    \label{dz_wd_polluted}
\end{figure*}

Gaia DR3 557277267992311808 (Rad1; $d50=91.5$~pc), a $pm=112$~mas~yr$^{-1}$ probable white dwarf candidate in GCNS100pc and \citet{2019MNRAS.482.4570G}, appears as a probable, unique optical counterpart to the relatively faint NVSS~J053905+820343 radio source, with a coordinate offset of only 5.6~arcsec at the NVSS epoch. Within the positional uncertainty region of $(\Delta RA, \Delta Dec)=\pm(58.4, 6.7)$~arcsec, there are no other optical--infrared source. The 1.4~GHz radio flux density is of $2.3\pm0.5$~mJy\footnote{The Jansky unit corresponds to the amount of energy that reaches the Earth per unit collecting area and frequency, i.e. the brightness per unit frequency or flux density, measured in 10$^{-26}$~W~m$^{-2}$~Hz$^{-1}$ or 10$^{-23}$~erg~s$^{-1}$~cm$^{-2}$~Hz$^{-1}$.} (Table~\ref{tab:radiocan}). NVSS~J053905+820343 was observed at 1.4~GHz five times in about 1995 \citep{2011ApJ...737...45O}. As shown in Table~\ref{tab:radiopulsar}, the radio flux density is 147$\pm$33 times greater than the $G$-band broad-optical flux density, which is represented in Fig.~\ref{dz_wd_polluted} (10$^{-27.80}$~erg~s$^{-1}$~cm$^{-2}$~Hz$^{-1}$).

\begin{table}
	\centering
	\caption{Comparison of Rad1 and the AR Scorpii pulsar white dwarf (psrWD).}
	\label{tab:radiopulsar}
	\begin{tabular}{l@{\hspace{1.2\tabcolsep}}l@{\hspace{1.2\tabcolsep}}l@{\hspace{1.2\tabcolsep}}l@{\hspace{1.2\tabcolsep}}l@{\hspace{1.2\tabcolsep}}l}
		\hline
Source   & $G$  & F$_\mathrm{radio}$ & F$_G$ & F$_\mathrm{radio}$/F$_G$  \\
             & (mag) &   (mJy)              & (mJy)  &       \\
		\hline
Rad1     & 20.784~$\pm$~0.012  &  2.3~$\pm$~0.5    & ~0.01568~$\pm$~0.00017    & 147~~~~~~$\pm$~33\\
psrWD   & $>$14.990~$\pm$~0.031  &  5.0~$\pm$~0.5    & $<$3.259~~~$\pm$~0.093  & ~$>$1.53~$\pm$~0.23 \\
              \hline
	\end{tabular}
\end{table}

Significant radio emission in white dwarfs is not frequent and it could be produced by magnetic interaction with a close companion, such as for the AR Scorpii pulsar white dwarf (116~pc) that has a close M-type dwarf companion of orbital period of 3.6~h \citep{2016Natur.537..374M,2018A&A...611A..66S}. AR Scorpii has an NVSS 1.4 GHz flux density of $8.4\pm0.5$~mJy, but it includes flux of a neighbour object at 14~arcsec and of 3.49$\pm$0.08~mJy at 1.5~GHz \citep{2018A&A...611A..66S}; we assumed that AR Scorpii has a deblended 1.4~GHz flux density of 5.0$\pm$0.5~mJy. Also, the $G$-band flux of AR Scorpii (Gaia DR3 6050296829033196032) includes a significant contribution of the M-type companion, which is brighter at the red-optical \citep{2016Natur.537..374M}. Thus the radio-to-optical flux ratio of the pulsar white dwarf is probably greater than 1.53 (Table~\ref{tab:radiopulsar}). The \texttt{ipd\_gof\_harmonic\_amplitude}, \texttt{ipd\_frac\_multi\_peak}, \texttt{ipd\_frac\_odd\_win}, and \texttt{ruwe} (see Section~\ref{disc:bin}) values of Rad1 are of 0.065, 0 per cent, 0 per cent, and 0.977. These are very close to the median values of \textit{Gaia} DR3 sources with six-parameter solutions at $G=21$~mag (Table~\ref{tab:all1}), suggesting no obvious signature of a companion. In the case of AR Scorpii, these parameters are also unrevelatory, except possibly for \texttt{ruwe} that has a value of 1.393, greater than the median value of 1.104 at $G=15$~mag.

Alternatively, if the 1.4~GHz radio flux of NVSS~J053905+820343 were to be emitted by a neutron-star pulsar, then it is comparable to the fluxes of a few mJy of low-luminosity radio millisecond pulsars located at a few hundred pc \citep[see e.g.][]{1997ApJ...481..386B}. Assuming that at $G$ band, the pulsar flux is lesser than the white dwarf flux, then the $F_\mathrm{radio}/F_G$ ratio would be greater than 147$\pm$33 and compatible with a power law of spectral index $\alpha_{\nu}\ge1$ (assuming a non-broken power law from radio to optical).

The PS1 fluxes at epoch 2013.40 of Rad1 are not reproduced by the theoretical pure He- or H atmosphere white-dwarf fluxes nor a blackbody (Fig.~\ref{dz_wd_polluted}). In fact, the PS1 SED has a feature in common with the SED of Gaia DR3 4353607450860305024 (`DZ-E', right panel in Fig.~\ref{dz_wd_polluted}), a DZ white dwarf of $T_\mathrm{eff}=3830\pm230$~K with an atmosphere enriched in lithium, sodium, potassium, and calcium ascribed to planetesimal accretion \citep{2021Sci...371..168K}. The feature in common is the significantly depressed $r$-band flux, reflected in slightly unequal colours $g-r<r-i$. The PS1 $r$ band is defined at 5500--6890~\AA, with an effective wavelength of 6170~\AA. In the spectrum of Gaia DR3 4353607450860305024, the feature is caused by the intense Na~\textsc{i}~D (5893 \AA) absorption line \citep[fig.~1 of][]{2021Sci...371..168K}. Rad1 is completely isolated within 19~arcsec and its PS1 DR1 and DR2 photometric flags indicate good-quality measurements, suggesting that these are reliable. The SDSS fluxes at epoch 2006.23 have larger uncertainties at $riz$ bands, and the SDSS bandpasses are broader and overlapping, possibly causing the absorption line to not being resolved in the SDSS flux profile. Rad1 could thus be a white dwarf with significant sodium absorption. 

Atmospheric enrichment in sodium could stem either from surfacing (e.g. through convective dredge-up, such as the dredge-up of carbon from the interior proposed for some white dwarfs that are enriched in carbon, \citealt{2019MNRAS.490.4166B,2022ApJ...930....8B}) or from accretion or collection of external matter. In the broader context of the origin of the elements \citep{2019Sci...363..474J}, sodium is expected to originate in exploding massive stars, that is supernova of type II, which are progenitors of neutron stars. Indications of on-going accretion of planetesimals through photometric transit observations have been found in WD~1145+017, a white dwarf with a metal-enriched atmosphere and a debris disc (\citealt{2015Natur.526..546V}; see also introduction in \citealt{2022MNRAS.511...71H}). In the case of Rad1, radio observations and optical spectroscopy are required to confirm whether there is radio emission (furthermore of synchrotron-, pulsating type) and chemical enrichment of the atmosphere, respectively.

The seven remaining \textit{Gaia} sources are towards the Galactic plane ($|b|<12\degr$), except one of them, and have low counterpart-likelihoods. We note that the $\sigma_\mathrm{pos}$ of some radio sources could be lower values, as for VLSSr J192721.2+252627 at 74 MHz, where it does not include ionospheric corrections, possibly of the order of tens of arcsec \citep{2014MNRAS.440..327L}. Gaia DR3 2185585922087166720 (Rad\_a) is in a sparse field and 13~arcsec west of NVSS~J201013+540843, which has three closer optical-infrared sources but none is in its positional error ellipse of $(\Delta RA, \Delta Dec)=(5.1, 3.1)$~arcsec. Two \textit{Gaia} sources have radio cross-identifications in the SPECFIND V3.0 catalogue \citep{2021A&A...655A..17S}. Gaia DR3 4688982132605446784 (Rad\_b; $b=-45\degr$) is at 1.6~arcsec of PMN J0050$-$7235 of position accuracy $(\Delta RA, \Delta Dec)=(36.2,30.6)$~arcsec using the relations in \citet{1994ApJS...91..111W}. In SPECFIND, there are three other radio identifications associated with PMN J0050$-$7235, and these are centred 1.2~arcmin to the west (survey beam sizes of 18--98~arcsec), the four radio flux measurements being consistent with a single power law. The optical--near-infrared SED of Rad\_b indicates blackbody emission and, as for Rad1, an $r$-band flux dip (Fig~\ref{fsed_radio_ir}, available as online supplementary material). The \textit{Gaia} parameters (e.g. \texttt{ruwe}) indicate that it could be a resolved binary. Gaia DR3 2024652913390061952 (Rad\_f) is close to multiple radio components, for example, B2.2 1925+25A \citep{1972A&AS....7....1C} at 4.2~arcsec, or NVSS J192721+252624, TGSSADR J192721.0+252623, and TXS 1925+253 at 6~arcsec, the three being consistent with the same coordinates within 2~arcsec and, in SPECFIND, with a radio power-law emission of $\alpha_{\nu}=0.7$. NVSS J174923$-$034832 (Rad\_d) and NVSS J075143$-$221953 (Rad\_e) appear to have PSO J267.3500$-$03.8089 and PSO J117.9316$-$22.3316 as optical counterparts within 1~$\sigma_\mathrm{pos}$.

Finally, we crossmatched the list of \textit{Gaia} sources within 30~arcsec with the full catalogue of transients of the Transient Name Server (TNS\footnote{\url{https://www.wis-tns.org}, developed by O. Yaron, A. Sass, N. Knezevic, L. Manulis, E. Ofek and A. Gal-Yam.}; as of 2022 September 22), which contains the complete catalogue of fast radio bursts (FRBs), and we find no match.

\subsection{Crossmatches with sources reported in the literature}\label{sec:lit}

We queried the Simbad data base (as revisited on 2023 February 1) on the \textit{Gaia} coordinates, propagated to epoch 2000 ($dyr=-16.0$~yr), and within $\delta=10$~arcsec. The visual check with the Aladin tool and the verification in the literature and catalogues yielded 97 matches. These have $M_G^{\pi}=16.0-17.4$~mag and $d50=32-98$~pc. From a crossmatch with the Montreal White Dwarf Database\footnote{\url{https://montrealwhitedwarfdatabase.org/}} \citep[as of 2022 December 20;][]{2017ASPC..509....3D}, we recovered additional spectral types estimated from spectra.

There are 86 probable white dwarf photometric candidates (\citealt{2019MNRAS.482.4570G, 2021MNRAS.508.3877G,2019MNRAS.485.5573T}; \citetalias{2021A&A...649A...6G}). Among these, 19 are spectroscopically confirmed and have $M_G^{\pi}=16.0-17.1$~mag. Fifteen have DC-type (continuous, feature-less) spectra, three have DZ-type (metal-polluted) spectra, and one ([IIB2000] F351-50) has a feature-less DA-type spectrum. Also, among these (and simultaneously in our subsample of 2464 sources), 25 are in the selection of \citet{2022RNAAS...6...36S} of faint, $pm>30$~mas~yr$^{-1}$ white dwarf candidates, of which some could be ultra-massive ($\gtrapprox$1.05~M$_{\odot}$; \citealt{2022A&A...668A..58A}). Finally, one source has an uncertain spectroscopic classification, sdA: subdwarf (SDSS J120918.83+470940.3, \citealt{2016MNRAS.455.3413K}); its SED based on PS1, SDSS, and NSC fluxes indicates a blackbody profile of $\sim$7500~K.

In the colour--magnitude and colour--colour diagrams forwards, such as Figs.~\ref{mz_iz_msbd}~and~\ref{gaia_can_bpg_vs_g_cmd}, we plot the white dwarfs that are confirmed spectroscopically, with labels related to their spectral type. In Section~\ref{sec:radio}, we comment on Gaia DR3 4353607450860305024 (`DZ-E'), a cool DZ white dwarf that has an atmosphere enriched in lithium, sodium, potassium, and calcium \citep{2021Sci...371..168K}, because the SED of Rad1 bears resemblance to its SED. In Appendix~\ref{app:litextraWDs}, we comment on Gaia DR3 3905335598144227200 (`DC1'), an example of nearby DC white dwarf, and Gaia DR3 283928743068277376, 
a hot white-dwarf source with an apparent flux excess at $i$ band.

\subsection{Power-law profiles at the broad-optical and infrared}\label{sec:pl}

We identify power-law profiles using the PS1-, SDSS-, or NSC photometry of the \textit{Gaia} sources, and then we extend the search using infrared photometry.

For that purpose, we fitted the observed fluxes in the $\log(F_{\nu})- \log(\nu)$ plane with power law and blackbody functions by non-linear least-squares using the \textsc{scipy.optimize.curve\_fit} function. Magnitude uncertainties were previously converted to logarithmic-flux uncertainties and used as relative weights. The best fit minimizes the chi-square expression
\begin{equation}
\chi^2=\sum_{n=1}^{N}\left(\frac{\log(F_{\nu,i}) - Y_i}{\sigma_i}\right)^2,
\end{equation}
where $N$ is the total number of bandpasses, $Y$ is the fitting function, and $\sigma_i$ are the uncertainties. The output variance of a parameter, for instance of the $\alpha_{\nu}$ spectral index, is the variance of the parameter as obtained from ${\chi^2}$ and then multiplied by $f_{\sigma}^2$, an uncertainty scaling factor equal to the reduced chi-square of the optimal fit, $\chi^2[\textrm{optimal parameters}]/(N-p)$, where $p$ is the number of parameters. The square root of the output variance (the standard deviation) informs us thus on the dispersion or deviation of the observed SED profile relative to the model profile. To avoid weight sensitivity at very small $\sigma_{i}$ uncertainties, these were artificially enlarged to 0.008~dex, equivalent to 0.02~mag or 2 per cent in flux.

In the SED figures, the power-law fits are represented by the coloured segments (spectral indices are given in the legend) and the blackbody fits are represented by the curves (effective temperatures are indicated in the plot,  in the same colours as the curves).

For the blackbody fitting of the observed fluxes in erg~s$^{-1}$~cm$^{-2}$~Hz$^{-1}$, we used the Planck flux density function $\pi B_{\nu} (\nu, T)$ in the form\\
\begin{equation}\label{eqn:BB}
  \pi  B_{\nu} (\nu, T) f_\mathrm{sc} = \pi  \frac{2 h \nu^3}{c^2} \frac{1}{e^{h\nu/(k_\mathrm{B}T)} - 1}  f_\mathrm{sc}
\end{equation}
where h [J/Hz], $k_\mathrm{B}$ [J/K], and c [m/s] are the constants of Planck, Boltzmann, and light velocity, respectively. We fitted for the parameters of temperature $T$ and scaling factor $f_\mathrm{sc}$. Then, using the Stefan--Boltzmann $\sigma T^4$ frequency-integrated blackbody flux [J/s/m$^2$] as normalization factor, we retrieved the bolometric flux in erg~s$^{-1}$~cm$^{-2}$: 
\begin{equation}\label{eqn:Fbol}
F_\mathrm{bol} =  f_\mathrm{sc} \sigma T^4.
\end{equation}
We verified that the resulting blackbody density function with this approach is identical to that using the BlackBody model from the \textsc{astropy} package, with relative $F_{\nu}$ differences lesser than 0.02 per cent.

For illustrative purpose and to guide in the interpretation, we also show in the SED figures the closest matches of the theoretical grid fluxes of white dwarfs of pure He and H atmospheres (brown and orange small stars, extended by brown and orange dashed lines, respectively) and masses of 0.4, 0.6, 0.8, 1.0, and 1.2~M$_\odot$ (same reference as above). The closest match is that having the flux-error-weighted mean distance that is smallest between the observed and theoretical fluxes at $griz$ bands. The default observed fluxes that are matched are those of PS1 (black dots), SDSS (blue upward triangles), or NSC (yellow downward triangles), depending on which has most $griz$ coverage. At effective temperatures cooler than $\sim$5000~K and compared to the He atmosphere fluxes, the H atmosphere fluxes tend in most cases to depart to less red optical- and infrared colours and thus to differ more from blackbody and power-law fluxes (see e.g. the colour--colour diagrams and the SEDs in the online supplementary material).

For the 1104 \textit{Gaia} sources that have PS1 PSF magnitudes in the four $griz$ bands, we fitted their $griz$-band fluxes as described above, with a power-law function of unconstrained $\alpha_{\nu}$ spectral index. We used the bandpass effective wavelengths listed in \citet{2012ApJ...756..158S} and an AB magnitude zero-point flux of 3631~Jy. We then selected the sources with $\alpha_{\nu}$ errors smaller than 0.1. Because the PS1 $grizy$ bandpasses do not overlap in wavelengths, these enable us to obtain a neat description of the broad-optical profiles and, to some extent, to distinguish power-law profiles from blackbody profiles. However, the profile identification becomes more efficient when considering also SDSS- or NSC $u$-band fluxes and infrared fluxes, as shown below. Besides this, the $G_\mathrm{BP}$ and $G_\mathrm{RP}$ fluxes of the faint sources can also support estimating the slopes and temperatures at the broad-optical, provided these fluxes are accurate enough.

\begin{table*}
	\centering
	\caption{Sources with power-law profiles at $griz$ bands or approximately straight profiles at the optical and infrared.}
	\label{tab:plcan}
	\footnotesize
	\begin{tabular}{l@{\hspace{1.2\tabcolsep}}r@{\hspace{1.2\tabcolsep}}r@{\hspace{1.2\tabcolsep}}r@{\hspace{1.2\tabcolsep}}r@{\hspace{1.2\tabcolsep}}r@{\hspace{1.2\tabcolsep}}r@{\hspace{1.2\tabcolsep}}r@{\hspace{1.2\tabcolsep}}r@{\hspace{1.2\tabcolsep}}r@{\hspace{1.2\tabcolsep}}r@{\hspace{1.2\tabcolsep}}l@{\hspace{1.2\tabcolsep}}l}
		\hline
Flag    &  $\mu_{\alpha^*}$ &  $\mu_{\delta}$ & $G$   &  $BP-G$ &$G-RP$  & $g-r$     & $r-i$       &  $i-z$  & P &  PL $\alpha_{\nu}$  &  Profile & Comment \\
        &  \multicolumn{2}{c}{(mas~yr$^{-1}$)}  & (mag) &  (mag) &(mag)    & (mag)    & (mag)    & (mag)   &    & at $griz$  & UV-Opt-IR &  \\
\hline
P0     &      6.1 &    -6.2 &  20.89 &   -0.16~$\pm$~0.23  &     0.02~$\pm$~0.20  &    -0.26 &    -0.20 &    -0.05 &   3 &  -0.89~$\pm$~0.02   &  S [$U,z$] or C & --            \\
P1X1   &     -0.2 &    -1.5 &  20.61 &   -0.33~$\pm$~0.08  &     0.64~$\pm$~0.17  &    -0.24 &    -0.21 &    -0.02 &   1 &  -0.86~$\pm$~0.07   &  C?             & --            \\
P2     &     10.8 &    -3.7 &  20.86 &   -0.17~$\pm$~0.08  &     0.20~$\pm$~0.11  &     0.04 &    -0.04 &     0.07 &   1 &   0.04~$\pm$~0.07   &  S [$g,z$] or C & --            \\
P3     &     -4.8 &    -3.3 &  20.81 &    0.09~$\pm$~0.16  &     0.33~$\pm$~0.17  &     0.29 &    -0.04 &     0.03 &   1 &   0.38~$\pm$~0.14   &  $\sim$S [$g,W2$] & Recovered from infrared         \\
P4     &     -2.6 &     1.4 &  20.90 &    0.27~$\pm$~0.19  &     0.74~$\pm$~0.17  &     0.05 &     0.08 &     0.32 &   2 &   0.52~$\pm$~0.22   &  $\sim$S [$g,W2$] & Recovered from infrared         \\
P5     &     -1.0 &    -2.3 &  20.84 &   -0.14~$\pm$~0.17  &     0.07~$\pm$~0.36  &     0.18 &     0.18 &     0.05 &   1 &   0.71~$\pm$~0.05   &  C             & --            \\
P6     &      0.2 &     0.4 &  20.69 &   -0.29~$\pm$~0.31  &     0.21~$\pm$~0.36  &     0.19 &     0.23 &     0.09 &   1 &   0.81~$\pm$~0.05   &  $\sim$S [$u,W2$] or C & Recovered from infrared         \\
P7     &     -2.9 &   -16.0 &  20.50 &    0.16~$\pm$~0.10  &     0.87~$\pm$~0.07  &     0.30 &     0.20 &     0.07 &   1 &   0.97~$\pm$~0.08   &  C             & --            \\
P8     &      4.6 &    -0.5 &  20.92 &    0.57~$\pm$~0.32  &     0.46~$\pm$~0.20  &     0.32 &     0.21 &     0.17 &   1 &   1.09~$\pm$~0.02   &  C             & --            \\
P9     &      2.3 &    -7.3 &  19.78 &    0.12~$\pm$~0.07  &     0.81~$\pm$~0.05  &     0.36 &     0.21 &     0.10 &   1 &   1.09~$\pm$~0.07   &  C             & --            \\
P10    &     -7.0 &   -10.5 &  20.85 &    0.10~$\pm$~0.14  &     0.47~$\pm$~0.11  &     0.35 &     0.23 &     0.21 &   1 &   1.22~$\pm$~0.03   &  C             & --            \\
P11    &     -3.9 &   -10.1 &  20.87 &   -0.05~$\pm$~0.17  &     0.51~$\pm$~0.25  &     0.36 &     0.27 &     0.18 &   1 &   1.24~$\pm$~0.03   &  S [$g,y$]     & --            \\
P12    &     -3.9 &    -7.3 &  20.67 &    0.49~$\pm$~0.23  &     0.72~$\pm$~0.15  &     0.33 &     0.27 &     0.21 &   1 &   1.25~$\pm$~0.02   &  C             & --            \\
P13    &     -4.2 &     4.1 &  20.95 &    0.20~$\pm$~0.17  &     0.82~$\pm$~0.17  &     0.38 &     0.28 &     0.25 &   1 &   1.39~$\pm$~0.03   &  C             & --            \\
P14    &     -8.3 &    -1.8 &  20.65 &    0.61~$\pm$~0.12  &     0.58~$\pm$~0.07  &     0.40 &     0.33 &     0.09 &   1 &   1.40~$\pm$~0.09   &  C             & --            \\
P15    &     33.0 &   -64.1 &  20.89 &    0.22~$\pm$~0.17  &     0.84~$\pm$~0.17  &     0.47 &     0.26 &     0.19 &   1 &   1.48~$\pm$~0.09   &  C             & --            \\
P16    &      3.0 &     4.5 &  20.69 &    0.13~$\pm$~0.14  &     0.84~$\pm$~0.09  &     0.45 &     0.28 &     0.23 &   1 &   1.49~$\pm$~0.05   &  C             & --            \\
P17    &     -3.9 &     1.3 &  20.20 &    0.16~$\pm$~0.06  &     0.89~$\pm$~0.04  &     0.52 &     0.28 &     0.20 &   1 &   1.56~$\pm$~0.09   &  C             & --            \\
P18    &    -88.6 &   -15.7 &  20.78 &    0.24~$\pm$~0.19  &     0.91~$\pm$~0.15  &     0.42 &     0.36 &     0.22 &   1 &   1.59~$\pm$~0.03   &  C             & Comoving with M dwarf \\
P19    &      3.5 &     7.2 &  20.77 &   -0.22~$\pm$~0.19  &     0.82~$\pm$~0.13  &     0.48 &     0.29 &     0.32 &   1 &   1.67~$\pm$~0.08   &  C             & Blend photometry \\
P20    &    -16.5 &     0.7 &  20.36 &    0.15~$\pm$~0.15  &     0.81~$\pm$~0.09  &     0.49 &     0.37 &     0.28 &   1 &   1.77~$\pm$~0.02   &  C?            & Blend photometry \\
P21    &     -3.6 &    -4.4 &  20.45 &    0.01~$\pm$~0.11  &     0.76~$\pm$~0.15  &     0.53 &     0.37 &     0.16 &   1 &   1.83~$\pm$~0.06   &  C             & --            \\
P22    &     -0.7 &     0.4 &  20.65 &    0.43~$\pm$~0.14  &     0.89~$\pm$~0.06  &     0.63 &     0.41 &     0.27 &   1 &   2.13~$\pm$~0.05   &  C             & --            \\
P23    &    -15.2 &    -9.2 &  20.85 &    0.08~$\pm$~0.15  &     0.82~$\pm$~0.11  &     0.62 &     0.44 &     0.29 &   1 &   2.13~$\pm$~0.05   &  C             & --            \\
P24    &    -10.1 &    -1.8 &  20.94 &    0.66~$\pm$~0.39  &     0.68~$\pm$~0.11  &     0.55 &     0.57 &     0.24 &   1 &   2.17~$\pm$~0.06   &  C             & --            \\
P25    &      8.3 &    -0.5 &  20.76 &    0.64~$\pm$~0.24  &     0.90~$\pm$~0.03  &     0.62 &     0.49 &     0.25 &   1 &   2.17~$\pm$~0.07   &  C             & --            \\
P26    &     -7.7 &    -0.3 &  20.69 &    0.40~$\pm$~0.09  &     0.65~$\pm$~0.13  &     0.65 &     0.42 &     0.37 &   1 &   2.17~$\pm$~0.07   &  C             & --            \\
P27    &     -1.5 &     1.1 &  20.39 &    0.69~$\pm$~0.07  &     0.68~$\pm$~0.05  &     0.64 &     0.46 &     0.29 &   1 &   2.18~$\pm$~0.07   &  C             & --            \\
P28    &     -2.7 &     9.6 &  20.94 &   -0.22~$\pm$~0.24  &     0.84~$\pm$~0.16  &     0.67 &     0.46 &     0.26 &   1 &   2.30~$\pm$~0.06   &  C             & --            \\
P29    &     -5.0 &     0.4 &  20.87 &    0.75~$\pm$~0.39  &     0.15~$\pm$~0.08  &     0.70 &     0.49 &     0.35 &   1 &   2.39~$\pm$~0.06   &  C             & --            \\
P30    &    -22.3 &   -26.9 &  20.84 &    0.67~$\pm$~0.23  &     0.82~$\pm$~0.11  &     0.69 &     0.54 &     0.35 &   1 &   2.41~$\pm$~0.05   &  C             & --            \\
P31    &     -0.7 &   -11.0 &  20.82 &    0.62~$\pm$~0.20  &     0.80~$\pm$~0.11  &     0.70 &     0.49 &     0.38 &   1 &   2.42~$\pm$~0.05   &  C             & --            \\
P32    &      8.2 &    -3.6 &  20.35 &    0.53~$\pm$~0.11  &     0.88~$\pm$~0.04  &     0.69 &     0.50 &     0.38 &   1 &   2.43~$\pm$~0.03   &  C             & --            \\
P33    &     -6.9 &     1.5 &  20.69 &    0.31~$\pm$~0.16  &     0.62~$\pm$~0.07  &     0.84 &     0.35 &     0.36 &   1 &   2.44~$\pm$~0.08   &  C             & --            \\
P34    &     -3.3 &    -3.9 &  20.83 &    0.33~$\pm$~0.26  &     0.87~$\pm$~0.14  &     0.80 &     0.60 &     0.37 &   1 &   2.75~$\pm$~0.07   &  C             & --            \\
P35    &     -5.4 &    -1.0 &  20.62 &    0.73~$\pm$~0.09  &     0.75~$\pm$~0.11  &     0.89 &     0.50 &     0.37 &   1 &   2.84~$\pm$~0.07   &  C             & --            \\
P36    &     -1.5 &    -3.6 &  20.87 &    0.64~$\pm$~0.20  &     0.86~$\pm$~0.14  &     0.83 &     0.68 &     0.40 &   1 &   2.95~$\pm$~0.04   &  C             & --            \\
P37    &     -5.5 &    -4.1 &  20.48 &    0.78~$\pm$~0.09  &     0.75~$\pm$~0.11  &     0.80 &     0.64 &     0.45 &   1 &   2.96~$\pm$~0.01   &  C             & --            \\
\\                                                                                               
UV7    &    -10.0 &     3.1 &  20.56 &   -0.03~$\pm$~0.10  &     0.81~$\pm$~0.10  &     0.08 &     0.22 &     0.00 &   2 &   0.54~$\pm$~0.12   &  $\sim$S [$NUV,W2$] & Recovered from infrared         \\
            \hline
	\end{tabular}
\vspace{-0.23cm}
\begin{flushleft}
Proper motion component errors are of 0.7--3.1 mas~yr$^{-1}$. The $griz$-band photometric (P) colours are from PS1 (1), NSC (2), and SDSS (3). SED profiles based on ultraviolet, optical, and infrared photometry are classified as straight (S) or curved (C); for the former, the bandpass range is given.
\end{flushleft}
\end{table*}

\begin{table*}
	\centering
	\caption{Power-law colours for \textit{GALEX}, \textit{Gaia} DR3, SDSS $u^*g^*$, PS1 $grizy$, WFCAM $JK$, and \textit{WISE} filters. BB stands for a blackbody of $7.37\times10^5$~K.}
	\label{tab:plcolours}
	\begin{tabular}{cccccccccccccccr} 
		\hline
		$\alpha_{\nu}$ & $FUV$      &   $NUV$   &   $BP$      &  $G$     & $G$     &  $u^*$    &  $g^*$   &  $g$     & $r$     &  $i$     &  $z$    & $y$    & $J$    & $K$     & $W1$ \\
		               & $-$$NUV$   &   $-$$G$  &   $-$$G$    &  $-$$RP$ & $-$$g$  &  $-$$g^*$ &  $-$$g$  &  $-$$r$  & $-$$i$  &  $-$$z$  &  $-$$y$ & $-$$J$ & $-$$K$ & $-$$W1$ & $-$$W2$ \\
	       & (mag)  &  (mag)  &  (mag)  & (mag)  & (mag)  &  (mag)  &  (mag)  &  (mag)  &  (mag)  & (mag)  & (mag)  &  (mag)  &  (mag)  & (mag)  & (mag)  \\
		\hline
		 BB  &     -0.81 &    -1.82 &    -0.21 &    -0.42 &     0.20 &    -0.58 &    -0.06 &    -0.54 &    -0.43 &    -0.31 &    -0.23 &     0.35 &    -0.27 &    -0.10 &    -0.05   \\
	    -2.0 &     -0.84 &    -1.84 &    -0.21 &    -0.42 &     0.20 &    -0.59 &    -0.06 &    -0.55 &    -0.43 &    -0.31 &    -0.23 &     0.35 &    -0.27 &    -0.11 &    -0.05   \\
	    -1.5 &     -0.63 &    -1.37 &    -0.14 &    -0.24 &     0.14 &    -0.44 &    -0.05 &    -0.41 &    -0.32 &    -0.24 &    -0.17 &     0.49 &     0.04 &     0.11 &     0.13   \\
	    -1.0 &     -0.42 &    -0.88 &    -0.07 &    -0.07 &     0.06 &    -0.30 &    -0.03 &    -0.27 &    -0.22 &    -0.15 &    -0.11 &     0.63 &     0.34 &     0.34 &     0.29   \\
	    -0.5 &     -0.22 &    -0.38 &     0.01 &     0.10 &    -0.03 &    -0.15 &    -0.02 &    -0.13 &    -0.11 &    -0.08 &    -0.05 &     0.77 &     0.65 &     0.57 &     0.46   \\
		 0.0 &      0.00 &     0.13 &     0.10 &     0.24 &    -0.13 &     0.00 &     0.00 &     0.00 &     0.00 &     0.00 &     0.00 &     0.92 &     0.95 &     0.80 &     0.64   \\
		 0.5 &      0.22 &     0.65 &     0.20 &     0.38 &    -0.24 &     0.15 &     0.02 &     0.13 &     0.11 &     0.08 &     0.05 &     1.06 &     1.26 &     1.03 &     0.81   \\
		 1.0 &      0.44 &     1.18 &     0.30 &     0.51 &    -0.37 &     0.30 &     0.04 &     0.27 &     0.21 &     0.15 &     0.12 &     1.19 &     1.58 &     1.25 &     0.98   \\
		 1.5 &      0.65 &     1.73 &     0.41 &     0.63 &    -0.51 &     0.45 &     0.05 &     0.40 &     0.32 &     0.23 &     0.17 &     1.34 &     1.88 &     1.48 &     1.16   \\
		 2.0 &      0.87 &     2.29 &     0.53 &     0.73 &    -0.67 &     0.61 &     0.06 &     0.54 &     0.42 &     0.31 &     0.22 &     1.49 &     2.18 &     1.72 &     1.32   \\
		 2.5 &      1.10 &     2.85 &     0.65 &     0.83 &    -0.83 &     0.76 &     0.08 &     0.67 &     0.52 &     0.38 &     0.29 &     1.62 &     2.50 &     1.95 &     1.49   \\
		 3.0 &      1.33 &     3.43 &     0.78 &     0.91 &    -1.01 &     0.92 &     0.10 &     0.79 &     0.64 &     0.45 &     0.34 &     1.77 &     2.81 &     2.18 &     1.66   \\
		\hline
	\end{tabular}\\
\end{table*}

\begin{figure}
    \includegraphics[width=\columnwidth]{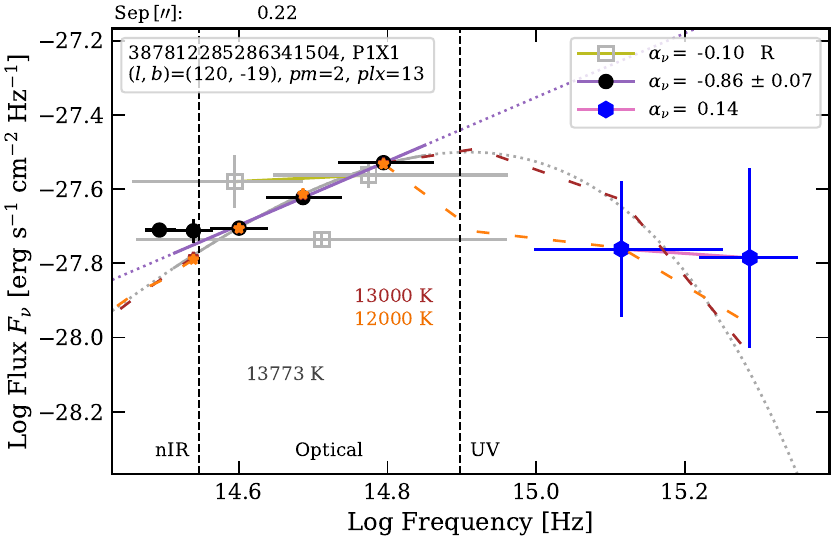}
    \includegraphics[width=\columnwidth]{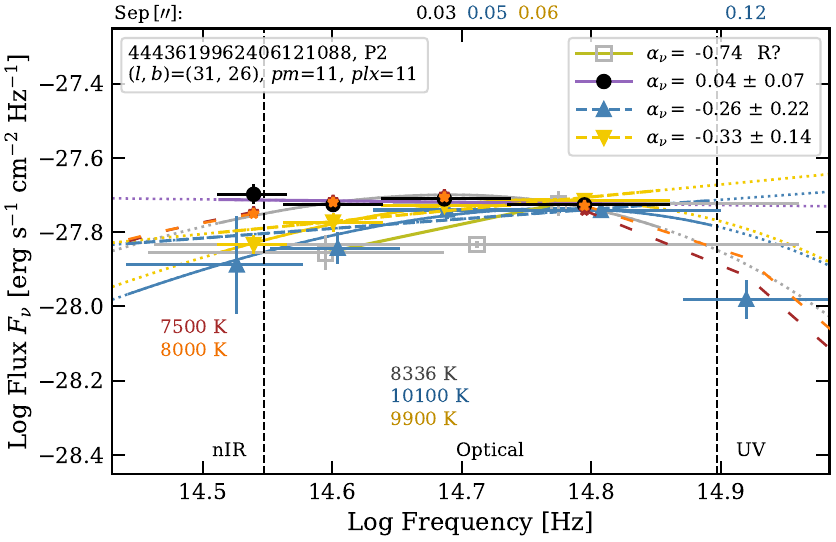}
    \caption{SEDs of P1X1 and P2, with PS1 $griz$-band power-law profiles. Same as in Fig.~\ref{fsed_uv1}.}
    \label{fsed_pl_can1}
\end{figure}

Among the PS1 $griz$-band power-law profile sources, 44 have $\alpha_{\nu}<3$ and five have $\alpha_{\nu}>3$. We were particularly interested in the former. One has $\alpha_{\nu}<0$, five have $\alpha_{\nu}=0.0$$-$0.8, and 38 have $\alpha_{\nu}=0.9$$-$3. Their minimum and maximum $griz$-band formal magnitude errors are of (0.003, 0.003, 0.005, and 0.006) and (0.139, 0.122, 0.112, and 0.446)~mag, respectively. We checked all the candidates in archival images and catalogues. From visual inspections of the SEDs, including ultraviolet and infrared fluxes, few have overall straight profiles and most have curved profiles. For illustrative and verification purposes, and to limit redundancy, we consider here explicitly only 35 of the 44 sources with $\alpha_{\nu}<3$. These are listed in Table~\ref{tab:plcan}, some are shown in Figs.~\ref{fsed_pl_can1} and \ref{fsed_pl_can2}, and all are shown in Fig.~\ref{fsed_pl_ir}, available as online supplementary material. The five sources with $\alpha_{\nu}>3$ are Gaia DR3 2022581399133091968, Gaia DR3 4109505149313851264, Gaia DR3 2022751273695068544, Gaia DR3 1827956227460470144, and Gaia DR3 2173177211819325568. These are towards the Galactic centre and plane, at $l=[-1, 97]\degr$ and $b=[-2, 8]\degr$, and have small proper motions $pm=3$$-$13~mas~yr$^{-1}$, suggesting that these are probably more distant and their optical power-law profiles stem from interstellar extinction and reddening. The PS1 source with a negative slope is P1X1 ($\alpha_{\nu}=-0.86\pm0.07$; Fig.~\ref{fsed_pl_can1}), at $d50=81$~pc. It is also an X-ray crossmatch candidate (Section~\ref{sec:xrays}), appears very faint in \textit{GALEX} ultraviolet images, and is fainter at ultraviolet than at $g$ band. The source has the highest \texttt{ruwe} value among the sources of Table~\ref{tab:plcan} (and those of Tables~\ref{tab:all1} and \ref{tab:all2}) and is possibly a binary (see Section~\ref{disc:bin}).

The sources with $\alpha_{\nu}\approx0$$-$0.9 are outwards of the Galactic plane, at latitudes $|b|=26-65\degr$. Gaia DR3 4443619962406121088 (P2, see Fig.~\ref{fsed_pl_can1}) has PS1 $griz$-band $\log F_{\nu}$ fluxes that are relatively close to one another, within 0.03 dex ($\alpha_{\nu}=0.04\pm0.07$). However, the NSC and SDSS profiles are slightly curved, with an $u$-band flux that is suppressed relative to the power-law and blackbody fits, suggesting a white-dwarf type. It is unclear whether, and to what extent, the difference between the flat and curved profiles is related to P2 possibly being a binary.

Among the sources of Table~\ref{tab:plcan}, Gaia DR3 3257308626125076480 (P15; $\alpha_{\nu}=1.48\pm0.09$) is a white dwarf photometric candidate in \citetalias{2021A&A...649A...6G}, with $P_\mathrm{WD}=0.81$, whereas the other sources have $P_\mathrm{WD}<0.35$ and a mean $<P_\mathrm{WD}>=0.04\pm0.08$. In \citet{2021MNRAS.508.3877G}, Gaia DR3 4125016578273139456 (P27; $\alpha_{\nu}=2.18\pm0.07$) and Gaia DR3 1833552363662146944 (P36; $\alpha_{\nu}=2.95\pm0.04$) are probable white dwarf photometric candidates, with $P_\mathrm{WD}=0.98$$-$1.00. None of the three sources is reported in Simbad. Six sources of Tables~\ref{tab:plcan} and \ref{tab:all1} are in \citet{2021MNRAS.508.3877G}, and P27 is the only source with a $G$-band \texttt{excess\_flux\_error} $>$ 4 (=7.404), indicating that it could be variable. 

Some sources have photometry affected by blends with second components and that might contribute to the $griz$ power-law profile. For example, Gaia DR3 4149950169090938240 (P20; $\alpha_{\nu}=1.8$) at $d50=48$~pc. It has a tight second component 0.37~arcsec north-east, Gaia DR3 4149950169090938112, which is 0.2~mag brighter at $G$ band, has no parallax nor proper motion (Table~\ref{tab:all2}), and appears as the slightly bluer part in their blend in the PS1 images. Both components are measured as a single source in the PS1 catalogue. The second component might be the cause of the high values of photoastrometric binarity parameters for P20, such as a $\texttt{ruwe}$ of 1.6 (see Tables~\ref{tab:all1} and \ref{tab:mul}). We define the parameters and discuss binarity further in Section~\ref{disc:bin}. Furthermore, the tight visual binary is located towards the crowded Galactic Centre, suggesting that if the distance is underestimated, the power-law profile might also be caused by reddening. Another case is Gaia DR3 309921197948007936 (P19; $\alpha_{\nu}=1.7$; $d50=67$~pc). Although, it is a well isolated, in a field at $(l, b)=(128, 33)\degr$ far from the Galactic plane, large parts of its red (and infrared) fluxes stem for a second component 1.0~arcsec north-west. Both components are almost resolved in the PS1 colour images and identified as a single source in the PS1 catalogue. In the case of Gaia DR3 3948582104883237632 (P18; $\alpha_{\nu}=1.6$), the photometry might be affected by its 5.1~mag brighter M-type neighbour, located 5.3~arcsec south (Gaia DR3 3948582104883968384); we found that these two objects form a large proper-motion, 886~au-wide stellar binary at 168~pc (Appendix~\ref{sec:comovingwidebinary}).

\begin{figure}
    \includegraphics[width=\columnwidth]{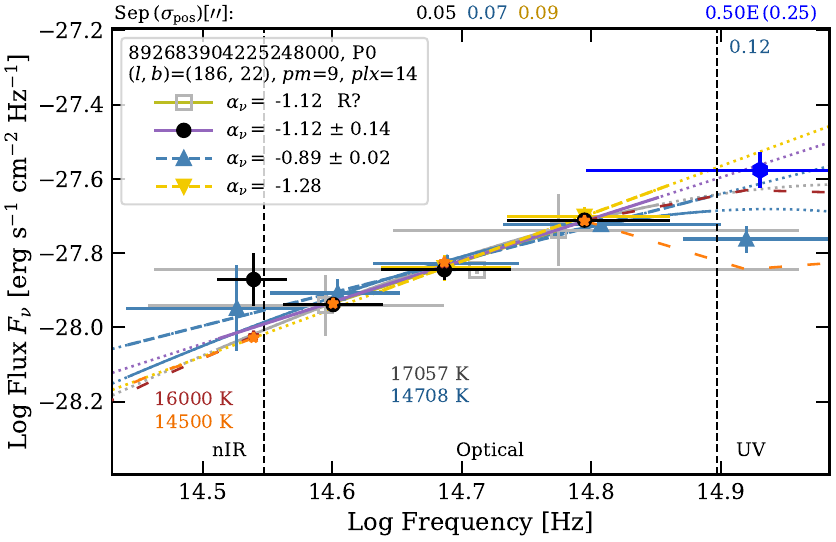}
    \includegraphics[width=\columnwidth]{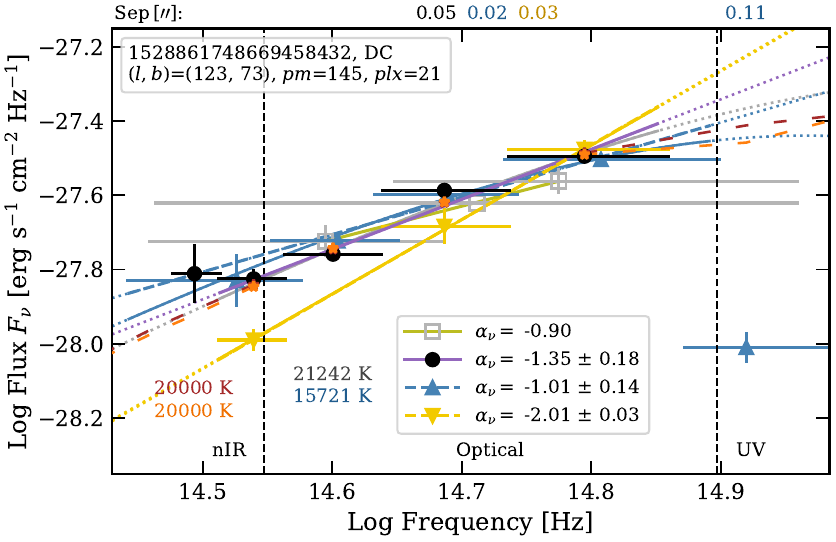}
    \caption{{\sl Top panel}: SED of P0, an SDSS $griz$-band power-law candidate. The $u$ and \textit{Swift}/UVOT $U$ bandpasses have their upper frequency-limit at $10^{15.0}$~Hz. {\sl Bottom panel}: SED of Gaia~DR3~1528861748669458432, an example of large proper-motion DC white dwarf with a $griz$-band flux profile that is almost straight (small $\alpha_{\nu}$ errors), peaking at $g$ band, and with significantly suppressed $u$-band flux. Same as in Fig.~\ref{fsed_uv1}.}
    \label{wd_blue}
\end{figure}

Five PS1 sources, P2, Gaia DR3 3910666820788919936 (P5), Gaia DR3 4486941991858372352 (P12), P15, and P19 have SDSS $u$-band magnitudes with uncertainties smaller than 0.5~mag in the range 0.12--040~mag. The SEDs (Figs.~\ref{fsed_pl_can1} and \ref{fsed_pl_ir}) indicate that the SDSS $u$-band fluxes are not only below the extrapolations of the optical power-law fits but also 0--3$\sigma$ below the extrapolations of the blackbody fits. We noted that 20~arcsec south of P5, there is a point-like radio source of 45.1$\pm$5.9~mJy at 150~MHz (TGSSADR J112929.6+073719), and 5.97$\pm$0.154~mJy (FIRST J112929.7+073718) and 8.9$\pm$0.5~mJy (NVSS 112929+073715) at 1.4~GHz, which coincides with the very faint mid-infrared source CWISE J112929.68+073717.9 ($W1W2\sim17$~mag). Gaia DR3 3622634116211982976 (P6) has an NSC $u$-band flux above the PS1 power-law fit extrapolation (Fig.~\ref{fsed_pl_can2}); however, its NSC $ugriz$-band fluxes are higher overall and have a wide spread, suggesting variability or other observational aspects. The source is well isolated in optical and infrared images, has a very small proper motion, and might be nearby or extragalactic.

Similarly, as for the PS1 sources, we proceeded with the power-law fitting of the 135 SDSS and 1195 NSC sources that have PSF and aperture $griz$-band magnitudes, respectively, with magnitude uncertainties smaller than 0.5~mag. Four SDSS sources have $\alpha_{\nu}$ errors smaller than 0.1. After visual check of their SEDs, we kept two as candidates, P19 and Gaia DR3 892683904225248000 (P0; $\alpha_{\nu}=-0.89\pm0.02$; Fig.~\ref{wd_blue}). Even though P0 has a $z$-band excess relative to the power-law fit, it has an $u$-band flux almost on par with the $g$-band flux. Based on the finding charts that we generated from survey images at different wavelengths, P0 is very bright and a point-like source in $U$ and $UVW1$-band images from the \textit{Swift} Ultraviolet and Optical Telescope \citep[UVOT;][]{2005SSRv..120...95R}. An $U$-band flux measurement (and $U_\mathrm{AB}=20.34\pm0.11$~mag) is provided for this source in the \textit{Swift}/UVOT Serendipitous Source Catalogue \citep{2014Ap&SS.354...97Y}, and we also represent it in the SED of P0. It indicates an even higher ultraviolet flux and possibly some ultraviolet variability. If the emission of P0 is thermal, then it would be relatively hot ($\sim$15\,000~K). We noted that about 2~arcsec west of P0, there is a very small, faint, and slightly red nebula that could be the origin of the mid-infrared $W1W2$-band fluxes. There are 41 NSC sources with power-law $\alpha_{\nu}$ errors smaller than 0.1. These include the P5 source. After visual check of their SEDs, we noted and concluded that most of the PS1, SDSS, and NSC optical power-law profile sources fall into two categories. Either these have slightly curved profiles with blue slopes of $\alpha_{\nu}<1$, and suppressed $u$-band flux, which can be ascribed to white dwarfs. As an indicative example, the SED of the Gaia DR3 1528861748669458432 DC white dwarf, with power-law slope errors of 0.1--0.2, is shown in Fig.~\ref{wd_blue}. Or these have straight profiles with red slopes of $\alpha_{\nu}>1$, and the sources are towards the plane and centre of the Galaxy, which suggests these sources are reddened and distant.

We also searched for power-law profiles among the 28 sources with hybrid catalogue coverage of the $griz$ bands, that is with (i) at most three of these bands covered in any of the PS1, SDSS, and NSC catalogues, and (ii) all four bands covered when combining these catalogues. We did not found any other interesting or unambiguous candidate.

There are 66 SDSS and 107 NSC sources with $u$-band magnitudes of uncertainties smaller than 0.45~mag. We did not find any other $ugri$-band power-law profiles among these. The sources typically have curved $griz$-band profiles with $u$-band fluxes that are either suppressed (as for the Gaia DR3 1505825635741455872 DC white dwarf, Fig.~\ref {dc_wd_op}) or in prolongation of blackbody profiles.

Interestingly, $griz$-band profiles that are straight or almost straight and with significantly negative slopes are very infrequent among the PS1-, SDSS-, and NSC sources in our subsample; there are only 14 with $\alpha_{\nu}<-0.8$ and $\sigma_{\alpha_{\nu}}<0.3$. By requiring that the straightness prolongs to the $u$ band, only one remains, P0. Neither the UV candidates with power-law components (Section~\ref{sec:uv}) nor P0 have the $\alpha_{\nu}=-2$ spectral index of a Rayleigh--Jeans tail, which would correspond to a very hot thermally emitting source ($\geq$10$^5$~K) such as a young neutron star. At wavelengths $\lambda$ greater than 1340~\AA -- the minimum wavelength of the \textit{GALEX} $FUV$ bandpass, the Rayleigh--Jeans approximation is valid if $T_\mathrm{eff}$ is equal or greater than for example $7.37\times10^5$~K (a comparison of colours is given in Table~\ref{tab:plcolours}), when $hc/(\lambda k_B T_\mathrm{eff})=0.15<<1$.

\begin{figure}
    \includegraphics[width=\columnwidth]{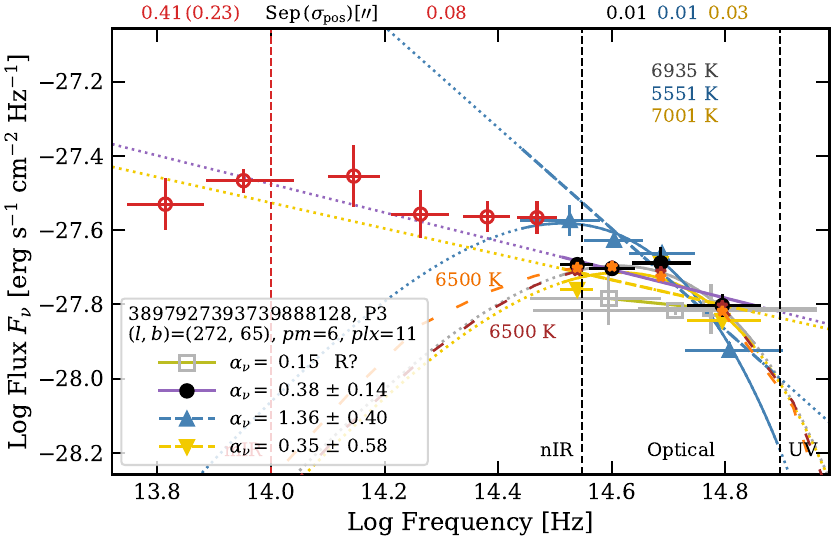}
    \includegraphics[width=\columnwidth]{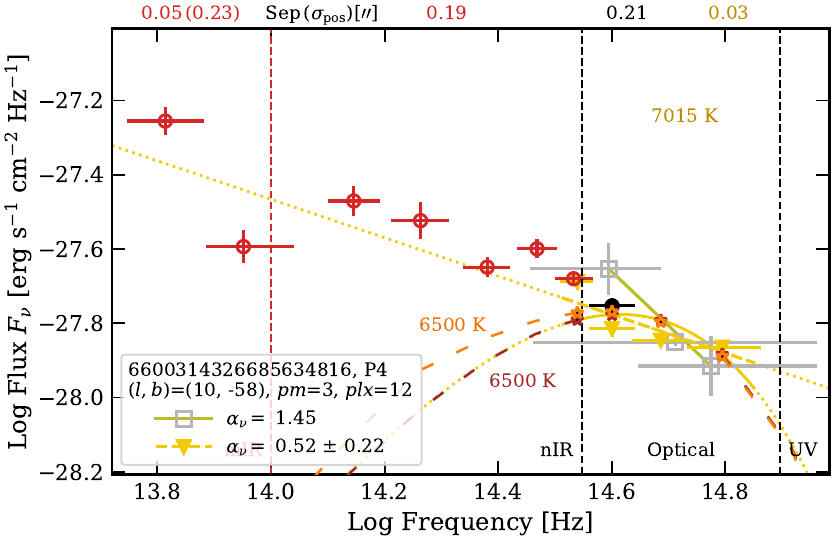}
    \includegraphics[width=\columnwidth]{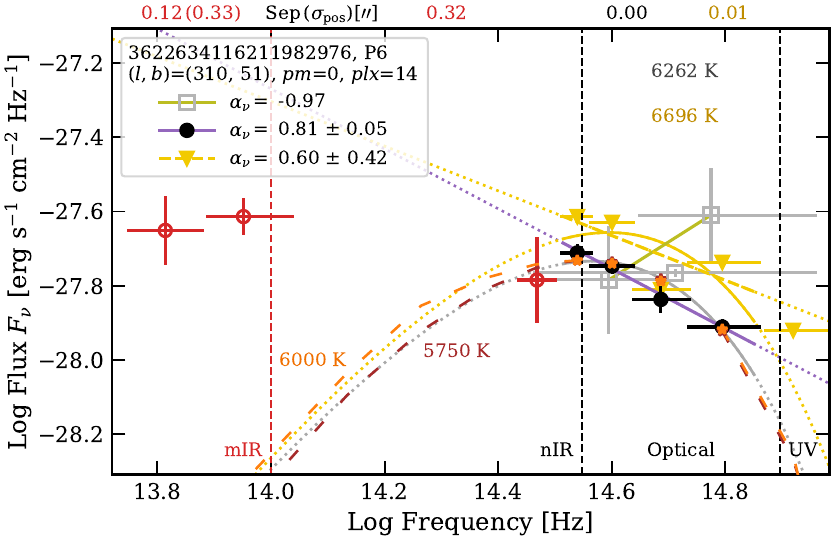}
     \caption{SEDs of P3, P4, and P6, which have approximately straight infrared--optical profiles (UV7 is shown in Fig.~\ref{fsed_uv8_uv9}). Same as in Fig.~\ref{fsed_uv1}.}
    \label{fsed_pl_can2}
\end{figure}

Finally, we searched for infrared power-law profiles among the 1936 sources that have photometry between the $z$ and $W2$ bands. First, independent of slope uncertainties, we looked at sources having infrared and optical slopes that coincide within 0.6, the optical ones being computed from photometry at at least two of the $griz$ bands. We recovered UV7 (which we discuss further in Section~\ref{disc:pl}) and P6, and we found two additional sources with approximately straight infrared--optical profiles, Gaia DR3 3897927393739888128 (P3) and Gaia DR3 6600314326685634816 (P4), whose SEDs are shown in Fig.~\ref{fsed_pl_can2}. P3 and P4 are both classified as point source and as probable quasar in the star-quasar-galaxy classification catalogue of \citet{2022A&A...668A..99H}, which is based on optical--infrared photometry and astrometry. Secondly, also independent of $\alpha_{\nu}$ uncertainties, we looked at the remaining sources having infrared slopes, with special attention to those having large proper motions or wide bandpass-coverage at the infrared, but we did not find any other candidate; many sources having overall curved profiles or mid-infrared flux excesses that can be ascribed to blends with nearby brighter sources.

For curiosity, we looked also at the sources with extremely blue or red slopes, independent of the slope uncertainties. Those with the bluest slopes are typically small-to-large proper-motion white dwarfs, having a curved flux-profile peaking at $gr$ bands and a significantly suppressed $u$-band flux, such as the profile of Gaia~DR3~1505825635741455872 (Fig.~\ref{dc_wd_op}). The five sources with the reddest slopes have M-type colours, are in sparse stellar fields, and have no indication of binarity; we describe these in Appendix~\ref{sec:serfind}, and their SEDs are shown in Fig.~\ref{fsed_superred_ir} (available as online supplementary material).

\subsection{Blueward and redward detectability of the power-law profile sources}\label{sec:plbr}

\begin{table*}
	\centering
	\caption{\textit{Gaia} DR3 parameters for the sources. The footnote is at Table~\ref{tab:all2}. \textcolor{gray}{(Preprint layout)}}
	\label{tab:all1}
	\scriptsize
	\begin{tabular}{l@{\hspace{1.2\tabcolsep}}r@{\hspace{1.2\tabcolsep}}r@{\hspace{1.2\tabcolsep}}r@{\hspace{1.2\tabcolsep}}r@{\hspace{1.2\tabcolsep}}r@{\hspace{1.2\tabcolsep}}r@{\hspace{1.2\tabcolsep}}r@{\hspace{1.2\tabcolsep}}r@{\hspace{1.2\tabcolsep}}r@{\hspace{1.2\tabcolsep}}r@{\hspace{1.2\tabcolsep}}l@{\hspace{1.2\tabcolsep}}l@{\hspace{1.2\tabcolsep}}l@{\hspace{1.2\tabcolsep}}l}
		\hline
Gaia DR3           &  RA         & Dec        & $Plx$              &  $pm$RA                &  $pm$Dec               &  gha   & fmp &  fow &  ruwe   &  ef$_{BR}$ &  $G$           &  $G_{BP}$         &  $G_{RP}$         &  Flag \\
                    &  (deg)      & (deg)      & (mas)              &  (mas~yr$^{-1}$)              &  (mas~yr$^{-1}$)              &        & (\%) & (\%) &         &            & (mag)          &  (mag)            & (mag)             &       \\
\hline
4043369078205765888  &     269.490361 &    -32.362352 &       19.4~$\pm$~2.2    &        -0.61~$\pm$~1.06     &        -5.18~$\pm$~1.08     &     0.072 &       6 &       5 &      1.515 &      0.078 &     19.863~$\pm$~0.010 &     20.425~$\pm$~0.346 &     19.032~$\pm$~0.017 &     $\gamma$\_a \\
4040994030072768896  &     265.851130 &    -35.777425 &       12.6~$\pm$~1.6    &        -3.61~$\pm$~2.50     &        -2.87~$\pm$~1.28     &     0.144 &       0 &       0 &      1.095 &     -0.103 &     20.697~$\pm$~0.009 &     21.233~$\pm$~0.444 &     20.177~$\pm$~0.106 &     $\gamma$\_b \\
5990998826943947520  &     241.492455 &    -45.738245 &       10.9~$\pm$~2.6    &        -1.70~$\pm$~2.01     &       -10.47~$\pm$~1.63     &     0.157 &       0 &       0 &      1.049 &     -0.072 &     20.822~$\pm$~0.011 &     21.598~$\pm$~0.177 &     20.064~$\pm$~0.091 &     $\gamma$\_c \\
4144738037302227072  &     267.657497 &    -17.354523 &       18.2~$\pm$~2.1    &       -19.26~$\pm$~1.25     &        -0.67~$\pm$~0.82     &     0.153 &       0 &       0 &      1.135 &     -0.005 &     20.146~$\pm$~0.011 &     20.869~$\pm$~0.133 &     19.325~$\pm$~0.072 &     $\gamma$\_d \\
2957941232274652032  &      81.037758 &    -24.498135 &       15.7~$\pm$~2.4    &         2.77~$\pm$~1.71     &         8.61~$\pm$~1.39     &     0.221 &       3 &       0 &      1.435 &      0.318 &     20.658~$\pm$~0.025 &     20.805~$\pm$~0.214 &     19.828~$\pm$~0.161 &     $\gamma$\_e \\
\\
6456175155513145344  &     315.733396 &    -58.644124 &       16.1~$\pm$~0.6    &        99.02~$\pm$~0.48     &      -164.89~$\pm$~0.58     &     0.023 &       0 &       0 &      0.996 &      0.120 &     20.512~$\pm$~0.006 &     20.695~$\pm$~0.083 &     19.949~$\pm$~0.085 &     X2   \\
5922780280824141056  &     258.890741 &    -54.828155 &       12.3~$\pm$~2.0    &         4.08~$\pm$~2.12     &        -1.71~$\pm$~1.69     &     0.161 &       2 &       0 &      1.466 &      0.296 &     20.809~$\pm$~0.020 &     21.055~$\pm$~0.131 &     19.924~$\pm$~0.099 &     X\_a \\
6027974578382721280  &     253.774144 &    -30.672080 &       21.9~$\pm$~2.1    &       -12.49~$\pm$~2.31     &        11.72~$\pm$~1.42     &     0.343 &       2 &       0 &      2.223 &      0.167 &     20.432~$\pm$~0.015 &     20.855~$\pm$~0.112 &     19.581~$\pm$~0.071 &     X\_b \\
6029621818614152576  &     254.404875 &    -29.838886 &       16.8~$\pm$~1.9    &        -6.22~$\pm$~1.60     &         3.66~$\pm$~1.29     &     0.207 &      12 &       6 &      1.568 &      0.454 &     20.280~$\pm$~0.013 &     20.322~$\pm$~0.121 &     19.369~$\pm$~0.136 &     X\_c \\
 384577010978265344  &       2.083260 &     42.597460 &       12.1~$\pm$~2.0    &        -4.13~$\pm$~1.14     &        -3.71~$\pm$~1.16     &     0.141 &       0 &       0 &      1.422 &      0.345 &     20.719~$\pm$~0.012 &     20.892~$\pm$~0.149 &     19.831~$\pm$~0.077 &     X\_d \\
5914009991961757952  &     259.742551 &    -60.530903 &       13.7~$\pm$~1.6    &        -8.12~$\pm$~1.62     &        -5.78~$\pm$~1.72     &     0.335 &       3 &       0 &      1.031 &      0.288 &     20.674~$\pm$~0.026 &     20.816~$\pm$~0.168 &     19.890~$\pm$~0.124 &     X\_e \\
\\
5386173659617460736* &     174.551694 &    -36.912644 &       10.6~$\pm$~1.7    &       -15.83~$\pm$~2.15     &        -3.88~$\pm$~2.17     &     0.362 &       3 &       0 &      1.236 &      0.429 &     20.885~$\pm$~0.026 &     20.537~$\pm$~0.128 &     20.531~$\pm$~0.151 &     UV1  \\
6829498931265990656* &     319.524762 &    -20.989340 &       17.3~$\pm$~1.5    &        -1.43~$\pm$~1.74     &        -7.08~$\pm$~1.36     &     0.428 &      30 &       0 &      1.995 &      0.543 &     20.166~$\pm$~0.018 &     19.921~$\pm$~0.080 &     19.454~$\pm$~0.096 &     UV2  \\
5777941133043108992* &     250.754760 &    -76.519108 &       12.2~$\pm$~2.0    &        -1.47~$\pm$~2.22     &         2.60~$\pm$~1.72     &     0.328 &       0 &       0 &      1.208 &      0.559 &     20.889~$\pm$~0.028 &     20.697~$\pm$~0.168 &     20.090~$\pm$~0.146 &     UV3  \\
6901051613248041216  &     307.001186 &    -11.663427 &       14.4~$\pm$~1.2    &         1.61~$\pm$~1.23     &        -9.00~$\pm$~1.07     &     0.844 &       9 &       0 &      1.769 &      0.504 &     20.235~$\pm$~0.016 &     20.129~$\pm$~0.058 &     19.406~$\pm$~0.053 &     UV4  \\
6585452983927662080  &     325.322938 &    -39.263001 &       11.9~$\pm$~3.2    &        -2.62~$\pm$~2.13     &        -2.26~$\pm$~2.45     &     0.070 &       0 &       0 &      0.998 &      0.238 &     20.901~$\pm$~0.016 &     20.845~$\pm$~0.158 &     20.427~$\pm$~0.262 &     UV5  \\
6858721824326705536  &     311.394787 &    -18.525829 &       17.3~$\pm$~1.6    &        -0.66~$\pm$~1.57     &       -13.42~$\pm$~1.07     &     0.299 &      33 &       3 &      2.048 &      0.421 &     20.119~$\pm$~0.016 &     20.080~$\pm$~0.048 &     19.330~$\pm$~0.048 &     UV6  \\
6529021095092491648  &     352.073550 &    -43.968727 &       13.4~$\pm$~1.4    &       -10.00~$\pm$~1.77     &         3.12~$\pm$~1.85     &     0.302 &       0 &       0 &      1.138 &      0.430 &     20.560~$\pm$~0.018 &     20.530~$\pm$~0.094 &     19.749~$\pm$~0.094 &     UV7  \\
3150537594577224960  &     118.499092 &     11.106838 &       13.0~$\pm$~2.3    &        -5.87~$\pm$~2.40     &        14.71~$\pm$~2.08     &     0.295 &       1 &       0 &      1.440 &      0.361 &     20.863~$\pm$~0.019 &     21.007~$\pm$~0.257 &     19.979~$\pm$~0.146 &     UV8  \\
 837736394742229248  &     168.202576 &     49.120746 &       10.5~$\pm$~3.8    &         2.31~$\pm$~1.62     &         7.81~$\pm$~1.44     &     0.036 &       0 &       0 &      1.091 &      0.267 &     20.895~$\pm$~0.013 &     21.150~$\pm$~0.146 &     20.039~$\pm$~0.127 &     UV9  \\
4240872805678917376* &     300.256515 &      2.242324 &       12.6~$\pm$~2.8    &        -6.31~$\pm$~2.73     &        -9.46~$\pm$~1.77     &     0.098 &       0 &       0 &      1.028 &     -0.215 &     20.894~$\pm$~0.018 &     21.157~$\pm$~0.166 &     21.017~$\pm$~0.304 &     UV10 \\
4438623270475014656  &     247.297403 &      5.872890 &       12.3~$\pm$~3.5    &       -11.68~$\pm$~2.22     &       -12.95~$\pm$~1.88     &     0.010 &       0 &       0 &      1.036 &      0.016 &     20.867~$\pm$~0.015 &     21.214~$\pm$~0.275 &     20.312~$\pm$~0.218 &     UV11 \\
4465431356821050496  &     247.534452 &     15.293884 &       13.7~$\pm$~2.9    &        -4.24~$\pm$~2.17     &       -14.60~$\pm$~2.29     &     0.300 &       3 &       6 &      1.418 &     -0.053 &     20.706~$\pm$~0.025 &     21.066~$\pm$~0.133 &     20.269~$\pm$~0.192 &     UV12 \\
2480784311444314880* &      23.825818 &     -4.664492 &       15.3~$\pm$~0.7    &       210.30~$\pm$~0.73     &       149.63~$\pm$~0.62     &     0.065 &       0 &       0 &      1.148 &     -0.067 &     20.191~$\pm$~0.006 &     20.670~$\pm$~0.087 &     19.664~$\pm$~0.073 &     UV13 \\
1900830823016998144  &     336.869018 &     31.233036 &       12.7~$\pm$~0.8    &       194.99~$\pm$~0.49     &        40.17~$\pm$~0.90     &     0.029 &       0 &       0 &      0.986 &      0.105 &     20.514~$\pm$~0.007 &     21.000~$\pm$~0.148 &     19.700~$\pm$~0.078 &     UV14 \\
2807663865638075264  &       8.667106 &     25.961247 &       11.6~$\pm$~1.3    &       100.22~$\pm$~1.46     &      -158.19~$\pm$~1.40     &     0.046 &       0 &       0 &      1.279 &     -0.024 &     20.750~$\pm$~0.010 &     21.326~$\pm$~0.163 &     20.060~$\pm$~0.111 &     UV15 \\
\\
 557277267992311808  &      84.762430 &     82.061317 &       11.3~$\pm$~1.1    &       -91.17~$\pm$~2.80     &        65.51~$\pm$~1.13     &     0.065 &       0 &       0 &      0.977 &      0.279 &     20.784~$\pm$~0.012 &     20.710~$\pm$~0.114 &     20.261~$\pm$~0.223 &     Rad1 \\
2185585922087166720  &     302.548637 &     54.145194 &       10.9~$\pm$~1.7    &        38.78~$\pm$~1.15     &        18.08~$\pm$~1.48     &     0.119 &       0 &       0 &      0.909 &      0.075 &     20.874~$\pm$~0.015 &     21.005~$\pm$~0.286 &     20.458~$\pm$~0.218 &     Rad\_a \\
4688982132605446784  &      12.552640 &    -72.595269 &       17.1~$\pm$~2.3    &        -2.76~$\pm$~2.40     &       -15.70~$\pm$~2.13     &     0.185 &       9 &       0 &      2.100 &      0.270 &     20.564~$\pm$~0.022 &     20.547~$\pm$~0.097 &     19.980~$\pm$~0.112 &     Rad\_b \\
6029611888623577088* &     253.732467 &    -29.553519 &       17.9~$\pm$~2.3    &       -14.08~$\pm$~1.96     &         7.17~$\pm$~1.29     &     0.294 &      14 &       6 &      2.313 &      0.484 &     20.096~$\pm$~0.017 &     20.084~$\pm$~0.129 &     19.205~$\pm$~0.049 &     Rad\_c \\
4176647926320238720  &     267.346904 &     -3.807569 &       11.9~$\pm$~3.1    &        -0.10~$\pm$~1.53     &         8.79~$\pm$~2.15     &     0.108 &       0 &       0 &      1.005 &      0.465 &     20.641~$\pm$~0.014 &     20.535~$\pm$~0.340 &     19.864~$\pm$~0.118 &     Rad\_d \\
5711114877897853696  &     117.929719 &    -22.333737 &       19.8~$\pm$~2.1    &        -0.28~$\pm$~1.75     &        -3.88~$\pm$~2.58     &     0.203 &       0 &       0 &      1.359 &      0.092 &     20.927~$\pm$~0.019 &     21.532~$\pm$~0.203 &     20.051~$\pm$~0.114 &     Rad\_e \\
2024652913390061952  &     291.838453 &     25.438382 &       12.9~$\pm$~2.0    &        -3.05~$\pm$~1.50     &        -1.69~$\pm$~2.04     &     0.055 &       0 &       0 &      1.154 &     -0.007 &     20.870~$\pm$~0.013 &     21.243~$\pm$~0.084 &     20.329~$\pm$~0.076 &     Rad\_f \\
4149153813464969856  &     266.106683 &    -14.689611 &       10.9~$\pm$~2.8    &        -0.99~$\pm$~2.14     &        -2.80~$\pm$~1.50     &     0.087 &       0 &       0 &      1.086 &     -0.016 &     20.879~$\pm$~0.016 &     21.052~$\pm$~0.209 &     20.601~$\pm$~0.256 &     Rad\_g \\
\\
 892683904225248000  &     113.064577 &     33.159563 &       14.2~$\pm$~4.3    &         6.06~$\pm$~2.35     &        -6.17~$\pm$~1.85     &     0.118 &       0 &       0 &      1.095 &      0.122 &     20.885~$\pm$~0.018 &     20.721~$\pm$~0.230 &     20.868~$\pm$~0.195 &     P0   \\
 387812285286341504  &       9.438664 &     44.117580 &       13.3~$\pm$~2.0    &        -0.24~$\pm$~1.79     &        -1.49~$\pm$~1.65     &     0.371 &       7 &       0 &      7.570 &      0.577 &     20.613~$\pm$~0.031 &     20.282~$\pm$~0.076 &     19.969~$\pm$~0.169 &     P1X1 \\
4443619962406121088* &     258.237430 &      9.819276 &       11.3~$\pm$~2.1    &        10.78~$\pm$~2.37     &        -3.68~$\pm$~2.13     &     0.173 &       0 &       0 &      1.309 &      0.199 &     20.856~$\pm$~0.014 &     20.685~$\pm$~0.074 &     20.659~$\pm$~0.108 &     P2   \\
3897927393739888128  &     180.320681 &      5.415227 &       11.1~$\pm$~4.0    &        -4.84~$\pm$~2.49     &        -3.27~$\pm$~1.32     &     0.181 &       0 &       0 &      0.991 &      0.064 &     20.815~$\pm$~0.016 &     20.905~$\pm$~0.158 &     20.479~$\pm$~0.168 &     P3   \\
6600314326685634816  &     336.849435 &    -34.658736 &       11.6~$\pm$~3.6    &        -2.57~$\pm$~1.93     &         1.41~$\pm$~1.51     &     0.064 &       0 &       0 &      1.038 &      0.184 &     20.899~$\pm$~0.019 &     21.164~$\pm$~0.189 &     20.152~$\pm$~0.165 &     P4   \\
3910666820788919936  &     172.374777 &      7.627222 &       12.5~$\pm$~3.2    &        -0.99~$\pm$~2.22     &        -2.35~$\pm$~1.31     &     0.343 &       0 &       0 &      0.926 &      0.120 &     20.840~$\pm$~0.018 &     20.702~$\pm$~0.169 &     20.772~$\pm$~0.360 &     P5   \\
3622634116211982976  &     197.465908 &    -11.356244 &       13.9~$\pm$~3.4    &         0.19~$\pm$~2.17     &         0.35~$\pm$~1.94     &     0.071 &       1 &       0 &      1.196 &      0.301 &     20.685~$\pm$~0.018 &     20.400~$\pm$~0.307 &     20.479~$\pm$~0.355 &     P6   \\
4234582018625351936* &     296.376750 &     -3.156466 &       14.0~$\pm$~1.5    &        -2.85~$\pm$~1.77     &       -15.95~$\pm$~1.07     &     0.300 &       0 &       0 &      1.387 &      0.339 &     20.501~$\pm$~0.015 &     20.665~$\pm$~0.097 &     19.634~$\pm$~0.067 &     P7   \\
5606206865714606208  &     107.879946 &    -29.719308 &       11.8~$\pm$~2.4    &         4.56~$\pm$~1.44     &        -0.51~$\pm$~2.61     &     0.051 &       0 &       0 &      1.008 &     -0.146 &     20.919~$\pm$~0.015 &     21.486~$\pm$~0.315 &     20.454~$\pm$~0.202 &     P8   \\
4085595994046913792* &     286.098462 &    -19.089506 &       21.1~$\pm$~1.2    &         2.35~$\pm$~1.10     &        -7.26~$\pm$~1.00     &     0.206 &       0 &       0 &      2.692 &      0.325 &     19.776~$\pm$~0.014 &     19.900~$\pm$~0.074 &     18.964~$\pm$~0.044 &     P9   \\
6041917107322758016* &     241.440552 &    -28.603047 &       10.8~$\pm$~2.9    &        -6.98~$\pm$~2.17     &       -10.49~$\pm$~2.16     &     0.097 &       0 &       0 &      1.083 &      0.128 &     20.847~$\pm$~0.011 &     20.948~$\pm$~0.140 &     20.375~$\pm$~0.108 &     P10  \\
1959239908639738240  &     329.732295 &     40.088230 &       11.2~$\pm$~2.4    &        -3.91~$\pm$~1.71     &       -10.12~$\pm$~1.94     &     0.113 &       0 &       0 &      1.054 &      0.256 &     20.874~$\pm$~0.017 &     20.821~$\pm$~0.165 &     20.362~$\pm$~0.251 &     P11  \\
4486941991858372352  &     263.143271 &      7.169085 &       15.5~$\pm$~2.3    &        -3.93~$\pm$~2.79     &        -7.26~$\pm$~1.56     &     0.137 &       0 &       0 &      1.135 &      0.040 &     20.670~$\pm$~0.021 &     21.163~$\pm$~0.227 &     19.945~$\pm$~0.152 &     P12  \\
4494124177899820800  &     263.879087 &     12.664908 &       10.8~$\pm$~3.9    &        -4.16~$\pm$~1.92     &         4.10~$\pm$~2.57     &     0.064 &       0 &       0 &      0.875 &      0.284 &     20.957~$\pm$~0.019 &     21.152~$\pm$~0.174 &     20.130~$\pm$~0.167 &     P13  \\
4112307735705947264  &     255.680142 &    -25.165337 &       12.6~$\pm$~1.9    &        -8.26~$\pm$~2.41     &        -1.79~$\pm$~1.49     &     0.112 &       0 &       0 &      1.093 &     -0.105 &     20.648~$\pm$~0.016 &     21.253~$\pm$~0.114 &     20.070~$\pm$~0.073 &     P14  \\
3257308626125076480  &      57.510592 &     -0.647049 &       12.2~$\pm$~3.0    &        32.98~$\pm$~1.85     &       -64.15~$\pm$~1.08     &     0.168 &       0 &       0 &      1.022 &      0.279 &     20.896~$\pm$~0.011 &     21.117~$\pm$~0.171 &     20.052~$\pm$~0.168 &     P15  \\
5648417357621486208  &     135.832152 &    -27.235757 &       16.0~$\pm$~2.2    &         3.03~$\pm$~2.87     &         4.52~$\pm$~1.68     &     0.127 &       2 &       0 &      1.486 &      0.339 &     20.691~$\pm$~0.016 &     20.817~$\pm$~0.142 &     19.852~$\pm$~0.087 &     P16  \\
5642345785702095616  &     131.226259 &    -29.453312 &       22.3~$\pm$~1.6    &        -3.90~$\pm$~0.99     &         1.29~$\pm$~1.68     &     0.387 &       3 &       0 &      2.274 &      0.354 &     20.206~$\pm$~0.012 &     20.366~$\pm$~0.054 &     19.317~$\pm$~0.033 &     P17  \\
3948582104883237632  &     187.599802 &     18.886339 &       13.5~$\pm$~1.7    &       -88.59~$\pm$~2.11     &       -15.68~$\pm$~1.15     &     0.061 &       0 &       0 &      1.040 &      0.318 &     20.781~$\pm$~0.014 &     21.021~$\pm$~0.193 &     19.873~$\pm$~0.153 &     P18  \\
 309921197948007936  &      17.766838 &     29.792370 &       17.4~$\pm$~3.0    &         3.49~$\pm$~1.61     &         7.25~$\pm$~2.26     &     0.172 &       0 &       0 &      1.201 &      0.599 &     20.770~$\pm$~0.018 &     20.550~$\pm$~0.193 &     19.942~$\pm$~0.127 &     P19  \\
4149950169090938240  &     266.931455 &    -12.899114 &       21.9~$\pm$~2.2    &       -16.46~$\pm$~1.90     &         0.67~$\pm$~1.23     &     0.390 &       5 &      18 &      1.639 &      0.306 &     20.368~$\pm$~0.030 &     20.513~$\pm$~0.148 &     19.555~$\pm$~0.089 &     P20  \\
6033410078502966784  &     255.370346 &    -26.810632 &       19.5~$\pm$~2.4    &        -3.65~$\pm$~2.32     &        -4.42~$\pm$~1.82     &     0.445 &      29 &      16 &      2.237 &      0.372 &     20.452~$\pm$~0.028 &     20.458~$\pm$~0.108 &     19.684~$\pm$~0.147 &     P21  \\
2928185797635630720  &     107.537105 &    -22.555094 &       12.6~$\pm$~1.6    &        -0.66~$\pm$~1.07     &         0.38~$\pm$~1.87     &     0.186 &       1 &       0 &      1.153 &      0.194 &     20.647~$\pm$~0.010 &     21.075~$\pm$~0.136 &     19.757~$\pm$~0.063 &     P22  \\
4137297229808521728  &     264.041050 &    -14.652376 &       12.2~$\pm$~2.6    &       -15.16~$\pm$~2.55     &        -9.24~$\pm$~2.07     &     0.050 &       0 &       0 &      1.029 &      0.359 &     20.857~$\pm$~0.016 &     20.938~$\pm$~0.147 &     20.032~$\pm$~0.112 &     P23  \\
4124968474672413440  &     264.497724 &    -15.685675 &       15.5~$\pm$~3.9    &       -10.08~$\pm$~2.76     &        -1.77~$\pm$~2.04     &     0.149 &       0 &       0 &      0.980 &     -0.067 &     20.937~$\pm$~0.016 &     21.591~$\pm$~0.389 &     20.256~$\pm$~0.110 &     P24  \\
4115860154696775168  &     258.698370 &    -21.012065 &       19.6~$\pm$~2.0    &         8.27~$\pm$~2.15     &        -0.53~$\pm$~1.52     &     0.067 &       0 &       4 &      1.197 &      0.089 &     20.756~$\pm$~0.014 &     21.401~$\pm$~0.242 &     19.858~$\pm$~0.031 &     P25  \\
4124806709024391424  &     263.996996 &    -16.524072 &       12.2~$\pm$~2.3    &        -7.74~$\pm$~2.02     &        -0.27~$\pm$~1.37     &     0.177 &       0 &       0 &      1.313 &      0.050 &     20.695~$\pm$~0.013 &     21.089~$\pm$~0.086 &     20.039~$\pm$~0.128 &     P26  \\
4125016578273139456  &     263.737777 &    -16.248787 &       13.4~$\pm$~1.3    &        -1.46~$\pm$~0.99     &         1.09~$\pm$~0.68     &     0.216 &       0 &       0 &      1.092 &     -0.083 &     20.392~$\pm$~0.009 &     21.086~$\pm$~0.073 &     19.708~$\pm$~0.054 &     P27  \\
5596729522158314752  &     122.234233 &    -29.593586 &       20.7~$\pm$~2.6    &        -2.70~$\pm$~1.99     &         9.59~$\pm$~2.82     &     0.346 &       0 &       0 &      1.287 &      0.611 &     20.948~$\pm$~0.029 &     20.725~$\pm$~0.235 &     20.107~$\pm$~0.157 &     P28  \\
4138259749157475200  &     261.579464 &    -14.165987 &       12.0~$\pm$~3.4    &        -4.96~$\pm$~2.87     &         0.44~$\pm$~2.46     &     0.001 &       0 &       0 &      1.415 &     -0.364 &     20.870~$\pm$~0.016 &     21.620~$\pm$~0.394 &     20.719~$\pm$~0.079 &     P29  \\
4297001190649825408  &     299.585459 &      6.173369 &       11.1~$\pm$~2.8    &       -22.35~$\pm$~2.39     &       -26.88~$\pm$~2.11     &     0.100 &       0 &       0 &      1.259 &      0.019 &     20.840~$\pm$~0.014 &     21.514~$\pm$~0.229 &     20.018~$\pm$~0.112 &     P30  \\
6034234441635923456  &     251.873459 &    -26.408243 &       11.5~$\pm$~2.6    &        -0.72~$\pm$~3.07     &       -11.01~$\pm$~2.72     &     0.282 &       0 &       0 &      1.246 &      0.028 &     20.821~$\pm$~0.015 &     21.446~$\pm$~0.199 &     20.018~$\pm$~0.107 &     P31  \\
4313639928323512064  &     285.198438 &     11.797166 &       17.1~$\pm$~2.6    &         8.20~$\pm$~1.57     &        -3.63~$\pm$~1.72     &     0.219 &       0 &       0 &      1.717 &      0.134 &     20.352~$\pm$~0.021 &     20.880~$\pm$~0.111 &     19.472~$\pm$~0.037 &     P32  \\
6030056675402064000  &     255.098205 &    -29.072116 &       13.1~$\pm$~2.4    &        -6.90~$\pm$~2.01     &         1.49~$\pm$~1.29     &     0.061 &       0 &       0 &      1.200 &      0.077 &     20.696~$\pm$~0.012 &     21.003~$\pm$~0.159 &     20.076~$\pm$~0.068 &     P33  \\
 423003361823693440  &       2.712613 &     59.015460 &       11.1~$\pm$~3.5    &        -3.29~$\pm$~1.30     &        -3.85~$\pm$~2.01     &     0.039 &       0 &       0 &      1.107 &      0.237 &     20.837~$\pm$~0.014 &     21.166~$\pm$~0.258 &     19.963~$\pm$~0.134 &     P34  \\
4124336221099985280  &     263.396696 &    -17.220216 &       13.2~$\pm$~2.3    &        -5.45~$\pm$~1.79     &        -0.96~$\pm$~1.09     &     0.022 &       0 &       0 &      1.180 &     -0.060 &     20.622~$\pm$~0.009 &     21.353~$\pm$~0.092 &     19.877~$\pm$~0.105 &     P35  \\
1833552363662146944  &     302.204178 &     23.678986 &       12.7~$\pm$~3.2    &        -1.46~$\pm$~1.45     &        -3.56~$\pm$~1.65     &     0.032 &       0 &       0 &      0.944 &      0.058 &     20.864~$\pm$~0.012 &     21.507~$\pm$~0.200 &     20.009~$\pm$~0.139 &     P36  \\
4108525450122867456  &     257.632522 &    -27.668680 &       16.1~$\pm$~2.3    &        -5.51~$\pm$~2.09     &        -4.09~$\pm$~1.46     &     0.303 &       0 &       0 &      1.639 &     -0.080 &     20.479~$\pm$~0.015 &     21.259~$\pm$~0.089 &     19.731~$\pm$~0.106 &     P37  \\
\\
Median DR3 6ps     &       --       &      --       &       --              &        --                 &          --               &     0.031 &     1.0 &     3.8 &      1.029 &      --    &     20               &         --           &        --            &     --   \\
Median DR3 6ps     &       --       &      --       &       --              &        --                 &          --               &     0.071 &     0.5 &     2.1 &      1.048 &      --    &     21               &         --           &        --            &     --   \\
\hline
	\end{tabular}
\end{table*}

\begin{table*}
	\centering
	\caption{\textit{Gaia} DR3 parameters for other sources of interest. \textcolor{gray}{(Preprint layout)}}
	\label{tab:all2}
	\scriptsize
	\begin{tabular}{l@{\hspace{1.2\tabcolsep}}r@{\hspace{1.2\tabcolsep}}r@{\hspace{1.2\tabcolsep}}r@{\hspace{1.2\tabcolsep}}r@{\hspace{1.2\tabcolsep}}r@{\hspace{1.2\tabcolsep}}r@{\hspace{1.2\tabcolsep}}r@{\hspace{1.2\tabcolsep}}r@{\hspace{1.2\tabcolsep}}r@{\hspace{1.2\tabcolsep}}r@{\hspace{1.2\tabcolsep}}r@{\hspace{1.2\tabcolsep}}r@{\hspace{1.2\tabcolsep}}r@{\hspace{1.2\tabcolsep}}l}
		\hline
Gaia DR3           &  RA         & Dec        & $Plx$              &  $pm$RA                &  $pm$Dec               &  gha   & fmp &  fow &  ruwe   &  ef$_{BR}$ &  $G$           &  $G_{BP}$         &  $G_{RP}$         &  Flag \\
                    &  (deg)      & (deg)      & (mas)              &  (mas~yr$^{-1}$)              &  (mas~yr$^{-1}$)              &        & (\%) & (\%) &         &            & (mag)          &  (mag)            & (mag)             &       \\
\hline
 283928743068277376  &      76.501845 &     59.055930 &       27.7~$\pm$~0.3    &        17.44~$\pm$~0.30     &      -345.56~$\pm$~0.27     &     0.045 &       0 &       0 &      1.025 &      0.139 &     19.635~$\pm$~0.004 &     19.503~$\pm$~0.048 &     19.506~$\pm$~0.079 &     OP   \\
4353607450860305024  &     251.071265 &     -4.830122 &       12.7~$\pm$~0.9    &        80.58~$\pm$~0.95     &       -45.40~$\pm$~0.75     &     0.051 &       1 &       0 &      1.042 &      0.041 &     20.522~$\pm$~0.007 &     21.171~$\pm$~0.152 &     19.686~$\pm$~0.057 &     DZ-E \\
3905335598144227200  &     185.201364 &      9.235241 &       26.7~$\pm$~0.3    &      -346.18~$\pm$~0.48     &      -388.23~$\pm$~0.28     &     0.065 &       0 &       2 &      1.006 &     -0.084 &     19.606~$\pm$~0.005 &     20.031~$\pm$~0.069 &     19.160~$\pm$~0.059 &     DC1 \\
 740483560857296768* &     152.922607 &     28.763239 &       67.9~$\pm$~0.1    &      -124.83~$\pm$~0.06     &      -713.24~$\pm$~0.04     &     0.019 &       0 &       0 &      0.997 &      0.012 &     16.547~$\pm$~0.004 &     17.352~$\pm$~0.007 &     15.649~$\pm$~0.006 &     DQpec \\
 900742499823605632  &     123.541222 &     30.313644 &       11.1~$\pm$~2.9    &        -4.49~$\pm$~2.17     &       -15.64~$\pm$~1.32     &     0.070 &       0 &       0 &      1.073 &      0.123 &     20.821~$\pm$~0.021 &     21.260~$\pm$~0.420 &     20.017~$\pm$~0.218 &     RC1 \\
4389594118549228416  &     263.707709 &      5.145780 &       10.8~$\pm$~2.8    &        -9.01~$\pm$~2.24     &        -9.89~$\pm$~1.39     &     0.094 &       0 &       0 &      1.005 &      0.153 &     20.942~$\pm$~0.016 &     21.301~$\pm$~0.287 &     20.159~$\pm$~0.128 &     RC2 \\
4780916942594976768  &      66.245093 &    -52.624239 &       11.1~$\pm$~2.6    &        29.24~$\pm$~2.44     &       -28.41~$\pm$~2.17     &     0.094 &       0 &       0 &      1.159 &      0.120 &     20.901~$\pm$~0.015 &     21.474~$\pm$~0.394 &     20.009~$\pm$~0.135 &     RC3 \\
3618788707439181440  &     208.721840 &     -8.839919 &       14.6~$\pm$~3.0    &        -8.85~$\pm$~1.90     &        10.41~$\pm$~1.81     &     0.139 &       0 &       0 &      1.012 &      0.103 &     20.861~$\pm$~0.018 &     21.421~$\pm$~0.416 &     20.000~$\pm$~0.169 &     RC4 \\
4906188219358053248  &       5.036097 &    -59.829282 &       10.7~$\pm$~2.0    &        53.79~$\pm$~2.67     &       -19.03~$\pm$~2.10     &     0.031 &       0 &       0 &      1.148 &      0.157 &     20.879~$\pm$~0.014 &     21.356~$\pm$~0.420 &     20.003~$\pm$~0.142 &     RC5 \\
6584418167391671808   &     326.985302 &    -40.591551 &       35.8~$\pm$~0.5    &       -84.11~$\pm$~0.42     &      -112.39~$\pm$~0.42     &     0.057 &       0 &       0 &      1.059 &     -0.028 &     19.959~$\pm$~0.008 &     20.949~$\pm$~0.089 &     19.024~$\pm$~0.046 &     DZQH    \\
5616323124310350976*  &     109.188531 &    -26.081808 &       33.6~$\pm$~0.2    &       -51.00~$\pm$~0.14     &       200.30~$\pm$~0.20     &     0.008 &       0 &       0 &      0.996 &     -0.008 &     18.910~$\pm$~0.003 &     20.143~$\pm$~0.040 &     17.829~$\pm$~0.012 &     VR    \\
6829498935560722560** &     319.524667 &    -20.989472 &            --         &              --           &              --           &     0.664 &      14 &      66 &        --  &      3.029 &     20.629~$\pm$~0.019 &     19.263~$\pm$~0.507 &     19.156~$\pm$~0.143 &     (UV2) \\
5777941133042341632** &     250.745035 &    -76.517860 &            --         &              --           &              --           &     0.395 &      52 &       1 &        --  &      0.494 &     16.089~$\pm$~0.014 &     16.129~$\pm$~0.004 &     15.137~$\pm$~0.004 &     (UV3) \\
6029611888623576960** &     253.732744 &    -29.553470 &            --         &              --           &              --           &     1.068 &      16 &      73 &        --  &      0.419 &     20.288~$\pm$~0.060 &     20.717~$\pm$~0.005 &     19.141~$\pm$~0.115 &     (Rad\_c) \\
3948582104883968384*  &     187.600219 &     18.884928 &        5.9~$\pm$~0.0    &       -90.56~$\pm$~0.04     &       -15.72~$\pm$~0.03     &     0.009 &       0 &       0 &      0.987 &      0.013 &     15.694~$\pm$~0.003 &     16.995~$\pm$~0.005 &     14.565~$\pm$~0.004 &     (P18) \\
4149950169090938112** &     266.931536 &    -12.899047 &            --         &              --           &              --           &     0.481 &       1 &      20 &        --  &     -0.067 &     20.188~$\pm$~0.022 &     20.720~$\pm$~0.081 &     19.612~$\pm$~0.106 &     (P20) \\
6033410082822657280** &     255.370155 &    -26.810740 &            --         &              --           &              --           &     0.642 &       8 &      33 &        --  &      0.612 &     20.404~$\pm$~0.039 &     20.278~$\pm$~0.128 &     19.466~$\pm$~0.148 &     (P21) \\
4115860154696775296** &     258.698092 &    -21.012248 &            --         &              --           &              --           &     0.114 &       0 &      30 &        --  &      0.256 &     20.985~$\pm$~0.028 &     22.030~$\pm$~0.337 &     19.713~$\pm$~0.218 &     (P25) \\
4313639928323511936** &     285.198566 &     11.797409 &            --         &              --           &              --           &     0.036 &       0 &      36 &        --  &      9.066 &     20.592~$\pm$~0.019 &     20.557~$\pm$~0.393 &     17.169~$\pm$~3.012 &     (P32) \\
\\
Median DR3  6ps     &    --       &   --       &    --              &     --                 &       --               &  0.031 &  1.0 &  3.8 &   1.029 &   --    &  20               &      --           &     --            &  --   \\
Median DR3  6ps     &    --       &   --       &    --              &     --                 &       --               &  0.071 &  0.5 &  2.1 &   1.048 &   --    &  21               &      --           &     --            &  --   \\
\hline
	\end{tabular}
\vspace{-0.23cm}
\begin{flushleft}
gha, fmp, fow, and ef$_{BR}$ stand for the \texttt{ipd\_gof\_harmonic\_amplitude}, \texttt{ipd\_frac\_multi\_peak}, \texttt{ipd\_frac\_odd\_win}, and \texttt{phot\_bp\_rp\_excess\_factor\_corrected} \textit{Gaia} DR3 parameters. * or **: 5- or 2- instead of 6-parameter solutions (6ps). For comparison, median values for \textit{Gaia} DR3 6ps sources are indicated at the bottom. Abbreviations: optically peculiar (OP), enriched DZ (DZ-E), red colour (RC), and very red (VR). Sources of interest near the main sources are indicated with the associated flag in parenthesis; except for \textit{Gaia} DR35777941133042341632, their angular separation and relative orientation are indicated in the binarity-indicator column `Vis.' in Table~\ref{tab:mul}.
\end{flushleft}
\end{table*}

The blueward and redward detectability of the power-law components of sources of interest can be assessed using the diagrams of Figs.~\ref{mg_vs_dlim1} and \ref{mg_vs_dlim2}, which show the maximum distances of average detectability for different catalogues or surveys and power-law slopes, as function of the $g$-band absolute magnitude. We represent P0, P1, P2, P3, P4, and P6 (Table~\ref{tab:plcan}) by vertical diamonds and UV3, UV4, UV7, UV8, and UV9 (Table~\ref{tab:uvcan}) by horizontal diamonds, in the plots of nearest spectral index (spectral index ranges are indicated at the bottom). Their colours are grey, blue, or red, depending on whether their power-law component was determined mostly from the optical ($griz$ bands), the ultraviolet, or the infrared, respectively. For P6, we adopt the infrared slope of $0.25\pm0.16$ in the $[y,W2]$ range, and we assume as anchor a $g$-band magnitude that is 0.15~mag brighter than that of PS1 (red symbol in $\alpha_{\nu}=0.5$ diagram). In the case of UV7, we represent it once with its slope and anchor magnitude from the optical (grey symbol in $\alpha_{\nu}=0.5$ diagram) and once with its slope as obtained when infrared- (and $NUV$-band) fluxes are also accounted for ($\alpha_{\nu}=1.25$; see Section~\ref{disc:pl} and Fig.~\ref{fsed_uv8_uv9}) and assuming an anchor magnitude that is 1.0~mag fainter ($\Delta \log F_{\nu}=-0.4$~dex) than that of NSC (red symbol in $\alpha_{\nu}=1.5$ diagram).

P1X1 (optical $\alpha_{\nu}=-0.86$) is represented in the upper right panel of Fig.~\ref{mg_vs_dlim1}. For its optical spectral index, it is indeed detected in RASS and at the limit in the $NUV$ and $FUV$ bands, but much fainter, for equivalent spectral indices of $\alpha_{\nu}=1$ and 0 at X rays and the ultraviolet, respectively (Figs.~\ref{xg_vs_mg}, \ref{x_vs_g} and \ref{uv_vs_g}). P2 (optical $\alpha_{\nu}=0.04$) is represented in the lower right panel of Fig.~\ref{mg_vs_dlim1}. It would be detectable at the limit at $NUV$ band, but there is no \textit{GALEX} coverage; however, its suppressed SDSS $u$-band flux and slightly curved SDSS and NSC $griz$-band SED peaking at $gr$ bands suggest a white dwarf profile (Fig.~\ref{fsed_pl_can1}). For its optical spectral index, it would also be detectable in RASS but it is not, suggesting that it may indeed have mostly a curved profile. Similar verifications can be done for the other sources that are represented in Figs.~\ref{mg_vs_dlim1} and \ref{mg_vs_dlim2}. Assuming unbroken power-law profiles, sources with slopes $\alpha_{\nu}>-0.5$ would be detectable at radio in NVSS, and those with $\alpha_{\nu}>1.0-1.5$ would be undetectable in RASS (as commented already in Section~\ref{sec:gaia_nsdetec}). This makes the latter type of sources also interesting for the search of nearby pulsars.

Besides this, the various colour--colour diagrams in the online supplementary material, starting with Fig.~\ref{gr_vs_ri} and based on $ugrizyJHK_\mathrm{s}W1W2$-band magnitudes, allow us to assess the colours of the sources with regard to the power-law spectral indices. The expected power-law colours involving PS1 and SDSS photometry are represented by the pink- and blue dashed lines. The assessment is relatively easy for sources having almost one single SED component at the involved bandpasses, such as UV7, P3, and P4 in the $i-W1$ versus $r-i$ (Fig.~\ref{ri_vs_iw1}) and $J-W1$ versus $r-J$ diagrams (Fig.~\ref{rj_vs_jw1}). Several of the plotted sources present a spread in their individual PS1- (pink), SDSS- (blue), and NSC (yellow) photometry (linked with solid lines), of up to 0.3--0.5~mag or more, as shown in the colour--colour diagrams involving the $g-r$ and $r-i$ (Fig.~\ref{gr_vs_ri}), $i-z$ (Fig.~\ref{gr_vs_iz}), and $z-y$ colours (Fig.~\ref{iz_vs_zy}). This could be caused by observational and calibration differences, variability, and binarity. For example, in the $z-J$ versus $i-z$ diagram of Fig.~\ref{iz_vs_zj}, several of the SDSS blue symbols connected with the blue lines to the PS1 pink symbols are to the right and lower. For some of these, this could be explained partly by the SDSS $z$-band filter being broader, redwards, than the PS1 $z$-band filter, implying that, in principle, for sources that are persistently red, the SDSS $i-z$ colour is redder and the $z-J$ colour is bluer than when using PS1 photometry (see e.g. the SED of UV13), and vice versa for persistently blue sources (the magnitude difference $m_1-m_2$ increases with the flux ratio $F_2/F_1$.).

\section{Discussion}

We find about two X-ray, 15 ultraviolet, one radio and about five power-law profile candidates. Their \textit{Gaia} parameters are in Table~\ref{tab:all1}, their PS1- and infrared photometry in Table~\ref{tab:opt_ir}, and SDSS and NSC photometry in Table~\ref{tab:opt_sdss_nsc}. Also, we find five isolated red-colour dwarfs with M-type colours, white dwarf distances, and no indication of binarity, whose classification depends on their true distances (Appendix~\ref{sec:serfind}). For these five objects and some other sources discussed in this study, the parameters and photometry are listed in Tables~\ref{tab:all2}, \ref{tab:opt_ir2}, and \ref{tab:opt_sdss_nsc2}.

Sources of Tables~\ref{tab:all1} and \ref{tab:all2} are represented in the following colour--absolute magnitude diagrams: $M_z$ versus $i-z$ (Fig.~\ref{mz_iz_msbd}), and $M_G$ versus $G_{BP}-G$ (Fig.~\ref{gaia_can_bpg_vs_g_cmd}), $G-G_{RP}$ (Fig.~\ref{gaia_can_grp_vs_g_cmd}), $G_{BP}-G_{RP}$ (Fig.~\ref{gaia_can_bprp_vs_g_cmd}), $G-J$ (Fig.~\ref{gaia_can_gj_vs_g_cmd}), $i-z$ (Fig.~\ref{mg_iz_msbd}), and $i-J$ (Fig.~\ref{mg_ij_msbd}). These are also represented in the colour--colour diagrams: $r-i$ versus $g-r$ (Fig.~\ref{gr_vs_ri}), $i-z$ versus $g-r$ (Fig.~\ref{gr_vs_iz}), $g-r$ versus $u-g$ (Fig.~\ref{ug_vs_gr}), $r-i$ versus $u-g$ (Fig.~\ref{ug_vs_ri}), $z-y$ versus $i-z$ (Fig.~\ref{iz_vs_zy}), $z-J$ versus $i-z$ (Fig.~\ref{iz_vs_zj}), $J-K_\mathrm{s}$ versus $i-z$ (Fig.~\ref{iz_vs_jks}), $J-H$ versus $z-J$ (Fig.~\ref{zj_vs_jh}), $i-W1$ versus $r-i$ (Fig.~\ref{ri_vs_iw1}), $J-W1$ versus $r-J$ (Fig.~\ref{rj_vs_jw1}), and $W1-W2$ versus $r-J$ (Fig.~\ref{rj_vs_w1w2}). The figures prepended with a letter are available as online supplementary material.

The SEDs comprising infrared, optical, and ultraviolet fluxes of the sources are available as online supplementary material and grouped per search category: gamma-ray (Fig.~\ref{fsed_gamma_ir}); X-ray (Fig.~\ref{fsed_x_ir}); ultraviolet (Fig.~\ref{fsed_uv_ir}); radio (Fig.~\ref{fsed_radio_ir}); close-to power-law profile (Fig.~\ref{fsed_pl_ir}); and red-colour dwarf (Fig.~\ref{fsed_superred_ir}).

\begin{figure}
    \includegraphics[width=\columnwidth]{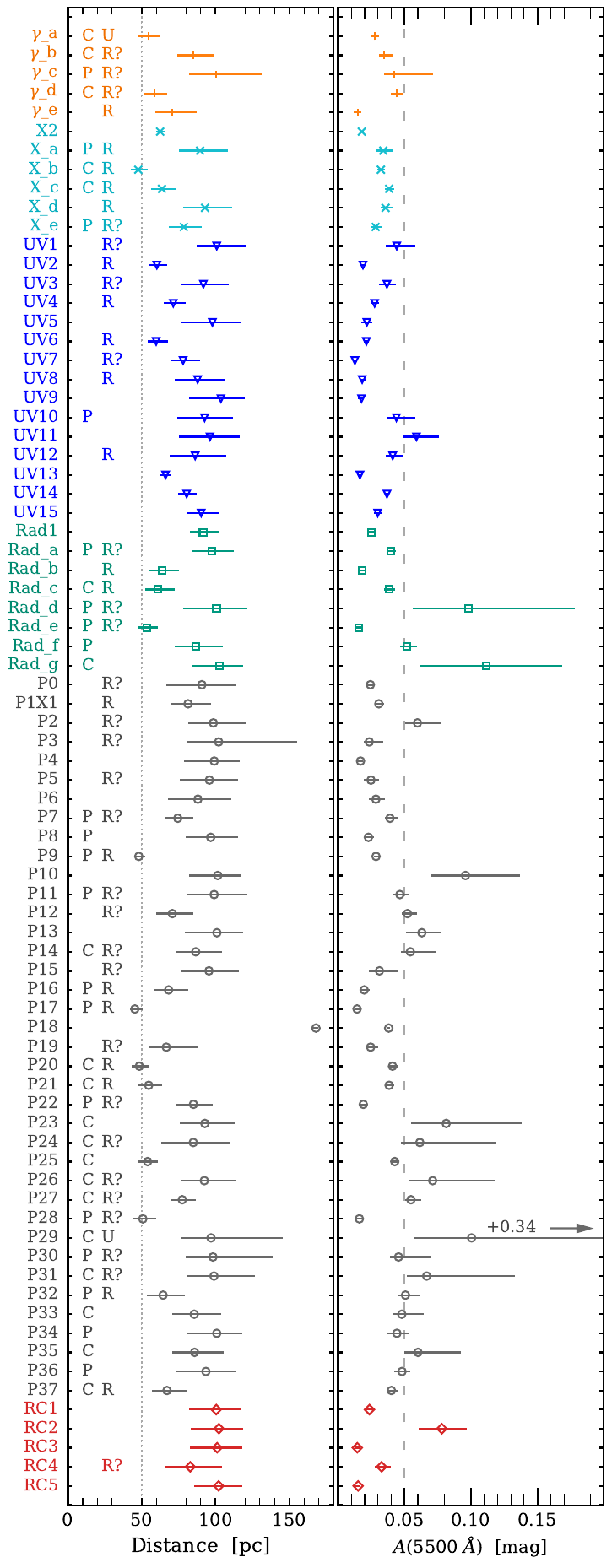}
    \caption{Distance (\textit{left-hand panel}) and cumulative $A_{5500~\AA}$ extinction (\textit{right-hand panel}) of the sources ordered as in Tables~\ref{tab:all1}, \ref{tab:all2}, and \ref{tab:mul}. Some sources are towards the Galactic plane ($|b|<15\degr$; `P') or centre ($|l|<15\degr$, $|b|<15\degr$; `C'), or have a photoastrometric indication (Table~\ref{tab:mul}) that these could be resolved (`R'), possibly resolved (`R?'), or unresolved (`U') binaries.}
    \label{extcplot}
\end{figure}

\subsection{Interstellar extinction and sky distribution}\label{disc:gen}

To characterize the sources further, we consider the amount of interstellar extinction these could have and their sky distribution.

\begin{figure*}
  \begin{minipage}[c]{0.92\textwidth}
   \includegraphics[width=\textwidth]{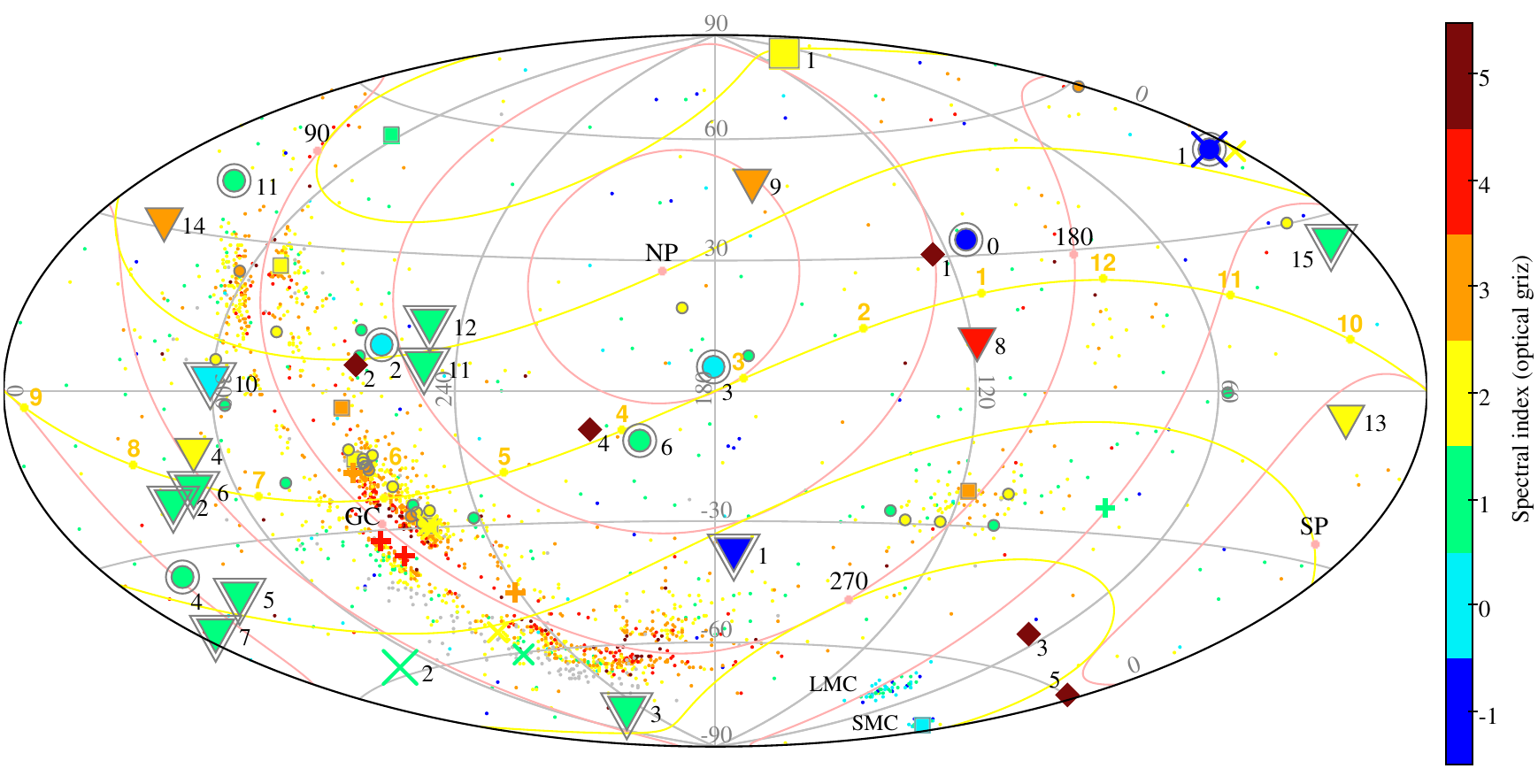}
   \includegraphics[width=\textwidth]{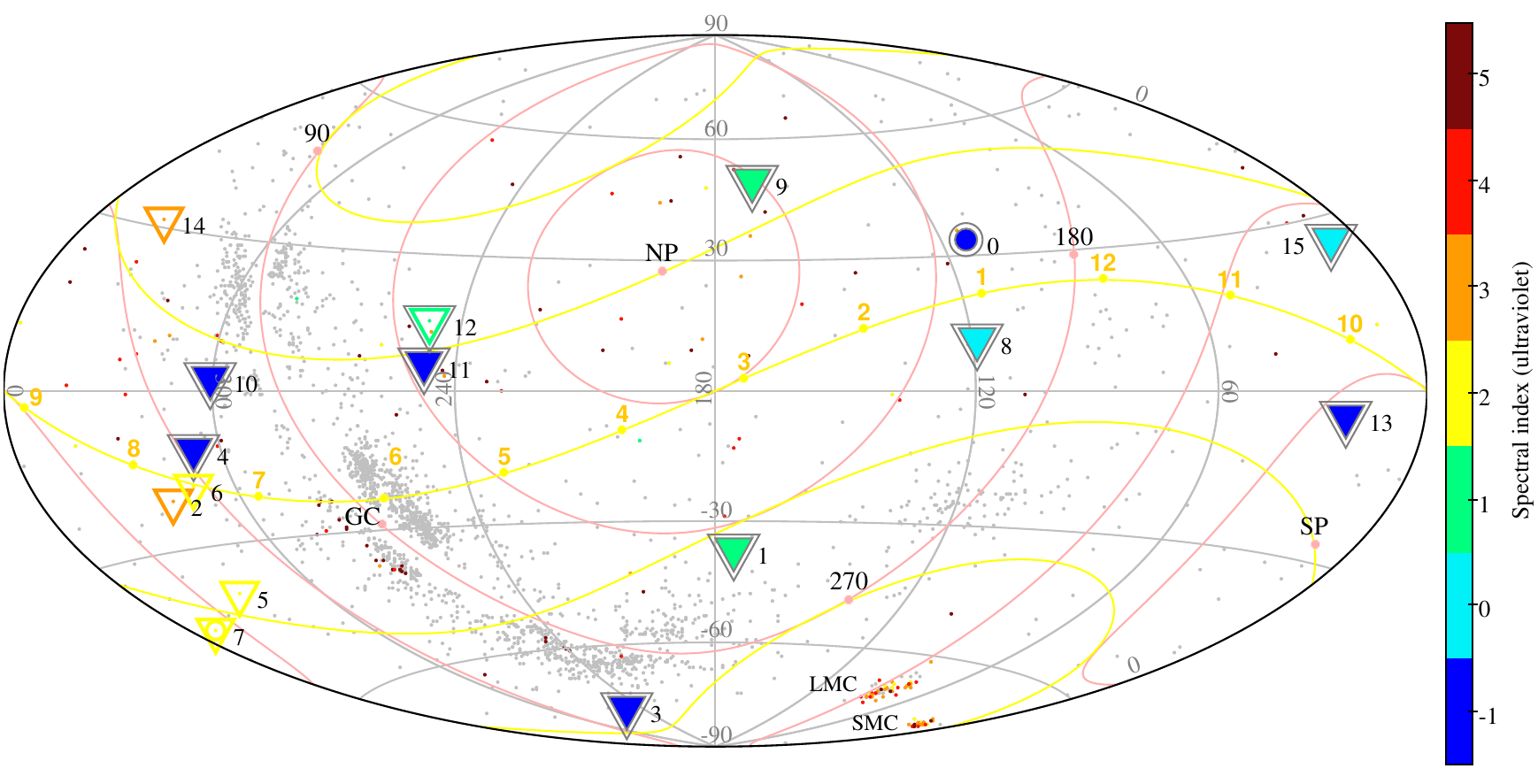}
  \end{minipage}
    \caption{Sky distribution in equatorial coordinates (Aeitoff projection). The 2464 \textit{Gaia} sources from the colour-selected subsample are represented by very small dots. Gamma-ray, X-ray, ultraviolet, radio, power-law profile, and very red-colour flagged sources are indicated by pluses, crosses, triangles, squares, circles, and lozenges, respectively. Probable candidates are represented by larger symbols and numbered. Sources are colour-coded as a function of the $\alpha_{\nu}$ spectral index (i.e slope) of their (red-)optical component at $griz$ bands ({\sl top panel}) and blue-optical--ultraviolet component ({\sl bottom panel}). The highest and lowest colour bins include spectral index values greater than 5.5 and smaller than $-$1.5, respectively. Candidates with low ultraviolet excesses are represented by open triangles in the bottom panel. Sources without available slope value are in grey colour. Numbered candidates with $\alpha_{\nu}<1.5$ are encased in a larger symbol of same shape (except X2). Galactic and ecliptic coordinate grids are shown in pink and yellow. The sky locations of the centre, north pole, and south pole of the Galaxy, and the Large and Small Magellanic Clouds are labelled as GC, NP, SP, LMC, and SMC, respectively. For each emphasized location on the ecliptic, the month (approximate mid-month date) of optimal night visibility from Earth is indicated.}
    \label{skydist}
\end{figure*}

We considered the high-resolution, spectro-photometrically calibrated, Gaia (E)DR3-2MASS 3D map of the Galactic extinction density from \citet{2022A&A...664A.174V}. We used the associated tool\footnote{\texttt{G-TOMO} at \url{https://explore-platform.eu/project/deliverables}} to retrieve the $A_{5500~\AA}$ cumulative monochromatic extinction as a function of distance, at the sky coordinates of the sources. In the left-hand and right-hand panels of Fig.~\ref{extcplot}, we show the ranges of distance and cumulative extinction for each of the sources as ordered and labelled in Tables~\ref{tab:all1} and \ref{tab:all2}. The $A_{5500~\AA}$ ranges do not include the uncertainties in the cumulative extinction, which, at the upper distance, are always smaller than 0.005~mag and thus negligible. In the left-hand panel, we indicate the sources that are towards the Galactic plane ($|b|<15\degr$; `P') or centre ($|l|<15\degr$, $|b|<15\degr$; `C'), and those that have a photoastrometric indication that these could be resolved (`R'), possibly resolved (`R?'), or unresolved (`U') binaries (see Section~\ref{disc:bin} and Table~\ref{tab:mul}). Most candidates of interest are at higher Galactic latitudes ($|b|>15\degr$) and their $A_{5500~\AA}$ values are smaller than 0.10~mag and textsl{a priori} negligible.

Nevertheless, the distances on which the cumulative extinctions are estimated depend primarily on the parallax measurements and whether the sources are binaries. The upward-corrected distance from 79~pc ($d50$) to 168~pc of P18, which is comoving with a brighter star (Appendix~\ref{sec:comovingwidebinary}), and the white-dwarf distances of the five isolated red-colour sources (Appendix~\ref{sec:serfind}), which, if these were M-type dwarfs would be 3--9 times farther, indicate that the distances of the sources with low $snr_{\pi}$ ($<$10) should be considered with caution.

In Fig.~\ref{skydist}, we show the sky distribution of the sources in equatorial coordinates. The 2464 \textit{Gaia} sources from the colour-selected subsample are represented by very small dots. Probable candidates are represented by larger symbols and numbered as in their labels. The sources are colour-coded as a function of the spectral index or slope $\alpha_{\nu}$. At the (red-)optical ({\sl top panel}), we derived PS1, SDSS, and NSC slopes separately and based on at least two of the $griz$ bands; we adopted the one with the largest number of bands covered and smallest uncertainty. For $\gamma$\_e and X\_e, we adopted the $G_\mathrm{BP}-G_\mathrm{RP}$ slope, and for UV3, the $r-G_\mathrm{RP}$ slope. At the blue-optical--ultraviolet ({\sl bottom panel}), we adopted preferentially the slopes based on contiguous bands at the shortest wavelengths, that is $NUV$ and $FUV$, else $u$ and $NUV$, else $g$ and $NUV$, else $g$ and $u$ (derived from SDSS and NSC separately; $gU$ in the case of P0). Only one source has a slope at optical--ultraviolet of $\alpha_{\nu}<-1.5$ (UV13 at $gu$). For the red components and as measured at the $griz$ bands, $\alpha_{\nu}$ values of $-1.5$ to 6.5 correspond roughly to $T_\mathrm{eff}$ of 25\,000 to 2000~K. For the blue components and as measured at varying blue-optical--ultraviolet bands, $\alpha_{\nu}$ values of about [$-1$, 1] correspond roughly to $T_\mathrm{eff}$ of 10\,000--50\,000~K.

From Figs.~\ref{extcplot} and \ref{skydist} and Table~\ref{tab:mul}, we see that P1X1--X2, UV1--UV15, Rad1, P0--P6, and RC1--RC5 are relatively far from the direction of the Galactic plane. Differently, $\gamma$\_a--$\gamma$\_d and most of P7--P37 are towards the Galactic plane or Bulge (though P15, P18, and P19 are clear exceptions); they have red spectral slopes of $\alpha_{\nu}=2.9$$-$4.4 and 1.0--3.0, respectively. Six of P7--P37 cluster near $(l, b)=(10,~9)\degr$, towards the Galactic centre. Gaia DR3 4124968474672413440 (P24), P26, and P27 cluster within 30~arcmin (all three have $\alpha_{\nu}$ of about 2.17), followed by Gaia DR3 4137297229808521728 (P23; $\alpha_{\nu}=2.13$) at about 1.4~deg and Gaia DR3 4138259749157475200 (P29; $\alpha_{\nu}=2.39$) and P20 ($\alpha_{\nu}=1.77$) at 3 and 4~deg. Generally, the gamma-ray and reddest P sources tend to appear towards crowded stellar regions, indicating that their current trigonometric distances might be spurious and probably underestimated. Their profiles might be shaped by interstellar extinction cumulated over large distances. The optical profiles of the outward-located sources are more likely to be intrinsic.

\subsection{Neutron stars as resolved emitters or unresolved companions}\label{disc:ns}

The 2464 \textit{Gaia} sources of the colour-selected subsample have practically no gamma- nor X-ray counterparts. Even the P1X1 and X2 candidates have each at least one optical neighbour source that could be a potential counterpart to the X-ray match. Assuming neither significant interstellar extinction nor self-absorption at X rays, these two candidates appear rather faint at X-rays for their bright $G$-band absolute magnitudes to resemble known thermally emitting neutron stars (Fig.~\ref{xg_vs_mg}). Alternatively, these could be, for example, interacting binaries, consisting of a white dwarf and a compact- or degenerate companion. A low occurrence of X-ray and gamma-ray pulsars is also apparent in the ATNF data base (although it excludes accretion-powered and X-ray-millisecond pulsars): out of the 3320 pulsars, 122 (4 per cent) are non-radio pulsars, emitting pulses at infrared or higher frequencies.

\subsubsection{Power-law emission}\label{disc:pl}

\begin{figure*}
    \includegraphics[width=0.70\textwidth]{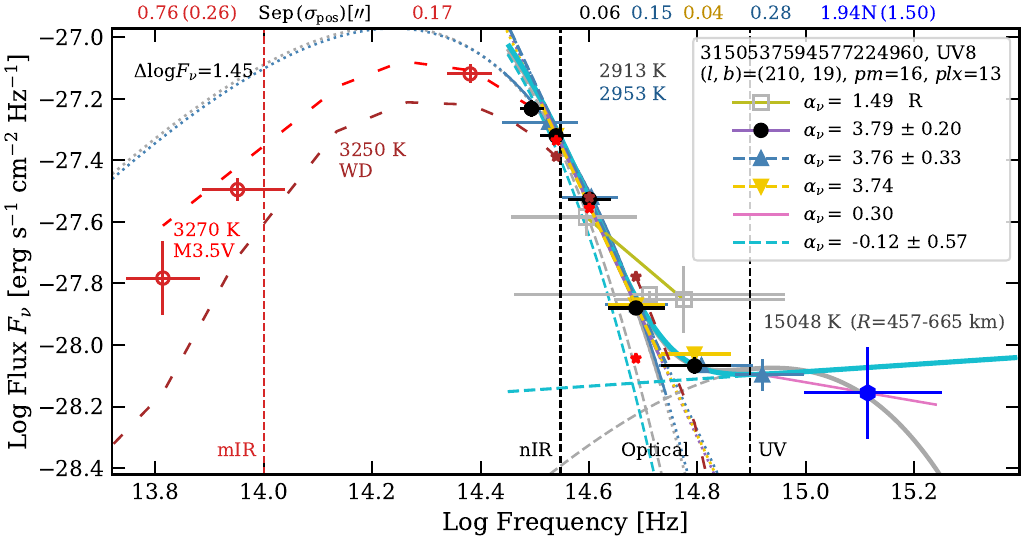}
    \includegraphics[width=0.70\textwidth]{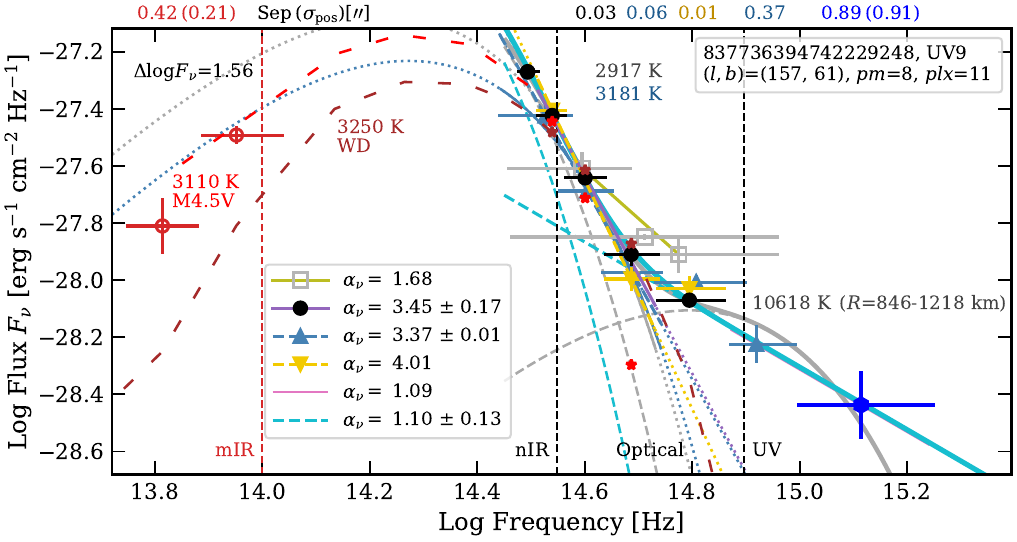}
    \includegraphics[width=0.70\textwidth]{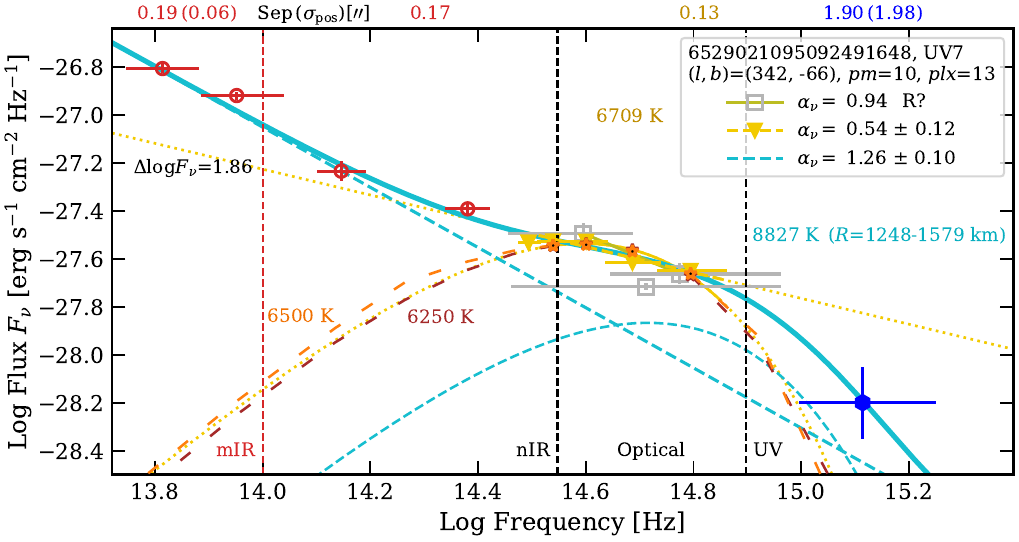}
     \caption{SEDs with ultraviolet, optical, and infrared fluxes of UV8, UV9, and UV7, from the top to bottom. Examples of two-component fits to the $NUVugriz$-band fluxes of UV8 and UV9 are shown: cool blackbody + hot blackbody (dark-grey lines) and cool blackbody + power law (cyan lines). Example fits with early M-type dwarfs (at $d\sim1000$~pc) to the red components of UV8 and UV9 are shown by the red dashed line. The fits with the coolest He-atmosphere white-dwarf model available (brown dashed line) suggests that a cooler or redder model is required to fit the optical--infrared fluxes. For UV7, an example of power law + hot blackbody fit to all of the fluxes is shown (cyan lines). Same as in Fig.~\ref{fsed_uv1}.}
    \label{fsed_uv8_uv9}
\end{figure*}

We find power-law components at the ultraviolet--optical in at least UV3 and UV4 (Figs.~\ref{fsed_uv1} and \ref{fsed_uv_ir}), UV8 and UV9 (Fig.~\ref{fsed_uv8_uv9}), and up to the mid-infrared $W2$ band in UV7 (Fig.~\ref{fsed_uv8_uv9}). We also find approximately straight profiles at the optical--infrared (up to $W2$ band) in P3, P4, and P6 (Fig.~\ref{fsed_pl_can2}), which have small proper motions of 0--6~mas~yr$^{-1}$. Finally, we find at least one possible transient power-law profile at the optical in P2 (Figs.~\ref{fsed_pl_can1} and \ref{fsed_pl_ir}).

The power-law extent to the infrared of the candidates can be contrasted with that of known pulsars. For instance, the Vela Pulsar has a flat power-law emission ($\alpha_{\nu}=0$) from the ultraviolet (10$^{15.4}$~Hz) to the near-infrared $K_\mathrm{s}$ band (10$^{14.1}$~Hz) (fig.~2 of \citealt{2013ApJ...775..101Z}). PSR B0656+14 has $JHK_\mathrm{s}$-band fluxes that tend to be along the main power-law fit \citep{2021MNRAS.502.2005Z}. The Geminga Pulsar has a power-law profile constrained up to the $H$ band and at optical, flux deviations of about 0.15~dex relative to the combined fit \citep[fig. 7 of][]{2006A&A...448..313S}. In the case of the Crab Pulsar, the power law is maintained up to the $L'$ band at the mid-infrared, and plausibly up to 8~$\micron$ depending on the flux contribution of the knot \citep{2009A&A...504..525S}. Still, not all of the pulsars detected in the optical and that we mention in this study are constrained in the infrared nor even entirely in the optical (see Section~\ref{sect:expphot} and Fig.~\ref{sfd}), mostly because of their larger distances and faintness.

Elucidating the true emission mechanisms and temporal variability of the sources is beyond the scope of the present study. However, we briefly mention some possible causes of the observed features.

In the most fortunate case(s), the power-law profile might indeed stem from synchrotron or gyro-synchrotron emission (from accelerated charged particles), as observed in pulsars \citep{1976MNRAS.177..109B,2021MNRAS.500.4530M}. The departure from a power-law profile at other wavelength ranges or epochs may indicate that:
\begin{enumerate}[label=(\roman*),align=left,itemindent=9pt]
\item the emission mechanism differs,
\item the observability of the emission mechanism differs, because of, for example, surrounding diffuse material, a binary companion, or self-absorption (cf. the synchrotron self-absorption roll-over; \citealt{2005ApJ...631..471O}),
\item the power-law emission is transient and either unique or of low occurrence (we note the pulse period of 18.18~min of the GLEAM-X~J162759.5$-$523504.3 radio transient, proposed to be a compact star; \citealt{2022Natur.601..526H}), or
\item the power-law profile is not caused by synchrotron emission but a combination of observed flux fluctuations from thermally emitting sources, unresolved or partially resolved blends of blue and red components in binaries, and reddening.
\end{enumerate}

The search for ultraviolet excesses and power-law profiles among the sources led us to identify intricate blackbody and power-law profiles, where the power-law component is substantiated by ultraviolet- or infrared measurements. This suggests that more power-law emission candidates could be found among colour-selected optical sources and using ultraviolet and infrared data. In continuation, we discuss UV7, UV8, and UV9, pushing the envelope of interpretation a bit, because these represent classes of sources that may hold neutron stars.

In the cases of UV8 and UV9 (Fig.~\ref{fsed_uv8_uv9}), we assume that (i) the two components in the SED have different origins, that is the SED divide does not result from opacity, (ii) interstellar extinction can be neglected, and (iii) the underlying $griz$-band part of the red component can be approximated by a blackbody curve. We then fit the $NUVugriz$-band fluxes ($griz$ from SDSS and PS1 for UV8 and UV9, respectively) by two components consisting of (a) a cool blackbody and a power law (cyan lines) and (b) a cool blackbody and a hot blackbody (dark-grey lines). We find power-law spectral indices of $\alpha_{\nu}=-0.12\pm0.57$ and $1.10\pm0.13$. Assuming unbroken power laws, UV8 and UV9 are expected to be detected well and marginally at X rays in RASS, as indicated in the corresponding panels in Figs.~\ref{mg_vs_dlim1} and \ref{mg_vs_dlim2}. However, these are not detected; the 2RXS exposure map at their coordinates indicate exposure times of about 357 and 439~s. An underlying, power-law component at ultraviolet--optical--infrared appears more plausible in UV8 than in UV9, because the power-law extension to the mid-infrared in UV8 is compatible with the $W1W2$ fluxes. In principle, independent of the stellar type and distance of the red component (for example, white dwarf at $\sim$100~pc or M-type dwarf at $\sim$1000~pc), the power-law component could stem from a neutron star pulsar. Alternatively, the blue component could be thermal emission and originate from a neutron star or its surroundings if its distance to us is small enough. The hot-blackbody parts of UV8 and UV9 imply effective temperatures of $T_\mathrm{eff}=15\,048$ and 10\,618~K, and, assuming the $d50$ distances, thermal radii of $R=552$ and 1060~km. These are 3.9 and 2.0 times smaller than the smallest known white dwarf radius of 2140~km (see Section~\ref{disc:bin}), which we use as a representative observational lower-limit. According to theoretical predictions of white dwarfs, including the effects of general relativity, at $T_\mathrm{eff}=10\,000$~K, a theoretical radius as small as 1000~km is only possible for 1.37~M$_{\odot}$, the highest mass for a white-dwarf and above which it becomes gravitationally unstable \citep{2022A&A...668A..58A}. Thus, taking the numbers and interpretation at face value, the blue components of UV8 and UV9 challenge the neutron star-white dwarf limit. To be less small than the smallest known white dwarf, the hot components would have to be at least 3.9 and 2.0 times farther than the $d50$ distances ($R=d\times\theta$, where $\theta$ is the constant angular radius), at $\ge$340 and 209~pc.

\begin{figure*}
\includegraphics[width=0.805\textwidth]{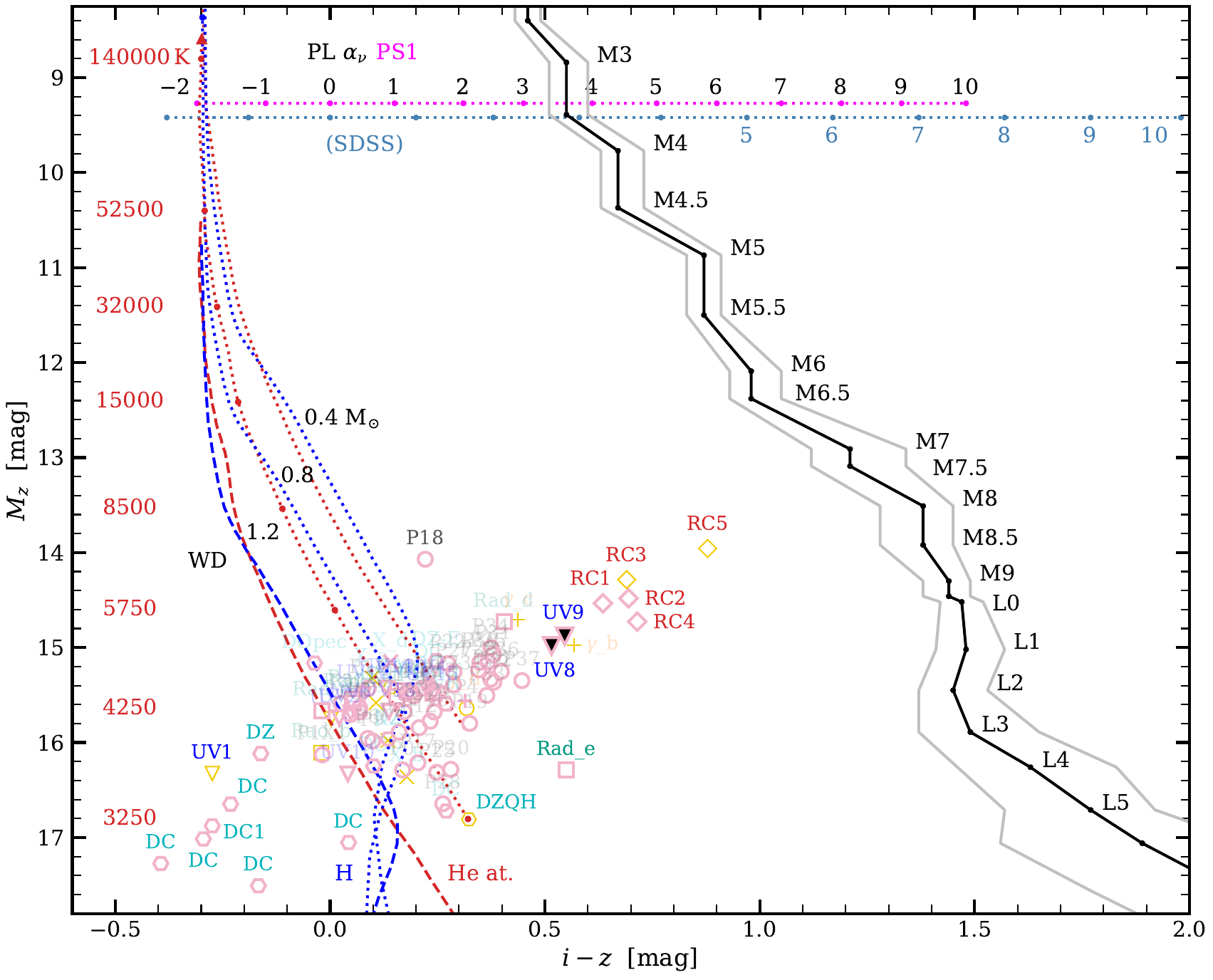}
    \caption{$M_z$ versus $i-z$ colour--absolute magnitude diagram, in PS1 photometry unless stated otherwise, and using the distances of Table~\ref{tab:mul}. Only photometry from PS1 (pink), else from NSC (yellow) are represented, for simplicity. A data point is plotted if each of the magnitude errors involved is smaller than 0.4~mag. Gamma-ray, X-ray, ultraviolet, radio, power-law profile, and red-colour flagged sources are represented by pluses, crosses, triangles, squares, circles, and lozenges and are labelled in orange, cyan, blue, green, grey, and red, respectively. Sources are labelled as in Tables~\ref{tab:all1} and \ref{tab:all2} and can be localized using the text search function in a PDF-document reader. UV8, UV9, RC1--RC5, and other outliers sources are highlighted; the symbols of UV8 and UV9 are filled in black. Spectroscopically confirmed and other literature white dwarfs mentioned in this study are represented by hexagons and labelled in turquoise. Isomasses of white-dwarf pure He and H atmospheres are represented by red and blue lines; those for 1.2~M$_{\odot}$ are dashed and those for 0.8 and 0.4~M$_{\odot}$ are dotted. Effective temperatures are indicated for the 0.8~M$_{\odot}$ He isomass. The stellar main sequence and its 68 per cent confidence limits are represented by the black and grey solid lines. Synthetic power-law colours are indicated for PS1 and SDSS photometry.
    } \label{mz_iz_msbd}
\end{figure*}

The red components of UV8 and UV9 have steep $riz$-band slopes ($\alpha_{\nu}=3.4$$-$3.8), red $i-z$ colours (0.5--0.6~mag), and curved profiles. Example fits in Fig.~\ref{fsed_uv8_uv9} suggest that these are either (i) very red and ultra-cool ($<$3250~K) white dwarfs at the \textit{Gaia} distances or (ii) early M-type dwarfs, which as shown in the $M_z$ versus $i-z$ colour--absolute magnitude diagram of Fig.~\ref{mz_iz_msbd}, would be about 5~mag brighter intrinsically. The ML-type dwarf sequence in this figure results from complementing the Modern Mean Dwarf Stellar Colour and Effective Temperature Sequence (hereafter MSC) compilation of Eric Mamajek\footnote{\url{https://www.pas.rochester.edu/~emamajek/EEM_dwarf_UBVIJHK_colors_Teff.txt}} with the MLT-type-dwarf mean PS1 colours from \citet{2018ApJS..234....1B}. For the SED fitting, we previously converted the MSC absolute magnitudes and colours to fluxes. After subtracting the blue components (power law or blackbody) from the SEDs, the red-optical slopes of UV8 and UV9 are slightly steeper than those of Gaia DR3 6584418167391671808 and Gaia DR3 5616323124310350976 \citep{2021RNAAS...5..229A}, two very red-, nearby (28--30~pc), and faint ($M_G\approx16.5$$-$17.7~mag) white-dwarf sources. The former is the coolest spectroscopically confirmed DZ white dwarf, of 3048$\pm$35~K and DZQH type \citep{2022MNRAS.517.4557E}. Their SEDs are shown in Fig.~\ref{fsed_superred_ir} of the online supplementary material. These two white-dwarf sources have $G-G_\mathrm{RP}=0.94$$-$1.08~mag and are thus slightly redder than our colour selection boundary. If UV8 and UV9 were M-type dwarfs, then these would be at much larger distances (cf. equation~(\ref{eqn:Mbol-d})), close to 1~kpc, without accounting for extinction. The low parallax signal-to-noise ratios, $snr_{\pi}$ of 5.7 (UV8) and 2.7 (UV9), and low \texttt{visibility\_periods\_used} of about 10 (Table~\ref{tab:mul}) indicate that the distances could be larger than the $d50$ ones. More accurate parallaxes or spectroscopy could solve the classification.

In the case of UV7, its power-law component (Fig.~\ref{fsed_uv8_uv9}) could be pulsar synchrotron emission. The source is well isolated, at $(l,b)=(342, -66)\degr$, has a relatively high $snr_{\pi}$ of 9.4, and a proper motion of $pm=10.5\pm2.6$~mas~yr$^{-1}$, which, given the short time-baseline of the available ground-based survey images, precludes an accurate verification of whether the source is moving transversely. The NSC $griz$-band slope of $\alpha_{\nu}=0.54\pm0.12$ is prolonged in the infrared by $JK_\mathrm{s}$-band fluxes and surpassed by about 0.3~dex by $W1W2$-band fluxes. In NSC-DES optical images, the PSF is stellar and almost circular. In near-infrared images of VHS DR6, the PSF is slightly broader than the average PSF of sources classified as stars, by a factor of about 1.5, this extension leading it to be classified as galaxy ($\texttt{pStar}=0.00$, $\texttt{pGalaxy}=0.99$). However, several pulsars are known to be in diffuse nebulae or knots. In the VISTA-DES-AllWISE (VEXAS-DESW) DR2 catalogue \citep{2021A&A...651A..69K}, its $grizyJK_\mathrm{s}W1W2$ magnitudes, colours, and VHS \texttt{pStar} parameter led it to be classified as a quasar ($\texttt{pQSO}=0.99$), among the three options, star, quasar, and galaxy. The proper motion amplitude of UV7 is 6.4 and 1.3 times the standard deviations of the proper motions of the \textit{Gaia} EDR3 crossmatches to the sources classified as quasars and galaxies in VEXAS-DESW (see table~7 of \citealt{2021A&A...651A..69K}), respectively, and suggest that UV7 could be classified as an extended AGN. However, its parallax is 17.8 and 10.0 times the standard deviations of the parallaxes of the two subsets. Unlike the SEDs of the P3 and P4 quasar candidates (Fig.~\ref{fsed_pl_can2}), the SED of UV7 appears to have a more obvious, outlined bulk at the optical. We note also that the bulk of UV7 has an $r$-band flux dip, as for Rad1 and Rad\_b, which in the description of Rad1, we proposed it could be caused by significant Na~\textsc{i}~D absorption (Section~\ref{sec:radio}). If we fit all of the fluxes by a power law + hot blackbody, assuming that the nominal value of the $NUV$-band flux is the true value, we obtain a power-law component of $\alpha_{\nu}=1.25\pm0.10$ and a blackbody of $T_\mathrm{eff}=8797$~K, and, assuming the $d50$ distance, a thermal radius of $R=1387$~km. With $\alpha_{\nu}>1.0$, UV7 is expected to be undetectable at X rays in RASS (see Fig.~\ref{mg_vs_dlim2} and Sections~\ref{sec:gaia_nsdetec}~and~\ref{sec:plbr}).

For illustrative purpose and for comparison with the ultraviolet--optical(--infrared) fluxes of known neutron stars, we represent the power-law component fits of $\alpha_{\nu}=-0.12$ and 1.25 of UV8 and UV7 by the grey dashed lines in the apparent- (top) and absolute (bottom) SED representations of Fig.~\ref{sfd}. These have relatively bright apparent fluxes, close to those of PSR~B0540$-$69, justifying the broad-optical search for synchrotron emission of pulsars in the solar neighbourhood.


Finally, we remark that the optical colours and faint magnitudes at which we searched for synchrotron emitters overlap with those expected for isolated stellar-mass black holes accreting interstellar matter \citep{2003ApJ...596..437C,2021ApJ...922L..15K}.

\subsubsection{Multiplicity}\label{disc:bin}

Knowing whether sources are isolated or binaries is important for searches of continuous gravitational wave emission, because binarity implies more modulations in the signal. The Gaia DR3 measurements for $G\ge19$$-$20~mag sources are single-source solutions and do not provide the masses of the components and the orbital elements. However, some of these measurements can give indications of binarity. In our search, we consider three types of binarity indicators: (i) \textit{Gaia} photoastrometric parameters, (ii), visual close neighbours or companions, and (iii) excesses or two components in SEDs.


(i) Useful \textit{Gaia} photoastrometric parameters for binarity are the parameters: \texttt{ruwe}, \texttt{ipd\_gof\_harmonic\_amplitude}, \texttt{ipd\_frac\_multi\_peak}, and \texttt{ipd\_frac\_odd\_win} \citep{2021A&A...649A...1G}. We paraphrase here briefly their definitions. \texttt{ruwe} is the renormalised unit weight error, for the astrometry; a value of \texttt{ruwe}~$\ga1.4$ is indicative of an ill-behaved astrometric solution and may indicate binarity or multiplicity. The \texttt{ipd\_gof\_harmonic\_amplitude} parameter measures the amplitude of the variation in the Image Parameters Determination (IPD) goodness of fit (GoF; reduced chi-square) as function of the position angle of the scan direction; the amplitude indicates the level of asymmetry of an image and a large amplitude indicates that the source is double (the associated \texttt{ipd\_gof\_harmonic\_phase} parameter indicates the orientation of the asymmetric image). The \texttt{ipd\_gof\_harmonic\_amplitude} parameter allows for the identification of spurious solutions, where values greater than 0.1 and a \texttt{ruwe} greater than 1.4 indicate resolved doubles, which are not correctly handled yet in the processing. \texttt{ipd\_frac\_multi\_peak} provides the percentage of windows (having a successful IPD result) for which the IPD algorithm has identified a double peak, meaning that the detection may be a visually resolved double star (either an optical pair or a physical binary). \texttt{ipd\_frac\_odd\_win} provides the percentage of transits having either truncation or multiple gates flagged in one or more windows and in general a non-zero value indicates that this source may be contaminated by another nearby source. In addition, there is the \texttt{duplicate\_source} parameter, for \textit{Gaia} pairs of sources closer together than 0.18~arcsec, for which only one source is retained in \textit{Gaia} DR3. Except Gaia DR3 5777941133042341632 (see below), none of the sources we list in Tables~\ref{tab:all1} and \ref{tab:all2} is flagged as duplicate.

In the bottom rows of Tables~\ref{tab:all1} and \ref{tab:all2}, we indicate for comparison the median values for sources with six-parameter solutions (accounting for the position, parallax, proper motion, and astrometrically estimated effective wavenumber) at $G=20$ and 21~mag\footnote{In section~4.5.2, Properties of the astrometric data, from Lennart Lindegren, in the \textit{Gaia} (E)DR3 documentation at \url{https://gea.esac.esa.int}.}. Five-parameter solutions account for the position, parallax, and proper motion, and two-parameter solutions account only for the position.

We assumed sources with (a) \texttt{ipd\_gof\_harmonic\_amplitude} $<$ 0.1 and \texttt{ruwe} $\ge$ 1.4 as unresolved, (b) \texttt{ipd\_gof\_harmonic-} \texttt{\_amplitude} $\ge$ 0.1 and \texttt{ruwe} $\ge$ 1.4 as resolved, and (c) \texttt{ipd\_gof\_harmonic\_amplitude} $\ge$ 0.1 and \texttt{ruwe} $<$ 1.4 as possibly resolved binaries. These are referred to by the abbreviations U, R, and R?, respectively, in the binary indicator column `Ast.' in Table~\ref{tab:mul}.
 
The \textit{Gaia} photoastrometric binarity criteria may be useful for identifying pairs that are too close ($\le$0.3~arcsec) to be resolved visually in the ground-based survey images. Most of our sources have $pm/plx$ ratios lesser than one (see Table~\ref{tab:mul}, where for simplicity the ratio error is computed from the linear sum of the relative errors in $pm$ and $plx$), implying that the astrometric motion is dominated by parallactic motion (uncertainty) or binarity. Exceptions with relatively large proper motions ($pm/plx>3)$ are X2, UV13, UV14, UV15, Rad1, Rad\_a, P15, P18, and P30.

\begin{table*}
	\centering
	\caption{Multiplicity of the sources. \textcolor{gray}{(Preprint layout)}}
	\label{tab:mul}
	\scriptsize
	\begin{tabular}{l@{\hspace{1.2\tabcolsep}}r@{\hspace{1.2\tabcolsep}}r@{\hspace{1.2\tabcolsep}}r@{\hspace{1.2\tabcolsep}}r@{\hspace{1.2\tabcolsep}}l@{\hspace{1.2\tabcolsep}}l@{\hspace{1.2\tabcolsep}}r@{\hspace{1.2\tabcolsep}}r@{\hspace{1.2\tabcolsep}}r@{\hspace{1.2\tabcolsep}}r@{\hspace{1.2\tabcolsep}}r@{\hspace{1.2\tabcolsep}}r@{\hspace{1.2\tabcolsep}}l@{\hspace{1.2\tabcolsep}}l@{\hspace{1.2\tabcolsep}}l@{\hspace{1.2\tabcolsep}}r@{\hspace{1.2\tabcolsep}}r@{\hspace{1.2\tabcolsep}}l}
		\hline
Gaia DR3           &  $l$~~~~~~~$b$    &  $snr_\pi$  & vp  &  $pm/\pi$          &  d    &  gha   &  fmp &  fow &  ruwe   & Flag   & \multicolumn{3}{c}{Binarity indicators$^a$} & $a_p^{d-}$ & \multicolumn{3}{l}{~~$a_p^{d+}$ (au)$^b$} & ~~~System$^c$ \\ 
                     &    (deg)~~   &      &  &      &  (pc)      &        & (\%) & (\%) &         &        & SED\,('')   & Ast.   & Vis.('') & 0.05''  & 0.3'' & Vis. & SED &  V1~V2~S/A \\ 
\hline
4043369078205765888  &  358~~~~-4&   8.8 &     9 &    0.3~$\pm$~0.1   &  ~~54.4~(~~6.0,~~7.7)  & 0.072 &    6 &    5 &   1.515 &  $\gamma$\_a & $\gamma448.9$    &  U   &  1.0 SW   &     2.4 &      19 &      62 &   27896 &  wd~~wd~~~? \\
4040994030072768896  &  354~~~~-3&   8.1 &     9 &    0.4~$\pm$~0.1   &  ~~84.9~(10.1,12.9)   & 0.144 &    0 &    0 &   1.095 &  $\gamma$\_b & $\gamma556.7$    &  R?  &  0.9 SW   &     3.7 &      29 &      88 &   54404 &  wd~~wd~~~? \\
5990998826943947520  &  335~~~~~~5&   4.1 &     9 &    1.0~$\pm$~0.2   &  100.5~(18.1,30.0)    & 0.157 &    0 &    0 &   1.049 &  $\gamma$\_c & $\gamma1109.4$   &  R?  &  --       &     4.1 &      39 &      -- &  144777 &  wd~~-----~~~? \\
4144738037302227072  &  ~~11~~~~~~5&   8.9 &     9 &    1.1~$\pm$~0.1   &  ~~58.5~(~~6.7,~~8.3)  & 0.153 &    0 &    0 &   1.135 &  $\gamma$\_d & $\gamma187.8$    &  R?  &  --       &     2.6 &      20 &      -- &   12551 &  wd~~-----~~~? \\
2957941232274652032  &  227~~-29&   6.6 &     9 &    0.6~$\pm$~0.1   &  ~~70.6~(11.0,15.7)   & 0.221 &    3 &    0 &   1.435 &  $\gamma$\_e & $\gamma1024.2$   &  R   &  --       &     3.0 &      26 &      -- &   88381 &  wd~~-----~~~? \\
\\
6456175155513145344  &  338~~-40&  24.9 &    24 &   11.9~$\pm$~0.3   &  ~~62.4~(~~2.3,~~2.9)  & 0.023 &    0 &    0 &   0.996 &  X2    & $X47.7$          &  --  &  --       &     3.0 &      20 &      -- &    3120 &  wd~~-----~~~? \\
5922780280824141056  &  335~~~~-9&   6.0 &    10 &    0.4~$\pm$~0.1   &  ~~89.5~(13.9,18.3)   & 0.161 &    2 &    0 &   1.466 &  X\_a  & $X35.0$          &  R   &  --       &     3.8 &      32 &      -- &    3769 &  wd~~-----~~~? \\
6027974578382721280  &  352~~~~~~8&  10.5 &    14 &    0.8~$\pm$~0.1   &  ~~47.6~(~~4.5,~~5.7)  & 0.343 &    2 &    0 &   2.223 &  X\_b  & $X37.5$          &  R   &  --       &     2.2 &      16 &      -- &    1999 &  wd~~-----~~~? \\
6029621818614152576  &  353~~~~~~8&   9.0 &    13 &    0.4~$\pm$~0.1   &  ~~63.6~(~~6.7,~~8.6)  & 0.207 &   12 &    6 &   1.568 &  X\_c  & $X42.7$          &  R   &  --       &     2.8 &      22 &      -- &    3086 &  wd~~-----~~~? \\
 384577010978265344  &  115~~-20&   6.0 &    16 &    0.5~$\pm$~0.1   &  ~~92.9~(14.1,17.8)   & 0.141 &    0 &    0 &   1.422 &  X\_d  & $X57.3$          &  R   &  --       &     3.9 &      33 &      -- &    6341 &  wd~~-----~~~? \\
5914009991961757952  &  331~~-13&   8.5 &    10 &    0.7~$\pm$~0.1   &  ~~78.5~(~~9.5,11.6)  & 0.335 &    3 &    0 &   1.031 &  X\_e  & $X54.3$          &  R?  &  --       &     3.5 &      27 &      -- &    4893 &  wd~~-----~~~? \\
\\
5386173659617460736* &  287~~~~24&   6.3 &    14 &    1.5~$\pm$~0.3   &  100.8~(12.7,19.6)    & 0.362 &    3 &    0 &   1.236 &  UV1   & $NF0.32$         &  R?  &  --       &     4.4 &      36 &      -- &      39 &  wd~~-----~~~? \\
6829498931265990656* &  ~~28~~-41&  11.6 &    12 &    0.4~$\pm$~0.1   &  ~~60.2~(~~5.1,~~6.1)  & 0.428 &   30 &    0 &   1.995 &  UV2   & $(N)1.55$        &  R   &  0.57 SW  &     2.8 &      20 &      38 &     103 &  wd~~wd~~~? \\
5777941133043108992* &  315~~-19&   6.0 &    13 &    0.2~$\pm$~0.1   &  ~~91.8~(14.2,16.5)   & 0.328 &    0 &    0 &   1.208 &  UV3   & $NF2.19$         &  R?  &  --       &     3.9 &      32 &      -- &     237 &  wd~~-----~~~? \\
6901051613248041216  &  ~~33~~-27&  12.1 &    15 &    0.6~$\pm$~0.1   &  ~~71.3~(~~5.7,~~7.8)  & 0.844 &    9 &    0 &   1.769 &  UV4   & $gNF0.72$        &  R   &  0.6 SW   &     3.3 &      24 &      47 &      57 &  ~~?~~~~wd~~~? \\
6585452983927662080  &  ~~~~3~~-49&   3.8 &    10 &    0.3~$\pm$~0.2   &  ~~97.9~(20.5,18.2)   & 0.070 &    0 &    0 &   0.998 &  UV5   & $(N)3.04$        &  --  &  --       &     3.9 &      35 &      -- &     353 &  wd~~-----~~~? \\
6858721824326705536  &  ~~28~~-33&  11.0 &    17 &    0.8~$\pm$~0.1   &  ~~59.8~(~~5.0,~~7.2)  & 0.299 &   33 &    3 &   2.048 &  UV6   & $(N)1.70$        &  R   &  --       &     2.7 &      20 &      -- &     114 &  wd~~-----~~~? \\
6529021095092491648  &  342~~-66&   9.4 &    10 &    0.8~$\pm$~0.1   &  ~~78.0~(~~7.9,10.7)  & 0.302 &    0 &    0 &   1.138 &  UV7   & $(N)1.90$        &  R?  &  --       &     3.5 &      27 &      -- &     168 &  ~~?~~~~-----~~~? \\
3150537594577224960  &  210~~~~19&   5.7 &    10 &    1.2~$\pm$~0.2   &  ~~87.8~(15.0,18.0)   & 0.295 &    1 &    0 &   1.440 &  UV8   & $guN0.28$        &  R   &  --       &     3.6 &      32 &      -- &      30 &  wd~~-----~~~? \\
 837736394742229248  &  157~~~~61&   2.7 &    11 &    0.8~$\pm$~0.2   &  103.6~(21.0,15.4)    & 0.036 &    0 &    0 &   1.091 &  UV9   & $guN0.37$        &  --  &  --       &     4.1 &      36 &      -- &      44 &  wd~~-----~~~? \\
4240872805678917376* &  ~~43~~-14&   4.6 &     9 &    0.9~$\pm$~0.2   &  ~~92.6~(17.7,18.3)   & 0.098 &    0 &    0 &   1.028 &  UV10  & $(N)F0.62$       &  --  &  --       &     3.7 &      33 &      -- &      69 &  wd~~-----~~~? \\
4438623270475014656  &  ~~21~~~~34&   3.5 &     9 &    1.4~$\pm$~0.3   &  ~~96.2~(20.5,19.3)   & 0.010 &    0 &    0 &   1.036 &  UV11  & $N1.54$          &  --  &  --       &     3.8 &      35 &      -- &     178 &  wd~~-----~~~? \\
4465431356821050496  &  ~~32~~~~38&   4.8 &    13 &    1.1~$\pm$~0.2   &  ~~86.2~(16.6,20.4)   & 0.300 &    3 &    6 &   1.418 &  UV12  & $(N)1.45$        &  R   &  0.6 EW   &     3.5 &      32 &      64 &     155 &  wd~~wd~~~? \\
2480784311444314880* &  150~~-65&  21.7 &    16 &   16.9~$\pm$~0.4   &  ~~66.1~(~~2.9,~~2.9)  & 0.065 &    0 &    0 &   1.148 &  UV13  & $u0.08$          &  --  &  --       &     3.2 &      21 &      -- &       6 &  wd~~-----~~~? \\
1900830823016998144  &  ~~90~~-22&  15.0 &    18 &   15.7~$\pm$~0.6   &  ~~80.4~(~~5.2,~~5.9)  & 0.029 &    0 &    0 &   0.986 &  UV14  & $(u)0.04$        &  --  &  --       &     3.8 &      26 &      -- &       3 &  wd~~-----~~~? \\
2807663865638075264  &  118~~-37&   8.9 &    14 &   16.1~$\pm$~1.0   &  ~~90.3~(~~9.4,11.7)  & 0.046 &    0 &    0 &   1.279 &  UV15  & $u0.58$          &  --  &  --       &     4.0 &      31 &      -- &      59 &  wd~~-----~~~? \\
\\
 557277267992311808  &  131~~~~24&  10.3 &    15 &    9.9~$\pm$~0.6   &  ~~91.5~(~~8.4,10.2)  & 0.065 &    0 &    0 &   0.977 &  Rad1  & $rad5.63$        &  --  &  --       &     4.2 &      31 &      -- &     573 &  wd~~-----~~~? \\
2185585922087166720  &  ~~89~~~~11&   6.4 &    12 &    3.9~$\pm$~0.4   &  ~~97.6~(12.5,14.1)   & 0.119 &    0 &    0 &   0.909 &  Rad\_a & $rad13.16$       &  R?  &  --       &     4.3 &      33 &      -- &    1470 &  wd~~-----~~~? \\
4688982132605446784  &  303~~-45&   7.4 &    14 &    0.9~$\pm$~0.2   &  ~~63.9~(~~8.6,10.5)  & 0.185 &    9 &    0 &   2.100 &  Rad\_b & $rad1.58$        &  R   &  --       &     2.8 &      22 &      -- &     118 &  wd~~-----~~~? \\
6029611888623577088* &  353~~~~~~9&   8.0 &    13 &    0.9~$\pm$~0.1   &  ~~60.9~(~~8.0,10.6)  & 0.294 &   14 &    6 &   2.313 &  Rad\_c & $rad11.56$       &  R   &  0.89 E   &     2.6 &      21 &      64 &     826 &  wd~~wd~~~? \\
4176647926320238720  &  ~~22~~~~12&   3.9 &     9 &    0.7~$\pm$~0.2   &  100.7~(21.8,20.2)    & 0.108 &    0 &    0 &   1.005 &  Rad\_d & $rad11.75$       &  R?  &  --       &     3.9 &      36 &      -- &    1420 &  wd~~-----~~~? \\
5711114877897853696  &  240~~~~~~2&   9.5 &     9 &    0.2~$\pm$~0.1   &  ~~53.4~(~~5.4,~~6.6)  & 0.203 &    0 &    0 &   1.359 &  Rad\_e & $rad11.15$       &  R?  &  --       &     2.4 &      18 &      -- &     668 &  wd~~-----~~~? \\
2024652913390061952  &  ~~60~~~~~~4&   6.3 &    11 &    0.3~$\pm$~0.1   &  ~~86.6~(13.7,17.5)   & 0.055 &    0 &    0 &   1.154 &  Rad\_f & $rad5.96$        &  --  &  0.7 W    &     3.6 &      31 &      73 &     621 &  wd~~wd~~~? \\
4149153813464969856  &  ~~12~~~~~~8&   3.9 &     9 &    0.3~$\pm$~0.2   &  102.7~(18.4,15.5)    & 0.087 &    0 &    0 &   1.086 &  Rad\_g & $rad12.14$       &  --  &  --       &     4.2 &      35 &      -- &    1435 &  wd~~-----~~~? \\
\\
 892683904225248000  &  186~~~~22&   3.3 &    10 &    0.6~$\pm$~0.2   &  ~~90.7~(23.4,22.1)   & 0.118 &    0 &    0 &   1.095 &  P0    & --               &  R?  &  --       &     3.4 &      34 &      -- &      -- &  ~~?~~~~-----~~~? \\
 387812285286341504  &  120~~-19&   6.7 &     9 &    0.1~$\pm$~0.1   &  ~~81.4~(11.4,15.2)   & 0.371 &    7 &    0 &   7.570 &  P1X1  & $X44.1$          &  R   &  --       &     3.5 &      29 &      -- &    4262 &  wd~~-----~~~? \\
4443619962406121088* &  ~~31~~~~26&   5.4 &    14 &    1.0~$\pm$~0.2   &  ~~98.5~(16.1,21.3)   & 0.173 &    0 &    0 &   1.309 &  P2    & --               &  R?  &  --       &     4.1 &      36 &      -- &      -- &  wd~~-----~~~? \\
3897927393739888128  &  272~~~~65&   2.8 &    10 &    0.5~$\pm$~0.2   &  102.1~(20.9,52.4)    & 0.181 &    0 &    0 &   0.991 &  P3    & --               &  R?  &  --       &     4.1 &      46 &      -- &      -- &  ~~?~~~~-----~~~? \\
6600314326685634816  &  ~~10~~-58&   3.2 &    10 &    0.3~$\pm$~0.1   &  ~~99.1~(20.0,16.5)   & 0.064 &    0 &    0 &   1.038 &  P4    & --               &  --  &  --       &     4.0 &      35 &      -- &      -- &  ~~?~~~~-----~~---- \\
3910666820788919936  &  254~~~~62&   3.9 &    10 &    0.2~$\pm$~0.1   &  ~~95.8~(19.6,18.9)   & 0.343 &    0 &    0 &   0.926 &  P5    & --               &  R?  &  --       &     3.8 &      34 &      -- &      -- &  wd~~-----~~~? \\
3622634116211982976  &  310~~~~51&   4.1 &     9 &    0.0~$\pm$~0.1   &  ~~88.0~(19.5,22.0)   & 0.071 &    1 &    0 &   1.196 &  P6    & --               &  --  &  --       &     3.4 &      33 &      -- &      -- &  ~~?~~~~-----~~---- \\
4234582018625351936* &  ~~36~~-13&   9.5 &    11 &    1.2~$\pm$~0.1   &  ~~74.4~(~~7.9,10.1)  & 0.300 &    0 &    0 &   1.387 &  P7    & --               &  R?  &  --       &     3.3 &      25 &      -- &      -- &  wd~~-----~~~? \\
5606206865714606208  &  242~~~~-9&   5.0 &     9 &    0.4~$\pm$~0.2   &  ~~96.7~(16.3,17.9)   & 0.051 &    0 &    0 &   1.008 &  P8    & --               &  --  &  --       &     4.0 &      34 &      -- &      -- &  wd~~-----~~---- \\
4085595994046913792* &  ~~17~~-11&  18.1 &    11 &    0.4~$\pm$~0.0   &  ~~48.0~(~~2.4,~~3.7)  & 0.206 &    0 &    0 &   2.692 &  P9    & --               &  R   &  0.8 NE   &     2.3 &      15 &      41 &      -- &  wd~~wd~~~? \\
6041917107322758016* &  347~~~~17&   3.8 &     9 &    1.2~$\pm$~0.3   &  101.5~(18.6,15.6)    & 0.097 &    0 &    0 &   1.083 &  P10   & --               &  --  &  --       &     4.1 &      35 &      -- &      -- &  wd~~-----~~---- \\
1959239908639738240  &  ~~91~~-12&   4.7 &     9 &    1.0~$\pm$~0.2   &  ~~99.0~(17.6,21.6)   & 0.113 &    0 &    0 &   1.054 &  P11   & --               &  R?  &  --       &     4.1 &      36 &      -- &      -- &  wd~~-----~~~? \\
4486941991858372352  &  ~~30~~~~21&   6.8 &    10 &    0.5~$\pm$~0.1   &  ~~70.7~(10.2,13.4)   & 0.137 &    0 &    0 &   1.135 &  P12   & --               &  R?  &  0.8 E    &     3.0 &      25 &      67 &      -- &  wd~~wd~~~? \\
4494124177899820800  &  ~~36~~~~22&   2.8 &     9 &    0.5~$\pm$~0.2   &  100.9~(21.0,17.3)    & 0.064 &    0 &    0 &   0.875 &  P13   & --               &  --  &  --       &     4.0 &      35 &      -- &      -- &  wd~~-----~~---- \\
4112307735705947264  &  358~~~~10&   6.7 &    10 &    0.7~$\pm$~0.2   &  ~~86.6~(12.7,17.2)   & 0.112 &    0 &    0 &   1.093 &  P14   & --               &  R?  &  0.8 NE   &     3.7 &      31 &      83 &      -- &  wd~~wd~~~? \\
3257308626125076480  &  189~~-40&   4.1 &    11 &    5.9~$\pm$~0.8   &  ~~95.5~(17.8,19.9)   & 0.168 &    0 &    0 &   1.022 &  P15   & --               &  R?  &  --       &     3.9 &      35 &      -- &      -- &  wd~~-----~~~? \\
5648417357621486208  &  253~~~~13&   7.3 &    13 &    0.3~$\pm$~0.1   &  ~~68.2~(~~9.5,12.6)  & 0.127 &    2 &    0 &   1.486 &  P16   & --               &  R   &  --       &     2.9 &      24 &      -- &      -- &  wd~~-----~~~? \\
5642345785702095616  &  252~~~~~~8&  14.0 &    24 &    0.2~$\pm$~0.1   &  ~~45.4~(~~2.8,~~4.8)  & 0.387 &    3 &    0 &   2.274 &  P17   & --               &  R   &  --       &     2.1 &      15 &      -- &      -- &  wd~~-----~~~? \\
3948582104883237632  &  271~~~~80&   7.9 &    12 &    6.7~$\pm$~0.5   &  168.0~(~~1.0,~~1.0)  & 0.061 &    0 &    0 &   1.040 &  P18   & --               &  --  &  5.282 S  &     8.3 &      51 &     893 &      -- &  wd~~ms~~---- \\
 309921197948007936  &  128~~-33&   5.8 &     9 &    0.5~$\pm$~0.1   &  ~~66.6~(11.6,20.3)   & 0.172 &    0 &    0 &   1.201 &  P19   & --               &  R?  &  0.9 NW   &     2.8 &      26 &      78 &      -- &  wd~~wd~~~? \\
4149950169090938240  &  ~~14~~~~~~8&  10.0 &    11 &    0.8~$\pm$~0.1   &  ~~48.4~(~~4.7,~~5.9)  & 0.390 &    5 &   18 &   1.639 &  P20   & --               &  R   &  0.37 NE  &     2.2 &      16 &      20 &      -- &  wd~~wd~~~? \\
6033410078502966784  &  356~~~~~~9&   8.1 &    11 &    0.3~$\pm$~0.1   &  ~~54.7~(~~6.2,~~8.4)  & 0.445 &   29 &   16 &   2.237 &  P21   & --               &  R   &  0.72 SW  &     2.4 &      19 &      45 &      -- &  wd~~wd~~~? \\
2928185797635630720  &  235~~~~-6&   7.8 &    13 &    0.1~$\pm$~0.1   &  ~~85.0~(11.0,12.7)   & 0.186 &    1 &    0 &   1.153 &  P22   & --               &  R?  &  --       &     3.7 &      29 &      -- &      -- &  wd~~-----~~~? \\
4137297229808521728  &  ~~11~~~~~~9&   4.8 &    10 &    1.5~$\pm$~0.3   &  ~~92.7~(16.3,19.7)   & 0.050 &    0 &    0 &   1.029 &  P23   & --               &  --  &  --       &     3.8 &      34 &      -- &      -- &  wd~~-----~~---- \\
4124968474672413440  &  ~~10~~~~~~8&   3.9 &    13 &    0.7~$\pm$~0.2   &  ~~84.9~(21.0,24.8)   & 0.149 &    0 &    0 &   0.980 &  P24   & --               &  R?  &  --       &     3.2 &      33 &      -- &      -- &  wd~~-----~~~? \\
4115860154696775168  &  ~~~~3~~~~10&  10.0 &    10 &    0.4~$\pm$~0.1   &  ~~54.0~(~~5.4,~~6.2)  & 0.067 &    0 &    4 &   1.197 &  P25   & --               &  --  &  1.15 SW  &     2.4 &      18 &      69 &      -- &  wd~~wd~~---- \\
4124806709024391424  &  ~~~~9~~~~~~8&   5.3 &    11 &    0.6~$\pm$~0.2   &  ~~92.4~(15.6,20.3)   & 0.177 &    0 &    0 &   1.313 &  P26   & --               &  R?  &  --       &     3.8 &      34 &      -- &      -- &  wd~~-----~~~? \\
4125016578273139456  &  ~~10~~~~~~9&  10.6 &    14 &    0.1~$\pm$~0.1   &  ~~77.4~(~~6.9,~~8.3)  & 0.216 &    0 &    0 &   1.092 &  P27   & --               &  R?  &  --       &     3.5 &      26 &      -- &      -- &  wd~~-----~~~? \\
5596729522158314752  &  248~~~~~~2&   7.9 &    10 &    0.5~$\pm$~0.1   &  ~~50.7~(~~5.6,~~8.2)  & 0.346 &    0 &    0 &   1.287 &  P28   & --               &  R?  &  --       &     2.3 &      18 &      -- &      -- &  wd~~-----~~~? \\
4138259749157475200  &  ~~10~~~~12&   3.5 &    12 &    0.4~$\pm$~0.2   &  ~~96.9~(19.5,48.0)   & 0.001 &    0 &    0 &   1.415 &  P29   & --               &  U   &  --       &     3.9 &      43 &      -- &      -- &  wd~~-----~~~? \\
4297001190649825408  &  ~~46~~-12&   4.0 &     9 &    3.1~$\pm$~0.5   &  ~~98.2~(17.7,39.7)   & 0.100 &    0 &    0 &   1.259 &  P30   & --               &  R?  &  --       &     4.0 &      41 &      -- &      -- &  wd~~-----~~~? \\
6034234441635923456  &  355~~~~12&   4.4 &    10 &    1.0~$\pm$~0.3   &  ~~98.9~(17.4,27.3)   & 0.282 &    0 &    0 &   1.246 &  P31   & --               &  R?  &  --       &     4.1 &      38 &      -- &      -- &  wd~~-----~~~? \\
4313639928323512064  &  ~~44~~~~~~3&   6.5 &    12 &    0.5~$\pm$~0.1   &  ~~64.4~(10.3,14.2)   & 0.219 &    0 &    0 &   1.717 &  P32   & --               &  R   &  0.98 NE  &     2.7 &      24 &      77 &      -- &  wd~~wd~~~? \\
6030056675402064000  &  354~~~~~~8&   5.6 &    10 &    0.5~$\pm$~0.1   &  ~~85.5~(14.1,17.4)   & 0.061 &    0 &    0 &   1.200 &  P33   & --               &  --  &  --       &     3.6 &      31 &      -- &      -- &  wd~~-----~~---- \\
 423003361823693440  &  118~~~~-3&   3.2 &    10 &    0.5~$\pm$~0.2   &  100.8~(19.6,16.5)    & 0.039 &    0 &    0 &   1.107 &  P34   & --               &  --  &  --       &     4.1 &      35 &      -- &      -- &  wd~~-----~~---- \\
4124336221099985280  &  ~~~~9~~~~~~9&   5.6 &    11 &    0.4~$\pm$~0.1   &  ~~85.8~(14.7,19.3)   & 0.022 &    0 &    0 &   1.180 &  P35   & --               &  --  &  1.0 NE   &     3.6 &      32 &     105 &      -- &  wd~~wd~~---- \\
1833552363662146944  &  ~~63~~~~-5&   4.0 &    10 &    0.3~$\pm$~0.1   &  ~~93.5~(19.2,20.1)   & 0.032 &    0 &    0 &   0.944 &  P36   & --               &  --  &  --       &     3.7 &      34 &      -- &      -- &  wd~~-----~~---- \\
4108525450122867456  &  357~~~~~~7&   6.9 &    14 &    0.4~$\pm$~0.1   &  ~~67.1~(~~9.3,12.4)  & 0.303 &    0 &    0 &   1.639 &  P37   & --               &  R   &  --       &     2.9 &      24 &      -- &      -- &  wd~~-----~~~? \\
\hline
	\end{tabular}
\vspace{-0.23cm}
\begin{flushleft}
gha, fmp, and fow, as in Table~\ref{tab:all2}. vp stands for \texttt{visibility\_periods\_used}. $^a$ Column SED: significant/low excess without/within parenthesis. $\gamma$, X, N, and F stand for Fermi, RASS 0.1--2.4~keV, \textit{GALEX} NUV and FUV bands, respectively. Angular separation in~arcsec. Column Ast.: \textit{Gaia} photoastrometric indicator (Unresolved -- U, Resolved -- R, and possibly resolved -- R?). Column Vis.: angular separation in~arcsec and relative location of visual companion. $^b$ Physical projected binary separation at $d_\mathrm{l}$ and 0.05~arcsec, and at $d_\mathrm{u}$ and an angular separations of 0.3~arcsec, of the visual companion, and of the SED-excess counterpart. $^c$ Possible system components (V1,V2,S/A): visible source exactly at \textit{Gaia} source position (V1), visual companion (V2), and SED excess- or photoastrometric companion (S/A).
\end{flushleft}
\end{table*}

Among the sources of Table~\ref{tab:mul}, all those that are photoastrometrically resolved (`R') have typically the greatest values of the \texttt{astrometric\_excess\_noise} parameter (a measure of the residuals in the astrometric solution), in the range of 5.4--11.3~mas. P29 is not photoresolved (`U'), suggesting that, if the source has a companion that causes the high \texttt{ruwe} value, then the companion is not visible. Considering the \texttt{astrometric\_sigma5d\_max} parameter (the five-dimensional equivalent to the semimajor axis of the \textit{Gaia} position error ellipse, including position, parallax, and proper motion), the sources have values in the range 0.81--4.54~mas, and P29 has the greatest value.

(ii) We checked all the sources of Table~\ref{tab:mul} for visual companions in single-band images from PS1, SDSS, NSC-NOIRLab, and \textit{GALEX}. This allowed us to constrain them beyond an angular separation of approximately 0.3~arcsec, above which binary components, especially of different colours, are easy to identify. The visual companions are either blends with the sources or fully resolved. Most have about the same colour and are as faint or fainter. None of the visual companions has proper motion nor parallax in \textit{Gaia} DR3, except the widely separated and bright M-type companion of P18 (Appendix~\ref{sec:comovingwidebinary}). Those that have \textit{Gaia} parameters are listed in Table~\ref{tab:all2}, with indication of associated candidate. \textit{Gaia} can resolve varying percentages of binaries of equal magnitudes down to separations of 0.05~arcsec \citep{2021A&A...649A...6G}; the completeness may start to decrease at separations $<$1.5~arcsec, drop below 80 per cent at $<$0.7~arcsec, and reduce to about 10 per cent or less at $<$0.4~arcsec \citep{2021A&A...649A...5F}. The visual companions are referred to by their angular separation (in arcsec) and approximate relative orientation in the column `Vis.' of binary indicators in Table~\ref{tab:mul}. One particular case is UV12 ($pm=15$~mas~yr$^{-1}$), where the PS1 flux counts decrease symmetrically by about 0.3~arcsec east-west from the source position, without clearly indicating two peaks. We assumed that the source is a 0.6~arcsec wide pair centred on the \textit{Gaia} source position, although it could also be an extended source. 

Finally, we checked the sources at wider separations within 10~arcsec, whether these have relatively bright neighbour stars that are comoving or do not have \textit{Gaia} proper motion nor parallax. At 9.4~arcsec north-west from UV3, there is a bright star, Gaia DR3 5777941133042341632 ($G=16.1$~mag), without proper motion nor parallax, with \texttt{ipd\_frac\_multi\_peak} $=$ 52 per cent and \texttt{duplicate\_source} $=$ 1, indicating that this bright star is itself probably a binary of separation $<$0.18~arcsec. It has a small proper motion, of $(-9.7\pm6.5, 2.6\pm7.0)$~mas~yr$^{-1}$ in CATWISE2020.

(iii) From the crossmatches with other catalogues and from the SEDs, we could identify sources with gamma-ray, X-ray, ultraviolet, and radio excesses, which we consider as probable- or possible indications of binarity. These are referred to in the binary indicator column `SED' in Table~\ref{tab:mul}. The abbreviations $\gamma$, X, N, and F stand for \textit{Fermi}, RASS 0.1--2.4~keV, \textit{GALEX} $NUV$ and $FUV$ bands, respectively. Significant and low ultraviolet-excesses in the SEDs are indicated without and within parenthesis. We provide the angular separation between the \textit{Gaia} coordinates and the counterpart centroid. For UV8, UV9, UV13, UV14, and UV15, we adopt the angular separation with SDSS $u$ band instead of \textit{GALEX}. We note that the gamma-ray, X-ray, and a few of the radio counterparts have large coordinates uncertainties; also, UV4, UV10, and UV11 have $NUV$- or $FUV$ counts that spread up to more than 2.5, 1, and 2~arcsec, respectively.

Out of the 72 sources of Table~\ref{tab:mul}, 49, 17, and 35 have (possible) indications of binarity from photoastrometry, a visual companion, and flux excess, respectively, and nine have no apparent indication of binarity. The later could still have for example tight companions of similar magnitudes and colours or substellar companions of LTY spectral types.

\begin{figure}
    \includegraphics[width=\columnwidth]{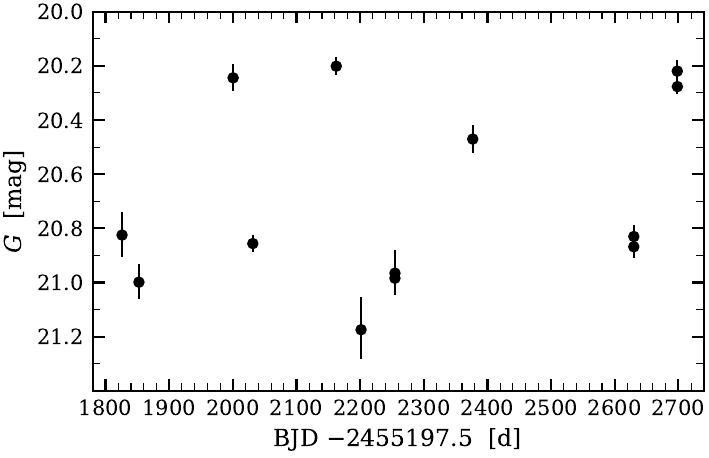}
    \caption{$G$-band magnitude versus time diagram of P1X1.}
    \label{time_series_G__Gaia_DR3_387812285286341504}
\end{figure}

In the whole \textit{Gaia} DR3 and compared to EDR3, many sources have additional measurements or parameters, for example related to binarity \citep{2022arXiv220605595G}. None of the sources of Table~\ref{tab:all1} and \ref{tab:all2} is flagged as non-single source, probably because these are fainter than the limits of $G=19$$-$20~mag for binarity analysis in DR3. One source has available epoch photometry and is significantly variable. P1X1 is in the \textit{Gaia} Andromeda Photometric Survey (pencil beam of 5.5~deg radius) and is located far away (more than 2~deg) from the spiral disc of the Andromeda Galaxy, as seen in the optical. The source has $G$-band variations of up to 1~mag within less than 40~d, in the 2014.9--2017.4 epoch range and considering measurements with $\texttt{flux\_over\_error}>7$. Fig.~\ref{time_series_G__Gaia_DR3_387812285286341504} shows the time series, in units of BJD. $G_\mathrm{BP}$ and $G_\mathrm{RP}$ bands measurements of $\texttt{flux\_over\_error}>7$ are too few and not shown.

If we combine these indications of binarity with the distances, we can infer physical projected binary separations. In Table~\ref{tab:mul}, we list the $d50$ distances with lower and upper errors. The one-sigma lower error was obtained by subtracting $d50$ by the lower distance $d_\mathrm{l}=d16$. The upper error $d_\mathrm{u}-d50$ depends on the photoastrometric binarity indicator (column Ast.). For unflagged sources, we adopted $d_\mathrm{u}=d84$. For flagged sources, we adopted $d_\mathrm{u}=\max(d84, 1/(plx-err_{plx}f))$, where $f$ is a correction factor for the parallax uncertainty underestimation. $f=1.5$ for sources with \texttt{ruwe}~$>1.4$ (flags `R' or `U') or $f=(1.5-1.2)/(1.4-1.04)*(\texttt{ruwe}-1.04)+1.2$ for sources with \texttt{ruwe}~$<1.4$ but IPD gha$\ge$0.1 (flag `R?'; $f>1$ for $\texttt{ruwe}>0.8$), allowing for a continuous transition to the \texttt{ruwe}~$>1.4$ case. The factors of 1.5 and 1.2 were adapted from the cases of binaries of $G=19$$-$20~mag and separations 0--2~arcsec shown in figs.~16 and 17 of \citet{2021MNRAS.506.2269E}. For P18, we adopted the distance of its brighter and comoving stellar companion. For orientation, we list in Table~\ref{tab:mul} the physical projected binary separations (i) $a_p^{d-}$ at the lower distance $d_\mathrm{l}$ and an angular separation of 0.05~arcsec, and (ii) $a_p^{d+}$ at the upper distance $d_\mathrm{u}$ and an angular separation of (a) 0.3~arcsec, (b) the visual companion, and (c) the SED-excess counterpart.

Finally, we indicate possible types of the system components, components that are: the visible source exactly at the \textit{Gaia} DR3 source position (V1), the visual companion (V2), and the SED excess- or photoastrometric companion (S/A). For UV12, we assume a V1+V2 pair centred on the \textit{Gaia} source position. In first approximation, we assume that all the V1 and V2 components are white dwarfs (WD; 0.1--1.3~M$_{\odot}$), except for (i) UV4 (V1; ultraviolet hot component), which at its current distance would have a too small radius to be a white dwarf (we thus indicate a question mark, `?'), (ii) UV7, P0, P3, P4, and P6 (V1; `?'), which have overall straight SED profiles, and (iii) the wide companion (V2) of P18, which probably is a main sequence star (MS) of 0.4~M$_{\odot}$. Besides this, the \textit{Gaia} sources of X2, Rad1, Rad\_a, UV13, UV14, UV15, P15, P27, and P36 are probable white dwarf candidates in \citetalias{2021A&A...649A...6G} or \citet{2021MNRAS.508.3877G}. Also, in first approximation, we assume that all of the gamma-ray, X-ray, ultraviolet, or radio SED excesses are of unknown (companion) type (question mark `?'). Similarly, we assume that all of the photoastrometric companions, independently of the visible companions, are of unknown type.

For guidance in the source classification, we derived radii from theoretical He and H atmosphere white-dwarf fitting of the optical fluxes (Section~\ref{sec:pl}) and blackbody fitting of the optical--ultraviolet fluxes. From the white-dwarf fitting, we have the theoretical white-dwarf mass and surface gravity (log~$g$ where $g$ is by default in cm s$^{-2}$ units), which with the surface gravity equation $g=GM / R^2$ provides us the white-dwarf radius. From the blackbody fitting [equations~(\ref{eqn:BB}) and (\ref{eqn:Fbol})], we have (i) the bolometric blackbody flux, which with the parallactic distance (Table~\ref{tab:mul}) and equation~(\ref{eqn:fluxlum}) allows us to obtain the bolometric luminosity (in erg~s$^{-1}$), and we have (ii) the effective temperature, which with the bolometric luminosity and equation~(\ref{eqn:BBlum}) provides us the thermal-emission radius. The derived radii are approximate because these depend on narrow spectral fitting regions (in most cases), opacity-related spectral redistributions of fluxes, distance, binarity, and reddening. The last one affects mostly sources towards the Galactic plane or centre. For simplicity, we account only for the uncertainty in the distance. 

The extrapolations at the infrared (see the SEDs in Appendix~\ref{fsed_ir} of the online supplementary material) of the blackbody and theoretical white-dwarf fits typically match or are lower than the observed fluxes, indicating unaccounted opacities and infrared fluxes in the fitting procedure. Despite the mismatches, most \textit{Gaia} (V1) components appear to have radii in the white dwarf range of 0.005--0.025~R$_{\odot}$ at the adopted distances. White dwarf radii from model-independent mass and radius measurements are in the range of 0.01--0.03~R$_{\odot}$ \citep[fig.~9 of][]{2017MNRAS.470.4473P}. Therefore, we preliminarily classified these components as white dwarfs.

\begin{figure*}
    \includegraphics[width=\textwidth]{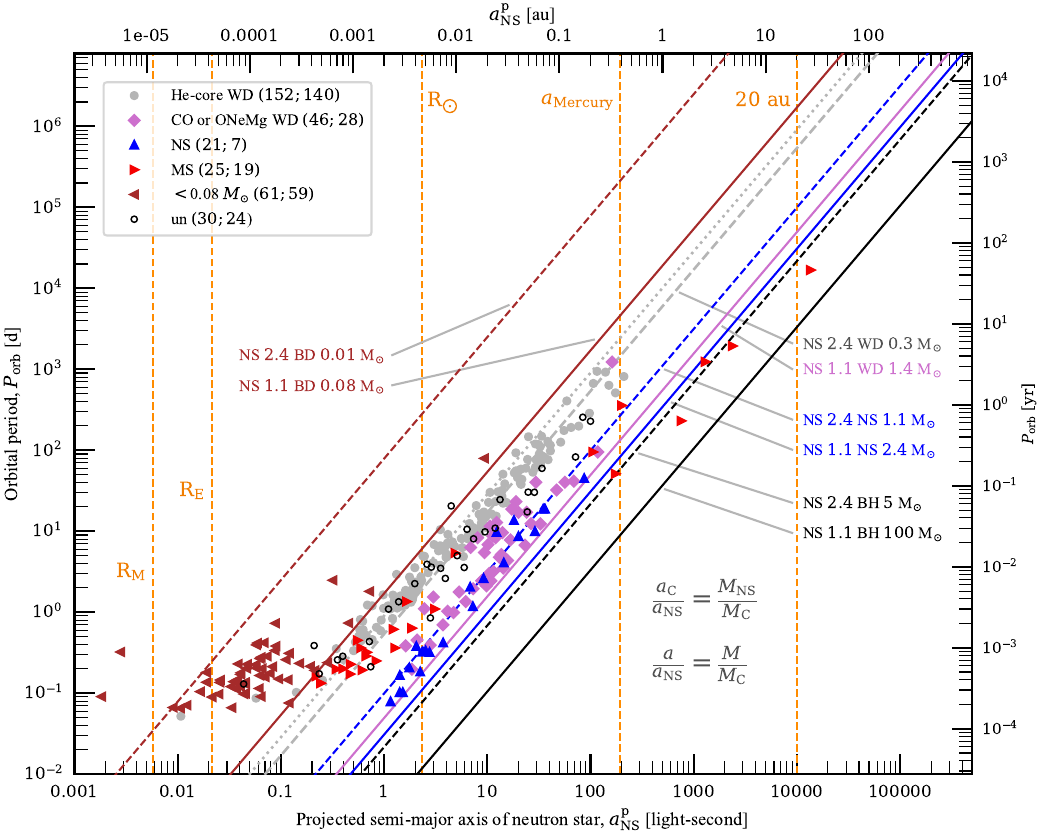}
     \caption{Orbital period versus projected semimajor axis of neutron stars with companions of different types. Parameter regions for low and high mass-ratio NS--BD (brown), NS--WD (grey or pink), NS--NS (blue), and NS--BH (black) binaries are delimited by the dashed and solid lines, where BD, WD, NS, and BH stand for brown dwarf, white dwarf, neutron star, and black-hole mass objects, respectively \citep[see also fig.~5 of][]{2019PhRvD.100b4058S}. These lines are for an inclination of 90~deg and thus indicate the full semimajor axis. The grey dotted line, as compared to the grey dashed line, is for an inclination of 45~deg and indicates a projected semimajor axis smaller by a factor 0.71. ATNF pulsars in binaries are shown with a colour according to the type or candidate-type of their companion: low-mass He-core white dwarf (grey dots), CO- or ONeMg-core white dwarf (violet diamonds), neutron star (blue upward triangles), main sequence star (red rightward triangles), substellar-mass object (brown leftward triangles), or unspecified (black open circles). The respective numbers and millisecond-pulsar (spin $\le$~30~ms) subsets are indicated in parenthesis. The PSR J0737$-$3039AB binary pulsar is counted once but both components are represented in the plot. For comparison, the semimajor axis of Mercury and the radii of the Sun, Earth, and Moon are represented by the vertical dashed lines. Semi-major axis -- mass relations are indicated, where $a=a_\mathrm{NS} + a_\mathrm{C}$ and $M=M_\mathrm{NS} + M_\mathrm{C}$ are the binary separation and total mass.}
    \label{porbsep}
\end{figure*}

\begin{figure*}
    \includegraphics[width=\textwidth]{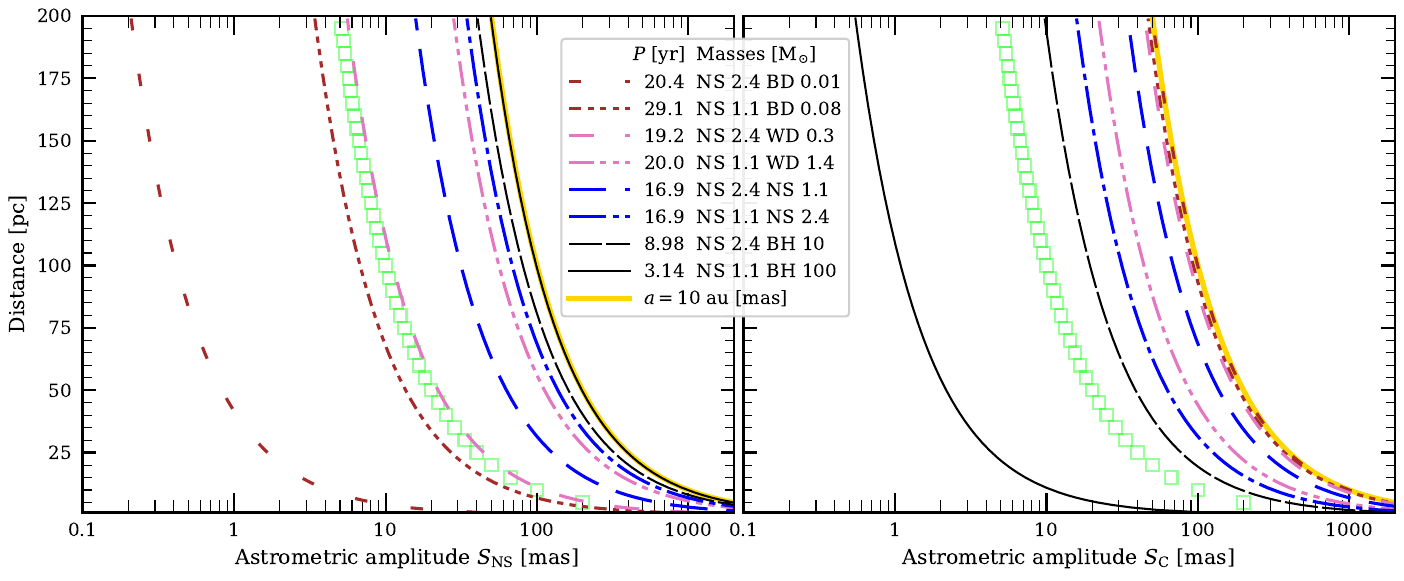}
    \includegraphics[width=\textwidth]{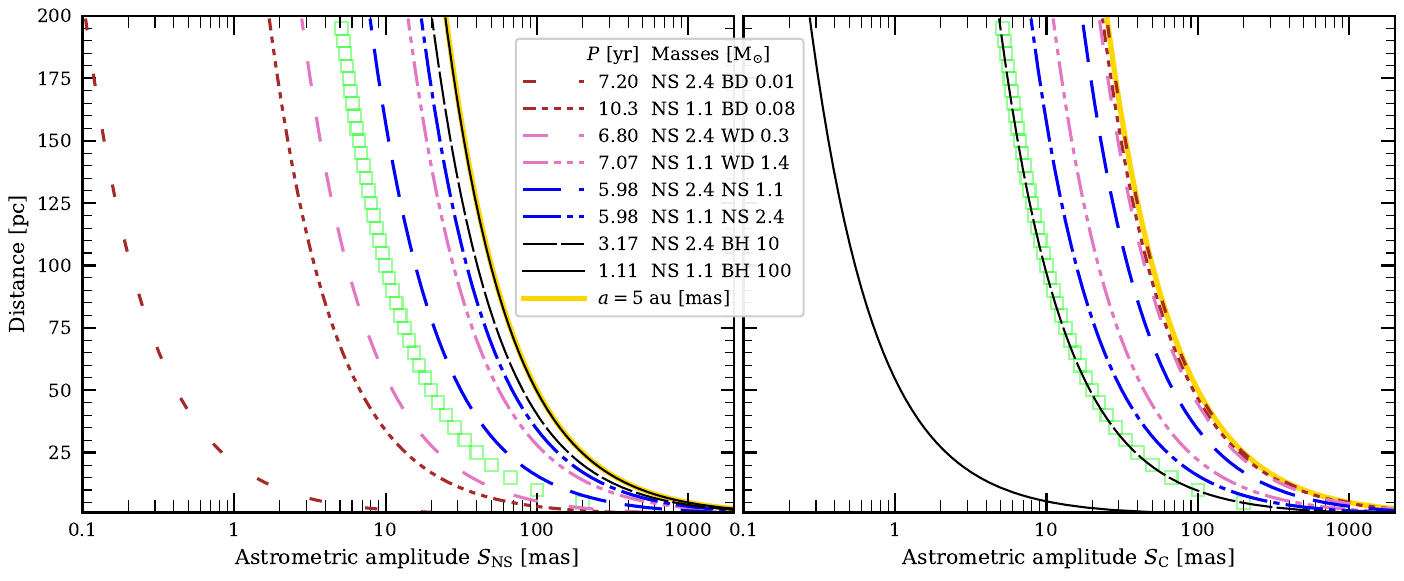}
    \includegraphics[width=\textwidth]{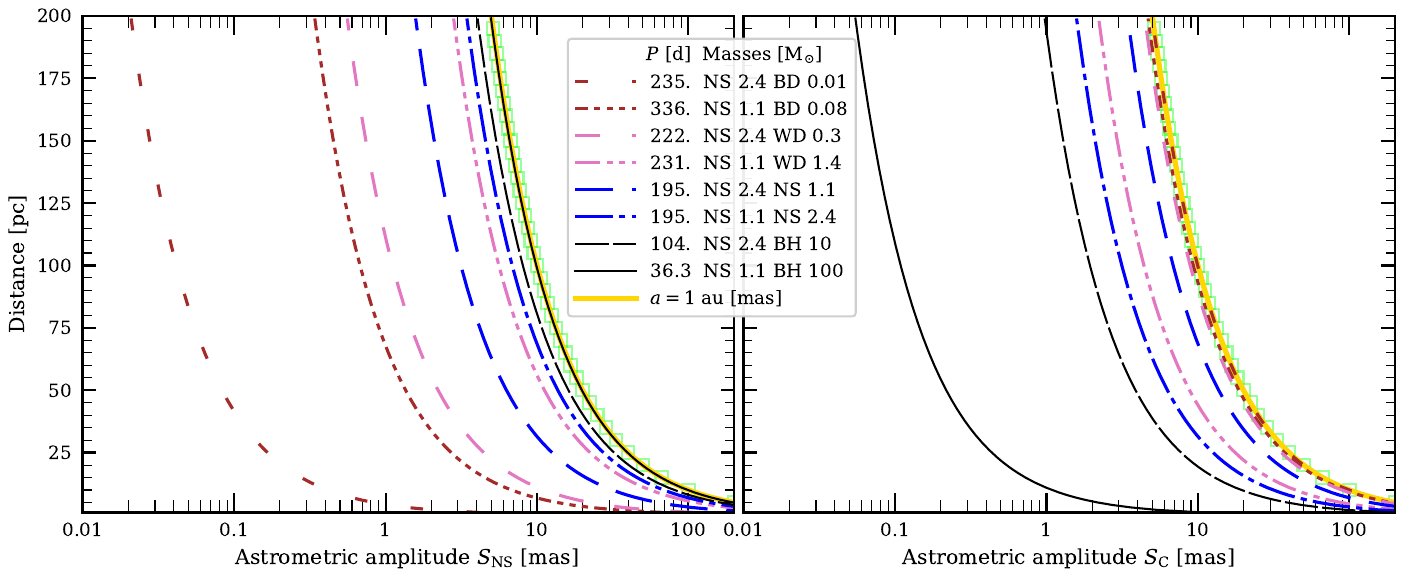}
      \caption{Astrometric orbital amplitude of a neutron star ({\sl left}) and its companion ({\sl right}), as a function of distance. Amplitudes for low and high mass-ratio NS--BD (brown), NS--WD (pink), NS--NS (blue), and NS--BH (black) binaries are represented by the dashed lines, and the dash-triple-dot, dash-double-dot, dash-single-dot, and solid lines, respectively. From top to bottom, the physical binary separations are of 10, 5, and 1~au and are represented as angular separations (yellow solid lines). The orbital period $P$ for each case of binary component masses is indicated. Parallaxes ($\pi=1/d$) are shown by green squares.}
    \label{aorbdis2_1}
\end{figure*}

The smallest known white dwarf is ZTF~J190132.9+145808.7 (Gaia DR3 4506869128279648512), with an estimated radius of $2140^{+160}_{-230}$~km (0.0031~R$_{\odot}$), an effective temperature of $\sim$50\,000~K, and a mass of 1.3~M$_{\odot}$ \citep{2021Natur.595...39C}. It is only slightly larger than the Moon, of 1737.5~km radius. In this context, the above 0.005--0.025~R$_{\odot}$ range corresponds to two to ten times the radius of the Moon. From the blackbody fitting at optical, we identified five V1 sources with very small radii at their $d50$ distances. These are P0, P1X1, UV10, P2, and Rad\_b, with $R=0.0012$, 0.0016, 0.0020, 0.0025, and 0.0031~R$_{\odot}$. The confirmation of these radius estimates would require the verification of the source type, distance, extinction, reddening, and binarity. Nevertheless, we can anticipate the effects, for example, of extinction and reddening. Assuming a low extinction of $<0.2$~mag over the $griz$-band range, then the unabsorbed flux is greater by a factor $<$10$^{0.4(0.2)}=1.2$, and the corrected radius is greater by a factor of only $<$1.2$^{0.5}=1.1$ [$R\propto F^{0.5}$, from equations~(\ref{eqn:BBlum}) and ~(\ref{eqn:fluxlum})]. Assuming a reddening that causes underestimating the effective temperature by 750~K, then the corrected radius is smaller by a factor $(T_\mathrm{eff}/(T_\mathrm{eff}+750))^2$, for example of 0.8--0.9 for $T_\mathrm{eff}=7\,000-13\,000$~K [$R\propto T_\mathrm{eff}^{-2}$, from equation~(\ref{eqn:BBlum})].

From the blackbody fitting at optical--ultraviolet, we derived emitting radii for UV1, UV3, UV4, and UV10 (Fig.~\ref{fsed_uv1}), which have $NUV$ and $FUV$-band fluxes, and for UV7, UV8, and UV9 (Fig.~\ref{fsed_uv8_uv9}), which have $NUV$-band fluxes and are already discussed in Section~\ref{disc:pl}. We indicate radius ranges accounting for distance uncertainties in the plots of the SEDs. For UV1, assuming that its optical and ultraviolet emission stems from a single blackbody source (which may not necessarily be true and the fit is approximate), the radius is of 891--1218~km. For the ultraviolet components of UV3 and UV4, the radii are of about 410--620~km (though in the case of UV4, the computation is purely formal, because the ultraviolet source is extended). For UV10, assuming that the ultraviolet emission stems only from a second component, the corresponding radius is of 95--140~km. In the cases of UV7, UV8, and UV9, the radii are of 1247--1578, 457--665, and 846--1218~km. The remaining eight ultraviolet candidates have too few near-ultraviolet measurements and do not allow us to determine whether these could have an ultraviolet blackbody component. From the perspective of our search, we tentatively interpret or propose that some of these ultraviolet excesses, of thermal or non-thermal origin, could relate to neutron stars, either from regions extending from these or from interaction processes with the system components. We recall that for instance for radio pulsars, because of the multipath propagation of their radio emission, these have emitting regions that can be much larger than the neutron star radius and from a few tens of km to several 10$^4$~km, comparable to the light-cylinder radius, which is proportional to the pulsar period \citep{2021ApJ...915...65M}.

As shown in the orbital-period versus projected-semimajor-axis diagram of Fig.~\ref{porbsep}, ATNF pulsars have detected companions typically of short orbital periods ($\la$10$^3$~d), low or lower masses, and mostly white-dwarf type. The two first aspects correlate with these pulsars having small projected semimajor axes. Besides the low-mass He-core and higher-mass CO/ONeMg-core white dwarfs, there are ultra light ($<$0.08~M$_{\odot}$), main-sequence, and neutron-star companions. The numbers of each types of companions and their millisecond-pulsar subsets are indicated in parenthesis next to the abbreviated labels in the legend box. We note the significant amount of millisecond pulsars. These have spin periods shorter than 30~ms and could emit continuous gravitational waves at frequencies detectable by ground-based detectors. Small (initial) binary separations are required for mass transfer, accretion, and thereby the spinning-up of neutron stars to rotation periods of milliseconds \citep{1982Natur.300..728A,1982CSci...51.1096R,1991PhR...203....1B,2008LRR....11....8L}. None of the  ATNF millisecond pulsars has a measured projected semimajor axis larger than 200 lt-s (about the semimajor axis of Mercury), implying that in the case of white dwarf companions the projected binary separation would not be larger than about ten times this value ($\lessapprox$4~au). However, subsequent mass loss of companions could increase the binary separation, and besides this, companions at relatively wide separations of the ATNF pulsars may not be easily detectable- nor systematically searched for with the pulsar-timing method.

Assuming that there are white dwarf -- neutron star systems among our sources, any real, curvilinear orbital motion captured by \textit{Gaia} DR3 data would concern those systems with significantly small angular separations. \textit{Gaia} is most sensitive to orbital periods close to its observing period, as shown in the study about hidden (sub)stellar companions by \citet{2019ApJ...886...68A}. For white dwarf -- neutron star systems (0.9+1.5~M$_{\odot}$) of orbital periods of once and twice the \textit{Gaia} DR3 observing period (2.83~yr), the binary separations are of $a=2.68$ and 4.26~au. For the sources of Table~\ref{tab:mul}, the lower and upper projected binary-separations are of $a_p^{d-}=2.1-8.3$ and $a_p^{d+}=2.5-8.5$~au at 0.05~arcsec angular separation ($a_p^{d+}=15-51$~au at 0.3~arcsec), implying that 0.05~arcsec is the level of angular separation at which these can be probed. From its own or complemented with other astrometric data, \textit{Gaia} can also enable for detecting accelerations at larger separations, where the direction of acceleration may coincide with the relative location of a companion \citep[][]{2019ApJ...877...60B,2021AJ....162..230B,2022MNRAS.513.5588B}.

Because the visual companions of our sources are at very large projected separations ($>$0.3~arcsec) of a few 10$^1$~au, too large to cause any significant binary interaction with the main visual component and excess emission, these do not hinder the existence of neutron stars in the systems. On the other hand, photoastrometric companions (R, R?, or U flags) with small projected separations of $\lessapprox$5~au ($\lessapprox$0.05~arcsec) could cause noticeable binary interaction on short time scales and excess emission.

In Fig.~\ref{aorbdis2_1}, we show the astrometric orbital amplitudes of a neutron star (left-hand panel) and its companion (right-hand panel) as a function of the distance, for physical binary separations of 10, 5, and 1~au. For a circular orbit, the astrometric orbital amplitude is the angular semimajor axis
\begin{equation}\label{eqn:astorbamp}
S_\mathrm{vis}\mathrm{[mas]}= \frac{1000}{d}~a_\mathrm{vis}=\frac{M_c}{M_\mathrm{vis}+ M_c}\frac{1000}{d}~a,
\end{equation} where $M_\mathrm{vis}$ and $M_c$ are the masses of the visible and companion objects, $d\mathrm{[pc]}$ is the distance, and $a\mathrm{[au]}=a_\mathrm{vis}+a_c$ the binary separation and sum of the component semimajor axes. The amplitude is obtained by subtracting the proper and parallactic motion components from the time-varying astrometry of the photocentre. The figures show the astrometric signature of each component separately; in reality both components may be visible and cause a blend photocentre or only one component may be visible.

The amplitudes for low and high mass-ratio NS--BD (brown), NS--WD (pink), NS--NS (blue), and NS--BH (black) binaries are represented by the dashed lines, and, the dash-triple-dot, dash-double-dot, dash-single-dot, and solid lines, respectively, where NS, WD, BD, and BH stand for neutron star, white dwarf, brown dwarf, and black-hole mass objects, respectively (the same binaries are also represented in Fig.~\ref{porbsep}). The physical binary separation is represented as angular separation by the yellow solid line. The orbital period $P$ is indicated for each case of binary component masses. In arithmetic average, the astrometric amplitude is covered in $P/4$. For comparison, we show parallaxes ($d=1/\pi$) by the green squares. Per definition, the parallax is covered in 1/4~yr~$\approx$~90~d. In this context, we note that most sources of Table~\ref{tab:mul} have $pm/plx$ ratios close to or lesser than one, that is their parallactic motion (uncertainty) appears to be a significant cause of the astrometric motion; nine sources are exceptions with relatively large proper motions ($pm/plx>3$). From the right-hand panels of Fig.~\ref{aorbdis2_1}, we see that for physical binary separations of 1--5~au, neutron stars orbiting white dwarfs can induce in white dwarfs significant astrometric jitter or modulation when compared to the parallactic motion. Quantitatively, from 1 to 5~au, it is of 44--89 to 220--444 per cent, on time-scales of 0.15--0.16 to 1.70--1.77~yr. This would thus concern potential binaries with angular separations $\lessapprox$0.05~arcsec in Table~\ref{tab:mul}. Finally and trivially, for a 0.1~au binary separation (not shown), orbital astrometric jitters induced in every of the above binary components are always smaller than 10 per cent of the parallax. Considering the inclination ($i$), the angle between the line of sight and the normal vector to the orbital plane, for inclinations of 90 and 0~deg, the astrometric displacements are a segment and a circle on the sky, respectively ($\cos(i) = b_\mathrm{vis}/S_\mathrm{vis}$, where $b_\mathrm{vis}$ is the semiminor axis of the visible object). For an inclination of 0~deg, it produces the largest binary wobble area in the sky, but no Doppler shifts in radial velocity. However, independent of inclination, an astrometric wobble for these binaries will always occur, and in general, binaries have inclinations distributed uniformly.

Regarding the minimal separations of the binary candidates, we can assume that these could be of 0.1~au or less. In the catalogue of low-mass X-ray binaries (LMXB) in the Galaxy and the Large and Small Magellanic Clouds \citep{2007A&A...469..807L}, the least intense LMXB sources are the transient ultra-soft X-ray sources. These have orbital periods of a few hours to about 150~h, which are shorter than for a white dwarf -- neutron star binary of 1.0+1.5~M$_{\odot}$ and 0.1~au separation, of $P=175.3~\mathrm{h}=7.30~\mathrm{d}$. Too close separations would imply intense accretion and intense X-ray emission, which is apparently not the case for our sources. At less close separations and up to possibly 5~au, weak interactions between the stellar components (including orbital material) through irradiation, particle winds, magnetic fields, and gravitation could occur, and possibly cause X-ray, ultraviolet, optical, infrared, or radio excesses. Discoveries of tight white dwarf -- neutron star binaries or double white dwarfs at separations $<$1--4~au are important for gravitational wave searches \citep[see e.g.][]{2022MNRAS.511.5936K}. In particular, the tightest binaries can be detected at 0.1--10~mHz with the Laser Interferometer Space Antenna \citep[LISA;][]{2017arXiv170200786A}.

We keep for a future work a more detailed characterization of the sources of this expandable search. This will benefit from updated parameters provided by the next \textit{Gaia} Data Releases, in particular distance- and binary parameters. For some of the sources, it would be interesting to confirm whether these have power-law emission (at $10^0$--$10^1~\mu$Jy flux level), and whether it is pulsed and relates to pulsars and neutron stars. In the affirmative case, it would be interesting to determine how their ultraviolet--optical--infrared emission relates to radio emission, the magnetic field, and spin-down emission flux \citep[see e.g.][]{2016MNRAS.459..427S}. Some of the sources with near and mid-infrared excesses could be investigated in relation to the possible existence of an extended pulsar wind nebula or a resolved supernova fallback disc (as considered for the RX~J0806.4$–$4123 neutron star, \citealt{2018ApJ...865....1P}) or discs of other types, which could be detected and resolved more easily because the sources have nearer distances. Because neutron stars have very high gravitational fields and are almost point-like with just 20--30~km in size, these can enable high fly-by velocities and potentially `heat up' orbital material shared with their stellar companions. Other indirect observational signatures of neutron stars, for example related to beam sweeping on orbital material \citep{2020A&A...644A.145M}, could occur.


\section{Conclusions}

These results are from a search for neutron stars among \textit{Gaia} broad-optical sources, based on magnitudes and colours derived from observations of known neutron stars and scaled to distances of the solar neighbourhood. Most nearby neutron stars are expected to be old and practically invisible electromagnetically, assuming rapid and unreversed dimming. These would only reveal themselves again, either as recycled pulsars or through interactions with companion stars and circumstellar material. In both cases, it is their -- past or present -- interactions with companions that would render these visible electromagnetically again.

Neutron stars with stellar companions can be divided into three categories: non-interacting, weakly interacting, and intensively interacting. The intensively interacting neutron stars appear to have already been identified among the accreting low-mass X-ray binaries and are few ($\sim$200) and relatively distant \citep[$\sim$kpcs;][]{2007A&A...469..807L}. Thus, the nearest neutron stars, which amount to 1000--2000 within 100 pc, would be weakly interacting or non interacting, isolated. The ones that can be found electromagnetically are probably those that were recycled in the past by a companion and are now (millisecond) pulsars and those that are presently weakly interacting with stellar companions and orbital material, through irradiation, particle winds, magnetic fields, and gravitation.

Because the sources are searched in a photometric region of faint and blue sources that is in common with white dwarfs, visible companions are likely to be white dwarfs. This is also in confluence with most known pulsars in binaries having white dwarf companions. The present, \textit{Gaia} DR3 measurements of the faint sources are single-star solutions and do not permit us to infer the masses of the components and the orbital elements yet. However, these measurements can give indications of binarity. With the complement of ultraviolet, optical, infrared fluxes, we do identify sources with two-component profiles (such as UV3, UV4, UV7, UV8, and UV10), involving power-law- or thermal emissions, with small thermal radii.

Further work is required to confirm that these sources hold very compact objects. In particular, the confirmation of the source emission profiles, multiplicity, and component masses, searches of changes in electromagnetic fluxes and properties over time, and the elucidation of the astrophysical processes involved are required. Pulses or modulations of periods shorter than a few seconds could hint at neutron stars.


\section*{Acknowledgements}


This research has made use of:
\begin{enumerate}[label=(\roman*),align=left,itemindent=9pt]
\item  the SkyView image service \citep{1998IAUS..179..465M} and data provided by the High Energy Astrophysics Science Archive Research Centre (HEASARC), which is a service of the Astrophysics Science Division at NASA/GSFC. 
\item  the Aladin Sky Atlas \citep{2000A&AS..143...33B}, SIMBAD data base \citep{2000A&AS..143....9W}, and VizieR catalogue access tool \citep[][DOI: 10.26093/cds/vizier]{2000A&AS..143...23O}, CDS, Strasbourg, France.
\item  the \textsc{saoimageds9} software, developed by the Smithsonian Astrophysical Observatory \citep{2003ASPC..295..489J}.
\item  the \textsc{iraf} software \citep{1986SPIE..627..733T,1993ASPC...52..173T}
\item  the \textsc{numpy} \citep{2020Natur.585..357H}, \textsc{matplotlib} \citep{2007CSE.....9...90H}, and \textsc{scipy} \citep{2020NatMe..17..261V} packages in the \textsc{python} programming language (\url{https://www.python.org/}).
\item  \textsc{astropy} (\url{http://www.astropy.org}), a community-developed core \textsc{python} package for Astronomy \citep{2013A&A...558A..33A, 2018AJ....156..123A}. 
\item  the \textsc{topcat} and \textsc{stilts} softwares \citep{2005ASPC..347...29T,2006ASPC..351..666T}. 
\item data, tools, or materials developed as part of the EXPLORE project that has received funding from the European Union's Horizon 2020 research and innovation programme under grant agreement No. 101004214.
\item  data from the European Space Agency (ESA) mission \textit{Gaia} (\url{https://www.cosmos.esa.int/gaia}), processed by the \textit{Gaia} Data Processing and Analysis Consortium (DPAC; \url{https://www.cosmos.esa.int/web/gaia/dpac/consortium}). Funding for the DPAC has been provided by national institutions, in particular the institutions participating in the \textit{Gaia} Multilateral Agreement.
\item  services or data provided by the Astro Data Lab at NSF's National Optical-Infrared Astronomy Research Laboratory. NOIRLab is operated by the Association of Universities for Research in Astronomy (AURA), Inc. under a cooperative agreement with the National Science Foundation.
\item  services or data of the DESI Legacy Imaging Surveys (\url{https://www.legacysurvey.org/acknowledgment}).
\item  the WFCAM Science Archive (WSA) holding the image and catalogue data products generated by the Wide Field Camera (WFCAM) on the United Kingdom Infrared Telescope (UKIRT).
\item  the VISTA Science Archive (VSA) holding the image and catalogue data products generated by the VISTA InfraRed CAMera (VIRCAM) on the Visible and Infrared Survey Telescope for Astronomy (VISTA).
\item  data products from the Wide-field Infrared Survey Explorer (\textit{WISE}), which is a joint project of the University of California, Los Angeles, and the Jet Propulsion Laboratory/California Institute of Technology, funded by the National Aeronautics and Space Administration. \item  NASA's Astrophysics Data System Bibliographic Services.
\item  the SVO Filter Profile Service \citep[\url{http://svo2.cab.inta-csic.es/theory/fps/};][]{2012ivoa.rept.1015R, 2020sea..confE.182R} supported from the Spanish MINECO through grant AYA2017-84089.
\end{enumerate}
The author expresses his appreciation to the Stack Exchange (\url{https://stackexchange.com/sites#}) Q\&A platform, askers, and answerers (e.g. ProfRob), for providing useful community inputs and hints for approaching and understanding topics.\\
The author thanks the anonymous referee and the editors for constructive comments that helped to improve the paper. The author acknowledges the useful suggestion and idea by the referee to explicit all the crossmatch separations of the sources, for a more precise counterpart- and SED analysis. The author thanks Maria Alessandra Papa, as well as Anjana Ashok, Banafsheh Beheshtipour, Maximillian Bensch, Pep Blai Covas Vidal, Vladimir Dergachev, Heinz-Bernd Eggenstein, Liudmila Fesik, Prasanna Mohan Joshi, Bernd Machenschalk, Thorben Menne, Jing Ming, Gianluca Pagliaro, Reinhard Prix, Gabriele Richardson, Avneet Singh, Benjamin Steltner, Yuanhao Zhang, Badri Krishnan, and Bruce Allen, for valuable exchanges. With funding from the Max-Planck-Gesellschaft (M.FE.A.QOP10004).

\section*{Data Availability}

No new data were generated or analysed in support of this research.
 



\bibliographystyle{mnras}
\bibliography{NS_ref_citeads}



\begin{appendix}

\section{FURTHER DETAILS ON THE $M_G>16$ SAMPLE, THE COLOUR-SELECTED SUBSAMPLE, AND PROBABLE WHITE-DWARF PHOTOMETRIC CANDIDATES}\label{app:method1}

 
For the 75\,073 sources in the $M_G>16$ sample, the relative flux uncertainties ($1/snr_\mathrm{flux}$) at $G$ band are lesser than 10 per cent, and for most of the 58\,135 and 58\,656 sources with $G_{BP}$ and $G_{RP}$ photometry, the uncertainties are lesser than 30 per cent and 20 per cent, that is 0.3 and 0.2~mag [magnitude uncertainty $\approx 2.5 \log (1\pm 1/snr_\mathrm{flux})$]. These uncertainties reflect the minimum numbers of observations for the bandpasses required for the \textit{Gaia} (E)DR3 catalogue: \texttt{phot\_g\_n\_obs} $\geq10$,  \texttt{phot\_bp\_n\_obs} $\geq2$, and \texttt{phot\_rp\_n\_obs} $\geq2$, for $G$, $G_{BP}$, and $G_{RP}$, respectively. The maximum apparent magnitudes at $G_{BP}$, $G$, and $G_{RP}$ in the sample are 24.5, 21.4, and 22.3~mag (sources without parallax in the whole \textit{Gaia} catalogue can have values as high as 25.3, 23.0, and 24.7~mag), whereas the mean apparent magnitudes are of 21, 20.5, and 19~mag, respectively. The maximum absolute magnitudes are of 23.3, 21.7, and 20.0~mag using either the inverse parallax or the $d50$ distance. The implied intrinsic limits are shown by the lower grey dotted lines in Fig.~\ref{cmd_gaia}. The absolute magnitude difference between the former and latter distance methods is always smaller than +0.8~mag (distance factor lesser than 1.5), with zero difference at $snr_{\pi}>10$, and an increasing difference towards $M_G=16$~mag for $snr_{\pi}<5$--10. For our search, these differences are not significant.


In the colour-selected subsample, two sources have very blue $G_{BP}-G$ colours of $-$2.7~mag and are not shown in Fig.~\ref{cmd_gaia}. These also have the highest value of 8 in the subsample, of the corrected colour-excess factor (\texttt{phot\_bp\_rp\_excess\_factor\_corrected}). The factor is defined as the sum of the $G_{BP}$ and $G_{RP}$-band fluxes divided by the $G$-band flux, corrected by a colour function. For an isolated stellar point source, it is expected to be close to zero, with positive values when the source has more flux at $G_{BP}$ and $G_{RP}$ bands than at $G$ band, and vice versa, for negative values \citep{2021A&A...649A...3R}. The $G$-band flux is obtained from profile fitting in a relatively small field, of $0.35\times2.1$~arcsec$^2$ for faint sources, and the $G_{BP}$ and $G_{RP}$ band fluxes are sums in an aperture of $3.5\times2.1$~arcsec$^2$, thus any high excess factor would indicate contamination from a nearby source. Gaia DR3 4661504581157717120 ($G=20.7$, $G-G_{RP}=0.5$~mag) is in a dense field of the Large Magellanic Cloud, clearly visible as a relatively blue source in $ugriz$-band stack images of NOIRLab. It blends with a star 2.2~arcsec south and of $G=18.6$~mag. It corresponds to NSC DR2 189095\_13039 with $u-g=21.235-21.219=0.02$~mag (only $ug$-band photometry are available) and the blending, redder star is NSC DR2 189095\_10634, with $u-g=20.643-18.854=1.79$, $g-r=0.49$, $r-i=0.26$, and $i-z=0.10$~mag. Gaia DR3 5853004242120032256 ($G=21.0$, $G-G_{RP}=0.8$~mag) is in the Galactic plane at $(l,b)=(313.17,-3.70)\degr$. It is clearly visible in images of the DECam Plane Survey (DECaPS) DR1 \citep{2018ApJS..234...39S}, where it appears almost as red as its nearest and brighter \textit{Gaia} neighbour 2.4~arcsec north. In fact, it consists of a 0.65~arcsec separation blend of $grizY$-band DECaPS 1028228089934061633 ($i=20.3$~mag) and $iz$-band DECaPS 1028228089938032219 ($i=21.2$~mag), which have red colours. Thus the unusually blue $G_{BP-G}$ colours of the two \textit{Gaia} sources and their very high colour-excess factors are probably caused by their faintness compared to sources blending within~arcsec. The remaining of the 2464 sources have corrected colour-excess factors in the range $-$0.6--1.6 and peaking at 0.1. There is also a flux bias towards greater values (overestimating the flux) at low fluxes close to the minimum thresholding in \textit{Gaia} DR3, and it affects mostly the $G_{BP}$ band for red sources \citep{2021A&A...649A...3R,2021A&A...649A...5F}. The flux bias is negligible for input $G_{BP}$ and $G_{RP}$-band magnitudes that are less faint than about 20.5 and 20.0~mag \citep{2021A&A...649A...5F}. This bias implies that in our colour selection of faint sources we may have taken in sources that are in fact redder than their colour values in the catalogue, as for example, RC1--RC5 (Appendix~\ref{sec:serfind}).


Regarding known probable white-dwarf photometric candidates among the sources of the colour-selected subsample, there are (i) 113 with $P_\mathrm{WD}>0.5$ in \citetalias{2021A&A...649A...6G} and (ii) 72 with $P_\mathrm{WD}\leq0.5$ in \citetalias{2021A&A...649A...6G} but $P_\mathrm{WD}>0.5$ in \citet[71 sources]{2021MNRAS.508.3877G} or \citet[Gaia DR3 1781440902671450880]{2019MNRAS.482.4570G}. Also, there are 35 in the 100~pc white-dwarf candidate sample of \citet{2019MNRAS.485.5573T}, of which two are not in the previous three references and 33 are with $P_\mathrm{WD}>0.5$ in \citetalias{2021A&A...649A...6G}. GCNS and non-GCNS probable white-dwarf photometric candidates are represented by yellow and orange small squares in Fig.~\ref{cmd_gaia} and have $M_G<17.7$~mag. Three of the former are identified as resolved multiple systems in GCNS. These are bound wide binaries and indicated by red circles. Gaia DR3 1570574230528214400 ($M_G=16.0$, $G_{BP}-G=0.7$~mag) is part of a double white dwarf binary of separation of 128~au, with Gaia DR3 1570574260593313152 ($M_G=15.5$) as the brighter component; it was previously identified as an unresolved large proper-motion star in the catalogue of \citet{2005AJ....129.1483L}. Gaia DR3 4476576831314323456 ($M_G=16.0$, $G_{BP}-G=0.5$~mag) is part of a double white dwarf binary of separation of 824~au, with Gaia DR3 4476576831311990144 ($M_G=15.5$) as the brighter component; both components were previously identified individually as white dwarf photometric candidates in the catalogue of \citet{2019MNRAS.482.4570G}. Gaia DR3 6178573689547383168 ($M_G=16.5$, $G_{BP}-G=-0.1$~mag) is part of a white-dwarf--main-sequence binary of separation of 2754~au, with Gaia DR3 6178573792627114496 ($M_G=7.7$, photometric spectral type $\sim$ K6V) as the brighter component; the components were previously identified as a wide binary in the catalogue of \citet{2020ApJS..247...66H}. In total, there are 143 sources with $P_\mathrm{WD}>0.5$ in \citet{2021MNRAS.508.3877G}: these have $A_V$ mean extinctions of 0.01--0.09~mag, $d50$ distances of 32--104~pc, $P_\mathrm{WD}$ of 0.91--1.00, $snr_{\pi}$ of 3.4--166, $M_G$ of 16.0--17.6~mag, numbers of \textit{Gaia} DR3 sources in their respective constant-area surrounding fields of 30--11\,355 (\texttt{DENSITY} parameter), H-atmosphere $T_\mathrm{eff}$ of 3734--6958~K (78 sources), He-atmosphere $T_\mathrm{eff}$ of 3956--7785~K (84 sources), and proper motions of 1.8--2354~mas~yr$^{-1}$.

\section{SEDS OF THREE NEARBY WHITE-DWARF SOURCES}\label{app:litextraWDs}

\begin{figure}
    \includegraphics[width=\columnwidth]{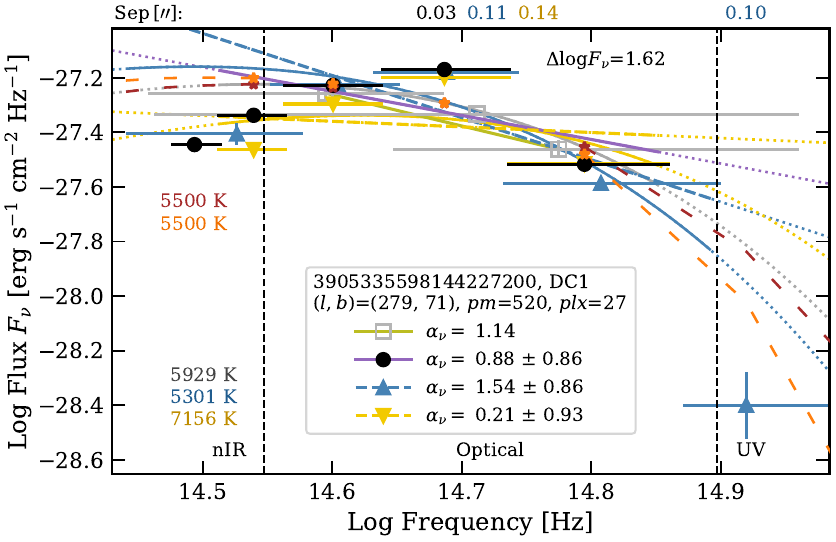}    
    \includegraphics[width=\columnwidth]{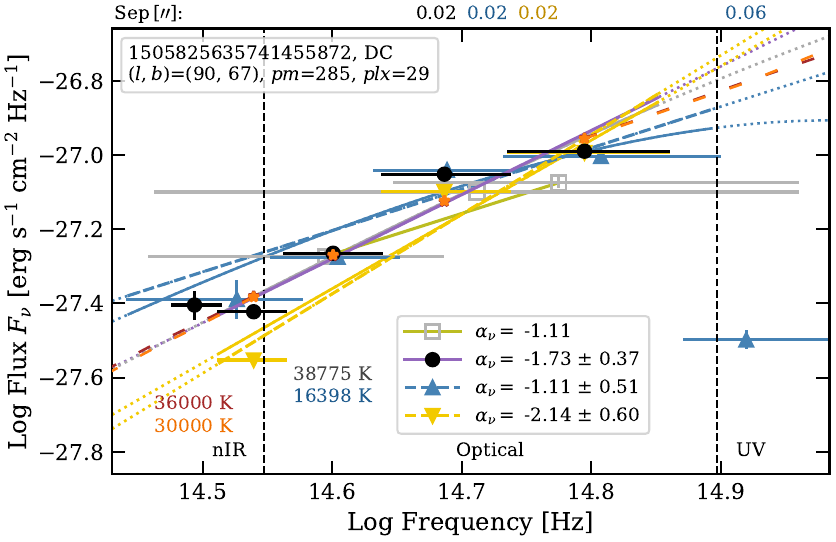}
    \includegraphics[width=\columnwidth]{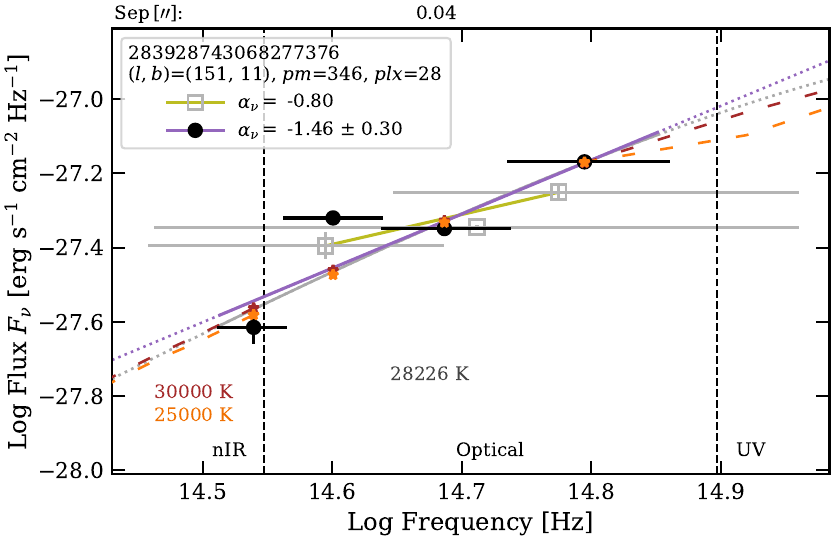}
    \caption{SEDs of three white-dwarf sources with $griz$-band flux profiles that are: (\textit{top panel}) very curved (`DC1'), ({\sl middle panel}) curved-up (large $\alpha_{\nu}$ errors), peaking at $gr$ bands, and with significantly suppressed $u$-band flux (Gaia~DR3~1505825635741455872), (\textit{bottom panel}) straight-up with an apparent $i$-band flux excess (Gaia~DR3~283928743068277376). Same as in Fig.~\ref{fsed_uv1}.}
    \label{dc_wd_op}
\end{figure}

The SEDs of three large proper-motion (285--520~mas~yr$^{-1}$), nearby ($d50=34$$-$37~pc) white-dwarf sources, are shown in Fig.~\ref{dc_wd_op}. Gaia DR3 3905335598144227200 (SDSS J122048.65+091412.1; `DC1') is a DC white dwarf of very cool- and mixed H/He atmosphere \citep{2004ApJ...612L.129G}. Its SED diverges from that of a blackbody and from those of theoretical white dwarf pure H and He atmospheres. The flux suppression at $iz$ bands, producing bluer colours relative to the $r$ band (i.e. $r-i$ and $r-z$), is caused by H$_2$ pressure-induced absorption; this absorption is greater in white dwarf H atmospheres that are cooler or richer in helium. The DC white dwarf has a mixed atmosphere of $N(\mathrm{H})/N(\mathrm{He})=10^{-1.22}$ and $T_\mathrm{eff}=3531\pm37$~K \citep{2018ApJS..239...26L}. Gaia~DR3~1505825635741455872 (SDSS~J140324.67+453332.7) is a DC white dwarf with a $griz$-band flux profile that is blue, curved (large $\alpha_{\nu}$ errors), peaking at $gr$ bands, and with significantly suppressed $u$-band flux.

Gaia DR3 283928743068277376 is a probable white dwarf candidate based on \textit{Gaia} photometry and astrometry (\citealt{2019MNRAS.482.4570G, 2021MNRAS.508.3877G}; \citetalias{2021A&A...649A...6G}). It is spectroscopically unconfirmed and noted as having peculiar \textit{Gaia} and PS1 colours \citep{2020MNRAS.499.1890M}. The SED in Figure~\ref{dc_wd_op} indicates that it is very blue ($\alpha_{\nu}=-1.5$) and has a peculiar $i$-band flux excess. The later probably stems from contamination by the glare of the bright and very small proper motion Gaia DR3 283928678644413824 star ($G=16.48$~mag, 2.6~mas~yr$^{-1}$). In PS1, the glare contamination is strongest redwards and at the $i$-band epoch (2013.8), when the separation of the sources is of 3.0~arcsec. (At the later \textit{Gaia} 2016.0 epoch, it is even closer, of 2.8~arcsec.) At epoch 2005.0, in $ugriz$-band images at the border of the SDSS DR9 survey (where no catalogue photometry is available), the white dwarf source is farther at 5.4~arcsec and well isolated. Its SDSS $ri$-band counts compared to other sources with PS1 magnitudes indicate fluxes consistent with a relatively straight $griz$-band profile.

\section{A LARGE PROPER-MOTION-, $\sim$900~AU-WIDE, WHITE-DWARF--M-TYPE-DWARF BINARY}\label{sec:comovingwidebinary}

\vspace{0.5cm}
\begin{figure*}
    \includegraphics[width=\textwidth]{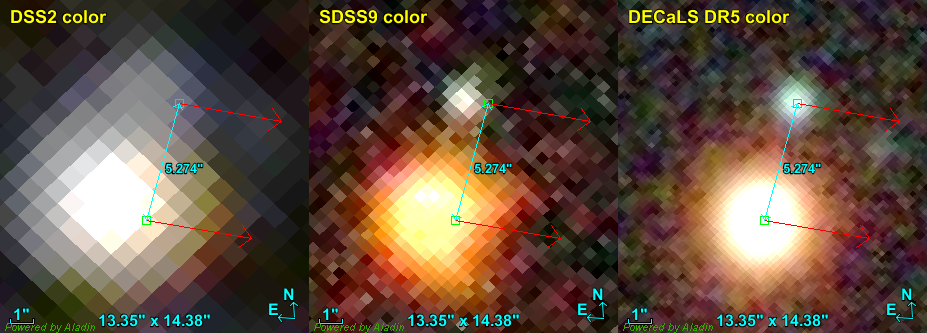}
     \caption{Sequence of DSS2- (1995--1996), SDSS- (2005), and DECaLS (2015--2017-epoch) colour images showing that P18 is comoving with the brighter and redder star, Gaia DR3 3948582104883968384. The small green squares indicate the \textit{Gaia} DR3 positions at epoch 2016.0, and the red arrows indicate the proper motions, magnified to show their equality.}
    \label{Gaia_EDR3_394858210488323763_comoving}
\end{figure*}

\begin{table}
	\centering
	\caption{Astrometry of the proper motion stellar binary of P18.}
	\label{tab:comoving}
	\footnotesize
	\begin{tabular}{l@{\hspace{1.2\tabcolsep}}r@{\hspace{1.2\tabcolsep}}c}
\hline
Parameter  &  Component A & Component B        \\
\hline
\textit{Gaia} DR3   &   3948582104883237632       & 3948582104883968384    \\
$pm$RA (mas~yr$^{-1}$)  & -88.586~$\pm$~2.108  & -90.558~$\pm$~0.044 \\
$pm$Dec (mas~yr$^{-1}$) & -15.676~$\pm$~1.150  & -15.720~$\pm$~0.035 \\
Plx (mas)       & (13.53~$\pm$~1.72)    & 5.882~$\pm$~0.041 \\
RA ICRS ($\degr$)   & 187.59980167314257   & 187.60021864187 \\
Dec ICRS ($\degr$)  &  18.886339386336545  & 18.88492845292  \\
Epoch (yr)     & 2016.0             & 2016.0            \\
$G$ (mag)      & 20.781~$\pm$~0.014   & 15.694~$\pm$~0.003  \\
$G_{BP}$ (mag) & 21.021~$\pm$~0.193   & 16.995~$\pm$~0.005  \\
$G_{RP}$ (mag) & 19.873~$\pm$~0.153   & 14.565~$\pm$~0.004  \\
$\rho$ (arcsec), PA (deg) & \multicolumn{2}{c}{5.274, 164.4} \\
$a_\mathrm{proj}$ (au) at 168 pc & \multicolumn{2}{c}{886} \\
\\
\multicolumn{3}{l}{NSC DR2}  \\
RA  J2000 ($\degr$)   & 187.599795417444    &  187.60020211637547  \\
Dec J2000 ($\degr$)  & 18.8863284791447    &  18.884931674173455  \\
Epoch (yr) & 2016.5714  &  2016.7941   \\
$\rho$ (arcsec), PA (deg) & \multicolumn{2}{c}{5.216, 164.6} \\
\\
\multicolumn{3}{l}{PS1 DR1}  \\
RA J2000 ($\degr$)  & 187.599828930   & 187.600258740 \\
Dec J2000 ($\degr$) &  18.886344310   &  18.884933650  \\
Epoch (yr) &  2011.5613  & 2011.8995 \\
$\rho$ (arcsec), PA (deg) & \multicolumn{2}{c}{5.285, 163.9} \\
\\
\multicolumn{3}{l}{SDSS DR 13} \\ 
RA ICRS ($\degr$)   &  187.600089120098    & 187.600512478144 \\
Dec ICRS ($\degr$)  &  18.8863672307873    & 18.8849507640562  \\
Epoch (yr)      &  2005.3562           & 2005.3562         \\
$\rho$ (arcsec), PA (deg) & \multicolumn{2}{c}{5.299, 164.2} \\
\\
\multicolumn{3}{l}{\textit{Gaia} - SDSS} \\ 
$pm$RA (mas~yr$^{-1}$)  & -92.0~$\pm$~1.6  & -94.0~$\pm$~0.1 \\
$pm$Dec (mas~yr$^{-1}$) & -9.4~$\pm$~1.9   & -7.5~$\pm$~0.1  \\
\multicolumn{3}{l}{NSC - SDSS} \\ 
$pm$RA (mas~yr$^{-1}$)  & -89.2~$\pm$~1.7  &  -92.4~$\pm$~0.3 \\
$pm$Dec (mas~yr$^{-1}$) & -12.4~$\pm$~2.0  &  -6.0~$\pm$~0.6  \\
\hline
	\end{tabular}
\begin{flushleft}
\end{flushleft}
\end{table}

Gaia DR3 3948582104883237632 (P18; $\alpha_{\nu}=1.6$) is located 5.3~arcsec north of a 5~mag brighter star, Gaia DR3 3948582104883968384 ($G=15.69$~mag). The two objects share the same \textit{Gaia} proper motion of ($-$90.6,~$-$15.7) within 2~mas~yr$^{-1}$. We corroborated this independently by measuring their \textit{Gaia}--SDSS and NSC-SDSS proper motions, assuming the source nominal coordinates in the catalogues (Table~\ref{tab:comoving}). The agreement between the declination components of the proper motions is corroborated less well, possibly because the PSFs of these sources blend along the declination. The displacement over time of the stellar binary can be seen in the survey colour images of Fig.~\ref{Gaia_EDR3_394858210488323763_comoving}, from the Second
Digitized Sky Survey (DSS2, \citealt{1996ASPC..101...88L}; blue epoch 1995.2 and red epoch 1996.4) to SDSS DR9 (epoch 2005.4) to the Dark Energy Camera Legacy Survey (DECaLS; \citealt{2018MNRAS.473.5113D}) DR5 (epoch 2015--2017), and relative to the \textit{Gaia} DR3 2016.0 positions. Possibly because of the glare of the bright star, the candidate has slightly larger than usual PS1 $griz$-band uncertainties (0.10, 0.08, 0.06, and 0.08~mag, respectively), its SDSS PSF magnitudes have unrealistic errors (we considered its SDSS model magnitudes instead), and its parallax (and $d50=79$~pc) disagrees significantly with that of the bright star. The parallax of the later is more accurate and implies a distance of about 168$\pm1$~pc \citep[from the \textit{Gaia} (E)DR3 distance catalogue of][]{2021AJ....161..147B} and thereby a tangential velocity of $v_\mathrm{t}=73$~km~s$^{-1}$. At the larger distance, the $G$-band absolute magnitude of P18 becomes $M_G=20.78 + 5 \times \log(5.8817)-10=14.63$~mag, which remains well below the stellar main sequence. The $G_{BP}-G=1.30$, $G-G_{RP}=1.13$, and $M_G=9.54$~mag of the bright star indicate consistently an M2.5V spectral type, corresponding to an effective temperature of 3500~K and a mass of 0.40~M$_{\odot}$, based on the MSC compilation. At 168~pc, the binary would have a projected physical separation of 886~au. The probability of chance alignment at this projected physical separation is likely to be less than 0.1 per cent, especially in this sparse stellar field well above the direction of the Galactic plane, and it is near the peak ($\sim$1000~au) of the distribution of separations of low-mass stellar wide binaries \citep{2018MNRAS.480.4884E,2021A&A...649A...6G,2021MNRAS.506.2269E}. Assuming a mass of 0.40~M$_{\odot}$ for each component, and multiplying the physical separation by 1.26 to account statistically for inclination angle and eccentricity of binary orbits \citep{1992ApJ...396..178F}, the binary would have an orbital period of $4.2\times10^4$~yr.

\section{Red-colour dwarfs}\label{sec:serfind}

\begin{table*}
	\centering
	\caption{Red-colour dwarfs.}
	\label{tab:superred}
       \footnotesize
	\begin{tabular}{l@{\hspace{1.2\tabcolsep}}r@{\hspace{1.2\tabcolsep}}r@{\hspace{1.2\tabcolsep}}r@{\hspace{1.2\tabcolsep}}r@{\hspace{1.2\tabcolsep}}r@{\hspace{1.2\tabcolsep}}r@{\hspace{1.2\tabcolsep}}r@{\hspace{1.2\tabcolsep}}r@{\hspace{1.2\tabcolsep}}r@{\hspace{1.2\tabcolsep}}r@{\hspace{1.2\tabcolsep}}r@{\hspace{1.2\tabcolsep}}r@{\hspace{1.2\tabcolsep}}r@{\hspace{1.2\tabcolsep}}r@{\hspace{1.2\tabcolsep}}r}
		\hline
Flag    & $Plx$         &  $\mu_{\alpha^*}$ &  $\mu_{\delta}$ & $G$   &  $BP-G$ &$G-RP$  & $g-r$     & $r-i$       &  $i-z$                 & P &  $z-J$  &  $J-H$   &  $J$$-$$K_\mathrm{s}$  &  $J$$-$$W1$  &  $W1$$-$$W2$  \\
          & (mas)        &  \multicolumn{2}{c}{(mas~yr$^{-1}$)}            & (mag) &  (mag) &(mag)    & (mag)    & (mag)    & (mag)   &   & (mag)  & (mag)  & (mag)               &  (mag)     & (mag)   \\
\hline
RC1    &     11.1~$\pm$~2.9    &     -4.5 &   -15.6 &  20.82 &  0.44~$\pm$~0.42  &   0.80~$\pm$~0.22  &   1.73 &   1.28 &   0.74 &   3 &  1.25 &    --  &    --  &   (1.18) &   (0.18) \\
RC2    &     10.8~$\pm$~2.8    &     -9.0 &    -9.9 &  20.94 &  0.36~$\pm$~0.29  &   0.78~$\pm$~0.13  &   1.76 &   1.43 &   0.77 &   3 &  1.67 &    --  &    --  &   0.98 &  -0.13 \\
RC3    &     11.1~$\pm$~2.6    &     29.2 &   -28.4 &  20.90 &  0.57~$\pm$~0.39  &   0.89~$\pm$~0.14  &   1.46 &   1.60 &   0.69 &   2 &  1.53 &   0.48 &   0.75 &   1.10 &  -0.44 \\
RC4    &     14.6~$\pm$~3.0    &     -8.8 &    10.4 &  20.86 &  0.56~$\pm$~0.42  &   0.86~$\pm$~0.17  &   1.49 &   1.66 &   0.71 &   2 &  1.64 &   0.38 &   0.82 &   1.02 &   0.03 \\
RC5    &     10.7~$\pm$~2.0    &     53.8 &   -19.0 &  20.88 &  0.48~$\pm$~0.42  &   0.88~$\pm$~0.14  &   1.65 &   2.15 &   0.88 &   2 &  1.64 &    --  &   0.71 &   0.92 &   0.01 \\
              \hline
	\end{tabular}
\scriptsize
\vspace{-0.23cm}
\begin{flushleft}
Proper motion component errors are of 1.3--2.7 mas~yr$^{-1}$. The griz-band photometric (P) colours are from 2-- NSC and 3-- SDSS.
\end{flushleft}
\end{table*}

The five sources with the reddest PS1, SDSS and NSC $griz$-band slopes, independent of slope uncertainty, are Gaia DR3 900742499823605632 (RC1), Gaia DR3 4389594118549228416 (RC2), Gaia DR3 4780916942594976768 (RC3), Gaia DR3 3618788707439181440 (RC4), and Gaia DR3 4906188219358053248 (RC5). These are listed in Table~\ref{sec:serfind}. Their parameters and photometry are listed in Tables~\ref{tab:all2}, \ref{tab:opt_ir2}, and \ref{tab:opt_sdss_nsc2}, and their SEDs are shown in Fig.~\ref{fsed_superred_ir} (available as online supplementary material).

In the $M_G$ versus $G-G_\mathrm{RP}$ colour--absolute magnitude diagram of Fig.~\ref{gaia_can_grp_vs_g_cmd} (available as online supplementary material), these have colours close to 0.9~mag, as red as the DQpec white dwarf LHS~2229 (Gaia DR3 740483560857296768, flagged `DQpec'). DQ white dwarfs have atmospheres enriched in carbon, and the peculiar DQ LHS~2229 has very strong shifted C$_2$ Swan bands in the blue optical \citep[see][]{2012ApJS..199...29G}, affecting mostly the $g$ band. Because of this, in the SED of LHS~2229, we fitted only the $riz$ bands (Figs.~\ref{fsed_uv2},~\ref{fsed_uv_ir},~and~\ref{fsed_superred_ir}). However, at the optical--infrared, the five sources are much redder than LHS~2229, as seen at the $G-J$, $i-z$ and $i-J$ (Figs.~\ref{gaia_can_gj_vs_g_cmd}, \ref{mg_iz_msbd}, and ~\ref{mg_ij_msbd}), $r-i$ (Fig.~\ref{gr_vs_ri}), $r-J$ (Figs.~\ref{rj_vs_jw1} and \ref{rj_vs_w1w2}), $z-J$ (Fig.~\ref{iz_vs_zj}), and $i-W1$ colours (Fig.~\ref{ri_vs_iw1}). In fact, their colours are those of early M-type dwarfs. Considering the increasing number of sources towards the $G-G_\mathrm{RP}$ selection boundary (Fig.~\ref{cmd_gaia}), similar red-colour sources could probably be found beyond the boundary, some of these merging with the ML-type dwarf sequence.

From least-square fitting with main-sequence dwarf photometry of the MSC compilation, we obtain the following spectral types: RC1 (M4V), RC2 (M4.5V), RC3 (M4.5V), RC4 (M5V), and RC5 (M5.5V). Because the sources are under the main sequence (3--4~mag under in the $M_G$ versus $i-z$ diagram of Fig.~\ref{mz_iz_msbd}), these could be of low-metallicity and subdwarfs. However, the $J-K$ versus $i-J$ colours of RC3, RC4, and RC5 overlap with the main sequence and do not suggest extreme- or ultra-subdwarf M types, which have less red $J-K$ colours (fig.~1 of \citealt{2017MNRAS.464.3040Z}). Their $J-K$ versus $J-W2$ and $J-H$ versus $z-J$ colours also overlap with the main sequence (fig.~18 of \citealt{2018MNRAS.480.5447Z}). Interestingly, the red-colour sources are located towards sparse stellar regions out of the direction of the Galactic plane (Section~\ref{disc:gen}), and their \textit{Gaia} photoastrometric parameters do not appear to indicate binarity. However, their $snr_{\pi}$ are low, of 4--5, and their \texttt{visibility\_periods\_used} are of about 10, suggesting that their distances could be larger. We note that RC1 has no $HK_\mathrm{s}$-band measurements and has a less coincidental mid-infrared counterpart, located 1.73~arcsec west. Supposing that RC1 is in fact relatively faint in the near and mid-infrared, then it could either be (i) an M-type subdwarf or (ii) a white dwarf cooler than 3250~K, such as the few ultra-cool white dwarfs known \citep{2020MNRAS.499.1890M}. Finally, we note that RC3 and RC5 have large proper motions of 41--57~mas~yr$^{-1}$.

\section{ADDITIONAL PHOTOMETRY}

\begin{table*}
	\centering
	\caption{PS1 and infrared photometry of the sources. \textcolor{gray}{(Preprint layout)}}
	\label{tab:opt_ir}
	\scriptsize
	\begin{tabular}{l@{\hspace{1.2\tabcolsep}}l@{\hspace{1.2\tabcolsep}}r@{\hspace{1.2\tabcolsep}}r@{\hspace{1.2\tabcolsep}}r@{\hspace{1.2\tabcolsep}}r@{\hspace{1.2\tabcolsep}}r@{\hspace{1.2\tabcolsep}}r@{\hspace{1.2\tabcolsep}}l@{\hspace{1.2\tabcolsep}}r@{\hspace{1.2\tabcolsep}}r@{\hspace{1.2\tabcolsep}}r@{\hspace{1.2\tabcolsep}}l@{\hspace{1.2\tabcolsep}}r@{\hspace{1.2\tabcolsep}}r}
		\hline
Flag     &  Sep & $g_\mathrm{~PS1}$   &  $r_\mathrm{~PS1}$   &  $i_\mathrm{~PS1}$   &  $z_\mathrm{~PS1}$   &  $y_\mathrm{~PS1}$    &  Sep    &  $J$              &  $H$              &  $K_\mathrm{s}$      &  P  &  Sep & $W1$             &  $W2$             \\
                        &  ($\prime\prime$)         &  (mag)            &  (mag)            &  (mag)            &  (mag)            &  (mag)    &  ($\prime\prime$)   &  (mag)            &  (mag)            &  (mag)            &     &  ($\prime\prime$)       & (mag)            &  (mag)            \\
\hline
$\gamma$\_a     &   --   &          --       &          --       &          --       &          --       &          --       &   --   &          --       &          --       &          --       &  6   &   --   &           --        &           --        \\
$\gamma$\_b$^a$ &   --   &          --       &          --       &          --       &          --       &          --       &  0.12  &  17.971~$\pm$~0.219 &  17.319~$\pm$~0.230 &  17.112~$\pm$~0.260 &  6   &   --   &           --        &           --        \\
$\gamma$\_c     &   --   &          --       &          --       &          --       &          --       &          --       &  0.24  &  18.647~$\pm$~0.160 &          --       &          --       &  1   &   --   &           --        &           --        \\
$\gamma$\_d     &  0.14  &  21.445~$\pm$~0.065 &  20.323~$\pm$~0.064 &  19.710~$\pm$~0.021 &  19.396~$\pm$~0.005 &  19.501~$\pm$~0.061 &  0.37  &  17.818~$\pm$~0.064 &          --       &  17.031~$\pm$~0.112 &  1   &   --   &           --        &           --        \\
$\gamma$\_e     &  0.20  &          --       &  19.642~$\pm$~0.043 &  20.052~$\pm$~0.017 &          --       &          --       &   --   &          --       &          --       &          --       &  --  &   --   &           --        &           --        \\
\hline
      X2     &   --   &          --       &          --       &          --       &          --       &          --       &  0.17  &  19.338~$\pm$~0.184 &  19.259~$\pm$~0.289 &          --       &  1   &   --   &           --        &           --        \\
    X\_a     &   --   &          --       &          --       &          --       &          --       &          --       &  0.30  &  18.735~$\pm$~0.070 &          --       &  18.165~$\pm$~0.208 &  1   &   --   &           --        &           --        \\
    X\_b     &  0.05  &  21.044~$\pm$~0.073 &  20.318~$\pm$~0.023 &  19.995~$\pm$~0.114 &          --       &  19.727~$\pm$~0.147 &  0.22  &  18.293~$\pm$~0.072 &          --       &  17.646~$\pm$~0.162 &  1   &   --   &           --        &           --        \\
    X\_c     &  0.05  &  20.855~$\pm$~0.032 &  20.192~$\pm$~0.032 &  19.935~$\pm$~0.450 &  19.568~$\pm$~0.136 &  19.780~$\pm$~0.087 &  0.29  &  18.777~$\pm$~0.181 &          --       &          --       &  1   &   --   &           --        &           --        \\
    X\_d     &  0.02  &  21.209~$\pm$~0.021 &  20.509~$\pm$~0.024 &  20.132~$\pm$~0.026 &  19.991~$\pm$~0.022 &  19.927~$\pm$~0.081 &  0.03  &  18.782~$\pm$~0.111 &          --       &          --       &  2   &  0.37  &   18.306~$\pm$~0.151  &   17.950            \\
    X\_e     &   --   &          --       &          --       &          --       &          --       &          --       &  0.24  &  19.035~$\pm$~0.095 &          --       &  18.375~$\pm$~0.304 &  1   &   --   &           --        &           --        \\
\hline
     UV1     &   --   &          --       &          --       &          --       &          --       &          --       &   --   &          --       &          --       &          --       &  --  &   --   &           --        &           --        \\
     UV2$^b$ &  0.05  &  20.031~$\pm$~0.021 &  19.767~$\pm$~0.012 &  19.660~$\pm$~0.009 &  19.639~$\pm$~0.023 &  19.534~$\pm$~0.024 &  0.17  &  18.688~$\pm$~0.089 &          --       &          --       &  1   &   --   &           --        &           --        \\
     UV3     &   --   &          --       &          --       &          --       &          --       &          --       &  0.26  &  18.533~$\pm$~0.072 &          --       &  17.634~$\pm$~0.159 &  1   &  0.84  &   17.440~$\pm$~0.068  &   17.923            \\
     UV4     &  0.05  &  20.127~$\pm$~0.018 &  20.262~$\pm$~0.017 &  19.926~$\pm$~0.020 &  19.702~$\pm$~0.019 &  19.293~$\pm$~0.032 &  0.51  &  18.137~$\pm$~0.078 &          --       &  17.466~$\pm$~0.156 &  1   &   --   &           --        &           --        \\
     UV5     &   --   &          --       &          --       &          --       &          --       &          --       &   --   &          --       &          --       &          --       &  --  &   --   &           --        &           --        \\
     UV6     &  0.07  &  20.060~$\pm$~0.020 &  19.727~$\pm$~0.016 &  19.620~$\pm$~0.008 &  19.572~$\pm$~0.031 &  19.497~$\pm$~0.057 &   --   &          --       &          --       &          --       &  --  &   --   &           --        &           --        \\
     UV7     &   --   &          --       &          --       &          --       &          --       &          --       &  0.17  &  18.951~$\pm$~0.050 &          --       &  17.646~$\pm$~0.087 &  1   &  0.19  &   16.020~$\pm$~0.026  &   15.099~$\pm$~0.028  \\
     UV8     &  0.06  &  21.570~$\pm$~0.056 &  21.099~$\pm$~0.033 &  20.216~$\pm$~0.033 &  19.701~$\pm$~0.026 &  19.481~$\pm$~0.056 &  0.17  &  18.275~$\pm$~0.065 &          --       &          --       &  2   &  0.76  &   17.465~$\pm$~0.082  &   17.544~$\pm$~0.288  \\
     UV9     &  0.03  &  21.575~$\pm$~0.040 &  21.174~$\pm$~0.091 &  20.505~$\pm$~0.020 &  19.959~$\pm$~0.029 &  19.576~$\pm$~0.052 &   --   &          --       &          --       &          --       &  --  &  0.42  &   17.457~$\pm$~0.066  &   17.613~$\pm$~0.229  \\
    UV10     &  0.06  &  20.895~$\pm$~0.035 &  20.974~$\pm$~0.025 &  21.204~$\pm$~0.068 &  21.163~$\pm$~0.056 &          --       &   --   &          --       &          --       &          --       &  --  &   --   &           --        &           --        \\
    UV11     &  0.07  &  21.242~$\pm$~0.020 &  20.736~$\pm$~0.025 &  20.539~$\pm$~0.016 &  20.369~$\pm$~0.038 &  20.283~$\pm$~0.089 &  0.27  &  19.276~$\pm$~0.161 &          --       &          --       &  2   &   --   &           --        &           --        \\
    UV12     &  0.07  &  20.751~$\pm$~0.076 &  20.419~$\pm$~0.011 &  20.247~$\pm$~0.026 &  20.169~$\pm$~0.042 &  19.811~$\pm$~0.142 &   --   &          --       &          --       &          --       &  --  &   --   &           --        &           --        \\
    UV13     &  0.59  &  21.141~$\pm$~0.028 &  20.155~$\pm$~0.016 &  19.912~$\pm$~0.013 &  19.775~$\pm$~0.018 &  19.477~$\pm$~0.040 &  0.19  &  18.345~$\pm$~0.045 &  17.983~$\pm$~0.059 &  17.948~$\pm$~0.117 &  1   &  0.75  &   17.818~$\pm$~0.097  &   17.811            \\
    UV14     &  0.52  &  21.364~$\pm$~0.051 &  20.471~$\pm$~0.026 &  20.083~$\pm$~0.014 &  19.952~$\pm$~0.033 &  19.858~$\pm$~0.066 &  0.05  &  18.954~$\pm$~0.116 &          --       &          --       &  2   &   --   &           --        &           --        \\
    UV15     &  0.49  &  21.338~$\pm$~0.033 &  20.610~$\pm$~0.017 &  20.520~$\pm$~0.017 &  20.282~$\pm$~0.038 &  19.822~$\pm$~0.136 &  0.19  &  19.251~$\pm$~0.170 &          --       &          --       &  2   &  1.00  &   17.915~$\pm$~0.113  &   18.388            \\
\hline
    Rad1     &  0.12  &  21.242~$\pm$~0.060 &  20.870~$\pm$~0.020 &  20.404~$\pm$~0.021 &  20.351~$\pm$~0.051 &          --       &   --   &          --       &          --       &          --       &  --  &   --   &           --        &           --        \\
  Rad\_a     &  0.05  &  21.347~$\pm$~0.091 &  20.865~$\pm$~0.043 &  20.583~$\pm$~0.027 &  20.529~$\pm$~0.182 &  20.250~$\pm$~0.186 &   --   &          --       &          --       &          --       &  --  &   --   &           --        &           --        \\
  Rad\_b$^c$ &   --   &          --       &          --       &          --       &          --       &          --       &  0.17  &  19.565~$\pm$~0.073 &          --       &  19.461~$\pm$~0.122 &  --  &   --   &           --        &           --        \\
  Rad\_c     &  0.03  &  20.591~$\pm$~0.043 &  19.949~$\pm$~0.046 &  19.569~$\pm$~0.038 &  19.588~$\pm$~0.076 &  19.550~$\pm$~0.028 &  0.39  &  18.469~$\pm$~0.161 &          --       &          --       &  1   &   --   &           --        &           --        \\
  Rad\_d     &  0.02  &  21.950~$\pm$~0.135 &  20.797~$\pm$~0.050 &  20.150~$\pm$~0.016 &  19.746~$\pm$~0.016 &  19.452~$\pm$~0.046 &  0.10  &  18.261~$\pm$~0.058 &          --       &  17.500~$\pm$~0.130 &  1   &   --   &           --        &           --        \\
  Rad\_e     &  0.03  &  21.967~$\pm$~0.172 &  20.858~$\pm$~0.069 &  20.479~$\pm$~0.085 &  19.928~$\pm$~0.052 &  19.510~$\pm$~0.065 &   --   &          --       &          --       &          --       &  --  &   --   &           --        &           --        \\
  Rad\_f     &  0.23  &  21.798~$\pm$~0.153 &  20.922~$\pm$~0.056 &  20.401~$\pm$~0.077 &  20.161~$\pm$~0.092 &  19.706~$\pm$~0.044 &  0.36  &  18.320~$\pm$~0.046 &  17.782~$\pm$~0.056 &  17.451~$\pm$~0.074 &  3   &   --   &           --        &           --        \\
  Rad\_g     &  0.03  &  21.934~$\pm$~0.161 &  21.057~$\pm$~0.064 &  20.727~$\pm$~0.065 &  20.571~$\pm$~0.178 &  19.953~$\pm$~0.172 &  0.09  &  19.075~$\pm$~0.164 &          --       &          --       &  1   &   --   &           --        &           --        \\
\hline
      P0     &  0.05  &  20.678~$\pm$~0.032 &  21.011~$\pm$~0.061 &  21.246~$\pm$~0.040 &  21.077~$\pm$~0.168 &          --       &   --   &          --       &          --       &          --       &  --  &  2.08  &  (17.156~$\pm$~0.059) &  (17.057~$\pm$~0.172) \\
    P1X1     &  0.22  &  20.223~$\pm$~0.018 &  20.458~$\pm$~0.019 &  20.663~$\pm$~0.028 &  20.682~$\pm$~0.071 &  20.676~$\pm$~0.012 &   --   &          --       &          --       &          --       &  --  &   --   &           --        &           --        \\
      P2     &  0.03  &  20.717~$\pm$~0.025 &  20.677~$\pm$~0.023 &  20.715~$\pm$~0.026 &  20.646~$\pm$~0.058 &          --       &   --   &          --       &          --       &          --       &  --  &   --   &           --        &           --        \\
      P3$^d$ &  0.01  &  20.908~$\pm$~0.067 &  20.621~$\pm$~0.098 &  20.659~$\pm$~0.019 &  20.633~$\pm$~0.035 &          --       &  0.08  &  19.387~$\pm$~0.095 &  18.922~$\pm$~0.152 &  18.201~$\pm$~0.199 &  4   &  0.41  &   17.394~$\pm$~0.070  &   16.913~$\pm$~0.166  \\
      P4$^e$ &  0.21  &          --       &          --       &  20.783~$\pm$~0.045 &          --       &          --       &  0.19  &  19.598~$\pm$~0.059 &  18.839~$\pm$~0.115 &  18.241~$\pm$~0.089 &  7   &  0.05  &   17.711~$\pm$~0.099  &   16.226~$\pm$~0.081  \\
      P5$^f$ &  0.02  &  21.020~$\pm$~0.029 &  20.838~$\pm$~0.047 &  20.660~$\pm$~0.029 &  20.613~$\pm$~0.058 &  19.631~$\pm$~0.160 &  0.21  &  20.081~$\pm$~0.262 &          --       &          --       &  4   &   --   &           --        &           --        \\
      P6$^g$ &  0.00  &  21.178~$\pm$~0.040 &  20.992~$\pm$~0.079 &  20.767~$\pm$~0.043 &  20.679~$\pm$~0.052 &          --       &  0.32  &          --       &          --       &          --       &  1   &  0.12  &   17.761~$\pm$~0.113  &   17.215~$\pm$~0.224  \\
      P7     &  0.05  &  20.569~$\pm$~0.017 &  20.267~$\pm$~0.033 &  20.062~$\pm$~0.035 &  19.994~$\pm$~0.043 &  20.002~$\pm$~0.093 &  0.12  &  19.089~$\pm$~0.181 &          --       &          --       &  1   &   --   &           --        &           --        \\
      P8     &  0.06  &  21.302~$\pm$~0.055 &  20.985~$\pm$~0.104 &  20.776~$\pm$~0.112 &  20.605~$\pm$~0.041 &  20.535~$\pm$~0.165 &   --   &          --       &          --       &          --       &  --  &   --   &           --        &           --        \\
      P9     &  0.38  &  20.057~$\pm$~0.021 &  19.701~$\pm$~0.073 &  19.488~$\pm$~0.018 &  19.389~$\pm$~0.026 &  19.492~$\pm$~0.057 &  0.42  &  18.422~$\pm$~0.090 &          --       &  17.785~$\pm$~0.174 &  1   &   --   &           --        &           --        \\
     P10$^h$ &  0.03  &  21.260~$\pm$~0.026 &  20.910~$\pm$~0.055 &  20.675~$\pm$~0.033 &  20.462~$\pm$~0.100 &  20.091~$\pm$~0.067 &  0.10  &  19.882~$\pm$~0.226 &          --       &          --       &  1   &   --   &           --        &           --        \\
     P11     &  0.04  &  21.267~$\pm$~0.036 &  20.909~$\pm$~0.012 &  20.640~$\pm$~0.022 &  20.464~$\pm$~0.006 &  20.296~$\pm$~0.117 &   --   &          --       &          --       &          --       &  --  &   --   &           --        &           --        \\
     P12     &  0.17  &  20.892~$\pm$~0.083 &  20.566~$\pm$~0.016 &  20.298~$\pm$~0.022 &  20.091~$\pm$~0.063 &  20.004~$\pm$~0.072 &  0.28  &  19.117~$\pm$~0.126 &          --       &          --       &  2   &   --   &           --        &           --        \\
     P13     &  0.01  &  21.400~$\pm$~0.032 &  21.017~$\pm$~0.076 &  20.737~$\pm$~0.030 &  20.486~$\pm$~0.072 &  20.744~$\pm$~0.175 &  0.06  &  19.884~$\pm$~0.228 &          --       &          --       &  2   &   --   &           --        &           --        \\
     P14     &  0.14  &  20.933~$\pm$~0.010 &  20.531~$\pm$~0.027 &  20.203~$\pm$~0.040 &  20.114~$\pm$~0.054 &          --       &  0.39  &  18.962~$\pm$~0.130 &          --       &          --       &  1   &   --   &           --        &           --        \\
     P15$^i$ &  0.03  &  21.336~$\pm$~0.003 &  20.866~$\pm$~0.024 &  20.604~$\pm$~0.020 &  20.409~$\pm$~0.035 &  20.035~$\pm$~0.185 &  0.11  &  19.588~$\pm$~0.135 &  19.324~$\pm$~0.215 &          --       &  4   &   --   &           --        &           --        \\
     P16     &  0.10  &  20.912~$\pm$~0.028 &  20.461~$\pm$~0.026 &  20.177~$\pm$~0.013 &  19.944~$\pm$~0.022 &  19.700~$\pm$~0.075 &  0.15  &  18.956~$\pm$~0.132 &          --       &  17.887~$\pm$~0.209 &  1   &   --   &           --        &           --        \\
     P17     &  0.01  &  20.497~$\pm$~0.022 &  19.979~$\pm$~0.030 &  19.701~$\pm$~0.019 &  19.496~$\pm$~0.010 &  19.666~$\pm$~0.043 &  0.07  &  18.457~$\pm$~0.079 &          --       &  17.985~$\pm$~0.213 &  1   &  1.75  &  (17.032~$\pm$~0.056) &  (16.835~$\pm$~0.128) \\
     P18     &  0.30  &  21.203~$\pm$~0.098 &  20.778~$\pm$~0.077 &  20.418~$\pm$~0.060 &  20.197~$\pm$~0.081 &          --       &  0.44  &  19.511~$\pm$~0.191 &          --       &          --       &  2   &   --   &           --        &           --        \\
     P19     &  0.51  &  21.011~$\pm$~0.030 &  20.534~$\pm$~0.027 &  20.241~$\pm$~0.037 &  19.916~$\pm$~0.037 &  19.703~$\pm$~0.027 &  1.05  &  18.436~$\pm$~0.051 &          --       &          --       &  2   &  1.14  &   17.599~$\pm$~0.084  &   17.605~$\pm$~0.264  \\
     P20     &  0.04  &  20.845~$\pm$~0.033 &  20.356~$\pm$~0.026 &  19.986~$\pm$~0.015 &  19.705~$\pm$~0.110 &  19.738~$\pm$~0.136 &   --   &          --       &          --       &          --       &  --  &   --   &           --        &           --        \\
     P21     &  0.47  &  20.637~$\pm$~0.036 &  20.107~$\pm$~0.075 &  19.741~$\pm$~0.027 &  19.581~$\pm$~0.157 &  19.419~$\pm$~0.176 &  0.30  &  18.502~$\pm$~0.089 &          --       &  18.143~$\pm$~0.281 &  1   &   --   &           --        &           --        \\
     P22     &  0.08  &  21.130~$\pm$~0.044 &  20.498~$\pm$~0.110 &  20.090~$\pm$~0.017 &  19.816~$\pm$~0.114 &  19.426~$\pm$~0.182 &  0.23  &  18.471~$\pm$~0.075 &          --       &  17.957~$\pm$~0.228 &  1   &  2.26  &  (16.792~$\pm$~0.057) &  (16.602~$\pm$~0.118) \\
     P23     &  0.18  &  21.583~$\pm$~0.027 &  20.962~$\pm$~0.052 &  20.517~$\pm$~0.049 &  20.230~$\pm$~0.035 &          --       &  0.17  &  18.987~$\pm$~0.126 &          --       &          --       &  1   &   --   &           --        &           --        \\
     P24     &  0.09  &  21.686~$\pm$~0.026 &  21.137~$\pm$~0.094 &  20.565~$\pm$~0.082 &  20.322~$\pm$~0.057 &  19.810~$\pm$~0.177 &  0.16  &  19.187~$\pm$~0.148 &          --       &          --       &  1   &   --   &           --        &           --        \\
     P25     &  0.25  &  21.334~$\pm$~0.025 &  20.715~$\pm$~0.036 &  20.223~$\pm$~0.059 &  19.975~$\pm$~0.044 &  19.674~$\pm$~0.146 &  0.13  &  18.452~$\pm$~0.084 &          --       &  17.620~$\pm$~0.145 &  1   &   --   &           --        &           --        \\
     P26     &  0.17  &  21.596~$\pm$~0.126 &  20.950~$\pm$~0.040 &  20.534~$\pm$~0.085 &  20.160~$\pm$~0.045 &  19.976~$\pm$~0.197 &  0.46  &  18.383~$\pm$~0.074 &          --       &  17.555~$\pm$~0.148 &  1   &   --   &           --        &           --        \\
     P27     &  0.02  &  21.096~$\pm$~0.035 &  20.453~$\pm$~0.073 &  19.990~$\pm$~0.024 &  19.701~$\pm$~0.015 &  19.757~$\pm$~0.096 &  0.14  &  18.427~$\pm$~0.076 &          --       &  17.771~$\pm$~0.178 &  1   &   --   &           --        &           --        \\
     P28     &  0.06  &  21.561~$\pm$~0.023 &  20.893~$\pm$~0.059 &  20.433~$\pm$~0.028 &  20.171~$\pm$~0.093 &          --       &   --   &          --       &          --       &          --       &  --  &  0.94  &   17.634~$\pm$~0.091  &   18.467~$\pm$~0.506  \\
     P29     &  0.03  &  21.703~$\pm$~0.079 &  21.001~$\pm$~0.056 &  20.513~$\pm$~0.026 &  20.168~$\pm$~0.084 &          --       &  0.05  &  19.013~$\pm$~0.136 &          --       &          --       &  1   &   --   &           --        &           --        \\
     P30     &  0.08  &  21.689~$\pm$~0.135 &  21.003~$\pm$~0.068 &  20.465~$\pm$~0.043 &  20.116~$\pm$~0.019 &  19.732~$\pm$~0.080 &  0.12  &  18.904~$\pm$~0.148 &          --       &          --       &  2   &   --   &           --        &           --        \\
     P31     &  0.04  &  21.613~$\pm$~0.075 &  20.916~$\pm$~0.060 &  20.429~$\pm$~0.019 &  20.049~$\pm$~0.046 &  19.580~$\pm$~0.141 &  0.14  &  18.795~$\pm$~0.095 &          --       &  17.805~$\pm$~0.177 &  1   &   --   &           --        &           --        \\
     P32     &  0.10  &  20.979~$\pm$~0.055 &  20.293~$\pm$~0.030 &  19.792~$\pm$~0.022 &  19.411~$\pm$~0.019 &  19.260~$\pm$~0.058 &  0.14  &  17.802~$\pm$~0.033 &  17.107~$\pm$~0.031 &  17.005~$\pm$~0.055 &  3   &   --   &           --        &           --        \\
     P33     &  0.18  &  21.712~$\pm$~0.006 &  20.875~$\pm$~0.085 &  20.529~$\pm$~0.019 &  20.164~$\pm$~0.087 &          --       &   --   &          --       &          --       &          --       &  --  &   --   &           --        &           --        \\
     P34     &  0.03  &  21.790~$\pm$~0.046 &  20.993~$\pm$~0.048 &  20.393~$\pm$~0.041 &  20.018~$\pm$~0.021 &  19.674~$\pm$~0.101 &  0.02  &  18.545~$\pm$~0.084 &          --       &          --       &  2   &  0.14  &   18.118~$\pm$~0.186  &   18.066            \\
     P35     &  0.23  &  21.575~$\pm$~0.037 &  20.686~$\pm$~0.122 &  20.188~$\pm$~0.005 &  19.820~$\pm$~0.118 &  19.800~$\pm$~0.142 &  0.34  &  18.261~$\pm$~0.067 &          --       &  17.851~$\pm$~0.193 &  1   &   --   &           --        &           --        \\
     P36     &  0.01  &  22.022~$\pm$~0.139 &  21.188~$\pm$~0.026 &  20.503~$\pm$~0.082 &  20.105~$\pm$~0.010 &  19.806~$\pm$~0.139 &  0.10  &  18.681~$\pm$~0.061 &  18.453~$\pm$~0.103 &  18.181~$\pm$~0.126 &  3   &   --   &           --        &           --        \\
     P37     &  0.09  &  21.368~$\pm$~0.075 &  20.568~$\pm$~0.033 &  19.927~$\pm$~0.110 &  19.480~$\pm$~0.058 &  19.343~$\pm$~0.026 &  0.19  &  18.062~$\pm$~0.069 &          --       &  17.484~$\pm$~0.164 &  1   &   --   &           --        &           --        \\
    \hline
	\end{tabular}
\vspace{-0.23cm}
\begin{flushleft}
Angular separations are in arcseconds. $JHK_\mathrm{s}$-band photometry (P) from (1) VHS, (2) UHS, (3) UKIDSS GPS, (4) UKIDSS LAS, (5) VMC, (6) VVV, and (7) VIKING (3--4: $K$- instead of $K_\mathrm{s}$ band). $W1W2$ magnitudes in parenthesis: less coincidental counterparts. $^a$ $Z=19.065$~$\pm$~0.231, $Y=18.538$~$\pm$~0.233 (6). $^b$ $Y=18.870$~$\pm$~0.047 (1). $^c$ $Y=19.676$~$\pm$~0.071 (5). $^d$ $Y=19.714$~$\pm$~0.102 (4). $^e$ $Z=20.088$~$\pm$~0.033, $Y=19.796$~$\pm$~0.053 (7). $^f$ $Y=19.959$~$\pm$~0.138 (4). $^g$ $Y=20.259$~$\pm$~0.277 (1). $^h$~$Z=20.286$~$\pm$~0.208 (UKIDSS GCS). $^i$ $Y=20.137$~$\pm$~0.146~mag (4).
\end{flushleft}
\end{table*}

\begin{table*}
	\centering
	\caption{SDSS and NSC photometry of the sources. \textcolor{gray}{(Preprint layout)}}
	\label{tab:opt_sdss_nsc}
	\scriptsize
	\begin{tabular}{l@{\hspace{1.2\tabcolsep}}l@{\hspace{1.2\tabcolsep}}l@{\hspace{1.2\tabcolsep}}l@{\hspace{1.2\tabcolsep}}l@{\hspace{1.2\tabcolsep}}l@{\hspace{1.2\tabcolsep}}l@{\hspace{1.2\tabcolsep}}l@{\hspace{1.2\tabcolsep}}l@{\hspace{1.2\tabcolsep}}l@{\hspace{1.2\tabcolsep}}l@{\hspace{1.2\tabcolsep}}l@{\hspace{1.2\tabcolsep}}l@{\hspace{1.2\tabcolsep}}r}
\hline
Flag     &  Sep  & $u_\mathrm{~SDSS}$      &  $g_\mathrm{~SDSS}$  &  $r_\mathrm{~SDSS}$  &  $i_\mathrm{~SDSS}$  &  $z_\mathrm{~SDSS}$  &   Sep  & $g_\mathrm{~NSC}$   &  $r_\mathrm{~NSC}$   &  $i_\mathrm{~NSC}$   &  $z_\mathrm{~NSC}$   &  $Y_\mathrm{~NSC}$   \\
                    &  ($\prime\prime$)         &  (mag)                &  (mag)            &  (mag)            &  (mag)            &  (mag)            & ($\prime\prime$)         &   (mag)            &  (mag)            &  (mag)            &  (mag)            &  (mag)            \\
\hline
$\gamma$\_a &   --   &          --           &          --       &          --       &          --       &          --       &  0.29  &          --       &  19.958~$\pm$~0.085 &  19.009~$\pm$~0.056 &          --       &          --       \\
$\gamma$\_b &   --   &          --           &          --       &          --       &          --       &          --       &  0.20  &  22.690~$\pm$~0.101 &  21.111~$\pm$~0.038 &  20.188~$\pm$~0.035 &  19.621~$\pm$~0.035 &  19.209~$\pm$~0.048 \\
$\gamma$\_c &   --   &          --           &          --       &          --       &          --       &          --       &  0.09  &  21.655~$\pm$~0.030 &  20.501~$\pm$~0.013 &  20.157~$\pm$~0.025 &  19.720~$\pm$~0.032 &  19.449~$\pm$~0.052 \\
$\gamma$\_d &   --   &          --           &          --       &          --       &          --       &          --       &   --   &          --       &          --       &          --       &          --       &          --       \\
$\gamma$\_e &   --   &          --           &          --       &          --       &          --       &          --       &   --   &          --       &          --       &          --       &          --       &          --       \\
\hline
X2       &   --   &          --           &          --       &          --       &          --       &          --       &  0.04  &  20.852~$\pm$~0.009 &  20.220~$\pm$~0.007 &  20.089~$\pm$~0.009 &  19.955~$\pm$~0.015 &  19.821~$\pm$~0.067 \\
X\_a     &   --   &          --           &          --       &          --       &          --       &          --       &  0.20  &  21.273~$\pm$~0.018 &  20.589~$\pm$~0.020 &  20.177~$\pm$~0.022 &  20.079~$\pm$~0.028 &  19.898~$\pm$~0.054 \\
X\_b     &   --   &          --           &          --       &          --       &          --       &          --       &  0.20  &  20.909~$\pm$~0.019 &  20.275~$\pm$~0.013 &  19.930~$\pm$~0.026 &  19.751~$\pm$~0.032 &  19.656~$\pm$~0.063 \\
X\_c     &   --   &          --           &          --       &          --       &          --       &          --       &  0.18  &  20.431~$\pm$~0.013 &  19.924~$\pm$~0.009 &  19.707~$\pm$~0.012 &  19.600~$\pm$~0.016 &  19.576~$\pm$~0.041 \\
X\_d     &   --   &          --           &          --       &          --       &          --       &          --       &   --   &          --       &          --       &          --       &          --       &          --       \\
X\_e     &   --   &          --           &          --       &          --       &          --       &          --       &  0.18  &          --       &  20.488~$\pm$~0.020 &          --       &          --       &          --       \\
\hline
UV1      &   --   &          --           &          --       &          --       &          --       &          --       &  0.09  &  20.449~$\pm$~0.007 &  20.818~$\pm$~0.005 &  21.070~$\pm$~0.033 &  21.344~$\pm$~0.091 &          --       \\
UV2      &   --   &          --           &          --       &          --       &          --       &          --       &  0.23  &  19.922~$\pm$~0.013 &          --       &  19.641~$\pm$~0.015 &          --       &          --       \\
UV3      &   --   &          --           &          --       &          --       &          --       &          --       &  0.02  &  20.747~$\pm$~0.017 &  20.939~$\pm$~0.020 &          --       &          --       &          --       \\
UV4      &   --   &          --           &          --       &          --       &          --       &          --       &  0.02  &  20.016~$\pm$~0.015 &          --       &          --       &          --       &          --       \\
UV5      &   --   &          --           &          --       &          --       &          --       &          --       &  0.10  &  21.109~$\pm$~0.080 &  20.812~$\pm$~0.040 &          --       &  20.688~$\pm$~0.106 &          --       \\
UV6      &   --   &          --           &          --       &          --       &          --       &          --       &  0.15  &  20.007~$\pm$~0.012 &          --       &  19.406~$\pm$~0.016 &          --       &          --       \\
UV7      &   --   &          --           &          --       &          --       &          --       &          --       &  0.13  &  20.518~$\pm$~0.006 &  20.435~$\pm$~0.007 &  20.213~$\pm$~0.008 &  20.211~$\pm$~0.018 &  20.224~$\pm$~0.040 \\
UV8      &  0.15  &  21.638~$\pm$~0.116     &  21.564~$\pm$~0.045 &  21.072~$\pm$~0.039 &  20.198~$\pm$~0.038 &  19.587~$\pm$~0.061 &  0.04  &  21.472~$\pm$~0.017 &  21.075~$\pm$~0.014 &          --       &  19.698~$\pm$~0.010 &          --       \\
UV9      &  0.06  &  21.954~$\pm$~0.154     &  21.416~$\pm$~0.044 &  21.328~$\pm$~0.048 &  20.619~$\pm$~0.043 &  19.956~$\pm$~0.087 &  0.01  &  21.476~$\pm$~0.099 &  21.390~$\pm$~0.091 &          --       &  19.912~$\pm$~0.014 &          --       \\
UV10     &   --   &          --           &          --       &          --       &          --       &          --       &  0.05  &          --       &          --       &          --       &          --       &          --       \\
UV11     &  0.04  &  22.563~$\pm$~0.030$^a$ &  21.647~$\pm$~0.100 &  20.960~$\pm$~0.071 &  20.722~$\pm$~0.073 &  20.447~$\pm$~0.153 &  0.12  &  21.106~$\pm$~0.006 &  20.660~$\pm$~0.008 &  20.490~$\pm$~0.005 &  20.351~$\pm$~0.008 &  20.098~$\pm$~0.035 \\
UV12     &  0.23  &  21.774~$\pm$~0.169     &  20.726~$\pm$~0.028 &  20.436~$\pm$~0.028 &  20.308~$\pm$~0.035 &  19.805~$\pm$~0.101 &  0.20  &  20.638~$\pm$~0.009 &  20.374~$\pm$~0.007 &          --       &  20.250~$\pm$~0.025 &          --       \\
UV13     &  0.15  &  20.792~$\pm$~0.091     &  21.440~$\pm$~0.054 &  20.137~$\pm$~0.038 &  19.891~$\pm$~0.024 &  19.591~$\pm$~0.056 &  0.11  &  21.101~$\pm$~0.011 &  20.159~$\pm$~0.005 &  19.966~$\pm$~0.007 &  19.731~$\pm$~0.009 &  19.475~$\pm$~0.027 \\
UV14     &  0.04  &  22.497~$\pm$~0.293     &  21.565~$\pm$~0.048 &  20.510~$\pm$~0.028 &  20.052~$\pm$~0.026 &  19.623~$\pm$~0.065 &  0.03  &  21.500~$\pm$~0.031 &  20.495~$\pm$~0.015 &          --       &  19.952~$\pm$~0.019 &          --       \\
UV15     &  0.09  &  21.699~$\pm$~0.190     &  21.699~$\pm$~0.057 &  20.661~$\pm$~0.038 &  20.459~$\pm$~0.052 &  19.999~$\pm$~0.109 &  0.03  &  21.408~$\pm$~0.027 &  20.772~$\pm$~0.021 &          --       &  20.199~$\pm$~0.029 &          --       \\
\hline
Rad1     &  0.18  &          --           &  21.424~$\pm$~0.049 &  20.867~$\pm$~0.055 &  20.624~$\pm$~0.067 &  20.553~$\pm$~0.180 &   --   &          --       &          --       &          --       &          --       &          --       \\
Rad\_a   &   --   &          --           &          --       &          --       &          --       &          --       &   --   &          --       &          --       &          --       &          --       &          --       \\
Rad\_b   &   --   &  21.337~$\pm$~0.129$^a$ &          --       &          --       &          --       &          --       &  0.15  &  20.198~$\pm$~0.012 &  20.208~$\pm$~0.012 &  20.113~$\pm$~0.011 &  20.133~$\pm$~0.021 &          --       \\
Rad\_c   &   --   &          --           &          --       &          --       &          --       &          --       &  0.36  &  20.406~$\pm$~0.010 &  19.752~$\pm$~0.010 &  19.482~$\pm$~0.017 &  19.337~$\pm$~0.020 &  19.252~$\pm$~0.042 \\
Rad\_d   &   --   &          --           &          --       &          --       &          --       &          --       &  0.02  &  22.000~$\pm$~0.039 &          --       &  20.312~$\pm$~0.028 &          --       &          --       \\
Rad\_e   &   --   &          --           &          --       &          --       &          --       &          --       &  0.04  &  21.891~$\pm$~0.092 &  20.609~$\pm$~0.055 &  20.234~$\pm$~0.052 &  19.781~$\pm$~0.066 &  18.828~$\pm$~0.138 \\
Rad\_f   &   --   &          --           &          --       &          --       &          --       &          --       &   --   &          --       &          --       &          --       &          --       &          --       \\
Rad\_g   &   --   &          --           &          --       &          --       &          --       &          --       &  0.07  &          --       &  21.212~$\pm$~0.023 &          --       &          --       &          --       \\
\hline
P0       &  0.07  &  20.804~$\pm$~0.080     &  20.705~$\pm$~0.030 &  20.968~$\pm$~0.048 &  21.165~$\pm$~0.086 &  21.213~$\pm$~0.278 &  0.09  &  20.651~$\pm$~0.053 &  20.998~$\pm$~0.075 &          --       &          --       &          --       \\
P1X1     &   --   &          --           &          --       &          --       &          --       &          --       &   --   &          --       &          --       &          --       &          --       &          --       \\
P2       &  0.05  &  21.350~$\pm$~0.121     &  20.754~$\pm$~0.031 &  20.751~$\pm$~0.048 &  21.007~$\pm$~0.096 &  21.077~$\pm$~0.321 &  0.06  &  20.688~$\pm$~0.011 &  20.721~$\pm$~0.020 &  20.833~$\pm$~0.054 &  20.984~$\pm$~0.054 &          --       \\
P3       &  0.01  &          --           &  21.208~$\pm$~0.042 &  20.555~$\pm$~0.031 &  20.467~$\pm$~0.035 &  20.324~$\pm$~0.095 &  0.03  &  21.010~$\pm$~0.016 &  20.626~$\pm$~0.016 &          --       &  20.801~$\pm$~0.024 &          --       \\
P4       &   --   &          --           &          --       &          --       &          --       &          --       &  0.03  &  21.062~$\pm$~0.024 &  21.016~$\pm$~0.029 &  20.938~$\pm$~0.043 &  20.620~$\pm$~0.045 &          --       \\
P5       &  0.14  &  22.571~$\pm$~0.349     &  21.195~$\pm$~0.043 &  20.800~$\pm$~0.043 &  20.719~$\pm$~0.055 &  20.865~$\pm$~0.218 &  0.02  &  21.068~$\pm$~0.015 &  20.911~$\pm$~0.021 &          --       &  20.636~$\pm$~0.032 &          --       \\
P6       &   --   &  21.198~$\pm$~0.022$^a$ &          --       &          --       &          --       &          --       &  0.01  &  20.742~$\pm$~0.009 &  20.926~$\pm$~0.007 &  20.472~$\pm$~0.012 &  20.434~$\pm$~0.022 &          --       \\
P7       &   --   &          --           &          --       &          --       &          --       &          --       &  0.18  &  20.571~$\pm$~0.013 &          --       &  19.515~$\pm$~0.025 &          --       &          --       \\
P8       &   --   &          --           &          --       &          --       &          --       &          --       &  0.01  &  21.485~$\pm$~0.017 &  21.031~$\pm$~0.022 &  20.842~$\pm$~0.024 &  20.726~$\pm$~0.064 &  20.781~$\pm$~0.198 \\
P9       &   --   &          --           &          --       &          --       &          --       &          --       &  0.42  &  19.788~$\pm$~0.012 &          --       &          --       &          --       &          --       \\
P10      &   --   &          --           &          --       &          --       &          --       &          --       &  0.01  &  21.350~$\pm$~0.019 &  20.932~$\pm$~0.008 &  20.753~$\pm$~0.017 &          --       &  20.721~$\pm$~0.073 \\
P11      &   --   &          --           &          --       &          --       &          --       &          --       &   --   &          --       &          --       &          --       &          --       &          --       \\
P12      &  0.17  &  22.497~$\pm$~0.302     &  21.162~$\pm$~0.036 &  20.518~$\pm$~0.029 &  20.334~$\pm$~0.034 &  20.350~$\pm$~0.113 &  0.24  &          --       &          --       &  20.136~$\pm$~0.019 &          --       &          --       \\
P13      &   --   &          --           &          --       &          --       &          --       &          --       &  0.01  &  21.484~$\pm$~0.037 &  21.049~$\pm$~0.022 &          --       &  20.682~$\pm$~0.037 &          --       \\
P14      &   --   &          --           &          --       &          --       &          --       &          --       &  0.33  &  21.065~$\pm$~0.037 &  20.430~$\pm$~0.015 &  20.081~$\pm$~0.032 &          --       &          --       \\
P15      &  0.11  &  23.056~$\pm$~0.403     &  21.552~$\pm$~0.048 &  20.975~$\pm$~0.039 &  20.583~$\pm$~0.039 &  20.486~$\pm$~0.117 &  0.02  &          --       &          --       &          --       &  20.562~$\pm$~0.050 &          --       \\
P16      &   --   &          --           &          --       &          --       &          --       &          --       &  0.13  &  20.906~$\pm$~0.027 &  20.359~$\pm$~0.018 &  20.141~$\pm$~0.022 &          --       &          --       \\
P17      &   --   &          --           &          --       &          --       &          --       &          --       &  0.05  &  20.531~$\pm$~0.008 &  19.922~$\pm$~0.008 &  19.662~$\pm$~0.013 &  19.601~$\pm$~0.016 &  19.446~$\pm$~0.033 \\
P18      &  0.08  &          --           &  21.418~$\pm$~0.050 &  20.961~$\pm$~0.047 &  20.893~$\pm$~0.065 &  20.564~$\pm$~0.172 &  0.04  &  21.264~$\pm$~0.019 &  20.745~$\pm$~0.018 &          --       &  20.214~$\pm$~0.034 &          --       \\
P19      &  0.20  &  22.194~$\pm$~0.364     &  21.111~$\pm$~0.050 &  20.512~$\pm$~0.044 &  20.110~$\pm$~0.040 &  19.635~$\pm$~0.104 &  0.48  &  20.940~$\pm$~0.017 &  20.397~$\pm$~0.015 &          --       &  19.594~$\pm$~0.015 &          --       \\
P20      &   --   &          --           &          --       &          --       &          --       &          --       &  0.31  &          --       &  20.265~$\pm$~0.009 &          --       &          --       &          --       \\
P21      &   --   &          --           &          --       &          --       &          --       &          --       &  0.53  &  20.090~$\pm$~0.018 &  19.970~$\pm$~0.008 &  19.294~$\pm$~0.019 &  19.239~$\pm$~0.035 &  19.458~$\pm$~0.078 \\
P22      &   --   &          --           &          --       &          --       &          --       &          --       &  0.21  &          --       &          --       &  19.900~$\pm$~0.015 &          --       &          --       \\
P23      &   --   &          --           &          --       &          --       &          --       &          --       &  0.18  &          --       &  20.727~$\pm$~0.014 &          --       &          --       &          --       \\
P24      &   --   &          --           &          --       &          --       &          --       &          --       &  0.16  &          --       &  20.774~$\pm$~0.019 &          --       &          --       &          --       \\
P25      &   --   &          --           &          --       &          --       &          --       &          --       &  0.48  &  20.939~$\pm$~0.018 &  20.243~$\pm$~0.009 &  20.009~$\pm$~0.026 &  19.830~$\pm$~0.032 &  20.016~$\pm$~0.213 \\
P26      &   --   &          --           &          --       &          --       &          --       &          --       &  0.06  &          --       &  20.837~$\pm$~0.020 &          --       &          --       &          --       \\
P27      &   --   &          --           &          --       &          --       &          --       &          --       &  0.09  &          --       &  20.389~$\pm$~0.012 &          --       &          --       &          --       \\
P28      &   --   &          --           &          --       &          --       &          --       &          --       &  0.03  &  21.857~$\pm$~0.031 &  20.846~$\pm$~0.023 &  20.234~$\pm$~0.023 &  19.982~$\pm$~0.027 &  19.789~$\pm$~0.049 \\
P29      &   --   &          --           &          --       &          --       &          --       &          --       &  0.02  &  21.807~$\pm$~0.045 &  20.982~$\pm$~0.013 &  20.484~$\pm$~0.031 &          --       &          --       \\
P30      &   --   &          --           &          --       &          --       &          --       &          --       &  0.10  &          --       &          --       &          --       &          --       &          --       \\
P31      &   --   &          --           &          --       &          --       &          --       &          --       &  0.04  &  21.673~$\pm$~0.030 &  20.979~$\pm$~0.019 &  20.448~$\pm$~0.022 &  20.188~$\pm$~0.033 &  20.045~$\pm$~0.058 \\
P32      &   --   &          --           &          --       &          --       &          --       &          --       &  0.51  &          --       &  19.997~$\pm$~0.019 &          --       &          --       &          --       \\
P33      &   --   &          --           &          --       &          --       &          --       &          --       &  0.12  &  21.619~$\pm$~0.041 &  20.688~$\pm$~0.019 &  20.247~$\pm$~0.021 &  20.007~$\pm$~0.025 &  20.089~$\pm$~0.069 \\
P34      &   --   &          --           &          --       &          --       &          --       &          --       &   --   &          --       &          --       &          --       &          --       &          --       \\
P35      &   --   &          --           &          --       &          --       &          --       &          --       &  0.40  &          --       &  20.060~$\pm$~0.018 &          --       &          --       &          --       \\
P36      &   --   &          --           &          --       &          --       &          --       &          --       &   --   &          --       &          --       &          --       &          --       &          --       \\
P37      &   --   &          --           &          --       &          --       &          --       &          --       &  0.04  &  21.809~$\pm$~0.025 &  20.601~$\pm$~0.013 &  19.978~$\pm$~0.026 &  19.604~$\pm$~0.028 &  19.352~$\pm$~0.049 \\
\hline
	\end{tabular}
\vspace{-0.23cm}
\begin{flushleft}
SDSS DR13 $u$ and $z$-band zero-point corrections of $-$0.04 and 0.02~mag are applied according to \url{https://www.sdss.org/dr13/algorithms/fluxcal/\#counts2mag}. Angular separations are in arcseconds. $^a$ $u$-band magnitude from NSC.
\end{flushleft}
\end{table*}

\begin{table*}
	\centering
	\caption{PS1 and infrared photometry of other sources. \textcolor{gray}{(Preprint layout)}}
	\label{tab:opt_ir2}
	\scriptsize
	\begin{tabular}{l@{\hspace{1.2\tabcolsep}}l@{\hspace{1.2\tabcolsep}}r@{\hspace{1.2\tabcolsep}}r@{\hspace{1.2\tabcolsep}}r@{\hspace{1.2\tabcolsep}}r@{\hspace{1.2\tabcolsep}}r@{\hspace{1.2\tabcolsep}}r@{\hspace{1.2\tabcolsep}}l@{\hspace{1.2\tabcolsep}}r@{\hspace{1.2\tabcolsep}}r@{\hspace{1.2\tabcolsep}}r@{\hspace{1.2\tabcolsep}}l@{\hspace{1.2\tabcolsep}}r@{\hspace{1.2\tabcolsep}}r}
		\hline
Flag     &  Sep & $g_\mathrm{~PS1}$   &  $r_\mathrm{~PS1}$   &  $i_\mathrm{~PS1}$   &  $z_\mathrm{~PS1}$   &  $y_\mathrm{~PS1}$    &  Sep    &  $J$              &  $H$              &  $K_\mathrm{s}$      &  P  &  Sep & $W1$             &  $W2$             \\
                        &  ($\prime\prime$)         &  (mag)            &  (mag)            &  (mag)            &  (mag)            &  (mag)    &  ($\prime\prime$)   &  (mag)            &  (mag)            &  (mag)            &     &  ($\prime\prime$)       & (mag)            &  (mag)            \\
\hline
 DZ-E     &  0.22  &  21.590~$\pm$~0.107 &  20.767~$\pm$~0.029 &  19.898~$\pm$~0.017 &  19.651~$\pm$~0.013 &  19.709~$\pm$~0.070 &  0.15  &  19.135~$\pm$~0.131 &          --       &          --       &  1   &   --   &           --        &           --        \\
  DC1     &  0.03  &  20.192~$\pm$~0.012 &  19.324~$\pm$~0.007 &  19.469~$\pm$~0.011 &  19.744~$\pm$~0.026 &  20.012~$\pm$~0.046 &   --   &          --       &          --       &          --       &  --  &   --   &           --        &           --        \\
DQpec     &  0.01  &  18.262~$\pm$~0.028 &  16.404~$\pm$~0.001 &  15.968~$\pm$~0.003 &  16.005~$\pm$~0.002 &  16.012~$\pm$~0.006 &  0.09  &  15.125~$\pm$~0.046 &  14.720~$\pm$~0.063 &  14.535~$\pm$~0.078 &  8   &  0.25  &   14.251~$\pm$~0.016  &   14.202~$\pm$~0.025  \\
     RC1     &  0.04  &  22.069~$\pm$~0.179 &  21.511~$\pm$~0.023 &  20.179~$\pm$~0.023 &  19.544~$\pm$~0.016 &  19.153~$\pm$~0.042 &  0.14  &  18.192~$\pm$~0.054 &          --       &          --       &  2   &  1.73  &   (17.017~$\pm$~0.057)  &   (16.832~$\pm$~0.132)  \\
     RC2     &  0.03  &          --       &  21.727~$\pm$~0.042 &  20.225~$\pm$~0.026 &  19.530~$\pm$~0.021 &  19.137~$\pm$~0.035 &  0.08  &  17.850~$\pm$~0.053 &          --       &          --       &  2   &  0.30  &   16.875~$\pm$~0.046  &   17.006~$\pm$~0.145  \\
  RC3     &   --   &          --       &          --       &          --       &          --       &          --       &  0.08  &  17.782~$\pm$~0.045 &  17.304~$\pm$~0.049 &  17.035~$\pm$~0.077 &  1   &  0.08  &   16.678~$\pm$~0.026  &   17.121~$\pm$~0.110  \\
  RC4$^a$ &  0.03  &          --       &  21.442~$\pm$~0.153 &  20.034~$\pm$~0.028 &  19.319~$\pm$~0.013 &  18.952~$\pm$~0.023 &  0.14  &  17.582~$\pm$~0.068 &  17.205~$\pm$~0.088 &  16.759~$\pm$~0.080 &  1   &  0.26  &   16.557~$\pm$~0.041  &   16.524~$\pm$~0.108  \\
  RC5     &   --   &          --       &          --       &          --       &          --       &          --       &  0.05  &  17.365~$\pm$~0.017 &          --       &  16.654~$\pm$~0.043 &  1   &  0.27  &   16.446~$\pm$~0.029  &   16.437~$\pm$~0.072  \\
  DZQH     &   --   &          --       &          --       &          --       &          --       &          --       &  0.23  &  17.581~$\pm$~0.015 &          --       &  17.625~$\pm$~0.091 &  1   &  0.13  &   16.965~$\pm$~0.051  &   16.717~$\pm$~0.126  \\
  VR     &  0.60  &  20.795~$\pm$~0.044 &  19.023~$\pm$~0.005 &  18.231~$\pm$~0.005 &  17.961~$\pm$~0.006 &  17.823~$\pm$~0.018 &  0.06  &  17.000~$\pm$~0.021 &          --       &  16.931~$\pm$~0.098 &  1   &   --   &           --        &           --        \\
\hline
  	\end{tabular}
\vspace{-0.23cm}
\begin{flushleft}
Angular separations are in arcseconds. $JHK_\mathrm{s}$ photometry (P) from (1) VHS, (2) UHS, (3) UKIDSS GPS, (4) UKIDSS LAS, (5) VMC, (6) VVV, (7) VIKING, and (8) 2MASS (3--4: $K$- instead of $K_\mathrm{s}$ band). $^a$ $Y=18.295$~$\pm$~0.073 (1).
\end{flushleft}
\end{table*}

\begin{table*}
	\centering
	\caption{SDSS and NSC photometry of other sources. \textcolor{gray}{(Preprint layout)}}
	\label{tab:opt_sdss_nsc2}
	\scriptsize
	\begin{tabular}{l@{\hspace{1.2\tabcolsep}}l@{\hspace{1.2\tabcolsep}}l@{\hspace{1.2\tabcolsep}}l@{\hspace{1.2\tabcolsep}}l@{\hspace{1.2\tabcolsep}}l@{\hspace{1.2\tabcolsep}}l@{\hspace{1.2\tabcolsep}}l@{\hspace{1.2\tabcolsep}}l@{\hspace{1.2\tabcolsep}}l@{\hspace{1.2\tabcolsep}}l@{\hspace{1.2\tabcolsep}}l@{\hspace{1.2\tabcolsep}}l@{\hspace{1.2\tabcolsep}}r}
\hline
Flag     &  Sep  & $u_\mathrm{~SDSS}$      &  $g_\mathrm{~SDSS}$  &  $r_\mathrm{~SDSS}$  &  $i_\mathrm{~SDSS}$  &  $z_\mathrm{~SDSS}$  &   Sep  & $g_\mathrm{~NSC}$   &  $r_\mathrm{~NSC}$   &  $i_\mathrm{~NSC}$   &  $z_\mathrm{~NSC}$   &  $Y_\mathrm{~NSC}$   \\
                    &  ($\prime\prime$)         &  (mag)                &  (mag)            &  (mag)            &  (mag)            &  (mag)            & ($\prime\prime$)         &   (mag)            &  (mag)            &  (mag)            &  (mag)            &  (mag)            \\
\hline

DZ-E     &   --   &          --           &          --       &          --       &          --       &          --       &  0.15  &  21.654~$\pm$~0.030 &          --       &  19.939~$\pm$~0.015 &          --       &          --       \\
DC1      &  0.11  &  22.382~$\pm$~0.292     &  20.366~$\pm$~0.028 &  19.349~$\pm$~0.028 &  19.458~$\pm$~0.023 &  19.909~$\pm$~0.084 &  0.14  &  20.189~$\pm$~0.007 &  19.395~$\pm$~0.006 &  19.641~$\pm$~0.022 &  20.059~$\pm$~0.014 &          --       \\
DQpec    &  0.08  &  18.210~$\pm$~0.022     &  18.242~$\pm$~0.025 &  16.411~$\pm$~0.017 &  15.962~$\pm$~0.018 &  16.012~$\pm$~0.014 &  0.11  &  18.205~$\pm$~0.003 &  16.331~$\pm$~0.002 &          --       &  16.101~$\pm$~0.002 &          --       \\
RC1      &  0.06  &          --           &  23.186~$\pm$~0.187 &  21.458~$\pm$~0.060 &  20.181~$\pm$~0.032 &  19.441~$\pm$~0.049 &  0.05  &  22.875~$\pm$~0.060 &  21.426~$\pm$~0.023 &          --       &  19.487~$\pm$~0.014 &          --       \\
RC2      &  0.04  &          --           &  23.481~$\pm$~0.282 &  21.724~$\pm$~0.114 &  20.296~$\pm$~0.048 &  19.523~$\pm$~0.070 &  0.06  &          --       &          --       &  20.079~$\pm$~0.017 &          --       &          --       \\
RC3      &   --   &          --           &          --       &          --       &          --       &          --       &  0.03  &  23.059~$\pm$~0.032 &  21.601~$\pm$~0.012 &  20.000~$\pm$~0.005 &  19.310~$\pm$~0.005 &  19.129~$\pm$~0.013 \\
RC4      &  0.15  &          --           &          --       &  21.739~$\pm$~0.091 &  20.115~$\pm$~0.039 &  19.113~$\pm$~0.064 &  0.01  &  23.078~$\pm$~0.177 &  21.588~$\pm$~0.045 &  19.926~$\pm$~0.006 &  19.219~$\pm$~0.019 &          --       \\
RC5      &   --   &          --           &          --       &          --       &          --       &          --       &  0.01  &  23.677~$\pm$~0.078 &  22.029~$\pm$~0.021 &  19.880~$\pm$~0.007 &  19.001~$\pm$~0.006 &  18.746~$\pm$~0.014 \\
DZQH      &   --   &          --           &          --       &          --       &          --       &          --       &  0.02  &  21.433~$\pm$~0.014 &  19.925~$\pm$~0.006 &  19.366~$\pm$~0.006 &  19.043~$\pm$~0.007 &  18.956~$\pm$~0.021 \\
VR      &   --   &          --           &          --       &          --       &          --       &          --       &  0.56  &  20.719~$\pm$~0.095 &  18.625~$\pm$~0.016 &  18.094~$\pm$~0.013 &          --       &          --       \\
\hline
	\end{tabular}
\vspace{-0.23cm}
\begin{flushleft}
SDSS DR13 $u$ and $z$-band zero-point corrections of $-$0.04 and 0.02~mag are applied according to \url{https://www.sdss.org/dr13/algorithms/fluxcal/\#counts2mag}. Angular separations are in arcseconds.
\end{flushleft}
\end{table*}

\section{Additional photometric diagrams of the sources (online supplementary material)}

\begin{figure*}
    \includegraphics[width=0.9\textwidth]{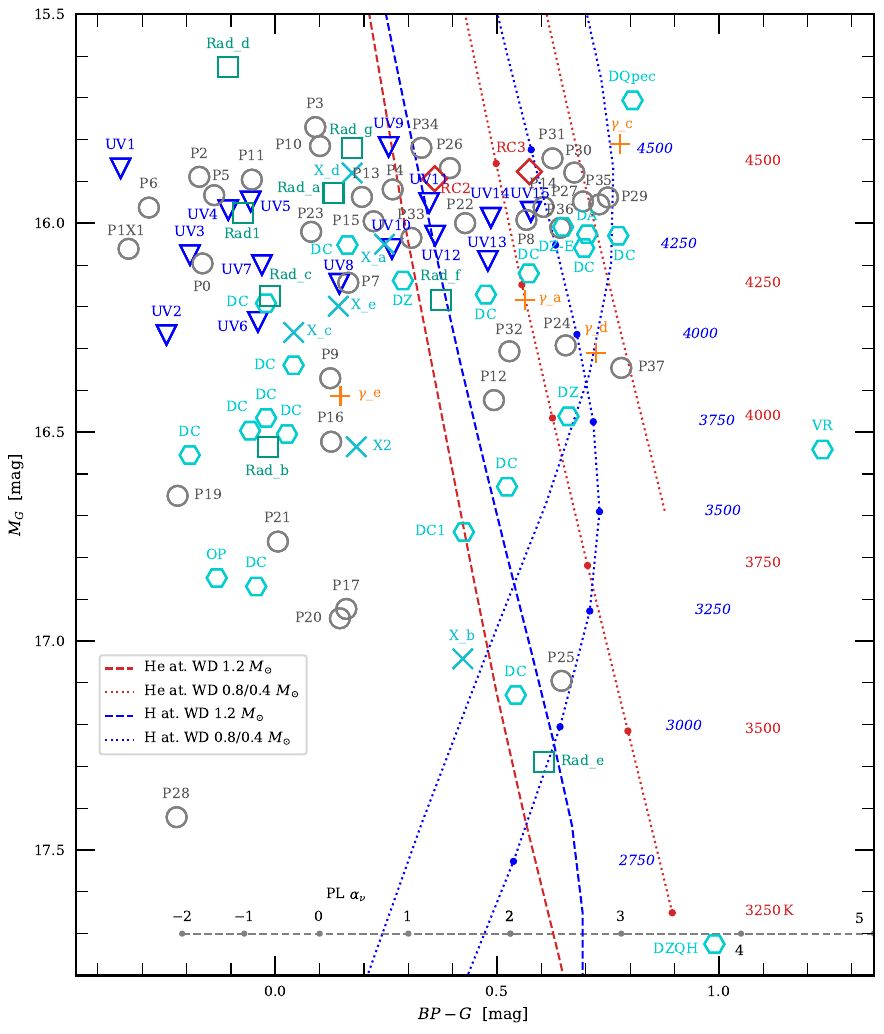}
    \caption{$M_G$ versus $G_{BP}-G$ colour--absolute magnitude diagram, using the distances of Table~\ref{tab:mul}. Gamma-ray, X-ray, ultraviolet, radio, optical--(infrared) power-law, and red-colour flagged sources are represented by pluses (orange), crosses (cyan), triangles (blue), squares (green), circles (grey), and lozenges (red), respectively. Candidates are labelled as in Tables~\ref{tab:all1} and \ref{tab:all2} and can be localised using the text search function in a PDF-document reader. Spectroscopically confirmed- and other literature white dwarfs mentioned in this study are represented by hexagons (turquoise) and labelled. In general, a data point is plotted if each of the involved magnitude errors is smaller than 0.4~mag. Isomasses of white-dwarf pure He and H atmospheres are represented by red and blue lines; those for 1.2~M$_{\odot}$ are dashed and those for 0.8 and 0.4~M$_{\odot}$ dotted. Effective temperatures are indicated for the 0.8~M$_{\odot}$ isomasses.}
    \label{gaia_can_bpg_vs_g_cmd}
\end{figure*}

\begin{figure*}
    \includegraphics[width=0.9\textwidth]{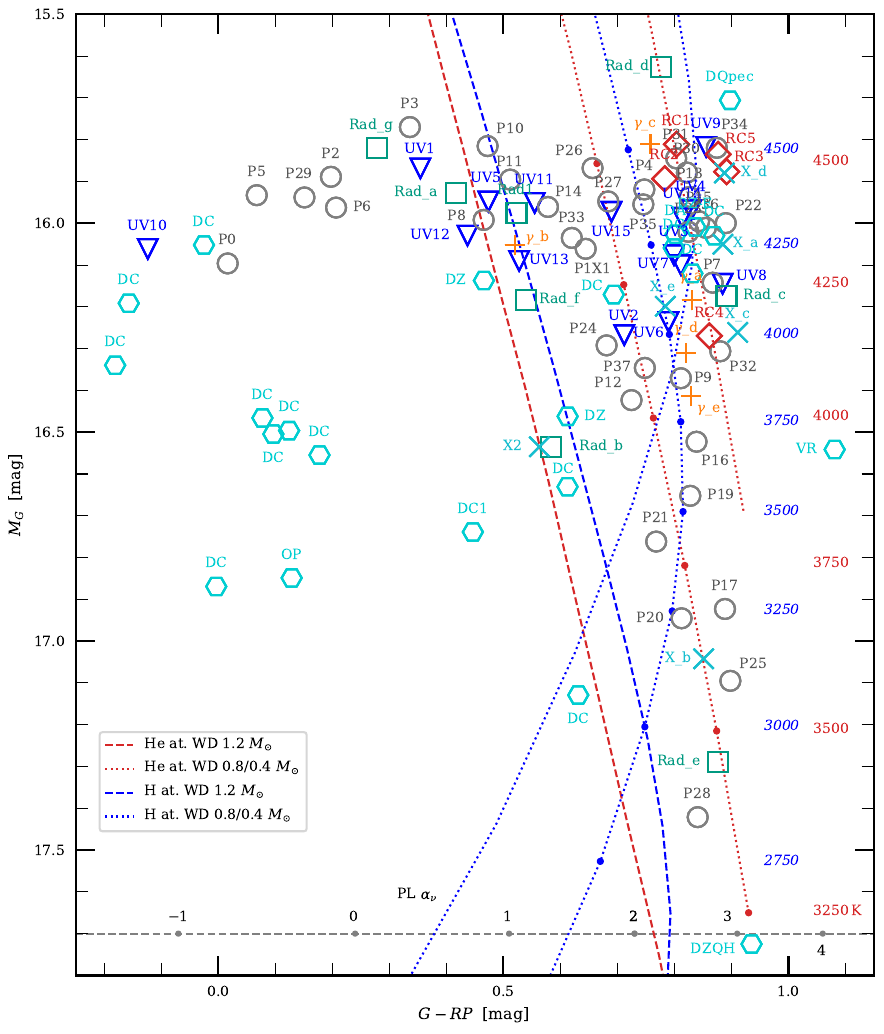}
    \caption{$M_G$ versus $G-G_{RP}$ colour--absolute magnitude diagram. Same as in Fig.~\ref{gaia_can_bpg_vs_g_cmd}.}
    \label{gaia_can_grp_vs_g_cmd}
\end{figure*}

\begin{figure*}
    \includegraphics[width=0.9\textwidth]{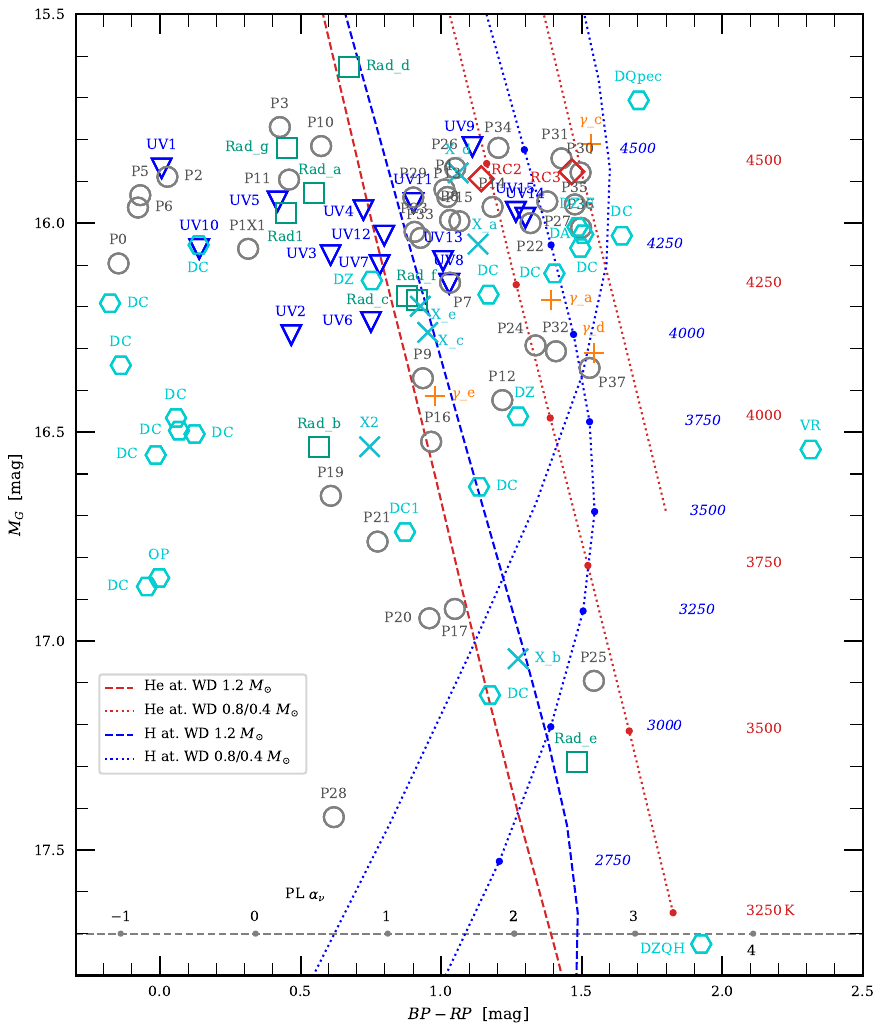}
    \caption{$M_G$ versus $G_{BP}-G_{RP}$ colour--absolute magnitude diagram. Same as in Fig.~\ref{gaia_can_bpg_vs_g_cmd}.}
    \label{gaia_can_bprp_vs_g_cmd}
\end{figure*}

\begin{figure*}
    \includegraphics[width=0.9\textwidth]{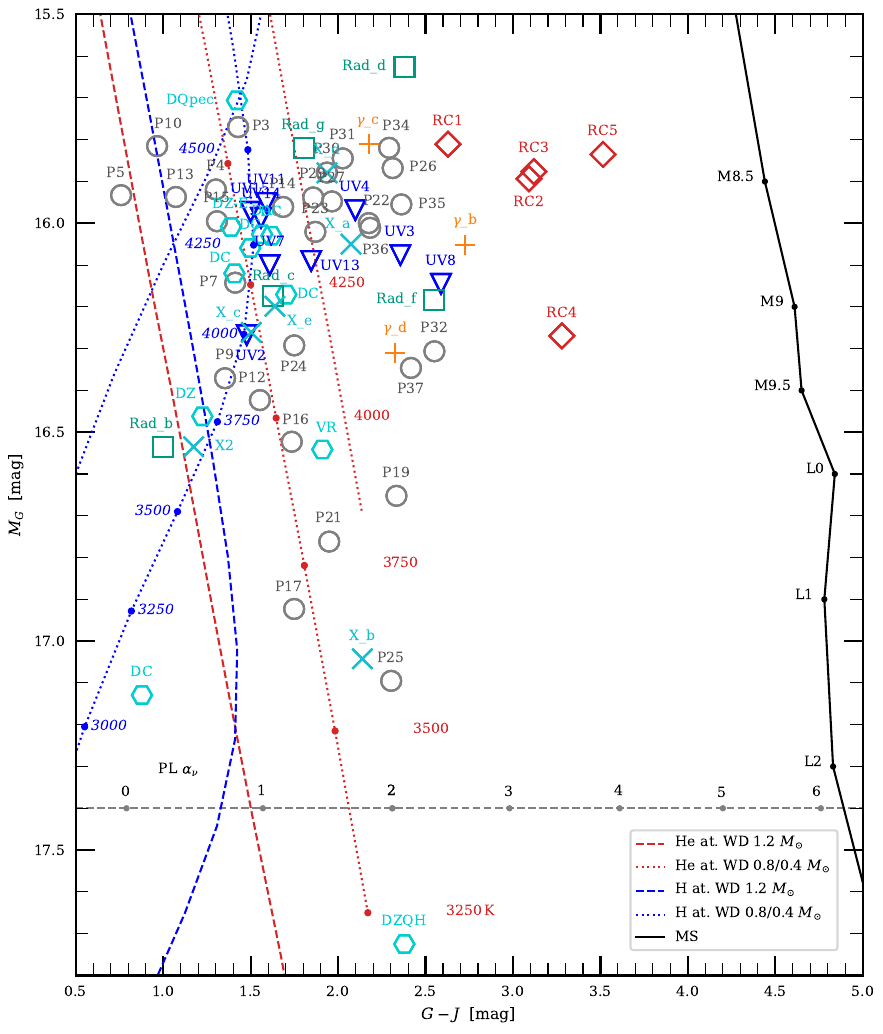}
    \caption{$M_G$ versus $G-J$ colour--absolute magnitude diagram. The stellar main sequence is represented by the black solid line. Same as in Fig.~\ref{gaia_can_bpg_vs_g_cmd}.}
    \label{gaia_can_gj_vs_g_cmd}
\end{figure*}
 
\begin{figure*}
    \includegraphics[width=0.9\textwidth]{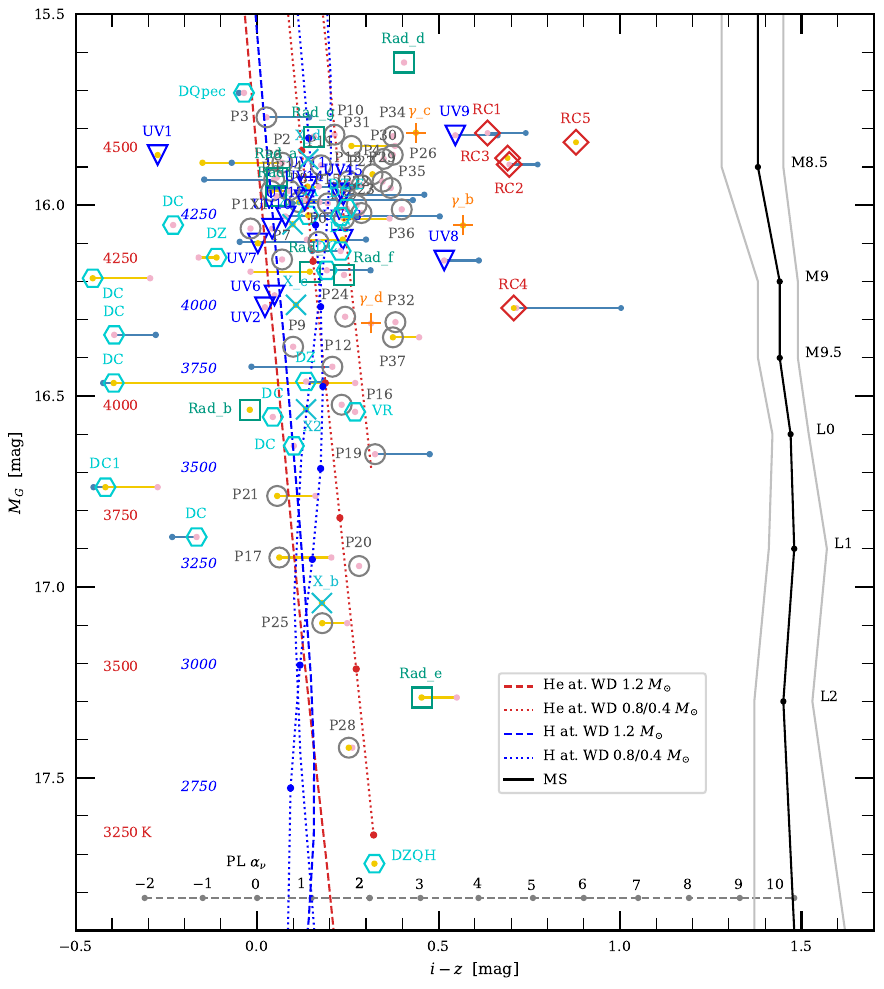}
    \caption{$M_G$ versus $i-z$ colour--absolute magnitude diagram. Same as in Fig.~\ref{gaia_can_bpg_vs_g_cmd}, except that symbols for a same object are linked with a straight line in yellow or blue colour, depending whether the additional photometry are from NSC or SDSS (see Tables~\ref{tab:opt_ir}, \ref{tab:opt_sdss_nsc}, \ref{tab:opt_ir2}, and \ref{tab:opt_sdss_nsc2}); to avoid overcrowding, additional symbols and links are plotted for an object only if the measurements per catalogue cover all the bands involved (among $ugrizy$). Symbols with pink, yellow, and blue small dots are for PS1, NSC, and SDSS photometry. The stellar main sequence in PS1 colour photometry and its 68 per cent confidence limits are represented by the black and grey solid lines.}
    \label{mg_iz_msbd}
\end{figure*}
 
\begin{figure*}
    \includegraphics[width=0.9\textwidth]{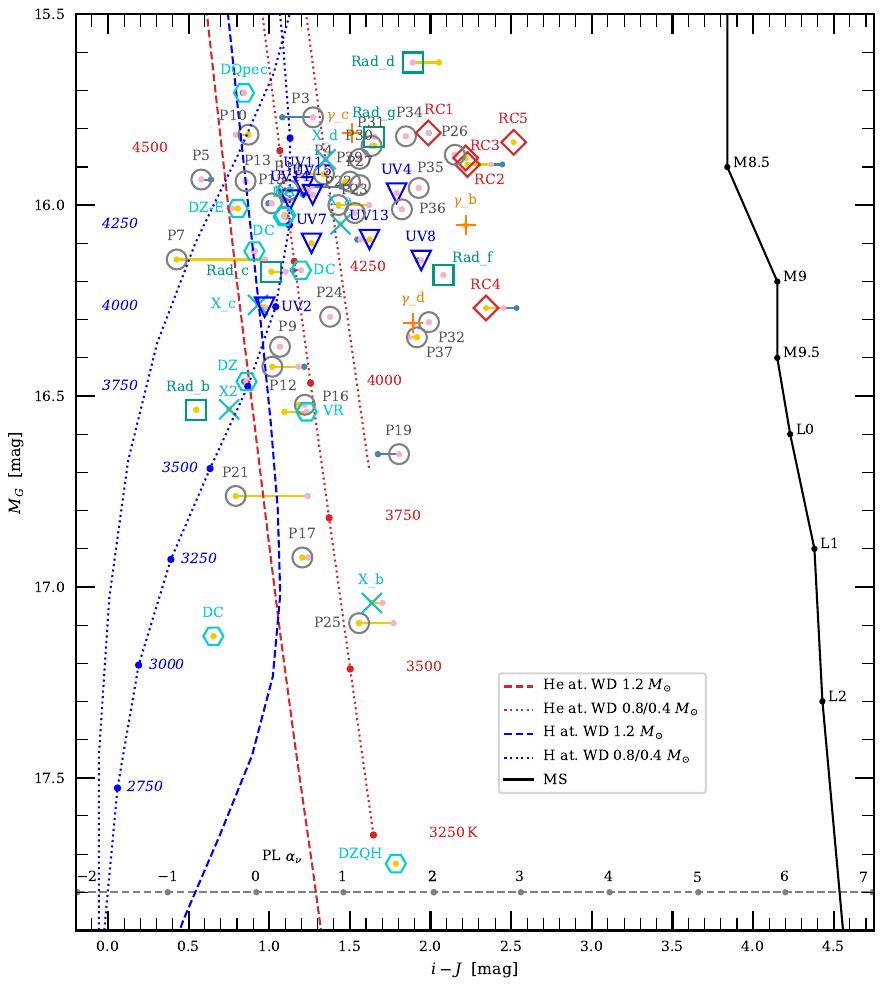}
    \caption{$M_G$ versus $i-J$ colour--absolute magnitude diagram. Same as in Fig.~\ref{mg_iz_msbd}.}
    \label{mg_ij_msbd}
\end{figure*}

\begin{figure*}
    \includegraphics[width=0.9\textwidth]{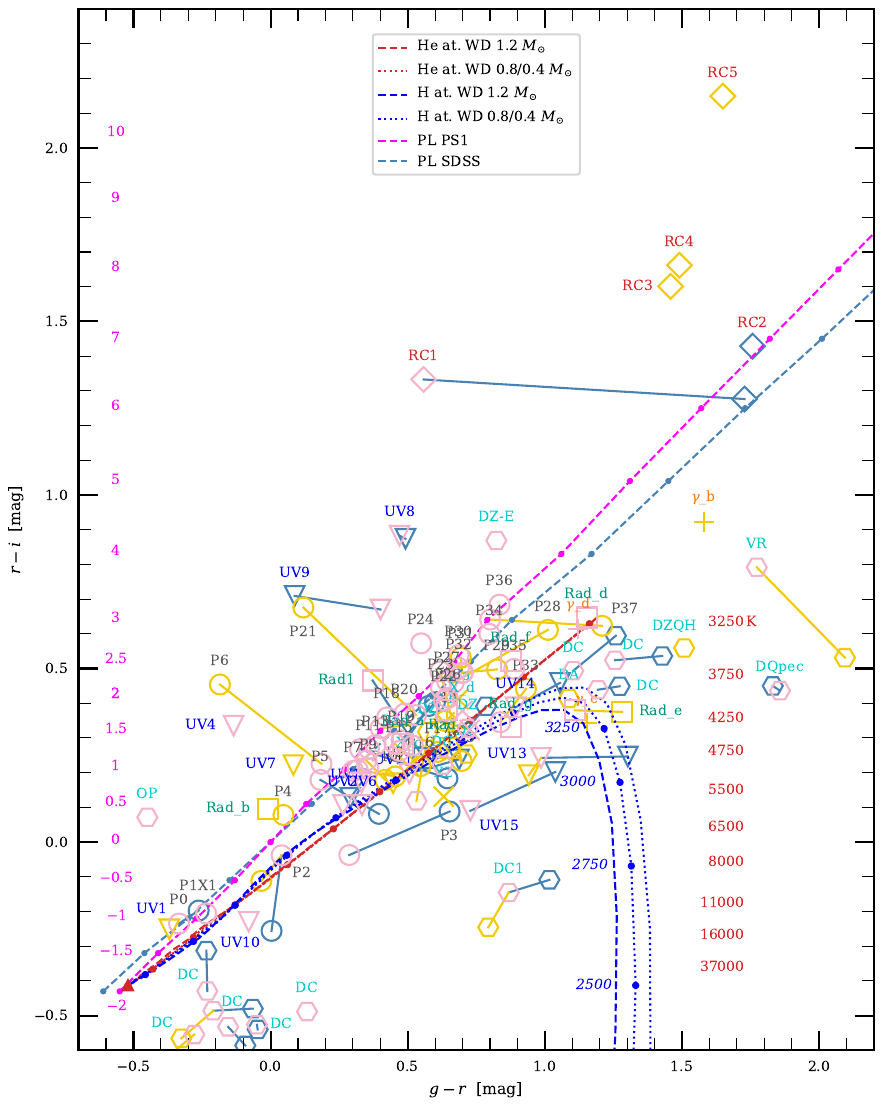}
    \caption{$r-i$ versus $g-r$ colour--colour diagram, in PS1 photometry unless stated otherwise. Same as in Fig.~\ref{mg_iz_msbd}, except that source symbols are in pink, yellow, or blue colour, depending whether the photometry are from PS1, NSC, or SDSS. The PS1 and SDSS colours for power laws of spectral index $\alpha_{\nu}=-2, -1.5, -1, -0.5$, 0, 0.5, 1, 1.5, 2, 2.5, and 3 are represented by pink and blue dots interconnected by dashed lines, respectively. For relatively high white-dwarf effective temperatures, the $T_\mathrm{eff}$ values are indicated for the 0.8~M$_{\odot}$ pure-He atmosphere track (small red dots); these apply also for the pure-H track where the data point locations (small blue dots) where obtained by interpolation using the $T_\mathrm{eff}$ values. For diverging He and H data points at lower temperatures, $T_\mathrm{eff}$ values are indicated separately. Both 0.8~M$_{\odot}$ He and H atmosphere tracks start at $T_\mathrm{eff}=150\,000$~K and at the same location (small red filled triangle).}
    \label{gr_vs_ri}
\end{figure*}

\begin{figure*}
    \includegraphics[width=0.9\textwidth]{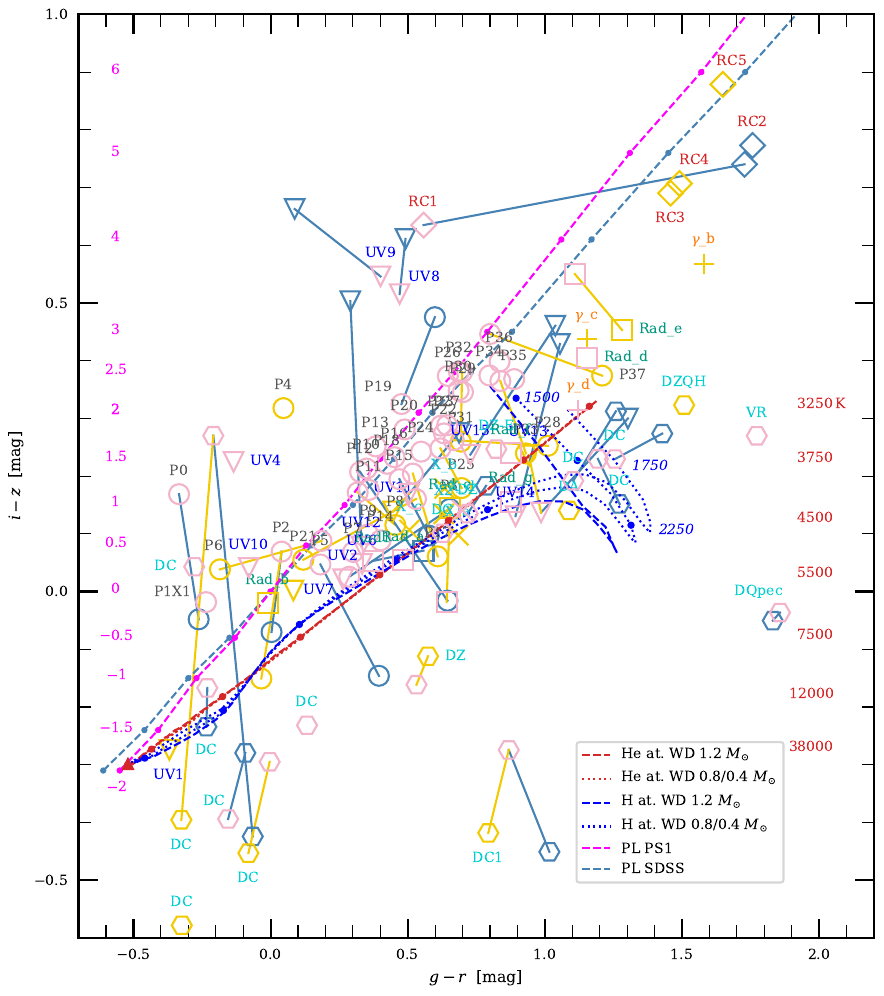}
    \caption{$i-z$ versus $g-r$ colour--colour diagram. Same as Fig.~\ref{gr_vs_ri}.}
    \label{gr_vs_iz}
\end{figure*}

\begin{figure*}
    \includegraphics[width=0.9\textwidth]{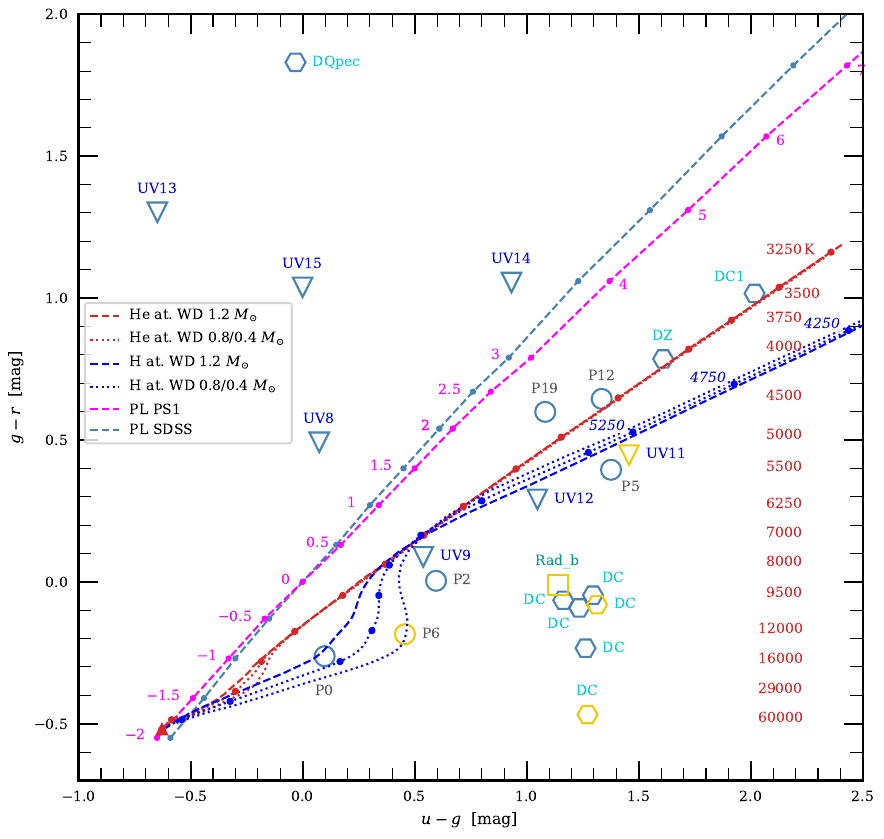}
    \caption{$g-r$ versus $u-g$ colour--colour diagram. Same as Fig.~\ref{gr_vs_ri}.}
    \label{ug_vs_gr}
\end{figure*}

\begin{figure*}
    \includegraphics[width=0.9\textwidth]{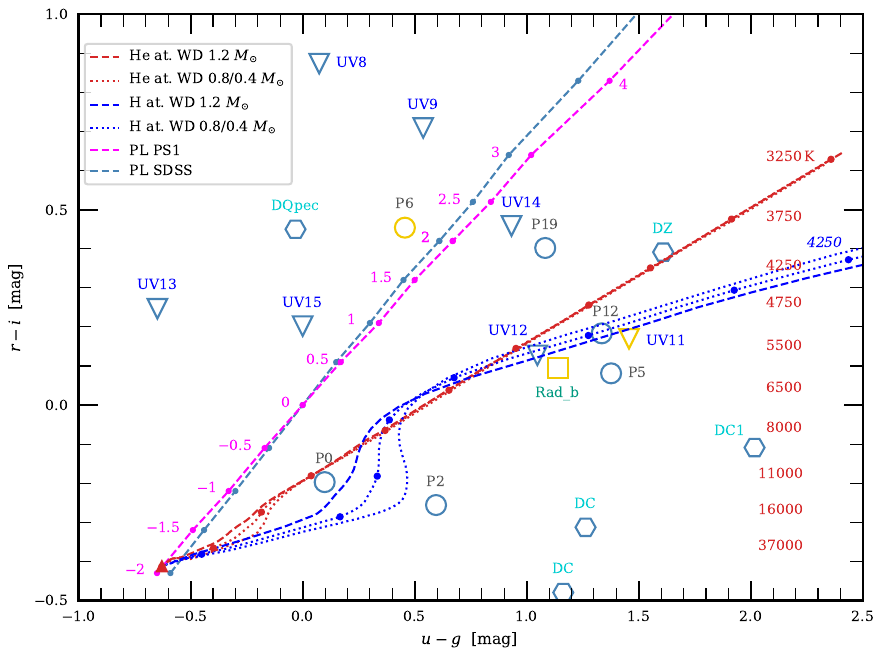}
    \caption{$r-i$ versus $u-g$ colour--colour diagram. Same as Fig.~\ref{gr_vs_ri}.}
    \label{ug_vs_ri}
\end{figure*}

\begin{figure*}
    \includegraphics[width=0.9\textwidth]{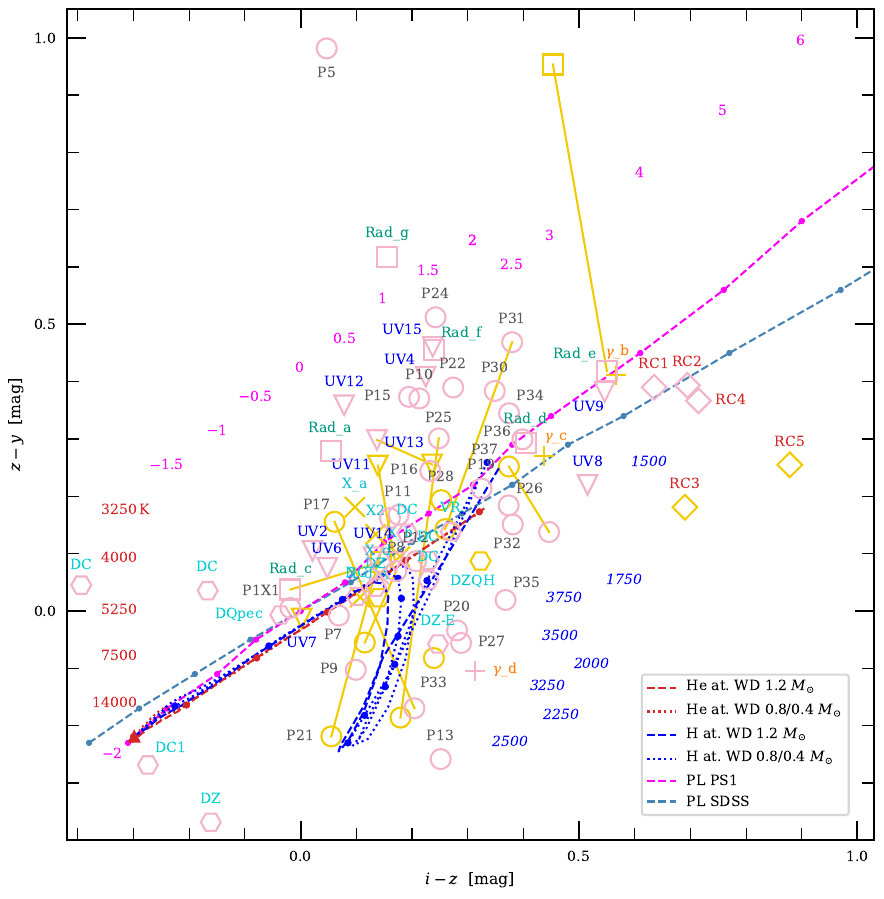}
    \caption{$z-y$ versus $i-z$ colour--colour diagram. Same as Fig.~\ref{gr_vs_ri}.}
    \label{iz_vs_zy}
\end{figure*}

\begin{figure*}
    \includegraphics[width=0.9\textwidth]{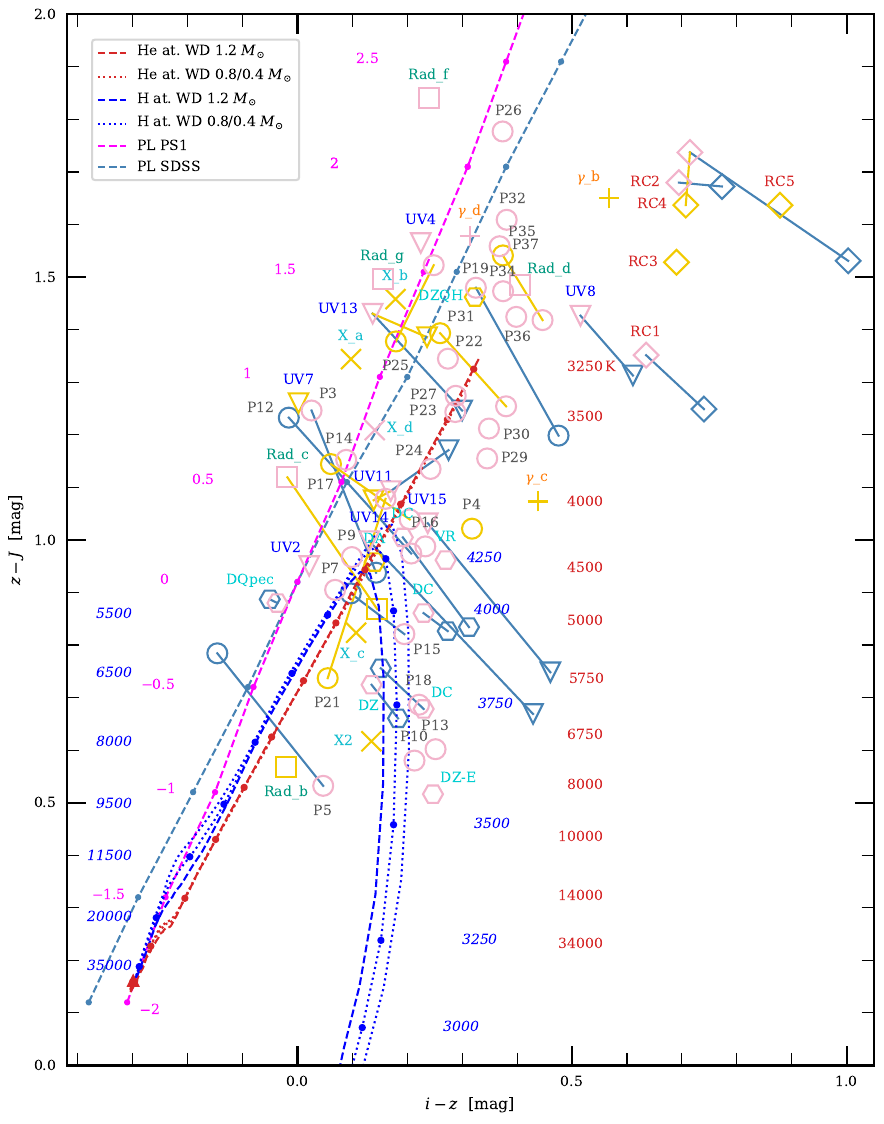}
    \caption{$z-J$ versus $i-z$ colour--colour diagram. Same as Fig.~\ref{gr_vs_ri}.}
    \label{iz_vs_zj}
\end{figure*}

\begin{figure*}
    \includegraphics[width=0.9\textwidth]{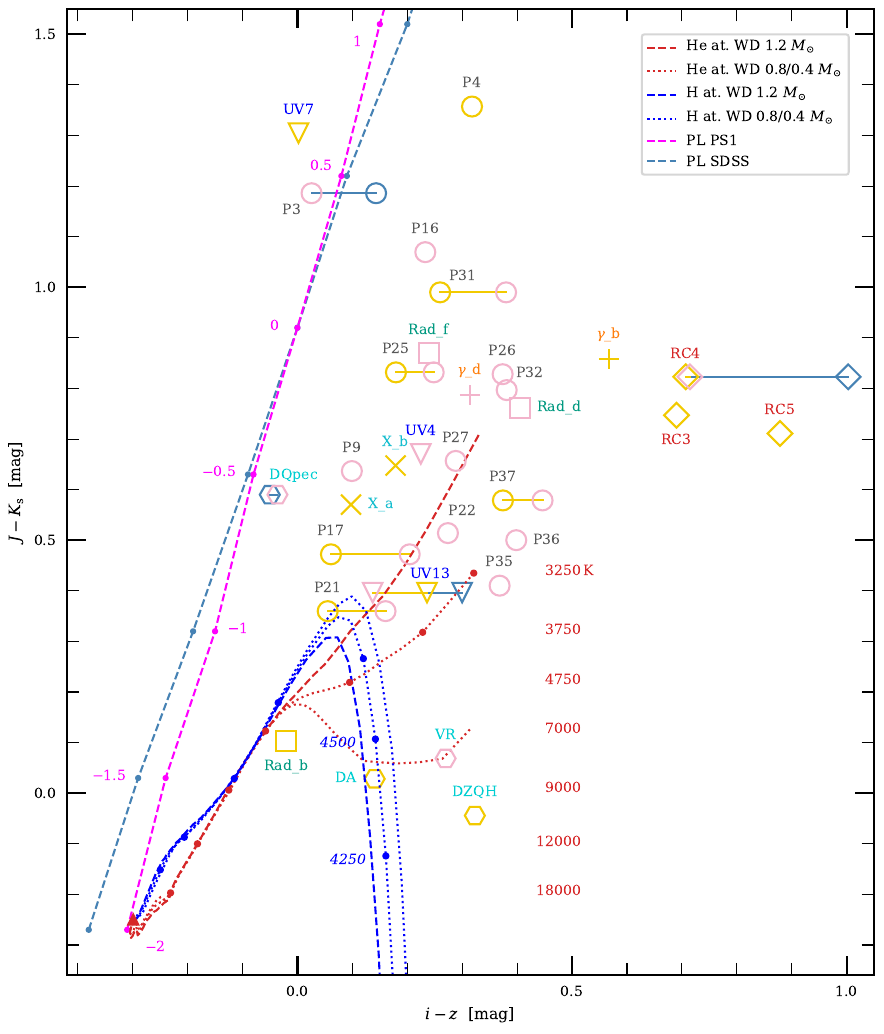}
    \caption{$J-K_\mathrm{s}$ versus $i-z$ colour--colour diagram. Same as Fig.~\ref{gr_vs_ri}.}
    \label{iz_vs_jks}
\end{figure*}

\begin{figure*}
    \includegraphics[width=0.9\textwidth]{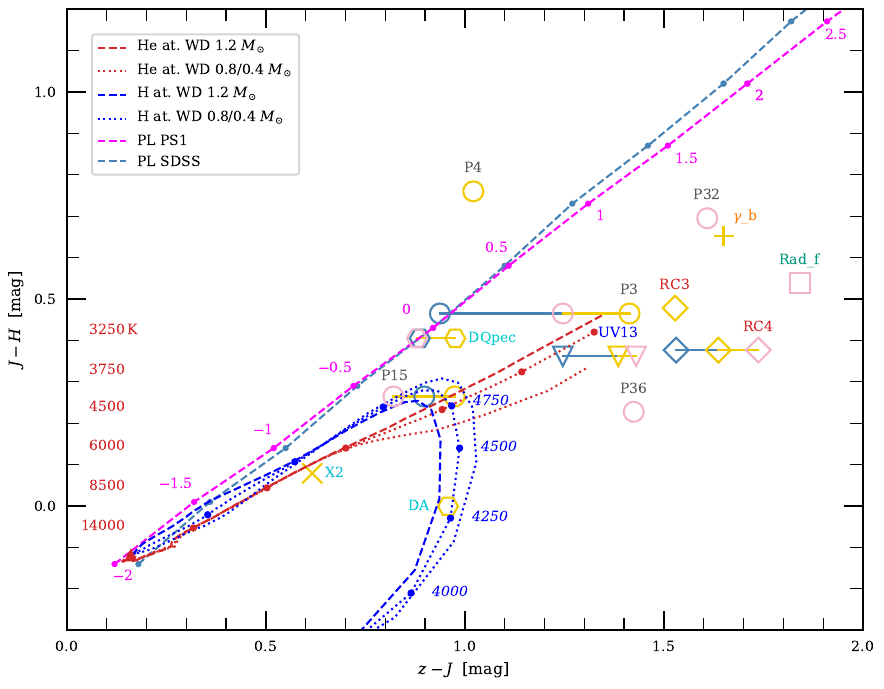}
    \caption{$J-H$ versus $z-J$ colour--colour diagram. Same as Fig.~\ref{gr_vs_ri}.}
    \label{zj_vs_jh}
\end{figure*}

\begin{figure*}
    \includegraphics[width=0.9\textwidth]{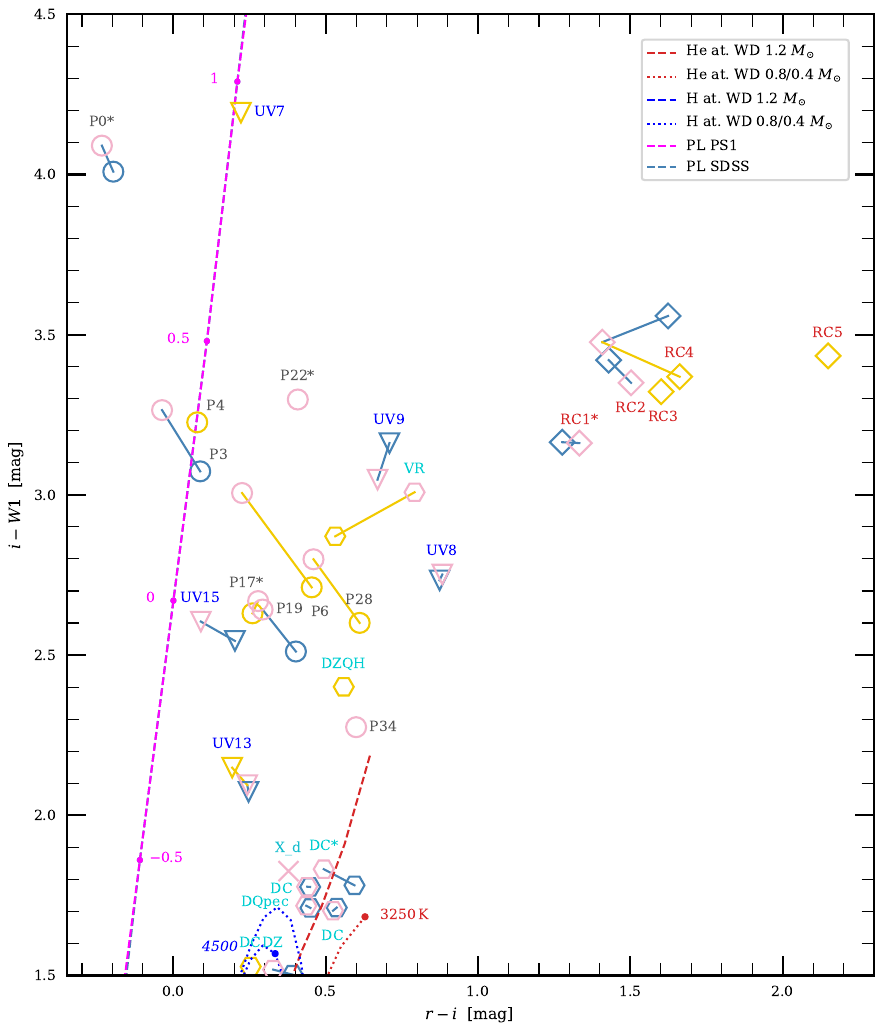}
    \caption{$i-W1$ versus $r-i$ colour--colour diagram. Same as Fig.~\ref{gr_vs_ri}. Sources with \textit{Gaia}--\textit{WISE} angular separations larger than 1.5~arcsec have an asterisk appended to their label.}
    \label{ri_vs_iw1}
\end{figure*}

\begin{figure*}
    \includegraphics[width=0.9\textwidth]{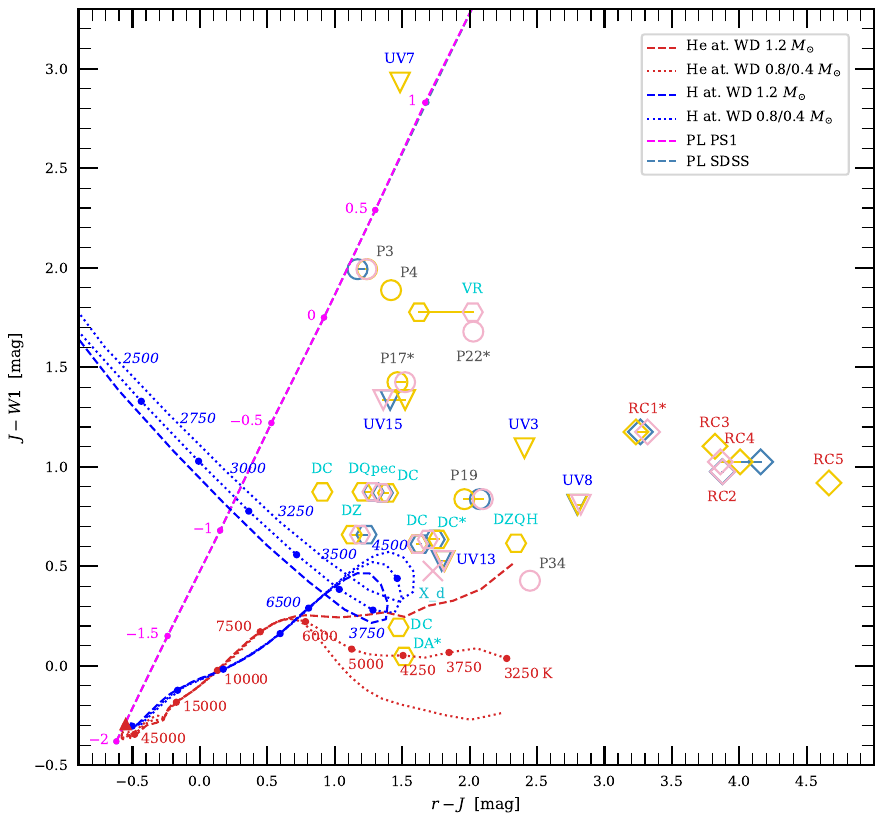}
    \caption{$J-W1$ versus $r-J$ colour--colour diagram. Same as Fig.~\ref{rj_vs_jw1}.}
    \label{rj_vs_jw1}
\end{figure*}

\begin{figure*}
    \includegraphics[width=0.9\textwidth]{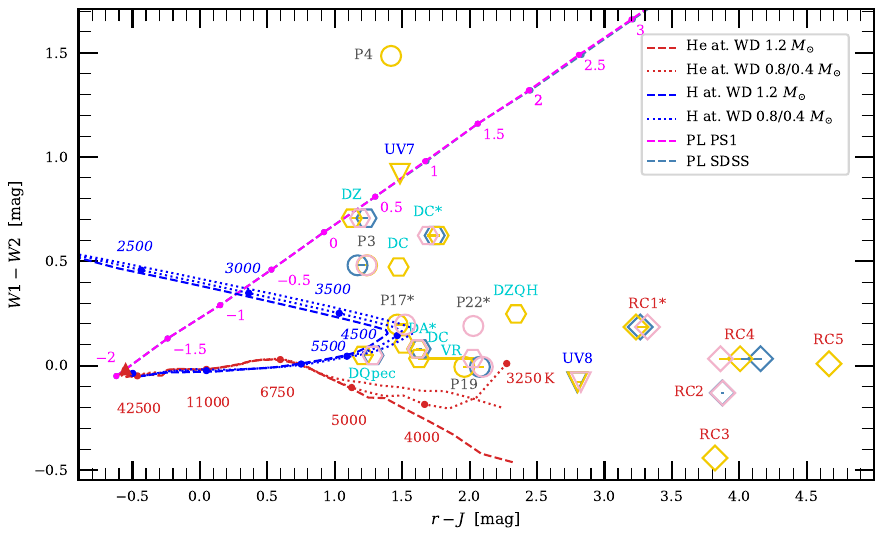}
    \caption{$W1-W2$ versus $r-J$ colour--colour diagram. Same as Fig.~\ref{rj_vs_w1w2}.}
    \label{rj_vs_w1w2}
\end{figure*}


\section{Spectral flux density distributions including infrared fluxes of the sources (online supplementary material)}\label{fsed_ir}

\begin{figure*}
    \includegraphics[width=\columnwidth]{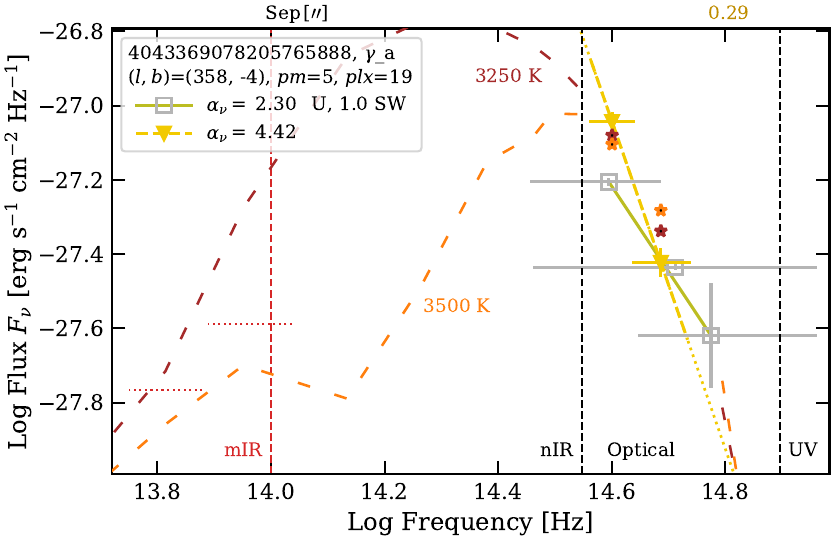}
    \includegraphics[width=\columnwidth]{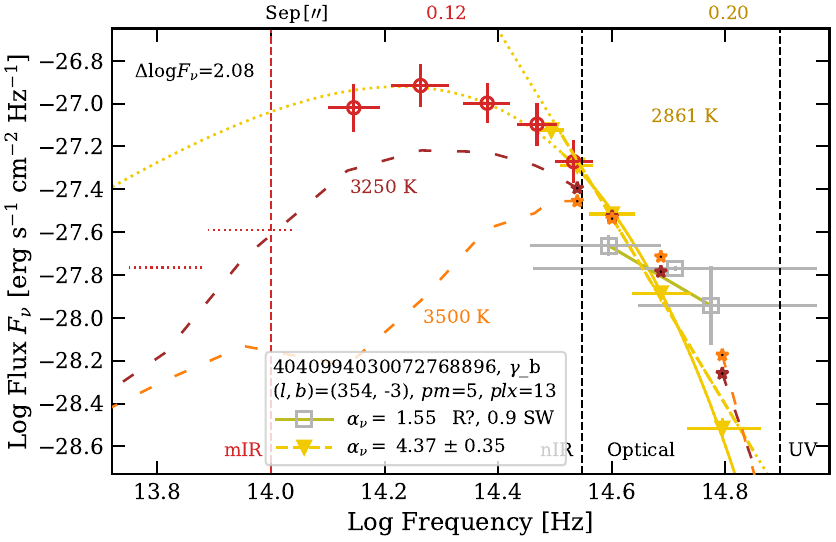}
    \includegraphics[width=\columnwidth]{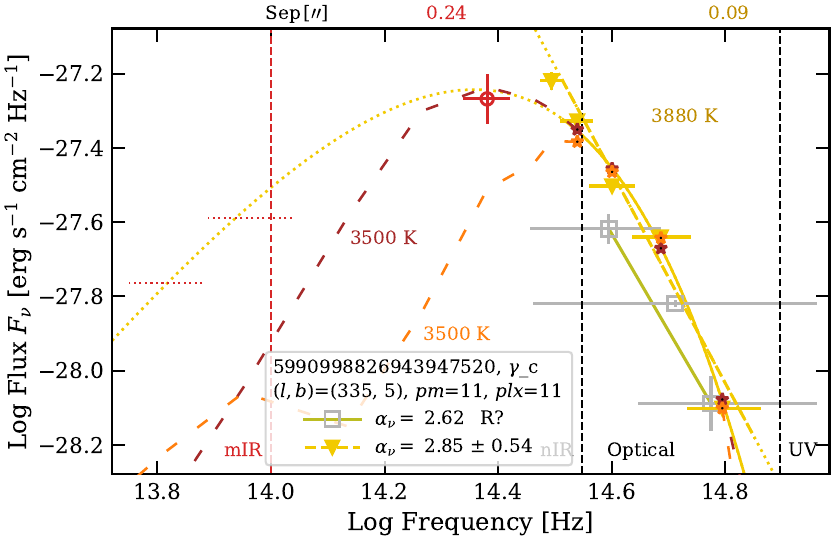}
    \includegraphics[width=\columnwidth]{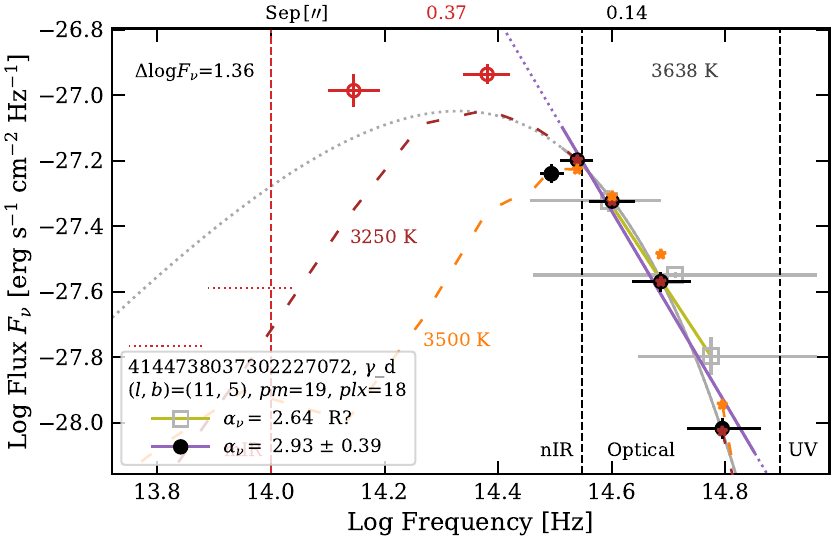}
    \includegraphics[width=\columnwidth]{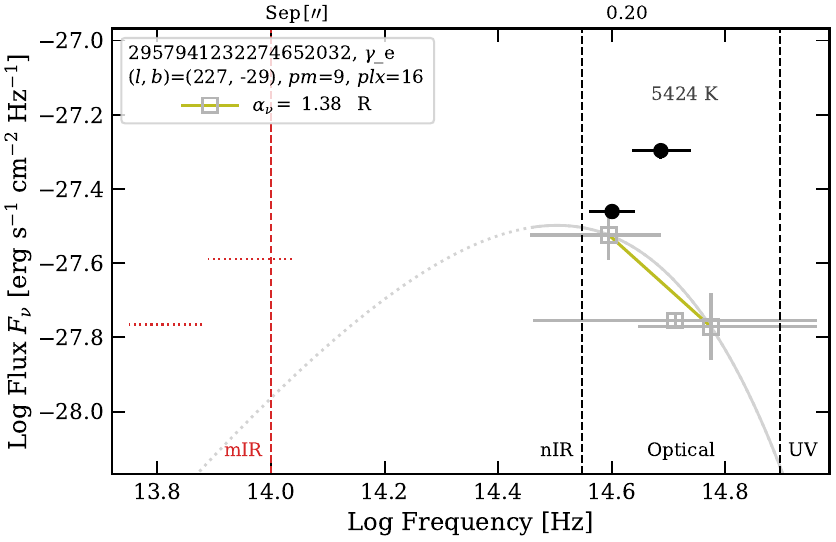}
     \caption{SEDs including infrared fluxes of the gamma-ray crossmatch sources. The $\Delta \log F_{\nu}$ ordinate span is of 1.2~dex, else if greater it is indicated. Infrared fluxes other than those of the broad-optical surveys are represented by red open circles.
     Same as in Fig.~\ref{fsed_uv1}.}
    \label{fsed_gamma_ir}
\end{figure*}

\begin{figure*}
    \includegraphics[width=\columnwidth]{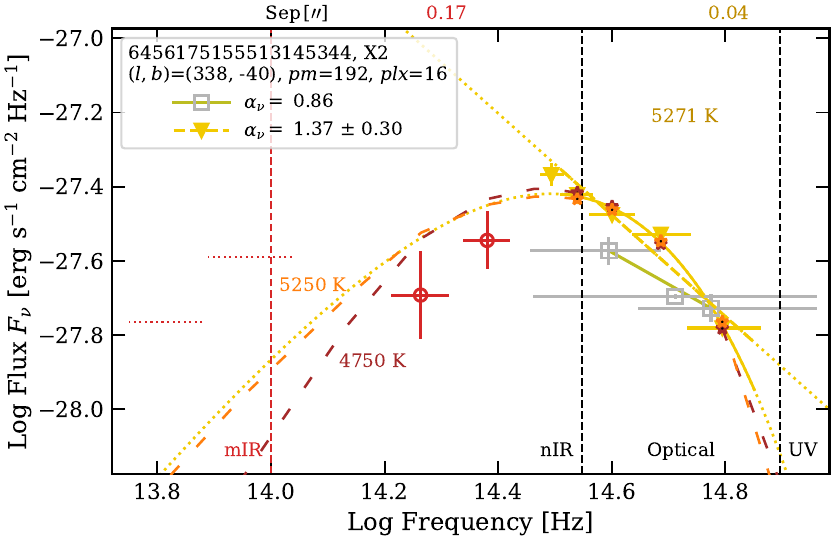}
    \includegraphics[width=\columnwidth]{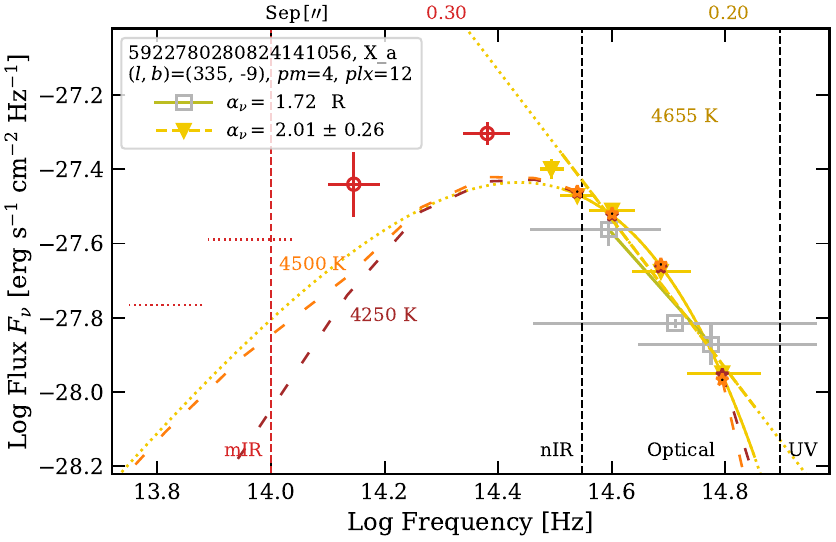}
    \includegraphics[width=\columnwidth]{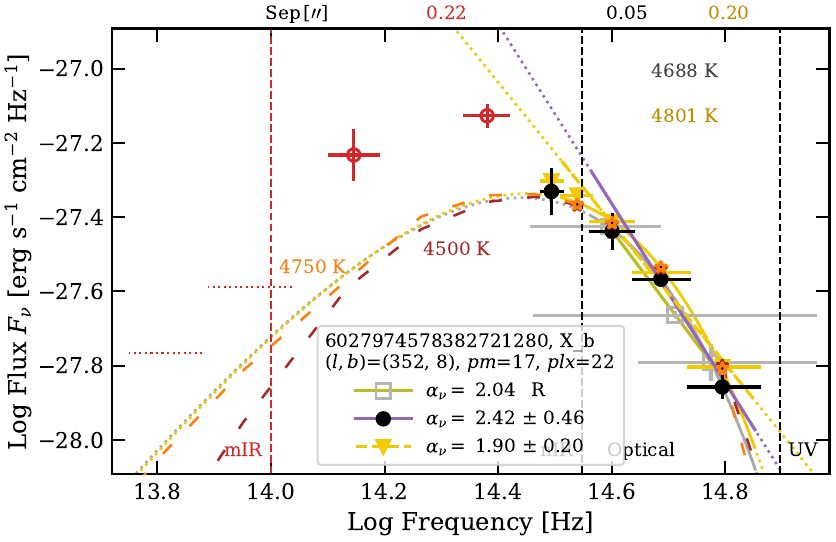}
    \includegraphics[width=\columnwidth]{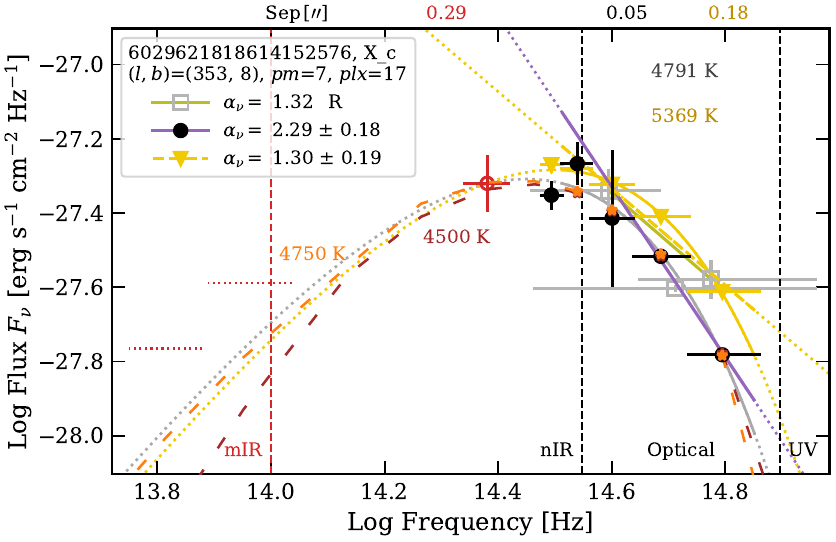}
    \includegraphics[width=\columnwidth]{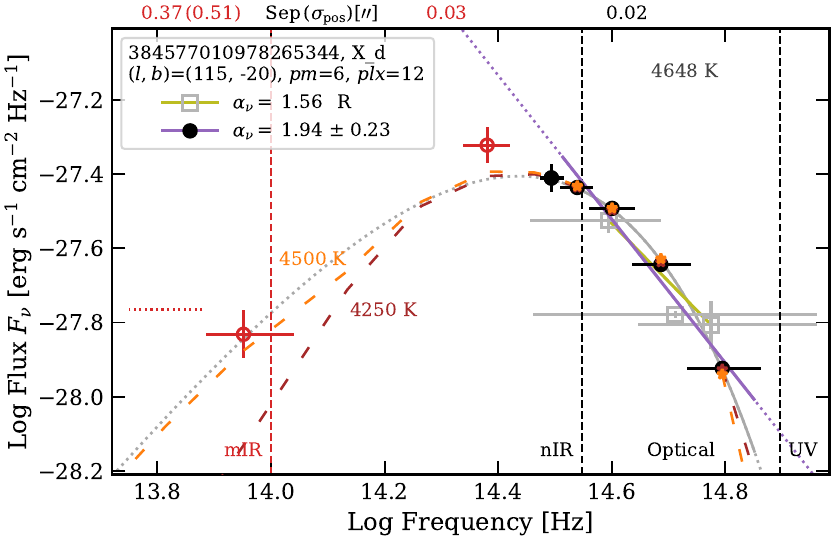}
    \includegraphics[width=\columnwidth]{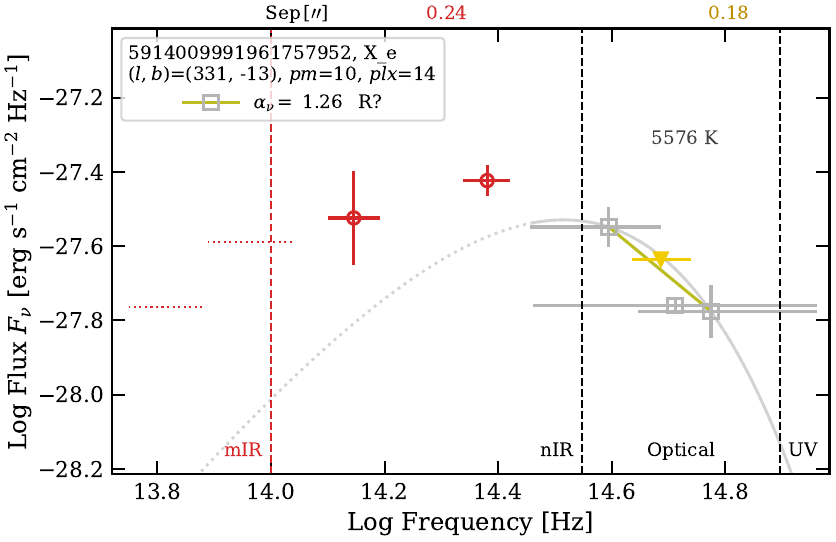}
     \caption{SEDs including infrared fluxes of the X-ray crossmatch sources.
     Same as in Fig.~\ref{fsed_uv1}.}
    \label{fsed_x_ir}
\end{figure*}

\begin{figure*}
    \includegraphics[width=0.60\textwidth]{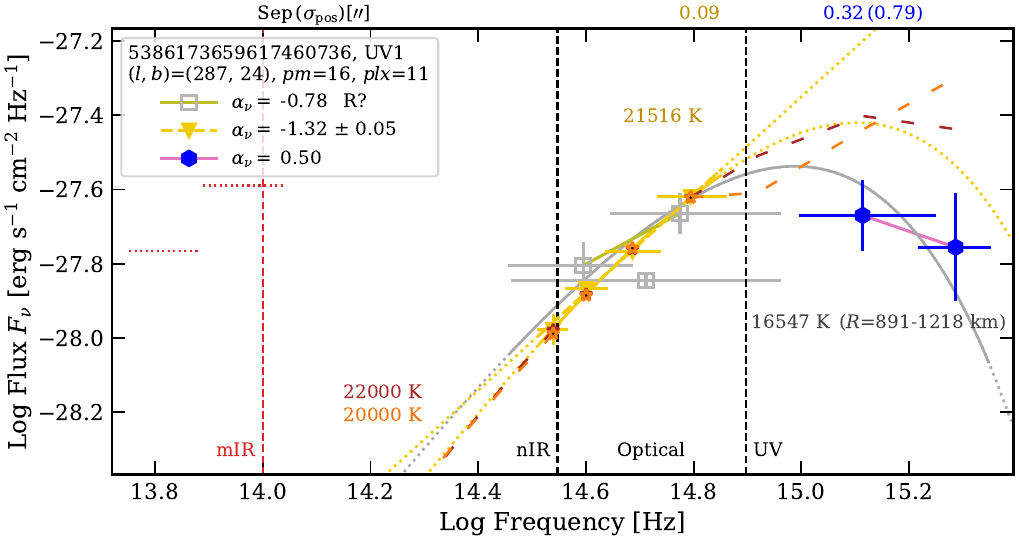}
    \includegraphics[width=0.60\textwidth]{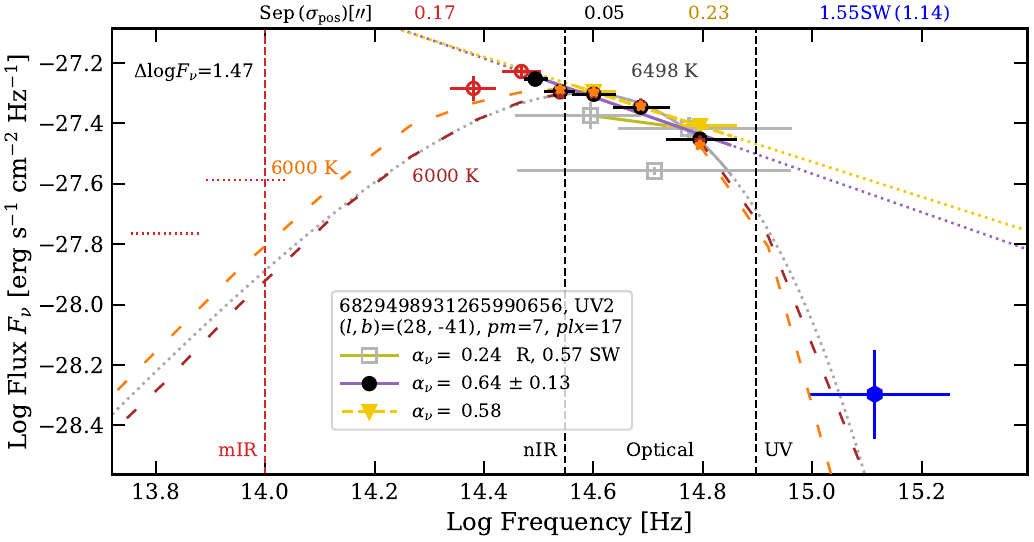}
    \includegraphics[width=0.60\textwidth]{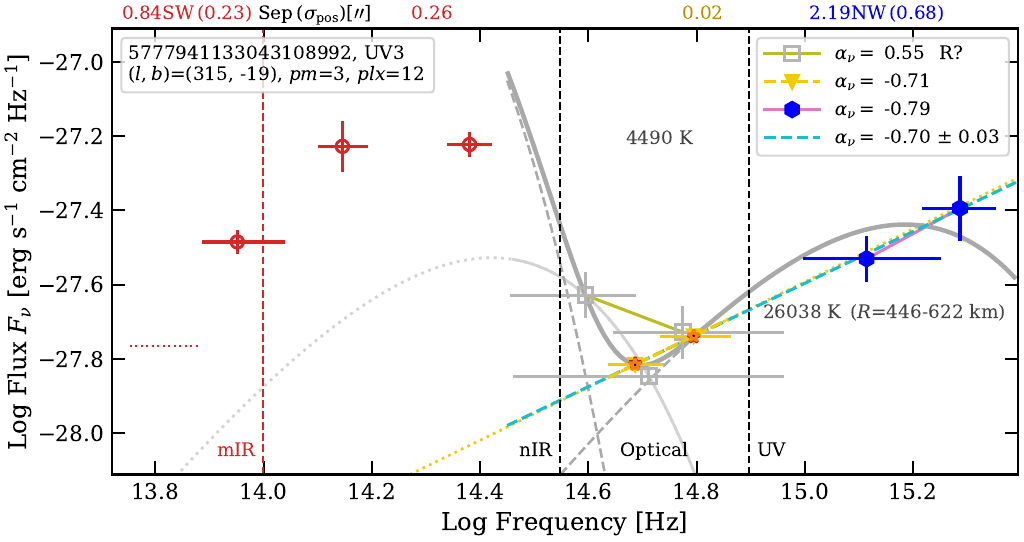}
    \includegraphics[width=0.60\textwidth]{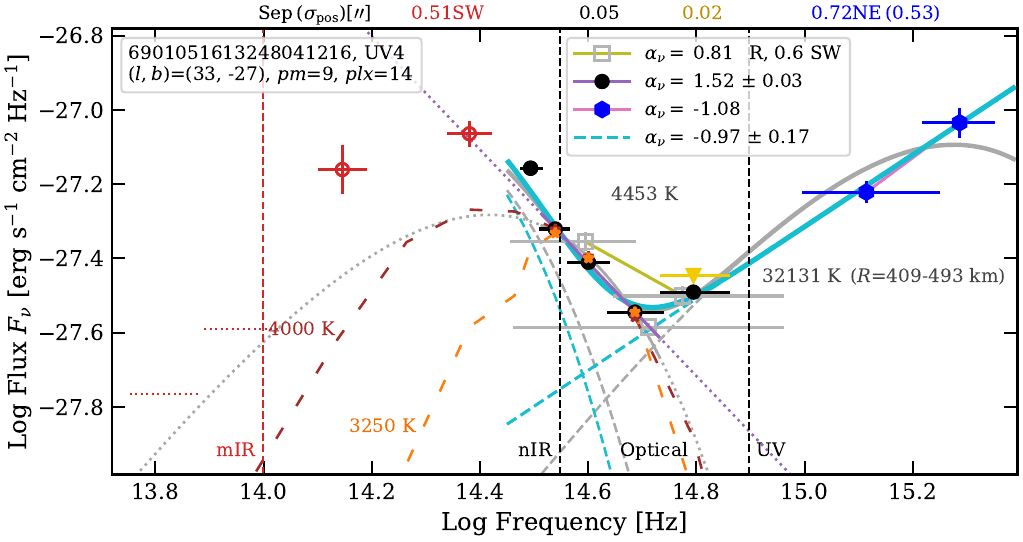}
     \caption{SEDs including infrared fluxes of the ultraviolet candidates.
     Same as in Fig.~\ref{fsed_uv1}.}
    \label{fsed_uv_ir}
\end{figure*}

\begin{figure*}
   \ContinuedFloat
    \includegraphics[width=0.60\textwidth]{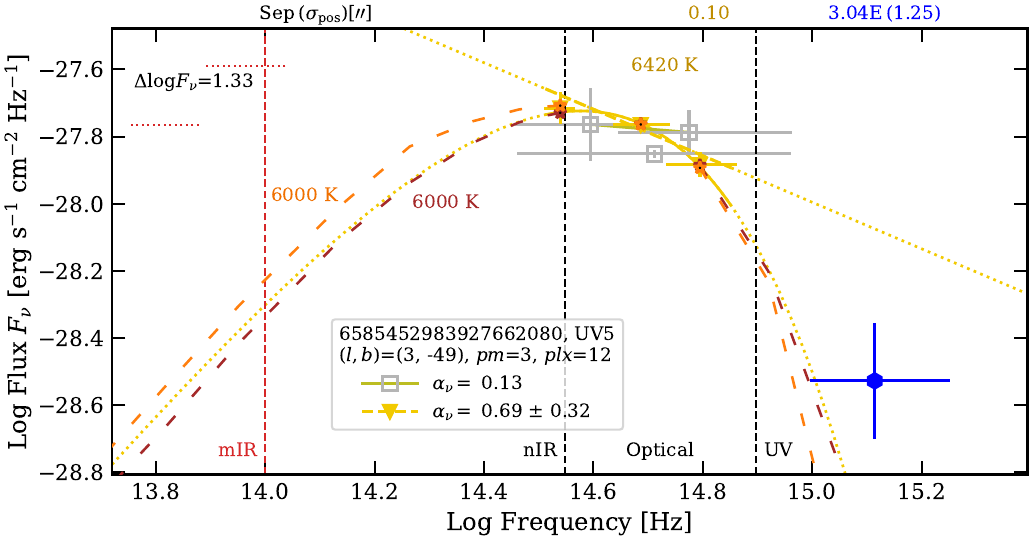}
    \includegraphics[width=0.60\textwidth]{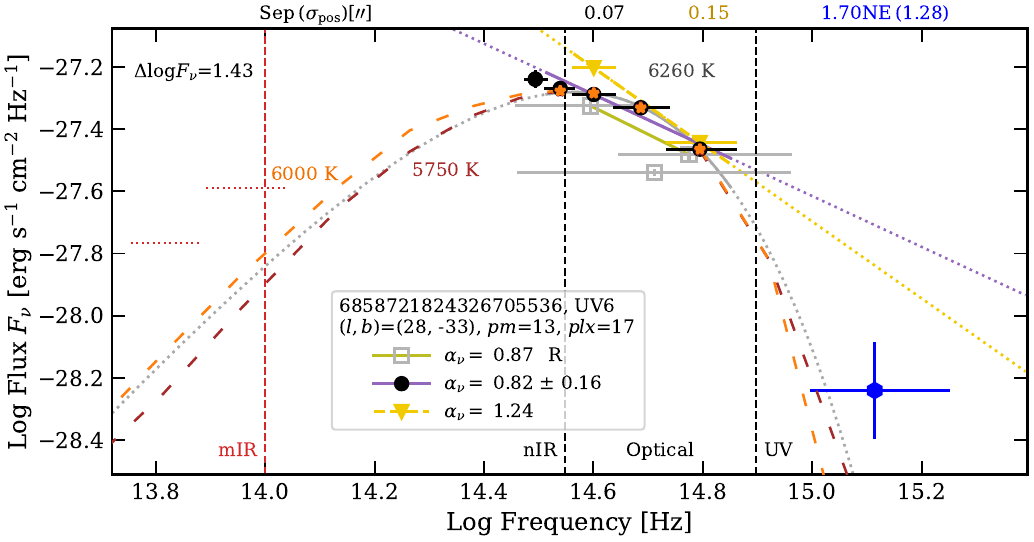}
    \includegraphics[width=0.60\textwidth]{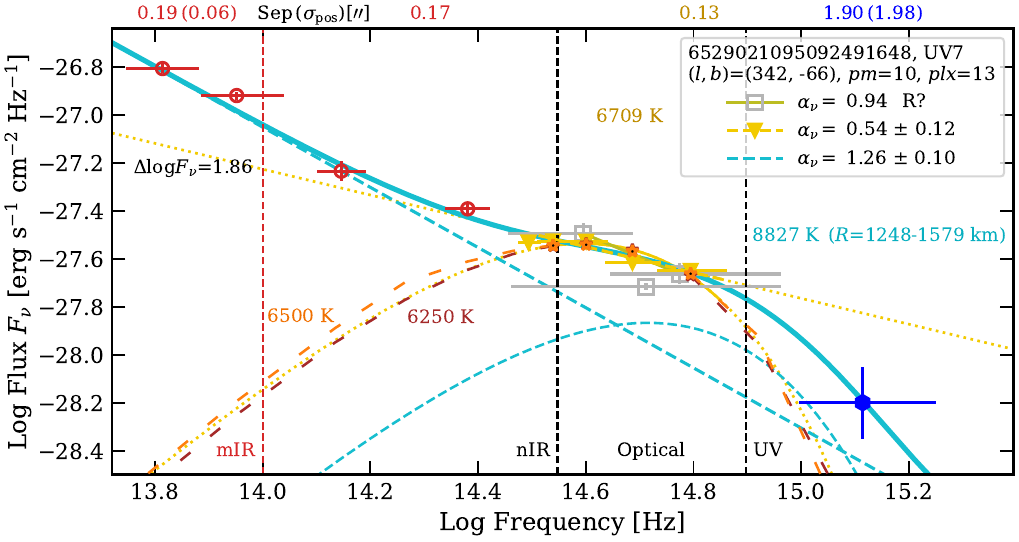}
    \includegraphics[width=0.60\textwidth]{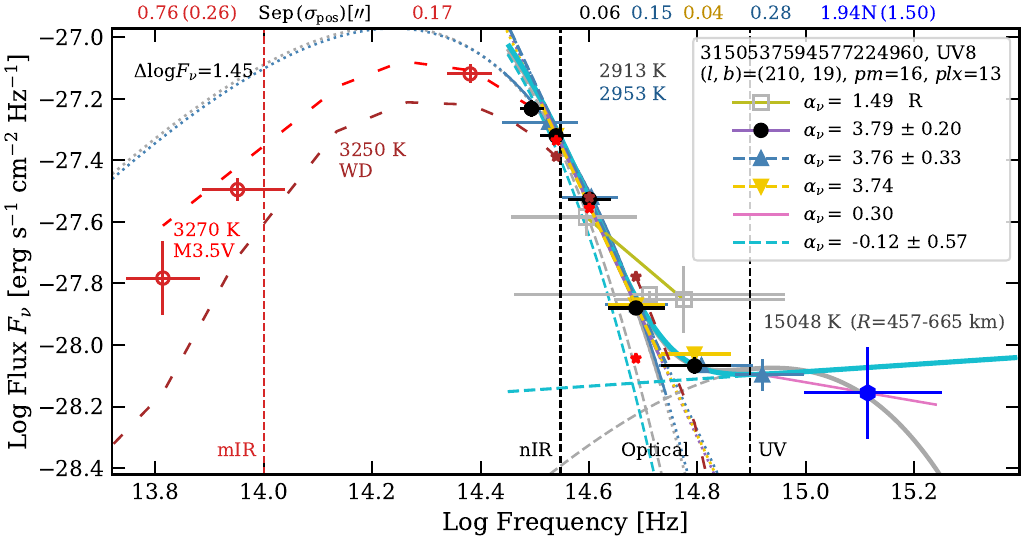}
     \caption{continued.}
    \label{fsed_uv_ir}
\end{figure*}

\begin{figure*}
   \ContinuedFloat
    \includegraphics[width=0.60\textwidth]{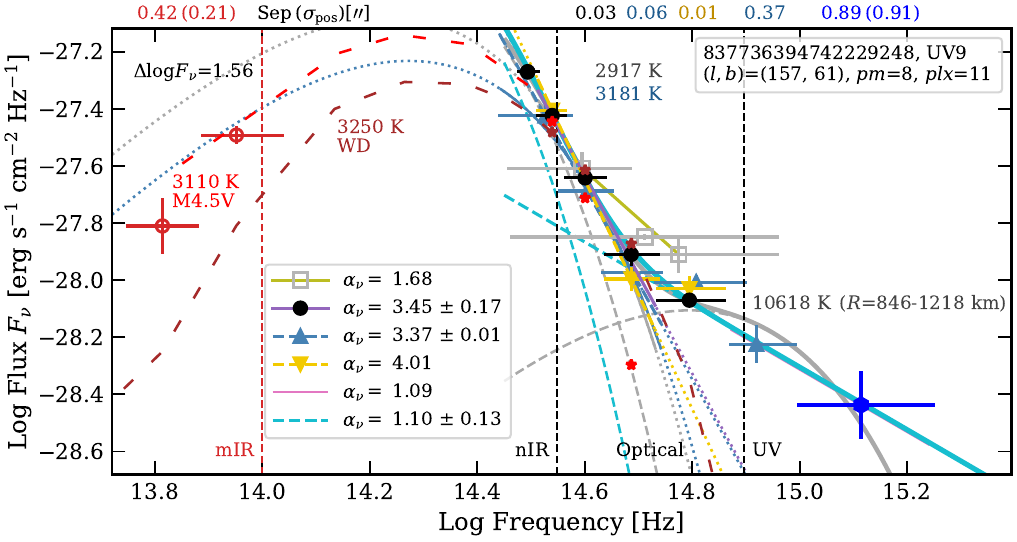}
    \includegraphics[width=0.60\textwidth]{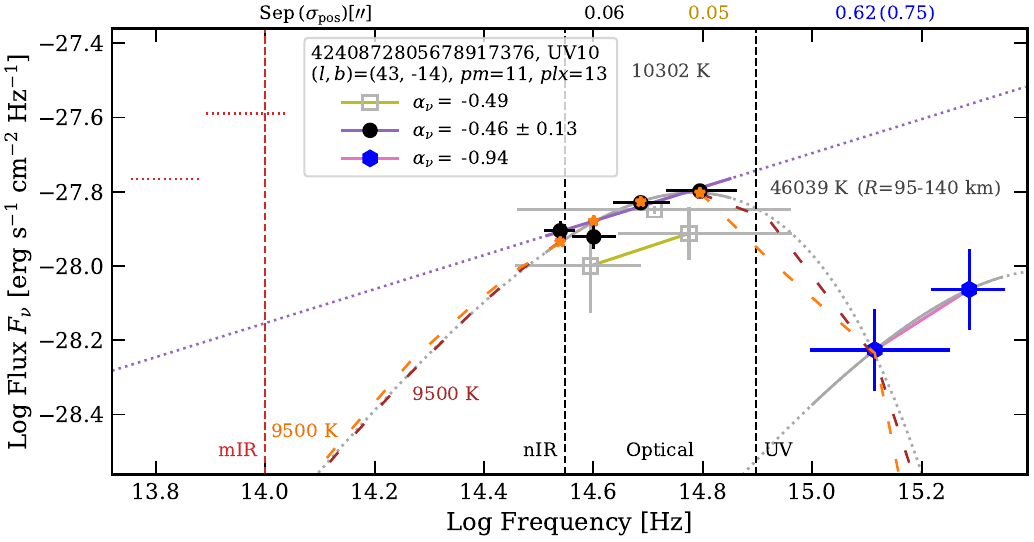}
    \includegraphics[width=0.60\textwidth]{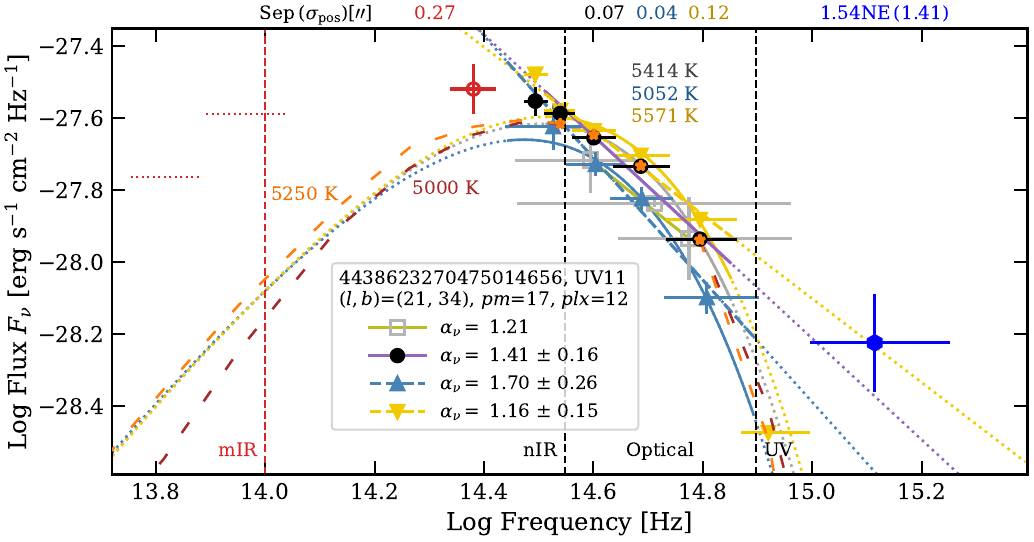}
    \includegraphics[width=0.60\textwidth]{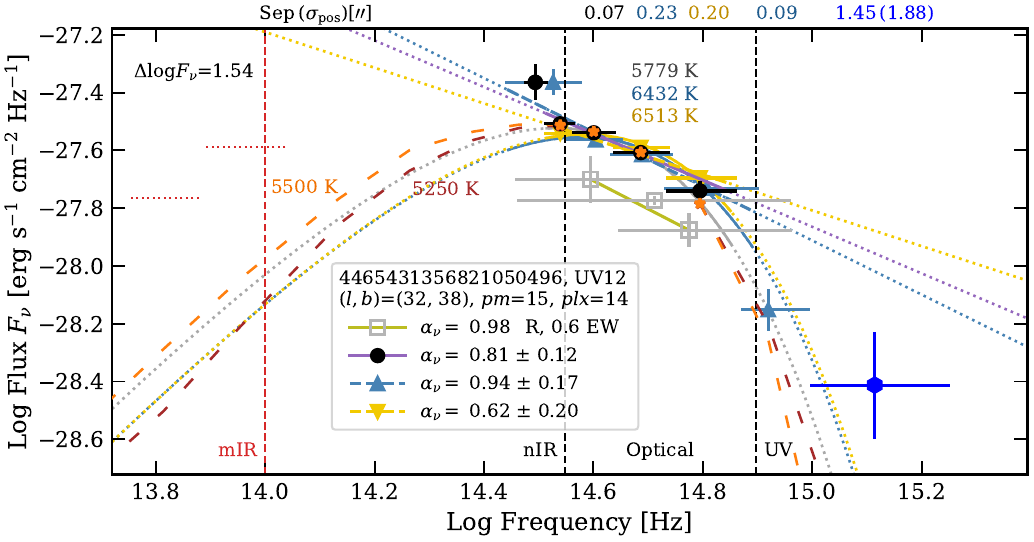}
     \caption{continued.}
    \label{fsed_uv_ir}
\end{figure*}

\begin{figure*}
   \ContinuedFloat
    \includegraphics[width=0.60\textwidth]{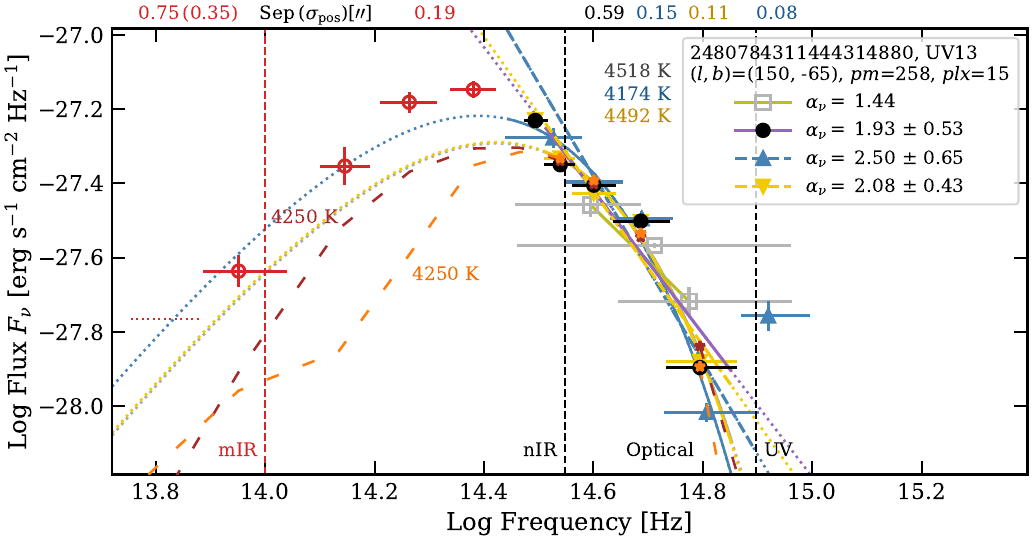}
    \includegraphics[width=0.60\textwidth]{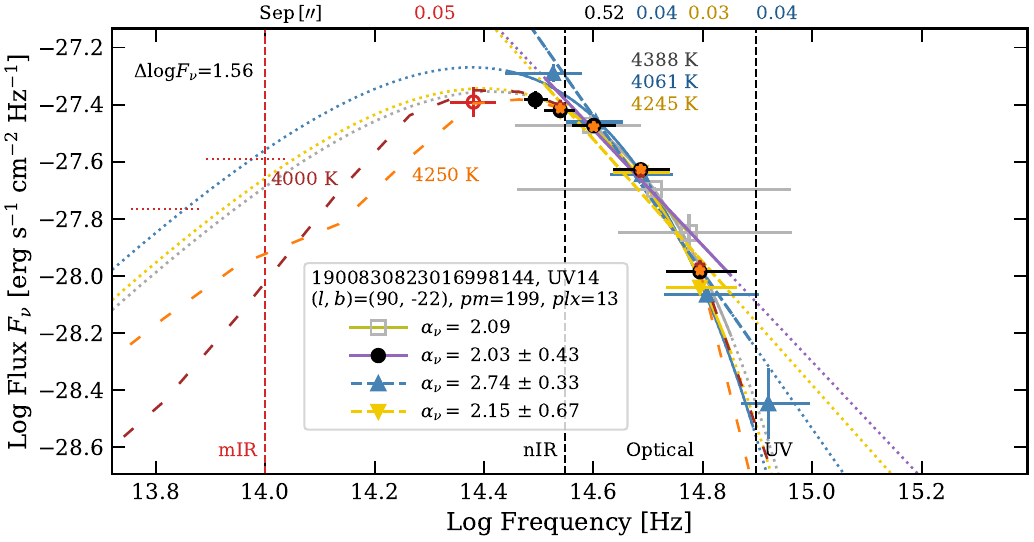}
    \includegraphics[width=0.60\textwidth]{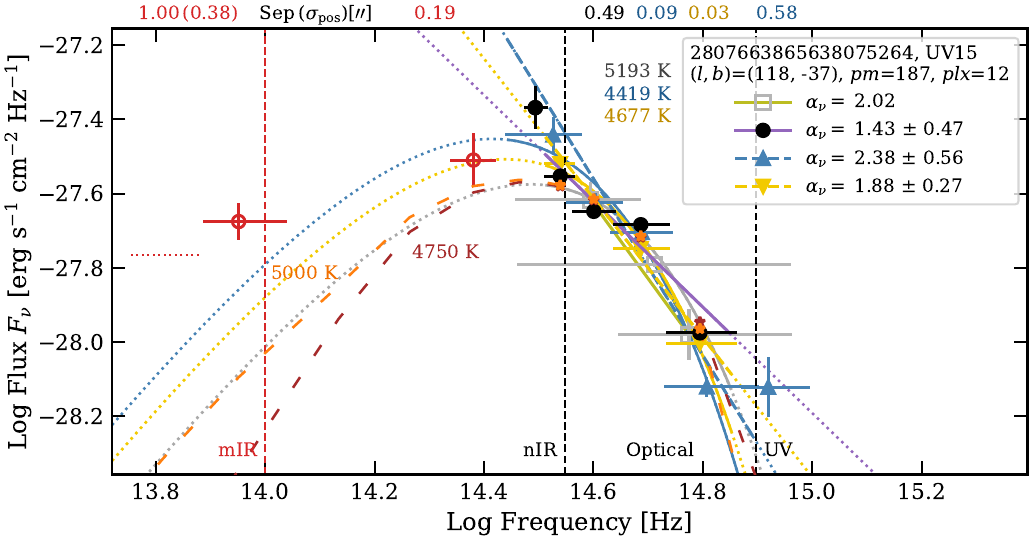}
    \includegraphics[width=0.60\textwidth]{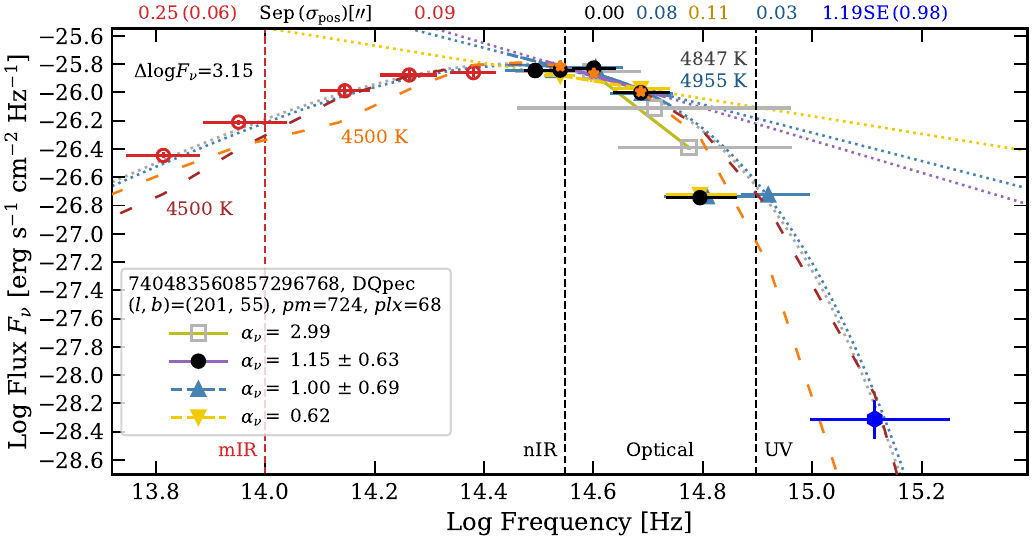}
     \caption{continued. The LHS 2229 DQpec white dwarf is shown for comparison in the bottom panel.}
    \label{fsed_uv_ir}
\end{figure*}

\begin{figure*}
    \includegraphics[width=\columnwidth]{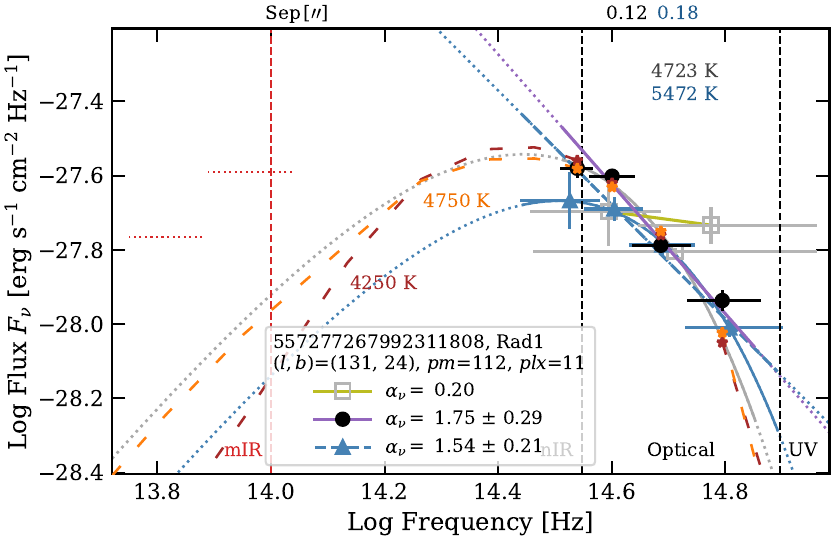}
    \includegraphics[width=\columnwidth]{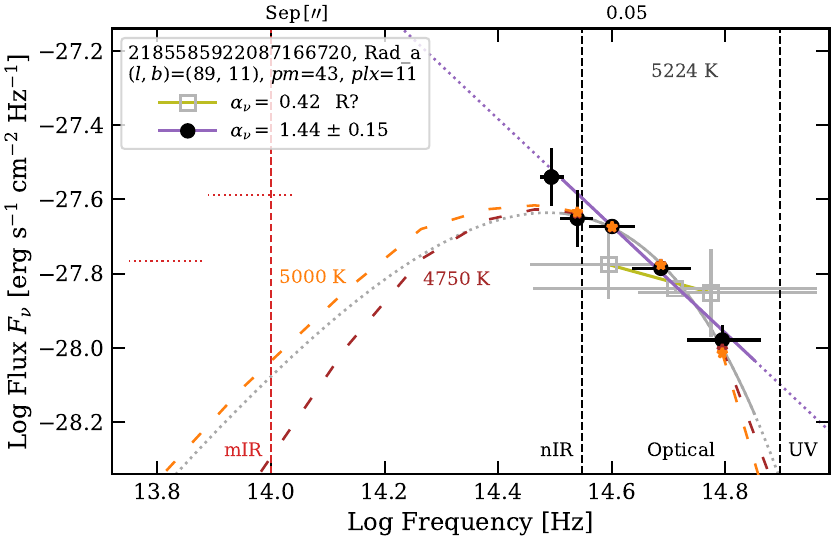}
    \includegraphics[width=\columnwidth]{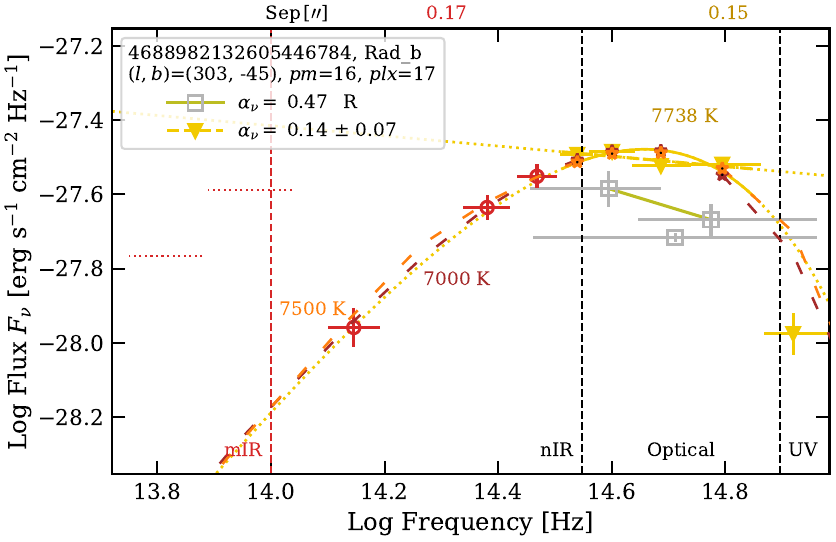}
    \includegraphics[width=\columnwidth]{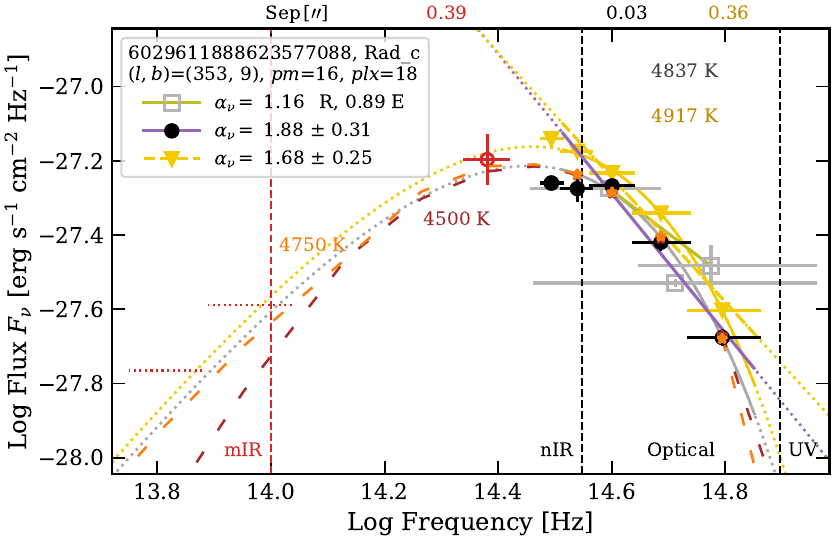}
    \includegraphics[width=\columnwidth]{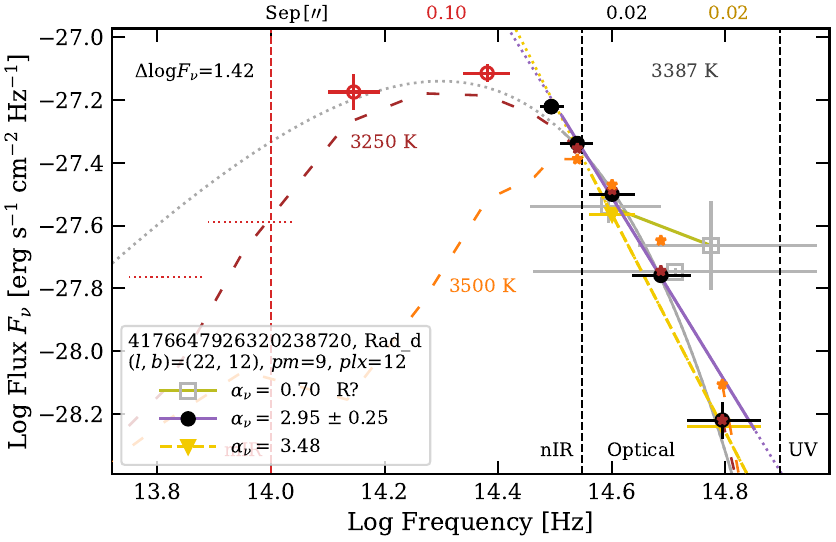}
    \includegraphics[width=\columnwidth]{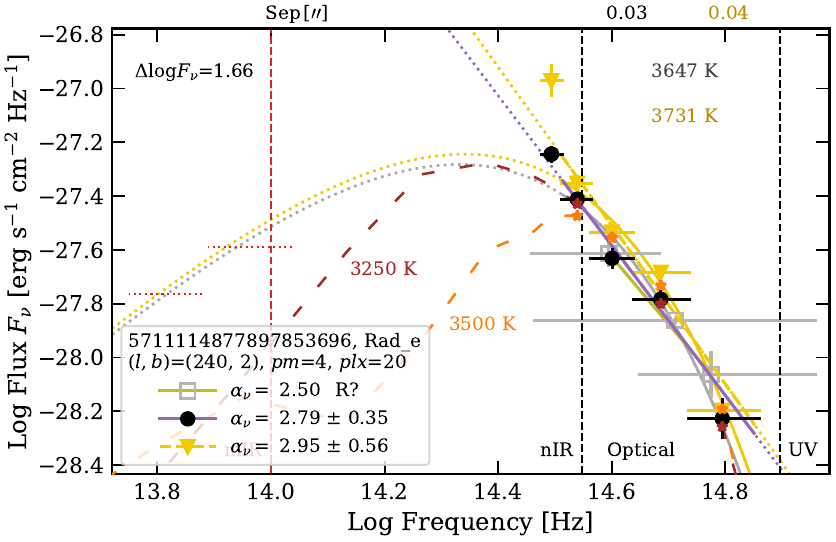}
    \includegraphics[width=\columnwidth]{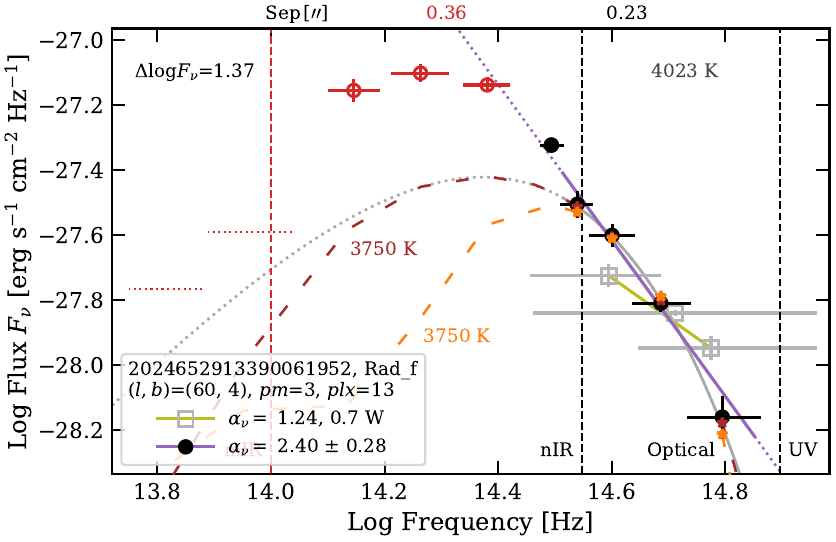}
    \includegraphics[width=\columnwidth]{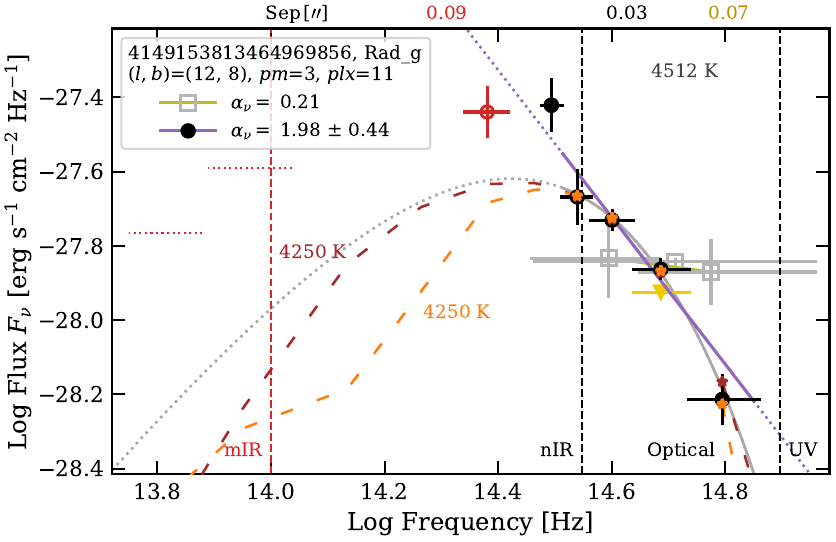}
     \caption{SEDs including infrared fluxes of the radio crossmatch sources.
     Same as in Fig.~\ref{fsed_uv1}.}
    \label{fsed_radio_ir}
\end{figure*}

\begin{figure*}
	\includegraphics[width=\columnwidth]{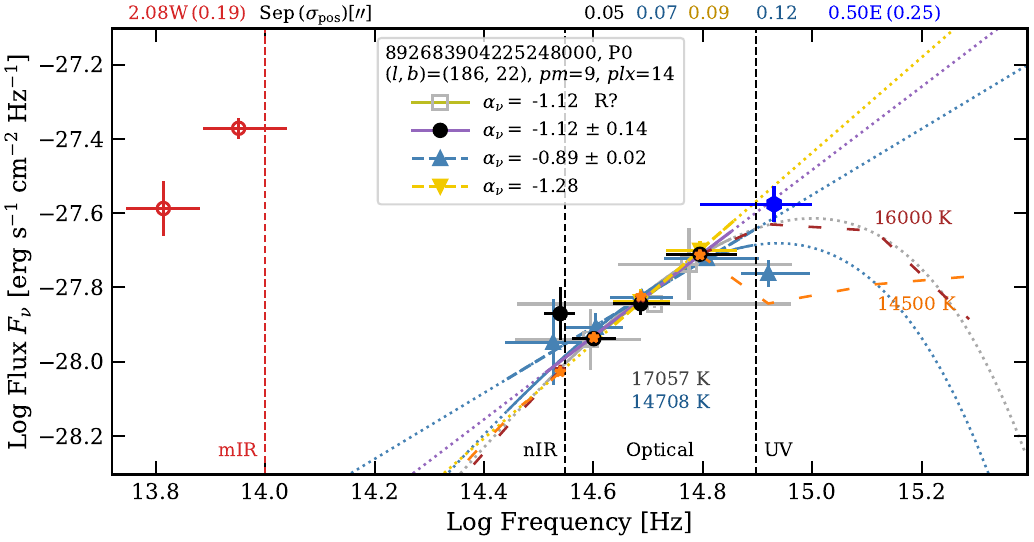}
	\includegraphics[width=\columnwidth]{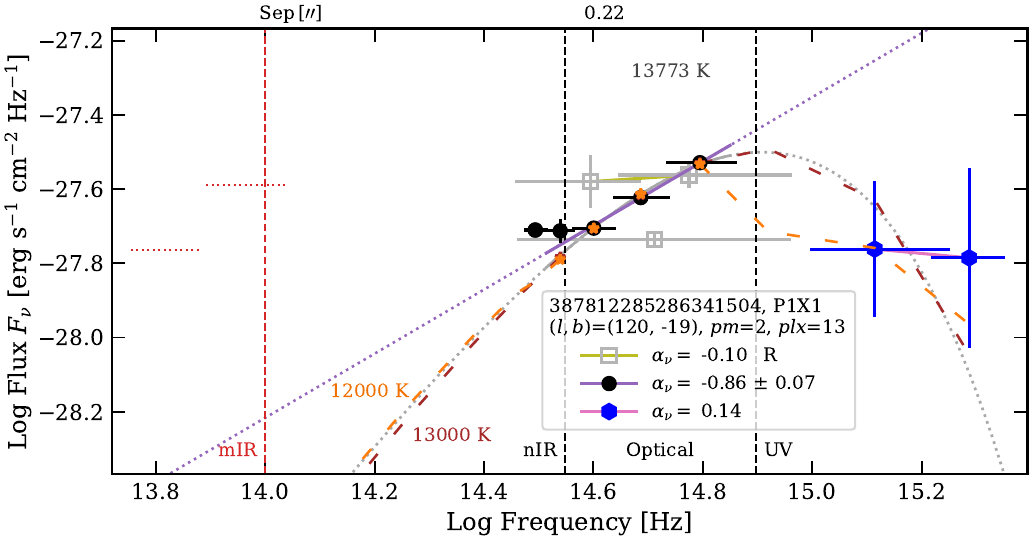}
	\includegraphics[width=\columnwidth]{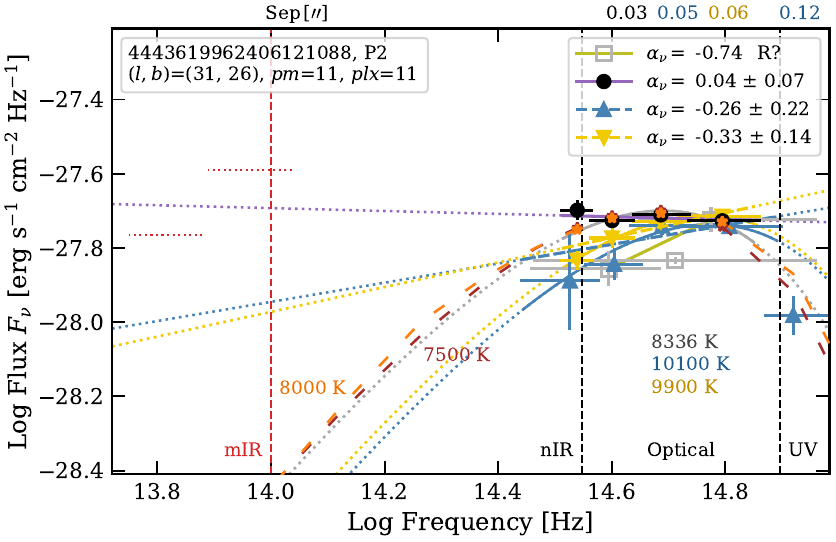}
	\includegraphics[width=\columnwidth]{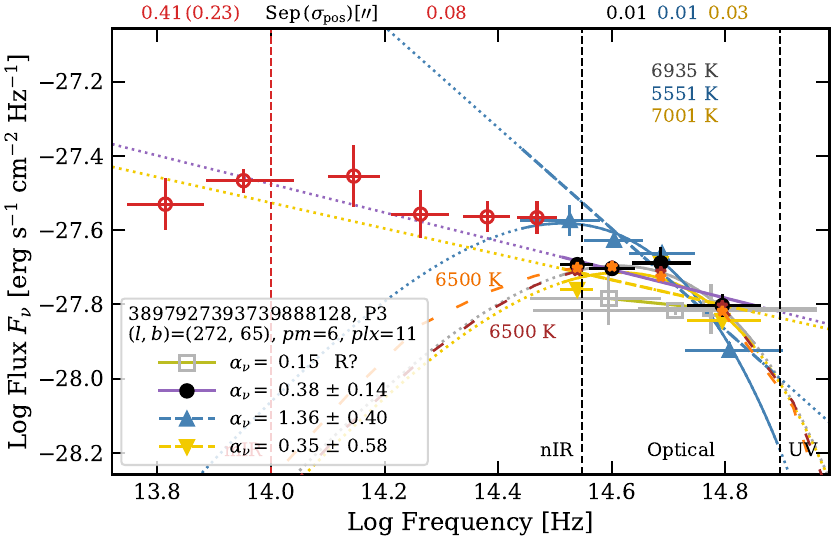}
	\includegraphics[width=\columnwidth]{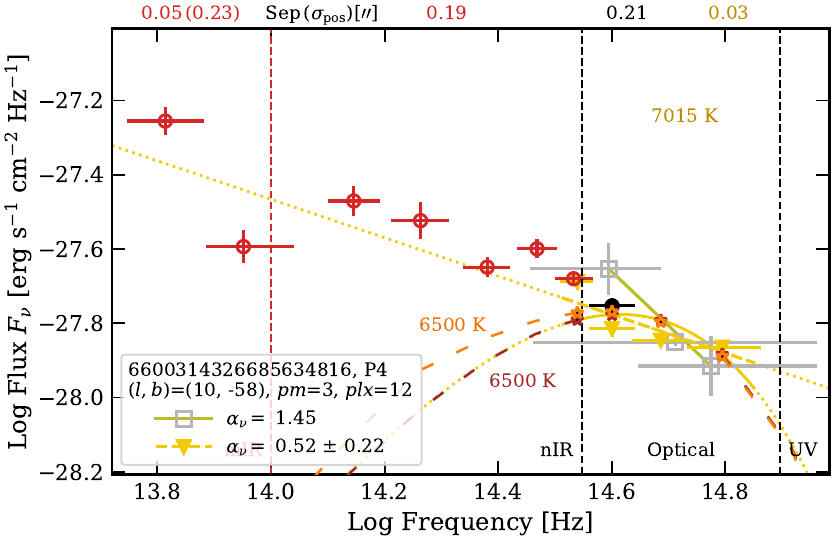}
	\includegraphics[width=\columnwidth]{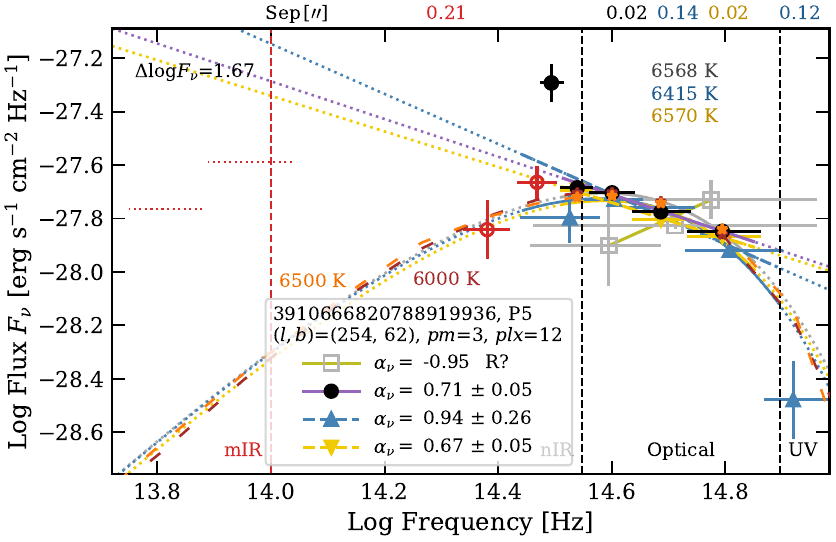}
	\includegraphics[width=\columnwidth]{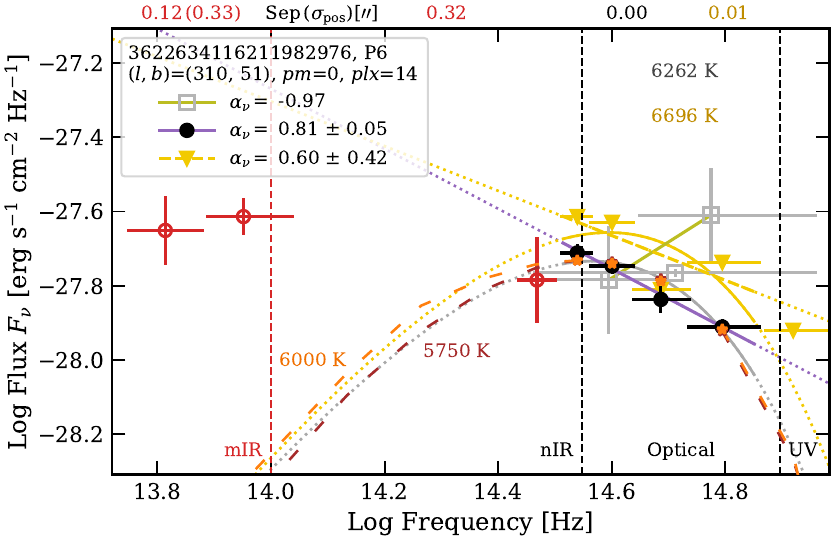}
	\includegraphics[width=\columnwidth]{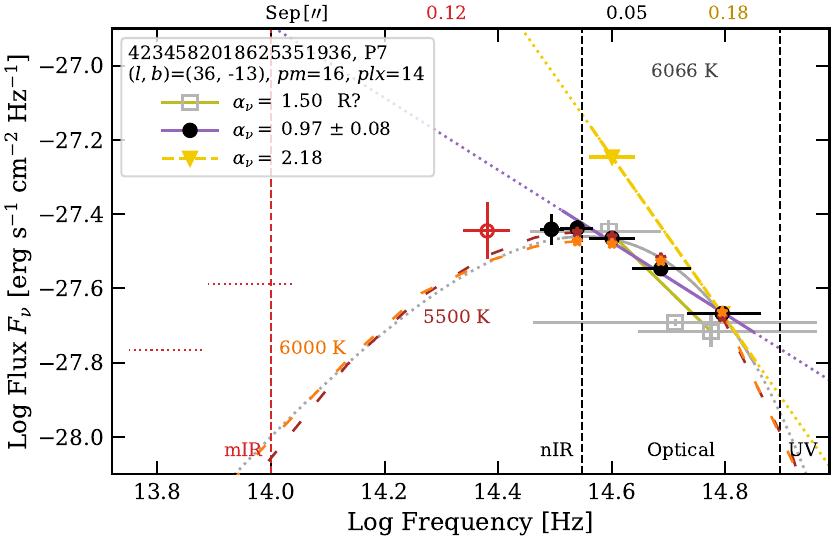}
     \caption{SEDs including infrared fluxes of the sources with power-law profiles at $griz$ bands or approximately-straight profiles at the optical and infrared (UV7 is shown in Fig.~\ref{fsed_uv_ir}).
     Same as in Fig.~\ref{fsed_uv1}.}
    \label{fsed_pl_ir}
\end{figure*}

\begin{figure*}
   \ContinuedFloat
	\includegraphics[width=\columnwidth]{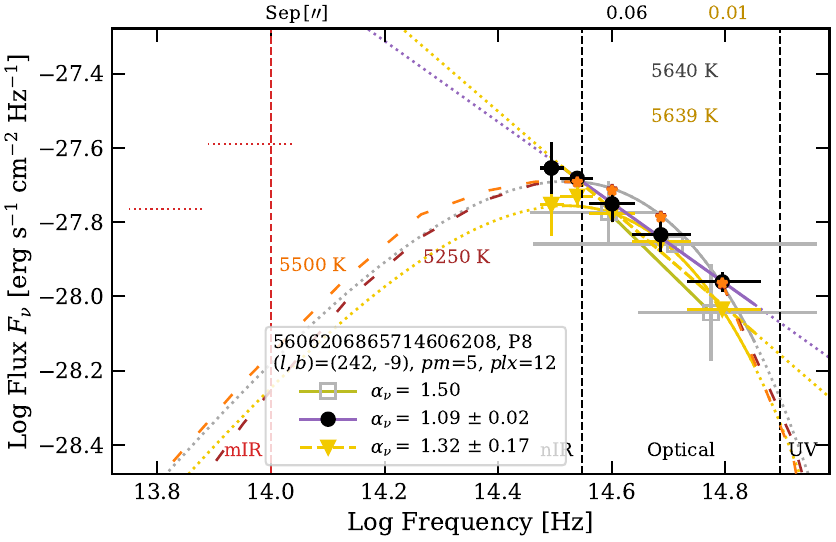}
	\includegraphics[width=\columnwidth]{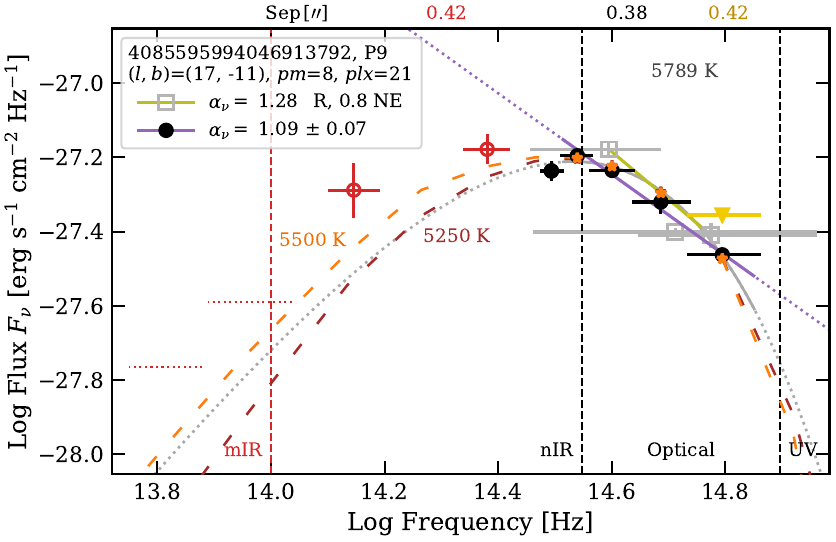}
	\includegraphics[width=\columnwidth]{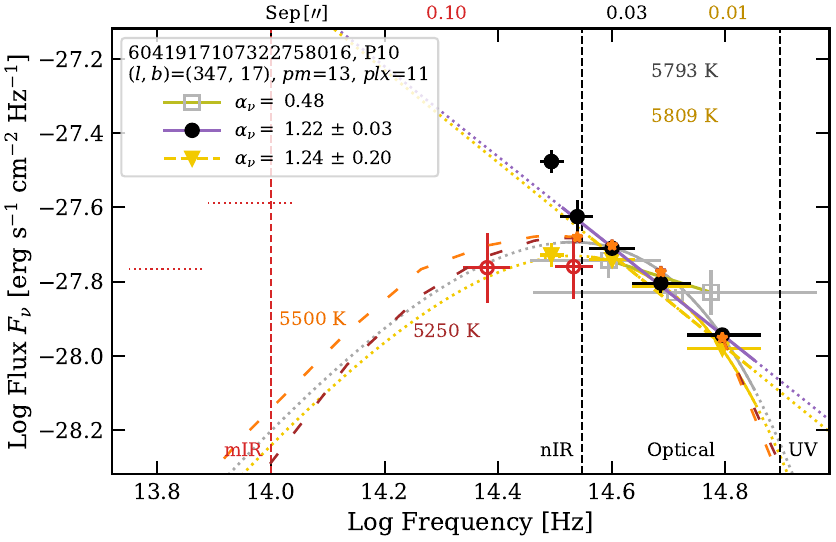}
	\includegraphics[width=\columnwidth]{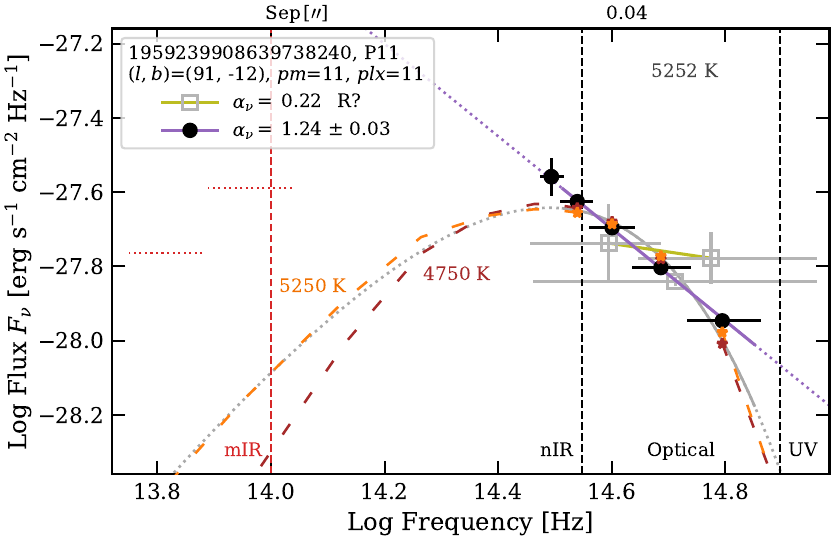}
	\includegraphics[width=\columnwidth]{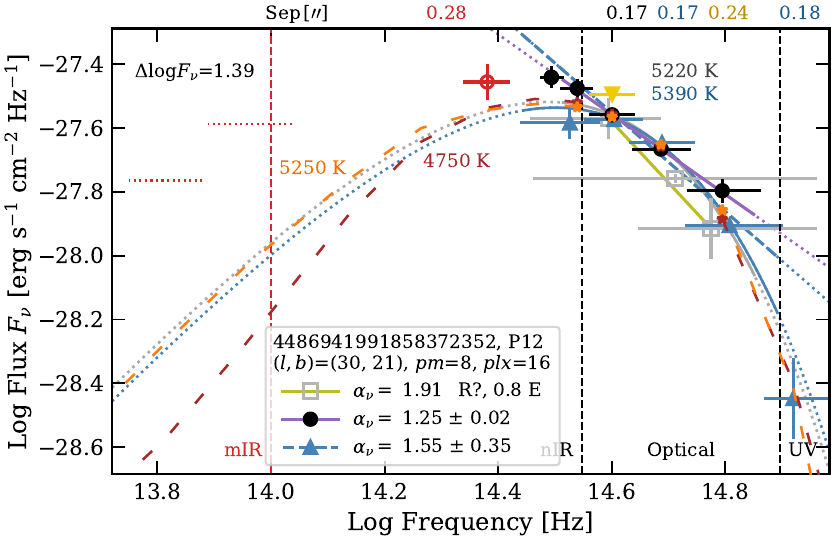}
	\includegraphics[width=\columnwidth]{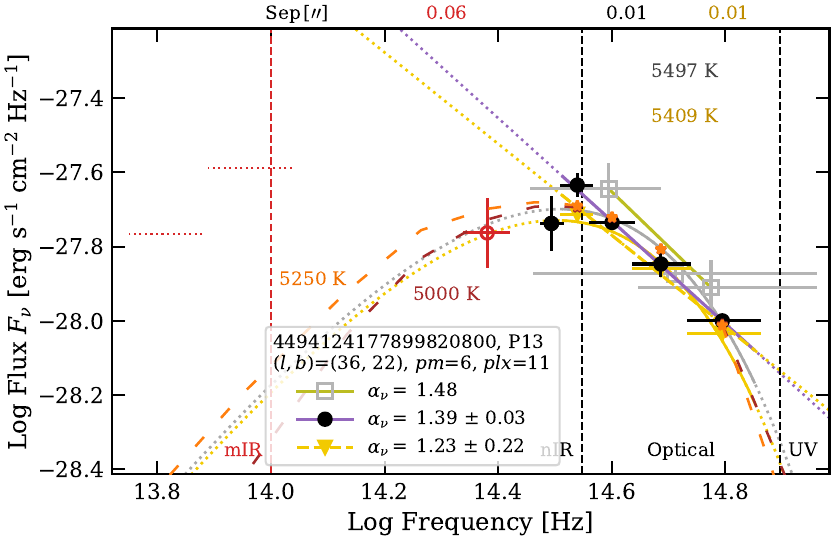}
	\includegraphics[width=\columnwidth]{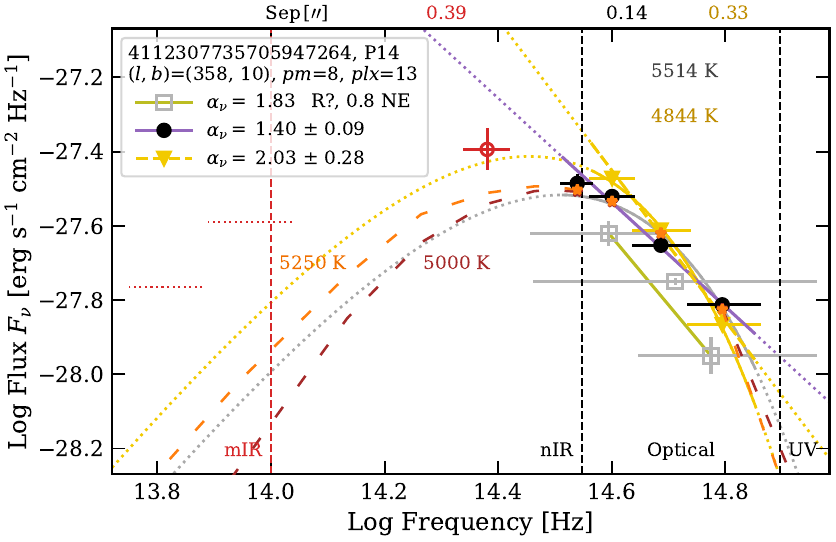}
	\includegraphics[width=\columnwidth]{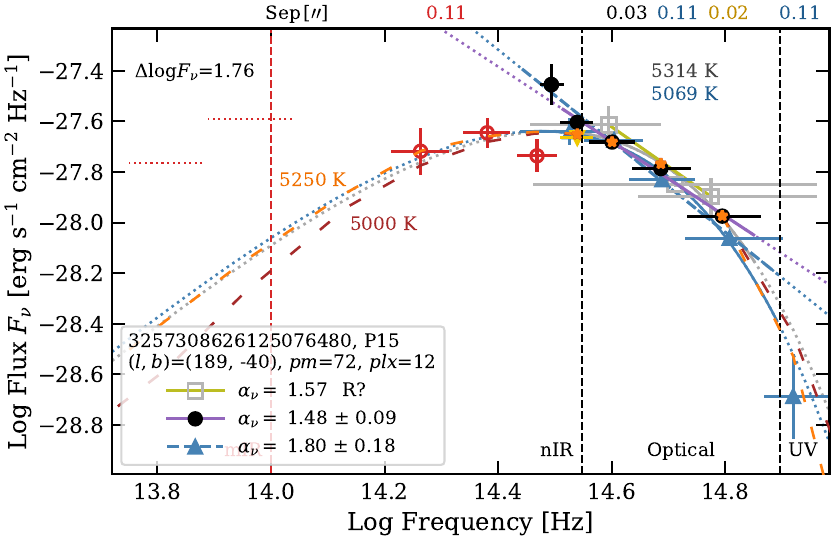}
     \caption{continued.}
    \label{fsed_pl_ir}
\end{figure*}

\begin{figure*}
   \ContinuedFloat
	\includegraphics[width=\columnwidth]{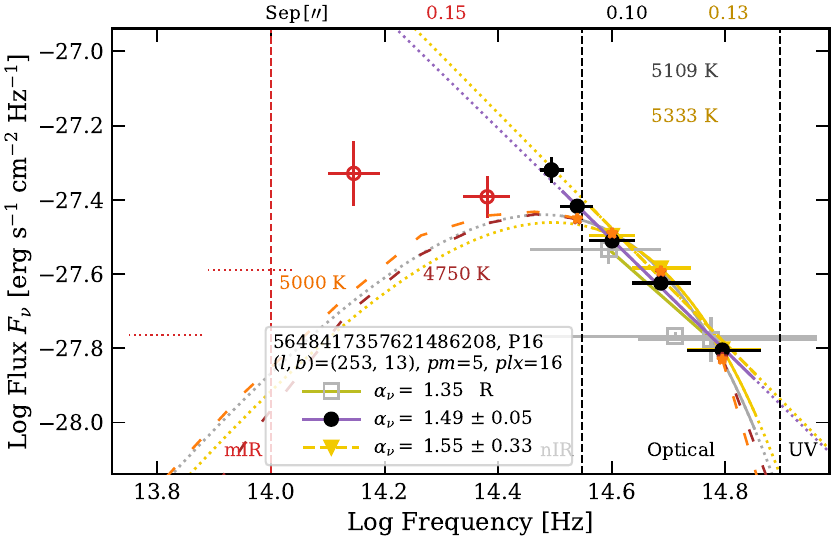}
	\includegraphics[width=\columnwidth]{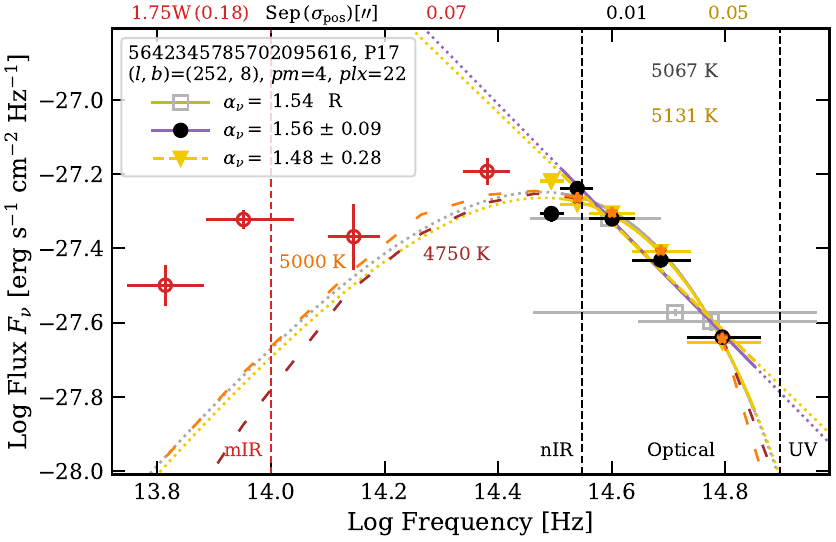}
	\includegraphics[width=\columnwidth]{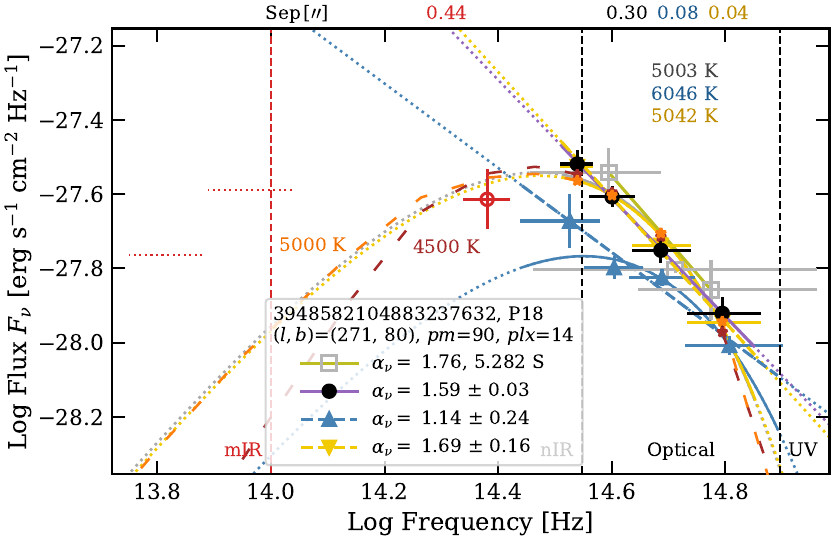}
	\includegraphics[width=\columnwidth]{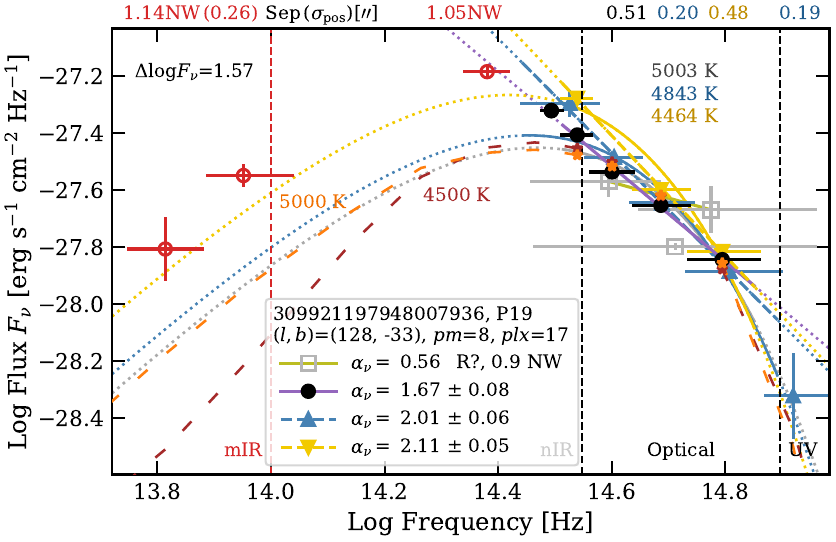}
	\includegraphics[width=\columnwidth]{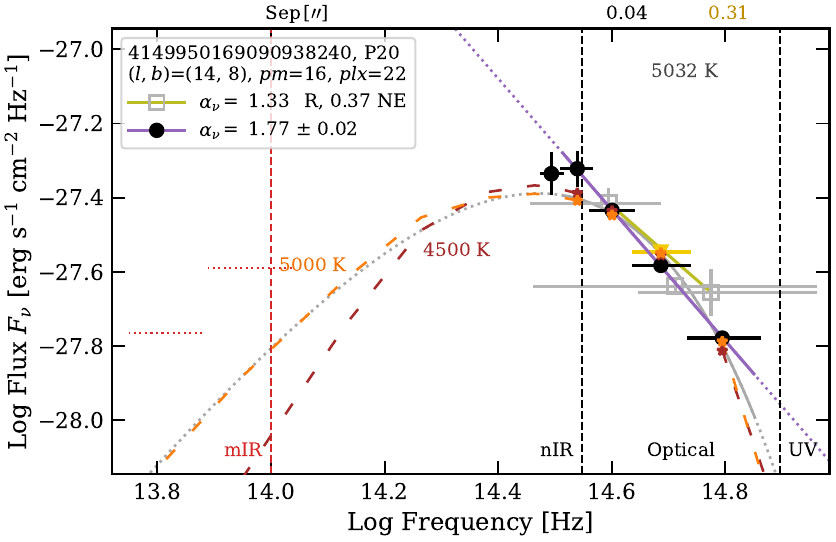}
	\includegraphics[width=\columnwidth]{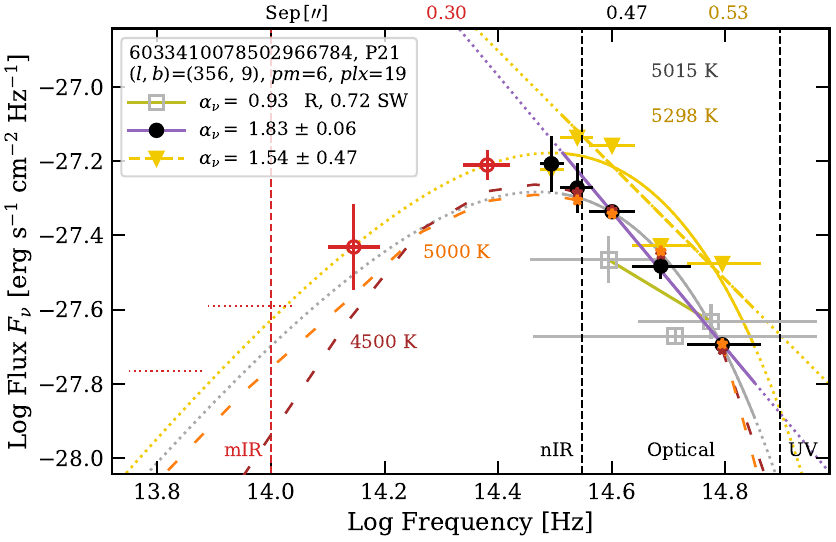}
	\includegraphics[width=\columnwidth]{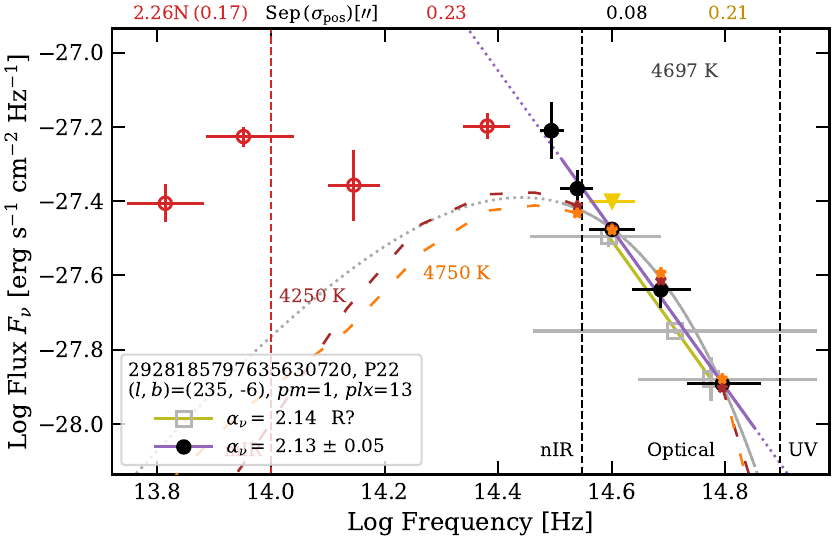}
	\includegraphics[width=\columnwidth]{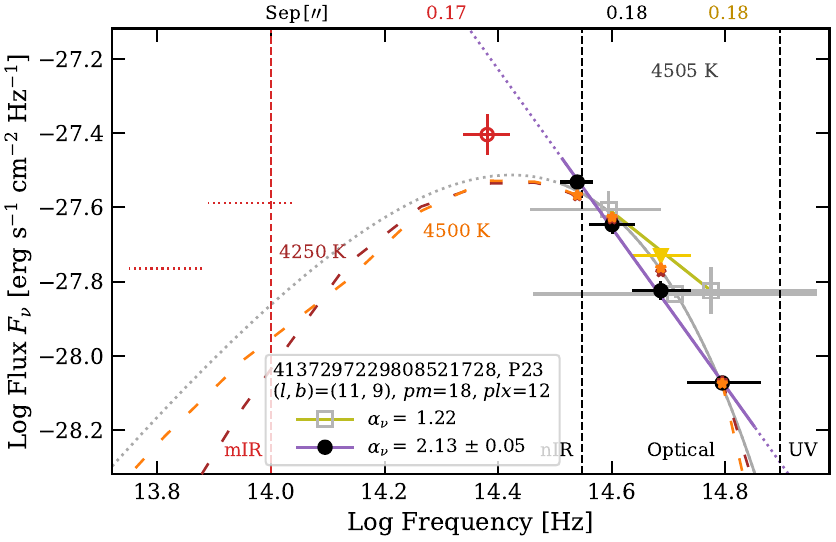}
     \caption{continued.}
    \label{fsed_pl_ir}
\end{figure*}

\begin{figure*}
   \ContinuedFloat
	\includegraphics[width=\columnwidth]{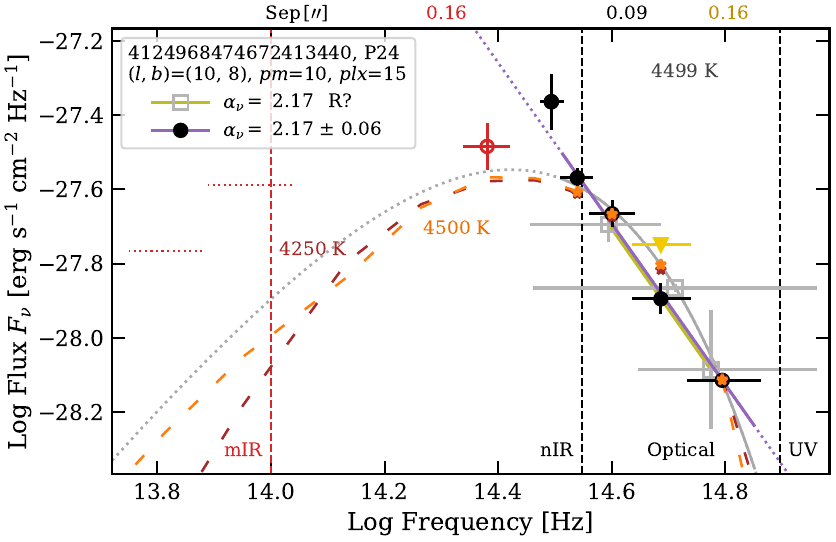}
	\includegraphics[width=\columnwidth]{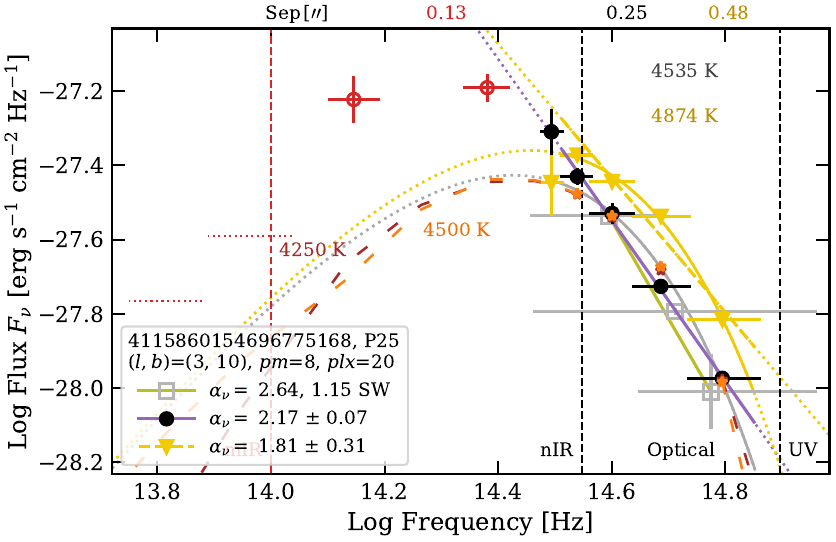}
	\includegraphics[width=\columnwidth]{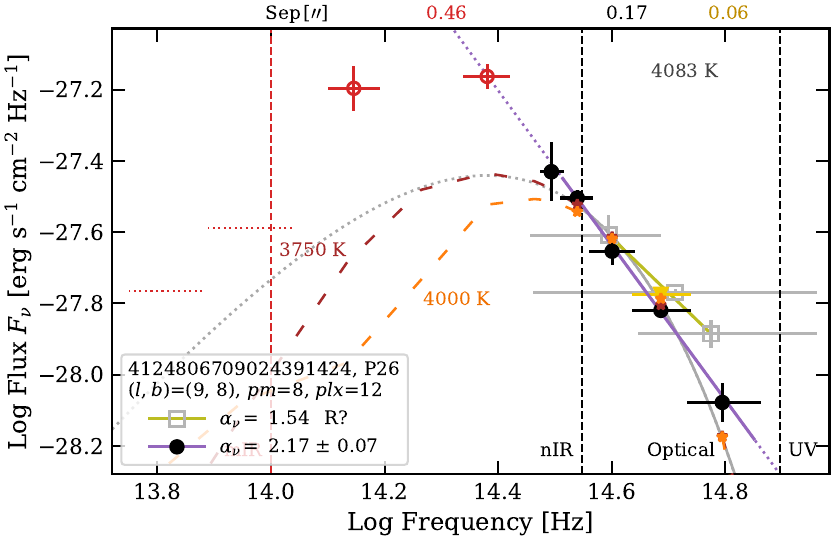}
	\includegraphics[width=\columnwidth]{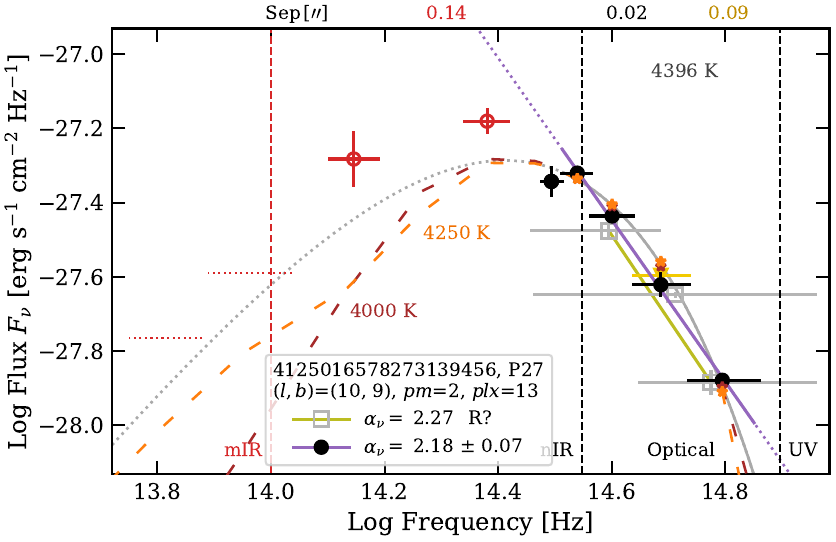}
	\includegraphics[width=\columnwidth]{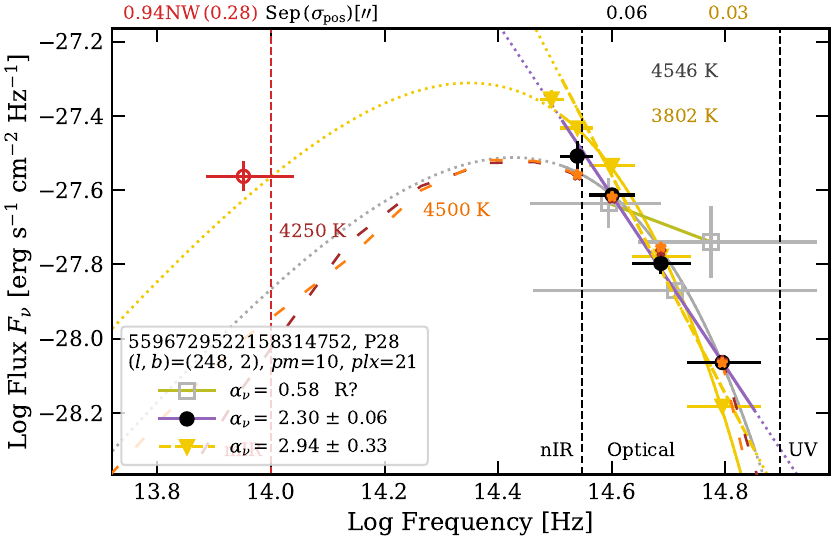}
	\includegraphics[width=\columnwidth]{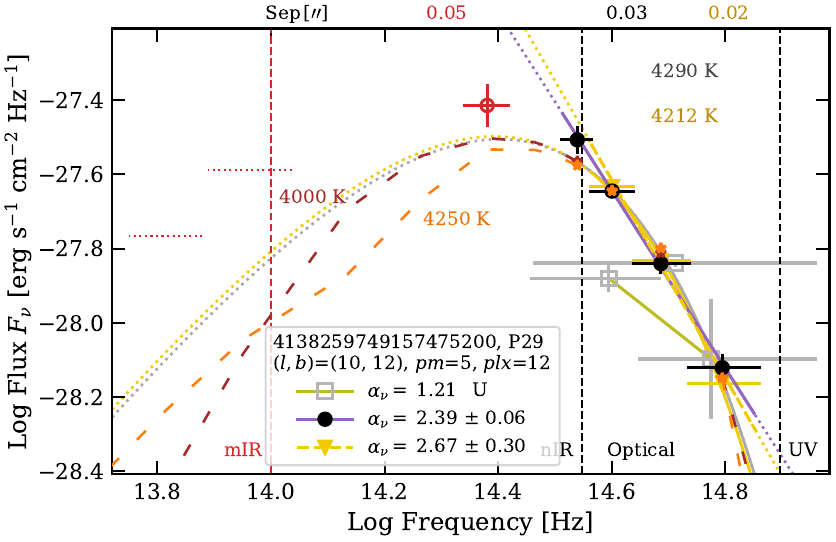}
	\includegraphics[width=\columnwidth]{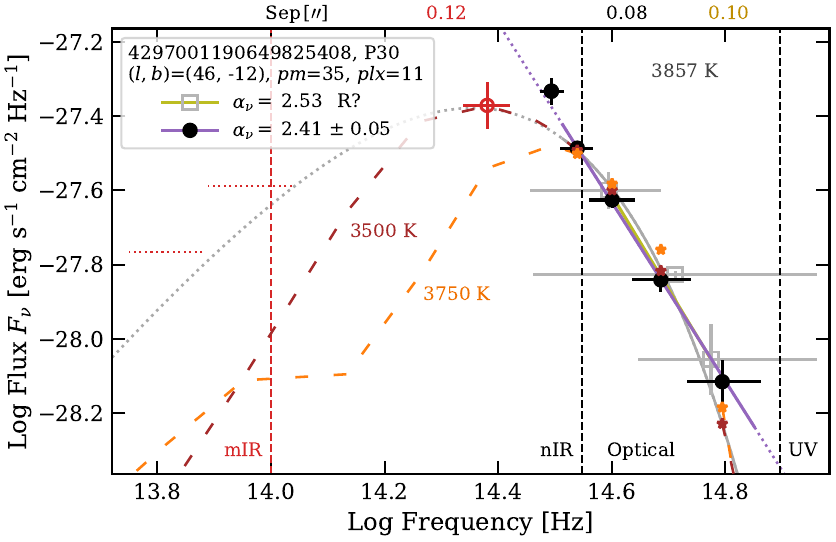}
	\includegraphics[width=\columnwidth]{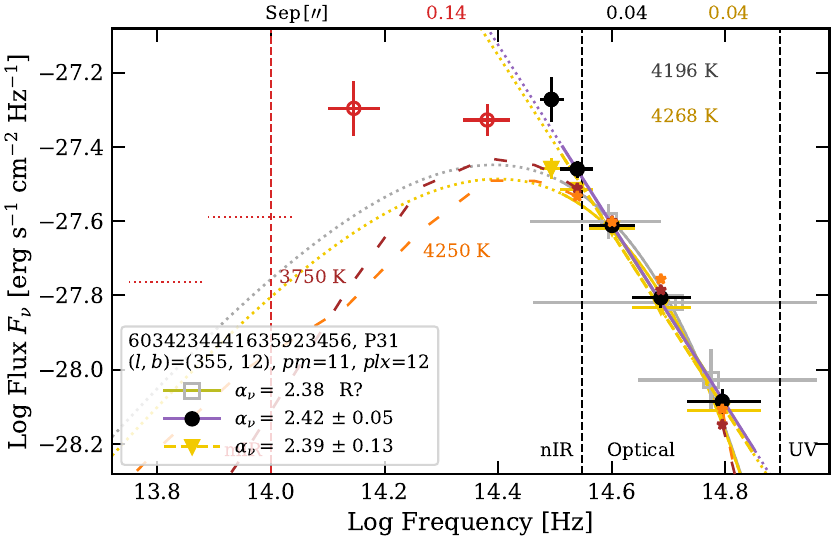}
     \caption{continued.}
    \label{fsed_pl_ir}
\end{figure*}

\begin{figure*}
   \ContinuedFloat
	\includegraphics[width=\columnwidth]{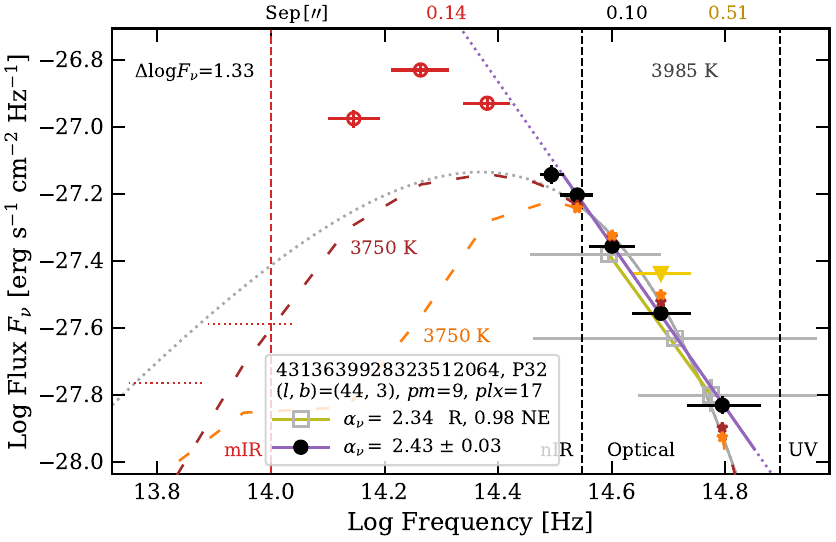}
	\includegraphics[width=\columnwidth]{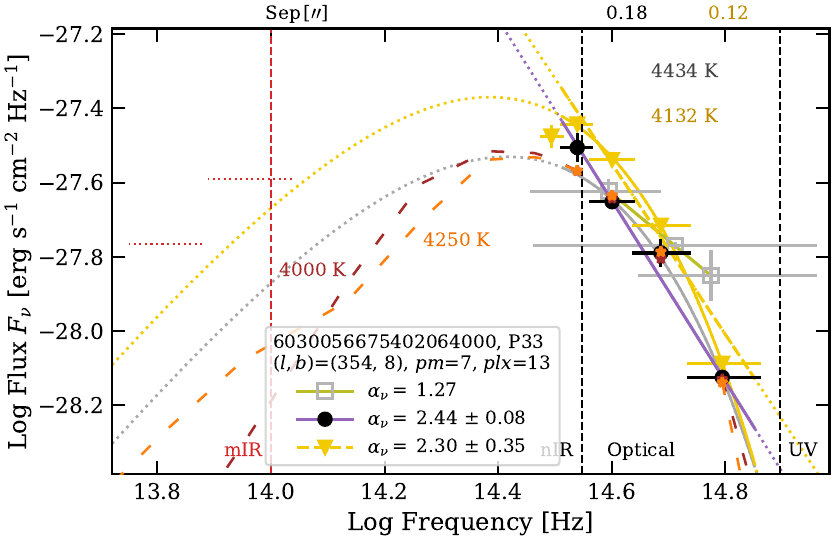}
	\includegraphics[width=\columnwidth]{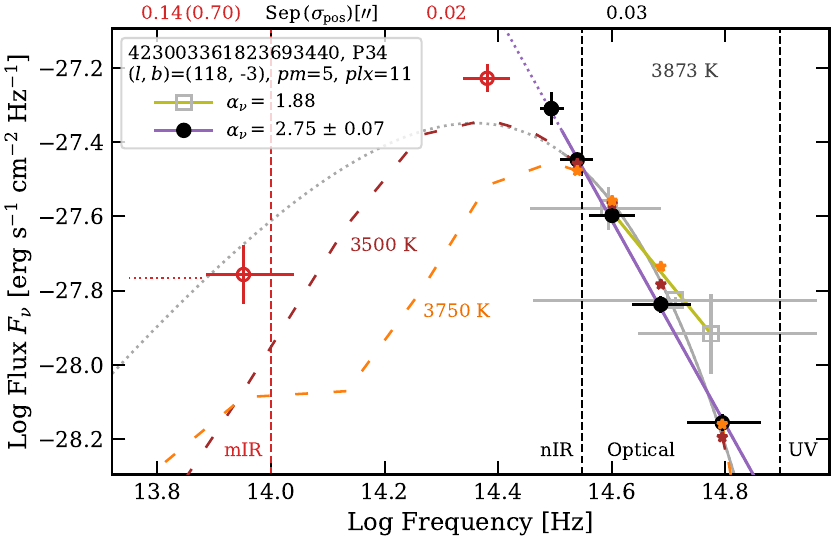}
	\includegraphics[width=\columnwidth]{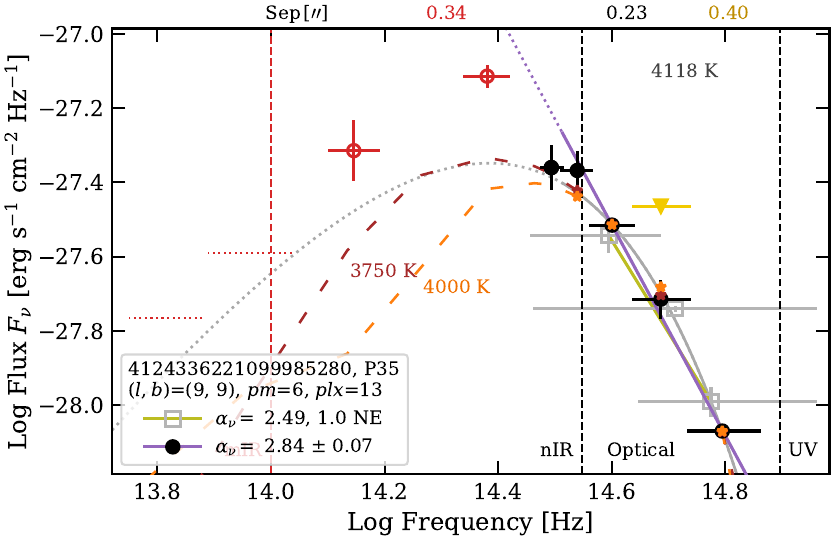}
	\includegraphics[width=\columnwidth]{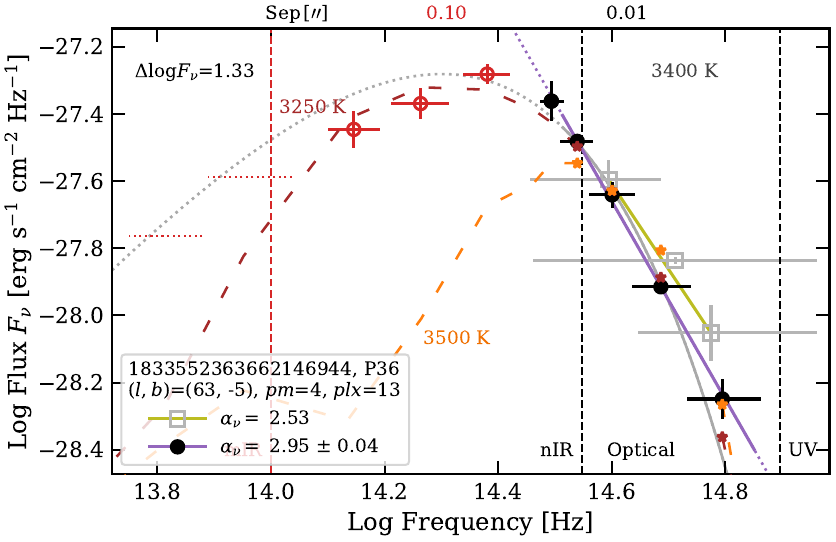}
	\includegraphics[width=\columnwidth]{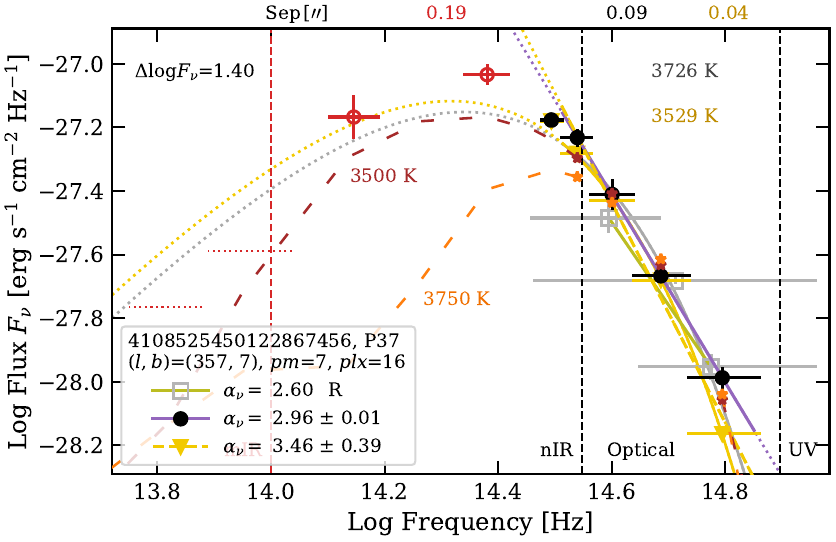}
     \caption{continued.}
    \label{fsed_pl_ir}
\end{figure*}

\begin{figure*}
    \includegraphics[width=\columnwidth]{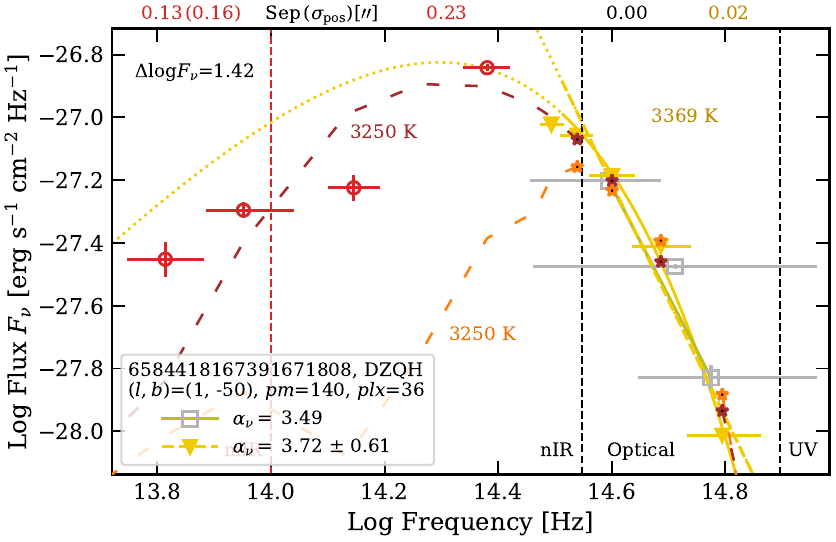}
    \includegraphics[width=\columnwidth]{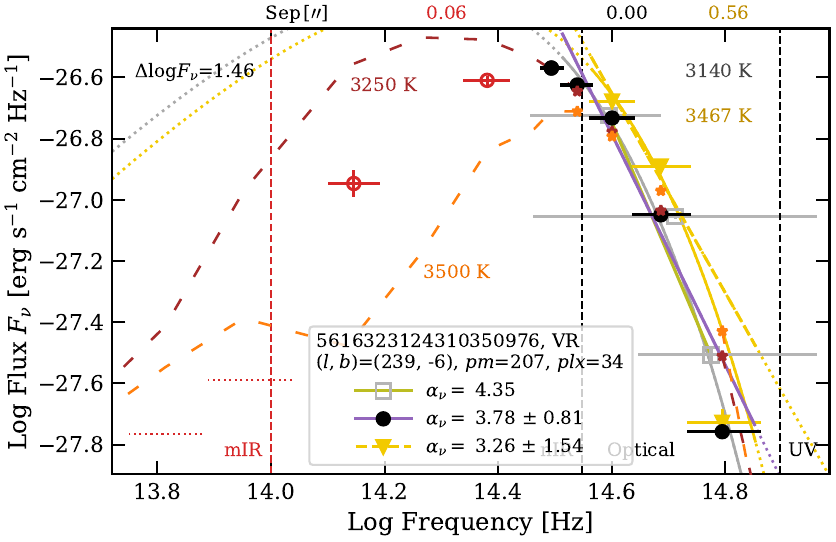}
    \includegraphics[width=\columnwidth]{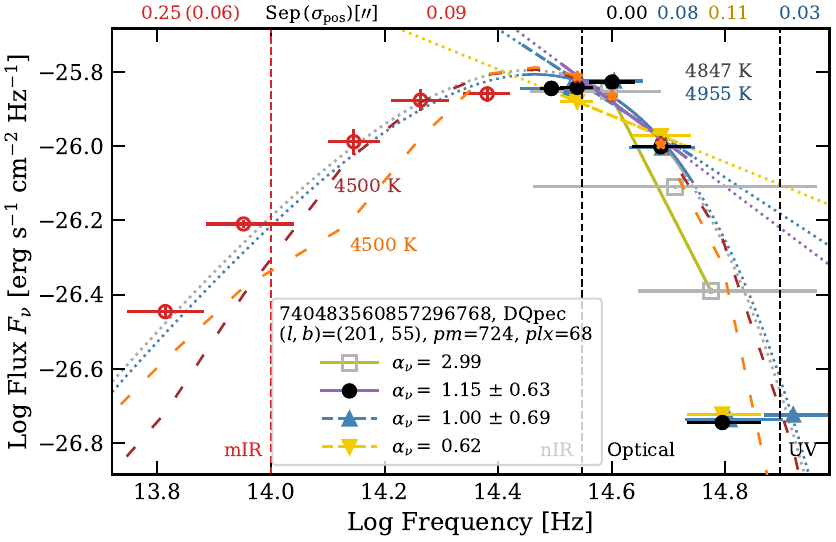}
    \includegraphics[width=\columnwidth]{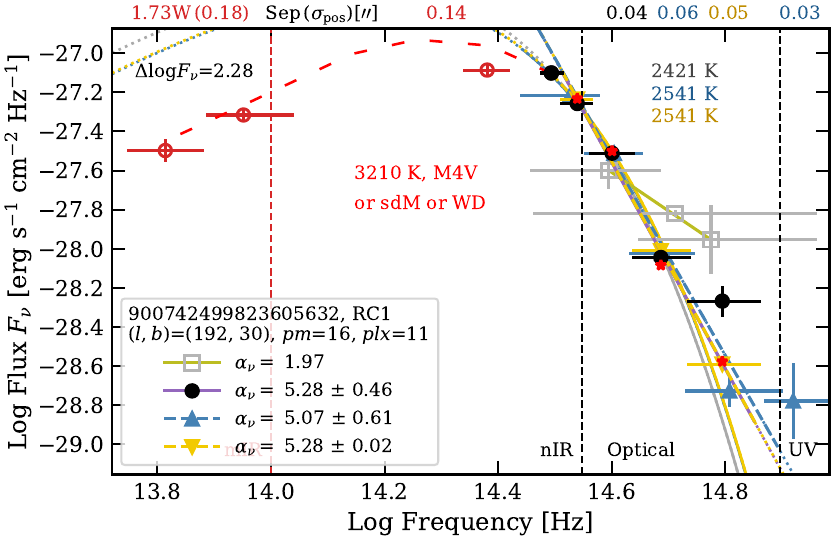}
    \includegraphics[width=\columnwidth]{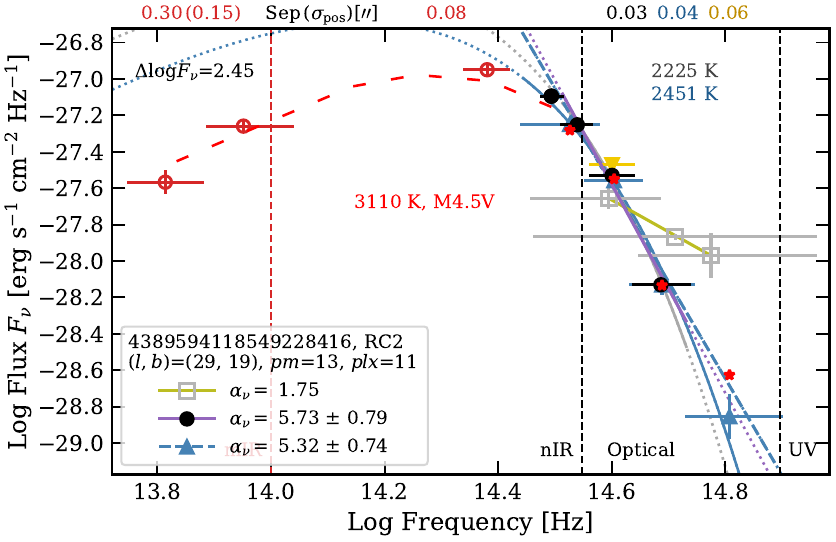}
    \includegraphics[width=\columnwidth]{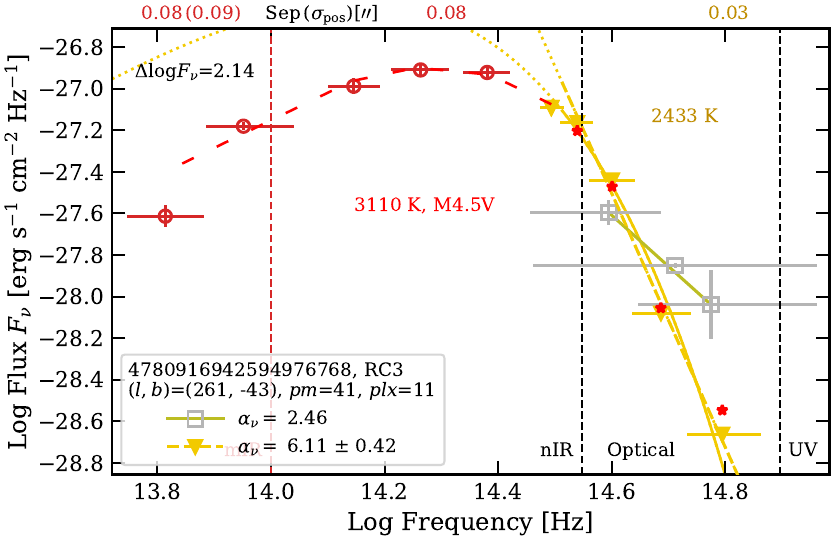}
    \includegraphics[width=\columnwidth]{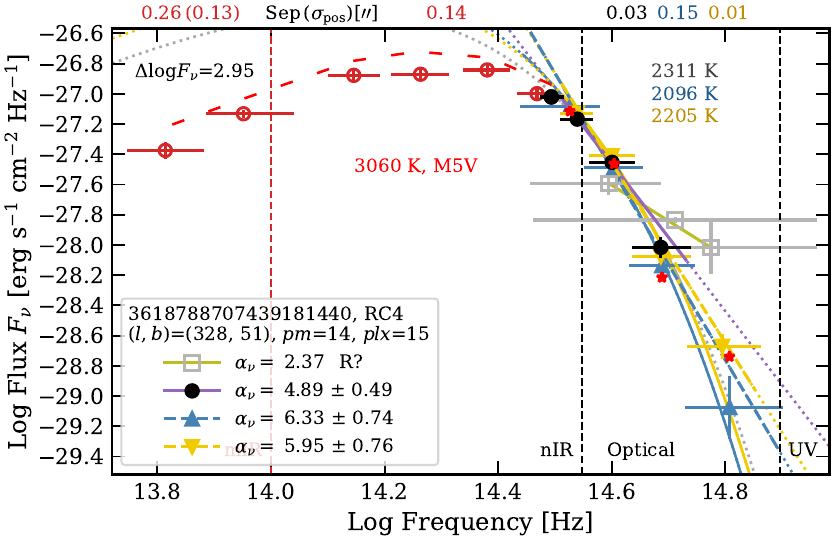}
    \includegraphics[width=\columnwidth]{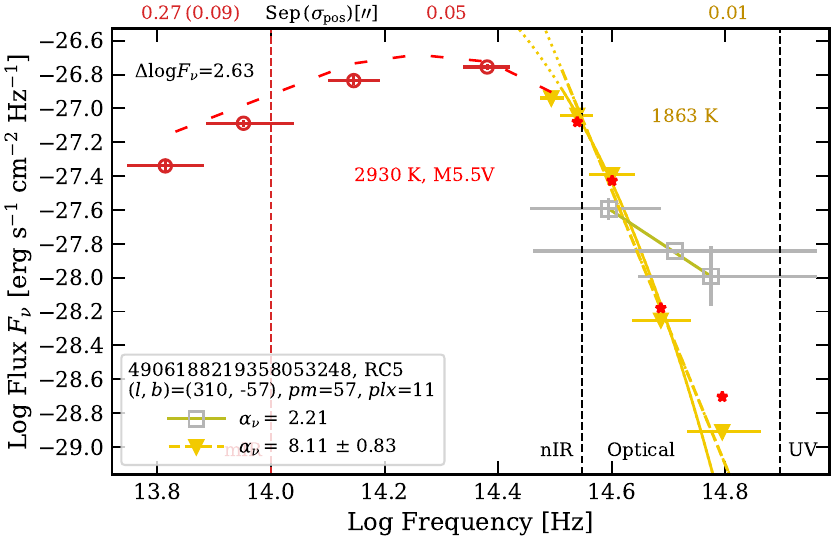}
    \caption{SEDs including infrared fluxes of the red-colour dwarf sources (RC1--RC5), having the reddest slopes in our subsample. Two red white-dwarf sources (flagged DZQH and VR) and the DQpec white dwarf LHS 2229 (Gaia DR3 740483560857296768) are shown for comparison.
     Same as in Fig.~\ref{fsed_uv1}.}
    \label{fsed_superred_ir}
\end{figure*}

\end{appendix}




\bsp	
\label{lastpage}
\end{document}